\pdfoutput=1

\documentclass[11pt,twoside,a4paper,cmspaper,final,collab]{cms-tdr}
\def\svnVersion{210297}\def\cmsCernNo{2013-129}\def\cmsCernDate{\today}\def\cmsMessage{Published in Physical Review Letters as \href{http://dx.doi.org/10.1103/PhysRevLett.111.101804}{\doi{10.1103/PhysRevLett.111.101804.}}}

\begin{document}\cmsNoteHeader{BPH-13-004}

\hyphenation{had-ron-i-za-tion}
\hyphenation{cal-or-i-me-ter}
\hyphenation{de-vices}

\RCS$Revision: 201989 $
\RCS$HeadURL: svn+ssh://svn.cern.ch/reps/tdr2/papers/BPH-13-004/trunk/BPH-13-004.tex $
\RCS$Id: BPH-13-004.tex 201989 2013-08-11 23:41:03Z alverson $
\newlength\cmsFigWidth
\ifthenelse{\boolean{cms@external}}{\setlength\cmsFigWidth{0.85\columnwidth}}{\setlength\cmsFigWidth{0.4\textwidth}}
\ifthenelse{\boolean{cms@external}}{\providecommand{\cmsLeft}{Top}}{\providecommand{\cmsLeft}{Left}}
\ifthenelse{\boolean{cms@external}}{\providecommand{\cmsRight}{Bottom}}{\providecommand{\cmsRight}{Right}}
\ifthenelse{\boolean{cms@external}}{\providecommand{\suppMaterial}{the supplemental material [URL will be inserted by publisher]}}{\providecommand{\suppMaterial}{Appendix~\ref{app:suppMat}}}
\newlength\figwid
\ifthenelse{\boolean{cms@external}}{\setlength\figwid{0.23\textwidth}}{\setlength\figwid{0.49\textwidth}}
\newcommand\vdef[1]{\expandafter\def\csname #1\endcsname}
\newcommand\vuse[1]{\csname #1\endcsname}
\newcommand{\base}{bdt2012}
\newcommand{\sample}{NoData}
\newcommand{\channel}{A}

\providecommand{\cPBds}{\ensuremath{\cmsSymbolFace{B_{d(s)}}\xspace}}
\newcommand{\cPhx}{\ensuremath{\cmsSymbolFace{h}}\xspace}
\newcommand{\bsdll}{\ensuremath{\cPBds\to\ell^+\ell^-}\xspace}
\newcommand{\bsmm}{\ensuremath{\PBzs\to\Pgmp\Pgmm}\xspace}
\newcommand{\bdmm}{\ensuremath{\PBz\to\Pgmp\Pgmm}\xspace}
\newcommand{\cbf}{\ensuremath{\mathcal{B}}}
\newcommand{\bupsik}{\ensuremath{\PBp \to \cPJgy \PKp}\xspace}
\newcommand{\bupsipi}{\ensuremath{\PBp \to \cPJgy \pi^+}\xspace}
\renewcommand{\bspsiphi}{\ensuremath{\PBzs \to \cPJgy \phi}\xspace}
\newcommand\fls      {\ensuremath{\ell_{3D}/\sigma(\ell_\mathrm{3D})}}
\newcommand\chidof   {\ensuremath{\chi^2/\mathrm{dof}}}
\newcommand\closetrk {\ensuremath{N_{\text{trk}}^{\text{close}}}}
\newcommand\docatrk  {\ensuremath{d^{\mathrm{0}}_\mathrm{ca}}}
\newcommand\dca      {\ensuremath{d_\mathrm{ca}}}
\newcommand\ip       {\ensuremath{\delta_\mathrm{3D}}}
\newcommand\ips      {\ensuremath{\delta_\mathrm{3D}/\sigma(\delta_\mathrm{3D})}}
\newcommand\lumiseven {\ensuremath{5}}
\newcommand\lumieight {\ensuremath{20}}
\newcommand\bdsignifAboveSM {\ensuremath{2.2}}
\newcommand\bsSMexclusion {\ensuremath{2.6}}
\newcommand\bdUL {\ensuremath{\vuse{20112012:ulObs:val:mu_d} \times 10^{\vuse{20112012:ulObs:exponent:mu_d} }}}
\newcommand\bsUL {\ensuremath{\vuse{20112012:ulObs:val:mu_s} \times 10^{\vuse{20112012:ulObs:exponent:mu_s} }}}
\newcommand\bdResultBF {\ensuremath{(3.5^{+2.1}_{-1.8})\times 10^{-10}}}
\newcommand\bdsignif {\ensuremath{2.0}}
\newcommand\bsResultBF {\ensuremath{(3.0^{+1.0}_{-0.9})\times 10^{-9}}}
\newcommand\bssignif {\ensuremath{4.3}}
\newcommand\bmm     {\ensuremath{\Bz\to \mu^+\mu^-}}
\newcommand\fufs {\ensuremath{f_{\cmsSymbolFace{u}}/f_{\cmsSymbolFace{s}}}}
\newcommand\fsfu {\ensuremath{f_{\cmsSymbolFace{s}}/f_{\cmsSymbolFace{u}}}}
\newcommand\red[1]{{\color{red}#1}}
\newcommand{\VERIFY}{\red{VERIFY}}

\vdef{2012:bdt:0}     {\ensuremath{{0.360 } } }
\vdef{2012:bdtMax:0}  {\ensuremath{{10.000 } } }
\vdef{2012:mBdLo:0}   {\ensuremath{{5.200 } } }
\vdef{2012:mBdHi:0}   {\ensuremath{{5.300 } } }
\vdef{2012:mBsLo:0}   {\ensuremath{{5.300 } } }
\vdef{2012:mBsHi:0}   {\ensuremath{{5.450 } } }
\vdef{2012:etaMin:0}   {\ensuremath{{0.0 } } }
\vdef{2012:etaMax:0}   {\ensuremath{{1.4 } } }
\vdef{2012:pt:0}   {\ensuremath{{6.5 } } }
\vdef{2012:m1pt:0}   {\ensuremath{{4.5 } } }
\vdef{2012:m2pt:0}   {\ensuremath{{4.0 } } }
\vdef{2012:m1eta:0}   {\ensuremath{{1.4 } } }
\vdef{2012:m2eta:0}   {\ensuremath{{1.4 } } }
\vdef{2012:iso:0}   {\ensuremath{{0.80 } } }
\vdef{2012:chi2dof:0}   {\ensuremath{{2.2 } } }
\vdef{2012:alpha:0}   {\ensuremath{{0.050 } } }
\vdef{2012:fls3d:0}   {\ensuremath{{13.0 } } }
\vdef{2012:docatrk:0}   {\ensuremath{{0.015 } } }
\vdef{2012:closetrk:0}   {\ensuremath{{2 } } }
\vdef{2012:pvlip:0}   {\ensuremath{{100.000 } } }
\vdef{2012:pvlips:0}   {\ensuremath{{100.000 } } }
\vdef{2012:pvlip2:0}   {\ensuremath{{-100.000 } } }
\vdef{2012:pvlips2:0}   {\ensuremath{{-100.000 } } }
\vdef{2012:maxdoca:0}   {\ensuremath{{100.000 } } }
\vdef{2012:pvip:0}   {\ensuremath{{0.008 } } }
\vdef{2012:pvips:0}   {\ensuremath{{2.000 } } }
\vdef{2012:doApplyCowboyVeto:0}   {no }
\vdef{2012:fDoApplyCowboyVetoAlsoInSignal:0}   {no }
\vdef{2012:bdt:1}     {\ensuremath{{0.380 } } }
\vdef{2012:bdtMax:1}  {\ensuremath{{10.000 } } }
\vdef{2012:mBdLo:1}   {\ensuremath{{5.200 } } }
\vdef{2012:mBdHi:1}   {\ensuremath{{5.300 } } }
\vdef{2012:mBsLo:1}   {\ensuremath{{5.300 } } }
\vdef{2012:mBsHi:1}   {\ensuremath{{5.450 } } }
\vdef{2012:etaMin:1}   {\ensuremath{{1.4 } } }
\vdef{2012:etaMax:1}   {\ensuremath{{2.4 } } }
\vdef{2012:pt:1}   {\ensuremath{{8.5 } } }
\vdef{2012:m1pt:1}   {\ensuremath{{4.5 } } }
\vdef{2012:m2pt:1}   {\ensuremath{{4.2 } } }
\vdef{2012:m1eta:1}   {\ensuremath{{2.0 } } }
\vdef{2012:m2eta:1}   {\ensuremath{{2.0 } } }
\vdef{2012:iso:1}   {\ensuremath{{0.80 } } }
\vdef{2012:chi2dof:1}   {\ensuremath{{1.8 } } }
\vdef{2012:alpha:1}   {\ensuremath{{0.030 } } }
\vdef{2012:fls3d:1}   {\ensuremath{{15.0 } } }
\vdef{2012:docatrk:1}   {\ensuremath{{0.015 } } }
\vdef{2012:closetrk:1}   {\ensuremath{{2 } } }
\vdef{2012:pvlip:1}   {\ensuremath{{100.000 } } }
\vdef{2012:pvlips:1}   {\ensuremath{{100.000 } } }
\vdef{2012:pvlip2:1}   {\ensuremath{{-100.000 } } }
\vdef{2012:pvlips2:1}   {\ensuremath{{-100.000 } } }
\vdef{2012:maxdoca:1}   {\ensuremath{{100.000 } } }
\vdef{2012:pvip:1}   {\ensuremath{{0.008 } } }
\vdef{2012:pvips:1}   {\ensuremath{{2.000 } } }
\vdef{2012:doApplyCowboyVeto:1}   {no }
\vdef{2012:fDoApplyCowboyVetoAlsoInSignal:1}   {no }
\vdef{2012:bdt:0}     {\ensuremath{{0.360 } } }
\vdef{2012:bdtMax:0}  {\ensuremath{{10.000 } } }
\vdef{2012:mBdLo:0}   {\ensuremath{{5.200 } } }
\vdef{2012:mBdHi:0}   {\ensuremath{{5.300 } } }
\vdef{2012:mBsLo:0}   {\ensuremath{{5.300 } } }
\vdef{2012:mBsHi:0}   {\ensuremath{{5.450 } } }
\vdef{2012:etaMin:0}   {\ensuremath{{0.0 } } }
\vdef{2012:etaMax:0}   {\ensuremath{{1.4 } } }
\vdef{2012:pt:0}   {\ensuremath{{6.5 } } }
\vdef{2012:m1pt:0}   {\ensuremath{{4.5 } } }
\vdef{2012:m2pt:0}   {\ensuremath{{4.0 } } }
\vdef{2012:m1eta:0}   {\ensuremath{{1.4 } } }
\vdef{2012:m2eta:0}   {\ensuremath{{1.4 } } }
\vdef{2012:iso:0}   {\ensuremath{{0.80 } } }
\vdef{2012:chi2dof:0}   {\ensuremath{{2.2 } } }
\vdef{2012:alpha:0}   {\ensuremath{{0.050 } } }
\vdef{2012:fls3d:0}   {\ensuremath{{13.0 } } }
\vdef{2012:docatrk:0}   {\ensuremath{{0.015 } } }
\vdef{2012:closetrk:0}   {\ensuremath{{2 } } }
\vdef{2012:pvlip:0}   {\ensuremath{{100.000 } } }
\vdef{2012:pvlips:0}   {\ensuremath{{100.000 } } }
\vdef{2012:pvlip2:0}   {\ensuremath{{-100.000 } } }
\vdef{2012:pvlips2:0}   {\ensuremath{{-100.000 } } }
\vdef{2012:maxdoca:0}   {\ensuremath{{100.000 } } }
\vdef{2012:pvip:0}   {\ensuremath{{0.008 } } }
\vdef{2012:pvips:0}   {\ensuremath{{2.000 } } }
\vdef{2012:doApplyCowboyVeto:0}   {no }
\vdef{2012:fDoApplyCowboyVetoAlsoInSignal:0}   {no }
\vdef{2012:bdt:1}     {\ensuremath{{0.380 } } }
\vdef{2012:bdtMax:1}  {\ensuremath{{10.000 } } }
\vdef{2012:mBdLo:1}   {\ensuremath{{5.200 } } }
\vdef{2012:mBdHi:1}   {\ensuremath{{5.300 } } }
\vdef{2012:mBsLo:1}   {\ensuremath{{5.300 } } }
\vdef{2012:mBsHi:1}   {\ensuremath{{5.450 } } }
\vdef{2012:etaMin:1}   {\ensuremath{{1.4 } } }
\vdef{2012:etaMax:1}   {\ensuremath{{2.4 } } }
\vdef{2012:pt:1}   {\ensuremath{{8.5 } } }
\vdef{2012:m1pt:1}   {\ensuremath{{4.5 } } }
\vdef{2012:m2pt:1}   {\ensuremath{{4.2 } } }
\vdef{2012:m1eta:1}   {\ensuremath{{2.0 } } }
\vdef{2012:m2eta:1}   {\ensuremath{{2.0 } } }
\vdef{2012:iso:1}   {\ensuremath{{0.80 } } }
\vdef{2012:chi2dof:1}   {\ensuremath{{1.8 } } }
\vdef{2012:alpha:1}   {\ensuremath{{0.030 } } }
\vdef{2012:fls3d:1}   {\ensuremath{{15.0 } } }
\vdef{2012:docatrk:1}   {\ensuremath{{0.015 } } }
\vdef{2012:closetrk:1}   {\ensuremath{{2 } } }
\vdef{2012:pvlip:1}   {\ensuremath{{100.000 } } }
\vdef{2012:pvlips:1}   {\ensuremath{{100.000 } } }
\vdef{2012:pvlip2:1}   {\ensuremath{{-100.000 } } }
\vdef{2012:pvlips2:1}   {\ensuremath{{-100.000 } } }
\vdef{2012:maxdoca:1}   {\ensuremath{{100.000 } } }
\vdef{2012:pvip:1}   {\ensuremath{{0.008 } } }
\vdef{2012:pvips:1}   {\ensuremath{{2.000 } } }
\vdef{2012:doApplyCowboyVeto:1}   {no }
\vdef{2012:fDoApplyCowboyVetoAlsoInSignal:1}   {no }
\vdef{bdt2012:N-CSBF-TNP-BS0:val}   {\ensuremath{{0.000024 } } }
\vdef{bdt2012:N-CSBF-TNP-BS0:err}   {\ensuremath{{0.000001 } } }
\vdef{bdt2012:N-CSBF-MC-BS0:val}   {\ensuremath{{0.000025 } } }
\vdef{bdt2012:N-CSBF-MC-BS0:err}   {\ensuremath{{0.000001 } } }
\vdef{bdt2012:N-CSBF-MC-BS0:syst}   {\ensuremath{{0.000002 } } }
\vdef{bdt2012:N-CSBF-BS0:val}   {\ensuremath{{0.000025 } } }
\vdef{bdt2012:N-CSBF-BS0:err}   {\ensuremath{{0.000001 } } }
\vdef{bdt2012:N-CSBF-BS0:syst}   {\ensuremath{{0.000002 } } }
\vdef{bdt2012:bgBd2K0MuMu:loSideband0:val}   {\ensuremath{{0.000 } } }
\vdef{bdt2012:bgBd2K0MuMu:loSideband0:err}   {\ensuremath{{0.000 } } }
\vdef{bdt2012:bgBd2K0MuMu:bdRare0}   {\ensuremath{{0.000000 } } }
\vdef{bdt2012:bgBd2K0MuMu:bdRare0E}  {\ensuremath{{0.000000 } } }
\vdef{bdt2012:bgBd2K0MuMu:bsRare0}   {\ensuremath{{0.000000 } } }
\vdef{bdt2012:bgBd2K0MuMu:bsRare0E}  {\ensuremath{{0.000000 } } }
\vdef{bdt2012:bgBd2K0MuMu:hiSideband0:val}   {\ensuremath{{0.000 } } }
\vdef{bdt2012:bgBd2K0MuMu:hiSideband0:err}   {\ensuremath{{0.000 } } }
\vdef{bdt2012:bgBd2KK:loSideband0:val}   {\ensuremath{{0.004 } } }
\vdef{bdt2012:bgBd2KK:loSideband0:err}   {\ensuremath{{0.005 } } }
\vdef{bdt2012:bgBd2KK:bdRare0}   {\ensuremath{{0.001967 } } }
\vdef{bdt2012:bgBd2KK:bdRare0E}  {\ensuremath{{0.002482 } } }
\vdef{bdt2012:bgBd2KK:bsRare0}   {\ensuremath{{0.000072 } } }
\vdef{bdt2012:bgBd2KK:bsRare0E}  {\ensuremath{{0.000091 } } }
\vdef{bdt2012:bgBd2KK:hiSideband0:val}   {\ensuremath{{0.000 } } }
\vdef{bdt2012:bgBd2KK:hiSideband0:err}   {\ensuremath{{0.000 } } }
\vdef{bdt2012:bgBd2KPi:loSideband0:val}   {\ensuremath{{0.177 } } }
\vdef{bdt2012:bgBd2KPi:loSideband0:err}   {\ensuremath{{0.125 } } }
\vdef{bdt2012:bgBd2KPi:bdRare0}   {\ensuremath{{0.455567 } } }
\vdef{bdt2012:bgBd2KPi:bdRare0E}  {\ensuremath{{0.322424 } } }
\vdef{bdt2012:bgBd2KPi:bsRare0}   {\ensuremath{{0.044416 } } }
\vdef{bdt2012:bgBd2KPi:bsRare0E}  {\ensuremath{{0.031435 } } }
\vdef{bdt2012:bgBd2KPi:hiSideband0:val}   {\ensuremath{{0.002 } } }
\vdef{bdt2012:bgBd2KPi:hiSideband0:err}   {\ensuremath{{0.001 } } }
\vdef{bdt2012:bgBd2MuMuGamma:loSideband0:val}   {\ensuremath{{0.028 } } }
\vdef{bdt2012:bgBd2MuMuGamma:loSideband0:err}   {\ensuremath{{0.006 } } }
\vdef{bdt2012:bgBd2MuMuGamma:bdRare0}   {\ensuremath{{0.002021 } } }
\vdef{bdt2012:bgBd2MuMuGamma:bdRare0E}  {\ensuremath{{0.000404 } } }
\vdef{bdt2012:bgBd2MuMuGamma:bsRare0}   {\ensuremath{{0.000204 } } }
\vdef{bdt2012:bgBd2MuMuGamma:bsRare0E}  {\ensuremath{{0.000041 } } }
\vdef{bdt2012:bgBd2MuMuGamma:hiSideband0:val}   {\ensuremath{{0.000 } } }
\vdef{bdt2012:bgBd2MuMuGamma:hiSideband0:err}   {\ensuremath{{0.000 } } }
\vdef{bdt2012:bgBd2Pi0MuMu:loSideband0:val}   {\ensuremath{{1.464 } } }
\vdef{bdt2012:bgBd2Pi0MuMu:loSideband0:err}   {\ensuremath{{0.732 } } }
\vdef{bdt2012:bgBd2Pi0MuMu:bdRare0}   {\ensuremath{{0.001312 } } }
\vdef{bdt2012:bgBd2Pi0MuMu:bdRare0E}  {\ensuremath{{0.000656 } } }
\vdef{bdt2012:bgBd2Pi0MuMu:bsRare0}   {\ensuremath{{0.000000 } } }
\vdef{bdt2012:bgBd2Pi0MuMu:bsRare0E}  {\ensuremath{{0.000000 } } }
\vdef{bdt2012:bgBd2Pi0MuMu:hiSideband0:val}   {\ensuremath{{0.001 } } }
\vdef{bdt2012:bgBd2Pi0MuMu:hiSideband0:err}   {\ensuremath{{0.000 } } }
\vdef{bdt2012:bgBd2PiMuNu:loSideband0:val}   {\ensuremath{{7.039 } } }
\vdef{bdt2012:bgBd2PiMuNu:loSideband0:err}   {\ensuremath{{3.537 } } }
\vdef{bdt2012:bgBd2PiMuNu:bdRare0}   {\ensuremath{{0.175771 } } }
\vdef{bdt2012:bgBd2PiMuNu:bdRare0E}  {\ensuremath{{0.088324 } } }
\vdef{bdt2012:bgBd2PiMuNu:bsRare0}   {\ensuremath{{0.016416 } } }
\vdef{bdt2012:bgBd2PiMuNu:bsRare0E}  {\ensuremath{{0.008249 } } }
\vdef{bdt2012:bgBd2PiMuNu:hiSideband0:val}   {\ensuremath{{0.004 } } }
\vdef{bdt2012:bgBd2PiMuNu:hiSideband0:err}   {\ensuremath{{0.002 } } }
\vdef{bdt2012:bgBd2PiPi:loSideband0:val}   {\ensuremath{{0.008 } } }
\vdef{bdt2012:bgBd2PiPi:loSideband0:err}   {\ensuremath{{0.008 } } }
\vdef{bdt2012:bgBd2PiPi:bdRare0}   {\ensuremath{{0.061578 } } }
\vdef{bdt2012:bgBd2PiPi:bdRare0E}  {\ensuremath{{0.061635 } } }
\vdef{bdt2012:bgBd2PiPi:bsRare0}   {\ensuremath{{0.025867 } } }
\vdef{bdt2012:bgBd2PiPi:bsRare0E}  {\ensuremath{{0.025891 } } }
\vdef{bdt2012:bgBd2PiPi:hiSideband0:val}   {\ensuremath{{0.001 } } }
\vdef{bdt2012:bgBd2PiPi:hiSideband0:err}   {\ensuremath{{0.001 } } }
\vdef{bdt2012:bgBs2KK:loSideband0:val}   {\ensuremath{{0.032 } } }
\vdef{bdt2012:bgBs2KK:loSideband0:err}   {\ensuremath{{0.032 } } }
\vdef{bdt2012:bgBs2KK:bdRare0}   {\ensuremath{{0.269032 } } }
\vdef{bdt2012:bgBs2KK:bdRare0E}  {\ensuremath{{0.272042 } } }
\vdef{bdt2012:bgBs2KK:bsRare0}   {\ensuremath{{0.113164 } } }
\vdef{bdt2012:bgBs2KK:bsRare0E}  {\ensuremath{{0.114430 } } }
\vdef{bdt2012:bgBs2KK:hiSideband0:val}   {\ensuremath{{0.002 } } }
\vdef{bdt2012:bgBs2KK:hiSideband0:err}   {\ensuremath{{0.002 } } }
\vdef{bdt2012:bgBs2KMuNu:loSideband0:val}   {\ensuremath{{4.517 } } }
\vdef{bdt2012:bgBs2KMuNu:loSideband0:err}   {\ensuremath{{2.270 } } }
\vdef{bdt2012:bgBs2KMuNu:bdRare0}   {\ensuremath{{0.208770 } } }
\vdef{bdt2012:bgBs2KMuNu:bdRare0E}  {\ensuremath{{0.104906 } } }
\vdef{bdt2012:bgBs2KMuNu:bsRare0}   {\ensuremath{{0.023368 } } }
\vdef{bdt2012:bgBs2KMuNu:bsRare0E}  {\ensuremath{{0.011742 } } }
\vdef{bdt2012:bgBs2KMuNu:hiSideband0:val}   {\ensuremath{{0.004 } } }
\vdef{bdt2012:bgBs2KMuNu:hiSideband0:err}   {\ensuremath{{0.002 } } }
\vdef{bdt2012:bgBs2KPi:loSideband0:val}   {\ensuremath{{0.002 } } }
\vdef{bdt2012:bgBs2KPi:loSideband0:err}   {\ensuremath{{0.001 } } }
\vdef{bdt2012:bgBs2KPi:bdRare0}   {\ensuremath{{0.012881 } } }
\vdef{bdt2012:bgBs2KPi:bdRare0E}  {\ensuremath{{0.009539 } } }
\vdef{bdt2012:bgBs2KPi:bsRare0}   {\ensuremath{{0.029181 } } }
\vdef{bdt2012:bgBs2KPi:bsRare0E}  {\ensuremath{{0.021610 } } }
\vdef{bdt2012:bgBs2KPi:hiSideband0:val}   {\ensuremath{{0.001 } } }
\vdef{bdt2012:bgBs2KPi:hiSideband0:err}   {\ensuremath{{0.000 } } }
\vdef{bdt2012:bgBs2MuMuGamma:loSideband0:val}   {\ensuremath{{1.166 } } }
\vdef{bdt2012:bgBs2MuMuGamma:loSideband0:err}   {\ensuremath{{0.233 } } }
\vdef{bdt2012:bgBs2MuMuGamma:bdRare0}   {\ensuremath{{0.221926 } } }
\vdef{bdt2012:bgBs2MuMuGamma:bdRare0E}  {\ensuremath{{0.044385 } } }
\vdef{bdt2012:bgBs2MuMuGamma:bsRare0}   {\ensuremath{{0.073130 } } }
\vdef{bdt2012:bgBs2MuMuGamma:bsRare0E}  {\ensuremath{{0.014626 } } }
\vdef{bdt2012:bgBs2MuMuGamma:hiSideband0:val}   {\ensuremath{{0.001 } } }
\vdef{bdt2012:bgBs2MuMuGamma:hiSideband0:err}   {\ensuremath{{0.000 } } }
\vdef{bdt2012:bgBs2PiPi:loSideband0:val}   {\ensuremath{{0.000 } } }
\vdef{bdt2012:bgBs2PiPi:loSideband0:err}   {\ensuremath{{0.000 } } }
\vdef{bdt2012:bgBs2PiPi:bdRare0}   {\ensuremath{{0.000292 } } }
\vdef{bdt2012:bgBs2PiPi:bdRare0E}  {\ensuremath{{0.000297 } } }
\vdef{bdt2012:bgBs2PiPi:bsRare0}   {\ensuremath{{0.002950 } } }
\vdef{bdt2012:bgBs2PiPi:bsRare0E}  {\ensuremath{{0.003004 } } }
\vdef{bdt2012:bgBs2PiPi:hiSideband0:val}   {\ensuremath{{0.000 } } }
\vdef{bdt2012:bgBs2PiPi:hiSideband0:err}   {\ensuremath{{0.000 } } }
\vdef{bdt2012:bgBu2KMuMu:loSideband0:val}   {\ensuremath{{0.000 } } }
\vdef{bdt2012:bgBu2KMuMu:loSideband0:err}   {\ensuremath{{0.000 } } }
\vdef{bdt2012:bgBu2KMuMu:bdRare0}   {\ensuremath{{0.000000 } } }
\vdef{bdt2012:bgBu2KMuMu:bdRare0E}  {\ensuremath{{0.000000 } } }
\vdef{bdt2012:bgBu2KMuMu:bsRare0}   {\ensuremath{{0.000000 } } }
\vdef{bdt2012:bgBu2KMuMu:bsRare0E}  {\ensuremath{{0.000000 } } }
\vdef{bdt2012:bgBu2KMuMu:hiSideband0:val}   {\ensuremath{{0.000 } } }
\vdef{bdt2012:bgBu2KMuMu:hiSideband0:err}   {\ensuremath{{0.000 } } }
\vdef{bdt2012:bgBu2PiMuMu:loSideband0:val}   {\ensuremath{{2.440 } } }
\vdef{bdt2012:bgBu2PiMuMu:loSideband0:err}   {\ensuremath{{0.634 } } }
\vdef{bdt2012:bgBu2PiMuMu:bdRare0}   {\ensuremath{{0.003499 } } }
\vdef{bdt2012:bgBu2PiMuMu:bdRare0E}  {\ensuremath{{0.000910 } } }
\vdef{bdt2012:bgBu2PiMuMu:bsRare0}   {\ensuremath{{0.001759 } } }
\vdef{bdt2012:bgBu2PiMuMu:bsRare0E}  {\ensuremath{{0.000457 } } }
\vdef{bdt2012:bgBu2PiMuMu:hiSideband0:val}   {\ensuremath{{0.000 } } }
\vdef{bdt2012:bgBu2PiMuMu:hiSideband0:err}   {\ensuremath{{0.000 } } }
\vdef{bdt2012:bgLb2KP:loSideband0:val}   {\ensuremath{{0.001 } } }
\vdef{bdt2012:bgLb2KP:loSideband0:err}   {\ensuremath{{0.001 } } }
\vdef{bdt2012:bgLb2KP:bdRare0}   {\ensuremath{{0.001964 } } }
\vdef{bdt2012:bgLb2KP:bdRare0E}  {\ensuremath{{0.001476 } } }
\vdef{bdt2012:bgLb2KP:bsRare0}   {\ensuremath{{0.021586 } } }
\vdef{bdt2012:bgLb2KP:bsRare0E}  {\ensuremath{{0.016223 } } }
\vdef{bdt2012:bgLb2KP:hiSideband0:val}   {\ensuremath{{0.008 } } }
\vdef{bdt2012:bgLb2KP:hiSideband0:err}   {\ensuremath{{0.006 } } }
\vdef{bdt2012:bgLb2PMuNu:loSideband0:val}   {\ensuremath{{33.730 } } }
\vdef{bdt2012:bgLb2PMuNu:loSideband0:err}   {\ensuremath{{37.711 } } }
\vdef{bdt2012:bgLb2PMuNu:bdRare0}   {\ensuremath{{8.518513 } } }
\vdef{bdt2012:bgLb2PMuNu:bdRare0E}  {\ensuremath{{9.523987 } } }
\vdef{bdt2012:bgLb2PMuNu:bsRare0}   {\ensuremath{{6.438319 } } }
\vdef{bdt2012:bgLb2PMuNu:bsRare0E}  {\ensuremath{{7.198259 } } }
\vdef{bdt2012:bgLb2PMuNu:hiSideband0:val}   {\ensuremath{{0.696 } } }
\vdef{bdt2012:bgLb2PMuNu:hiSideband0:err}   {\ensuremath{{0.778 } } }
\vdef{bdt2012:bgLb2PiP:loSideband0:val}   {\ensuremath{{0.000 } } }
\vdef{bdt2012:bgLb2PiP:loSideband0:err}   {\ensuremath{{0.000 } } }
\vdef{bdt2012:bgLb2PiP:bdRare0}   {\ensuremath{{0.000463 } } }
\vdef{bdt2012:bgLb2PiP:bdRare0E}  {\ensuremath{{0.000353 } } }
\vdef{bdt2012:bgLb2PiP:bsRare0}   {\ensuremath{{0.004296 } } }
\vdef{bdt2012:bgLb2PiP:bsRare0E}  {\ensuremath{{0.003277 } } }
\vdef{bdt2012:bgLb2PiP:hiSideband0:val}   {\ensuremath{{0.006 } } }
\vdef{bdt2012:bgLb2PiP:hiSideband0:err}   {\ensuremath{{0.005 } } }
\vdef{bdt2012:bsRare0}   {\ensuremath{{0.000 } } }
\vdef{bdt2012:bsRare0E}  {\ensuremath{{0.000 } } }
\vdef{bdt2012:bdRare0}   {\ensuremath{{0.000 } } }
\vdef{bdt2012:bdRare0E}  {\ensuremath{{0.000 } } }
\vdef{bdt2012:N-CSBF-TNP-BS1:val}   {\ensuremath{{0.000030 } } }
\vdef{bdt2012:N-CSBF-TNP-BS1:err}   {\ensuremath{{0.000000 } } }
\vdef{bdt2012:N-CSBF-MC-BS1:val}   {\ensuremath{{0.000030 } } }
\vdef{bdt2012:N-CSBF-MC-BS1:err}   {\ensuremath{{0.000001 } } }
\vdef{bdt2012:N-CSBF-MC-BS1:syst}   {\ensuremath{{0.000002 } } }
\vdef{bdt2012:N-CSBF-BS1:val}   {\ensuremath{{0.000031 } } }
\vdef{bdt2012:N-CSBF-BS1:err}   {\ensuremath{{0.000001 } } }
\vdef{bdt2012:N-CSBF-BS1:syst}   {\ensuremath{{0.000002 } } }
\vdef{bdt2012:bgBd2K0MuMu:loSideband1:val}   {\ensuremath{{0.000 } } }
\vdef{bdt2012:bgBd2K0MuMu:loSideband1:err}   {\ensuremath{{0.000 } } }
\vdef{bdt2012:bgBd2K0MuMu:bdRare1}   {\ensuremath{{0.020378 } } }
\vdef{bdt2012:bgBd2K0MuMu:bdRare1E}  {\ensuremath{{0.000000 } } }
\vdef{bdt2012:bgBd2K0MuMu:bsRare1}   {\ensuremath{{0.000000 } } }
\vdef{bdt2012:bgBd2K0MuMu:bsRare1E}  {\ensuremath{{0.000000 } } }
\vdef{bdt2012:bgBd2K0MuMu:hiSideband1:val}   {\ensuremath{{0.000 } } }
\vdef{bdt2012:bgBd2K0MuMu:hiSideband1:err}   {\ensuremath{{0.000 } } }
\vdef{bdt2012:bgBd2KK:loSideband1:val}   {\ensuremath{{0.001 } } }
\vdef{bdt2012:bgBd2KK:loSideband1:err}   {\ensuremath{{0.002 } } }
\vdef{bdt2012:bgBd2KK:bdRare1}   {\ensuremath{{0.000599 } } }
\vdef{bdt2012:bgBd2KK:bdRare1E}  {\ensuremath{{0.000756 } } }
\vdef{bdt2012:bgBd2KK:bsRare1}   {\ensuremath{{0.000082 } } }
\vdef{bdt2012:bgBd2KK:bsRare1E}  {\ensuremath{{0.000103 } } }
\vdef{bdt2012:bgBd2KK:hiSideband1:val}   {\ensuremath{{0.000 } } }
\vdef{bdt2012:bgBd2KK:hiSideband1:err}   {\ensuremath{{0.000 } } }
\vdef{bdt2012:bgBd2KPi:loSideband1:val}   {\ensuremath{{0.078 } } }
\vdef{bdt2012:bgBd2KPi:loSideband1:err}   {\ensuremath{{0.055 } } }
\vdef{bdt2012:bgBd2KPi:bdRare1}   {\ensuremath{{0.106973 } } }
\vdef{bdt2012:bgBd2KPi:bdRare1E}  {\ensuremath{{0.075710 } } }
\vdef{bdt2012:bgBd2KPi:bsRare1}   {\ensuremath{{0.029467 } } }
\vdef{bdt2012:bgBd2KPi:bsRare1E}  {\ensuremath{{0.020855 } } }
\vdef{bdt2012:bgBd2KPi:hiSideband1:val}   {\ensuremath{{0.002 } } }
\vdef{bdt2012:bgBd2KPi:hiSideband1:err}   {\ensuremath{{0.001 } } }
\vdef{bdt2012:bgBd2MuMuGamma:loSideband1:val}   {\ensuremath{{0.008 } } }
\vdef{bdt2012:bgBd2MuMuGamma:loSideband1:err}   {\ensuremath{{0.002 } } }
\vdef{bdt2012:bgBd2MuMuGamma:bdRare1}   {\ensuremath{{0.000788 } } }
\vdef{bdt2012:bgBd2MuMuGamma:bdRare1E}  {\ensuremath{{0.000158 } } }
\vdef{bdt2012:bgBd2MuMuGamma:bsRare1}   {\ensuremath{{0.000104 } } }
\vdef{bdt2012:bgBd2MuMuGamma:bsRare1E}  {\ensuremath{{0.000021 } } }
\vdef{bdt2012:bgBd2MuMuGamma:hiSideband1:val}   {\ensuremath{{0.000 } } }
\vdef{bdt2012:bgBd2MuMuGamma:hiSideband1:err}   {\ensuremath{{0.000 } } }
\vdef{bdt2012:bgBd2Pi0MuMu:loSideband1:val}   {\ensuremath{{0.382 } } }
\vdef{bdt2012:bgBd2Pi0MuMu:loSideband1:err}   {\ensuremath{{0.191 } } }
\vdef{bdt2012:bgBd2Pi0MuMu:bdRare1}   {\ensuremath{{0.003567 } } }
\vdef{bdt2012:bgBd2Pi0MuMu:bdRare1E}  {\ensuremath{{0.001783 } } }
\vdef{bdt2012:bgBd2Pi0MuMu:bsRare1}   {\ensuremath{{0.000474 } } }
\vdef{bdt2012:bgBd2Pi0MuMu:bsRare1E}  {\ensuremath{{0.000237 } } }
\vdef{bdt2012:bgBd2Pi0MuMu:hiSideband1:val}   {\ensuremath{{0.000 } } }
\vdef{bdt2012:bgBd2Pi0MuMu:hiSideband1:err}   {\ensuremath{{0.000 } } }
\vdef{bdt2012:bgBd2PiMuNu:loSideband1:val}   {\ensuremath{{2.029 } } }
\vdef{bdt2012:bgBd2PiMuNu:loSideband1:err}   {\ensuremath{{1.019 } } }
\vdef{bdt2012:bgBd2PiMuNu:bdRare1}   {\ensuremath{{0.080086 } } }
\vdef{bdt2012:bgBd2PiMuNu:bdRare1E}  {\ensuremath{{0.040243 } } }
\vdef{bdt2012:bgBd2PiMuNu:bsRare1}   {\ensuremath{{0.012728 } } }
\vdef{bdt2012:bgBd2PiMuNu:bsRare1E}  {\ensuremath{{0.006396 } } }
\vdef{bdt2012:bgBd2PiMuNu:hiSideband1:val}   {\ensuremath{{0.003 } } }
\vdef{bdt2012:bgBd2PiMuNu:hiSideband1:err}   {\ensuremath{{0.001 } } }
\vdef{bdt2012:bgBd2PiPi:loSideband1:val}   {\ensuremath{{0.005 } } }
\vdef{bdt2012:bgBd2PiPi:loSideband1:err}   {\ensuremath{{0.005 } } }
\vdef{bdt2012:bgBd2PiPi:bdRare1}   {\ensuremath{{0.016006 } } }
\vdef{bdt2012:bgBd2PiPi:bdRare1E}  {\ensuremath{{0.016021 } } }
\vdef{bdt2012:bgBd2PiPi:bsRare1}   {\ensuremath{{0.010246 } } }
\vdef{bdt2012:bgBd2PiPi:bsRare1E}  {\ensuremath{{0.010255 } } }
\vdef{bdt2012:bgBd2PiPi:hiSideband1:val}   {\ensuremath{{0.001 } } }
\vdef{bdt2012:bgBd2PiPi:hiSideband1:err}   {\ensuremath{{0.001 } } }
\vdef{bdt2012:bgBs2KK:loSideband1:val}   {\ensuremath{{0.021 } } }
\vdef{bdt2012:bgBs2KK:loSideband1:err}   {\ensuremath{{0.021 } } }
\vdef{bdt2012:bgBs2KK:bdRare1}   {\ensuremath{{0.064641 } } }
\vdef{bdt2012:bgBs2KK:bdRare1E}  {\ensuremath{{0.065364 } } }
\vdef{bdt2012:bgBs2KK:bsRare1}   {\ensuremath{{0.037683 } } }
\vdef{bdt2012:bgBs2KK:bsRare1E}  {\ensuremath{{0.038105 } } }
\vdef{bdt2012:bgBs2KK:hiSideband1:val}   {\ensuremath{{0.002 } } }
\vdef{bdt2012:bgBs2KK:hiSideband1:err}   {\ensuremath{{0.002 } } }
\vdef{bdt2012:bgBs2KMuNu:loSideband1:val}   {\ensuremath{{1.204 } } }
\vdef{bdt2012:bgBs2KMuNu:loSideband1:err}   {\ensuremath{{0.605 } } }
\vdef{bdt2012:bgBs2KMuNu:bdRare1}   {\ensuremath{{0.069817 } } }
\vdef{bdt2012:bgBs2KMuNu:bdRare1E}  {\ensuremath{{0.035082 } } }
\vdef{bdt2012:bgBs2KMuNu:bsRare1}   {\ensuremath{{0.012456 } } }
\vdef{bdt2012:bgBs2KMuNu:bsRare1E}  {\ensuremath{{0.006259 } } }
\vdef{bdt2012:bgBs2KMuNu:hiSideband1:val}   {\ensuremath{{0.001 } } }
\vdef{bdt2012:bgBs2KMuNu:hiSideband1:err}   {\ensuremath{{0.001 } } }
\vdef{bdt2012:bgBs2KPi:loSideband1:val}   {\ensuremath{{0.001 } } }
\vdef{bdt2012:bgBs2KPi:loSideband1:err}   {\ensuremath{{0.001 } } }
\vdef{bdt2012:bgBs2KPi:bdRare1}   {\ensuremath{{0.004780 } } }
\vdef{bdt2012:bgBs2KPi:bdRare1E}  {\ensuremath{{0.003540 } } }
\vdef{bdt2012:bgBs2KPi:bsRare1}   {\ensuremath{{0.007610 } } }
\vdef{bdt2012:bgBs2KPi:bsRare1E}  {\ensuremath{{0.005636 } } }
\vdef{bdt2012:bgBs2KPi:hiSideband1:val}   {\ensuremath{{0.001 } } }
\vdef{bdt2012:bgBs2KPi:hiSideband1:err}   {\ensuremath{{0.000 } } }
\vdef{bdt2012:bgBs2MuMuGamma:loSideband1:val}   {\ensuremath{{0.316 } } }
\vdef{bdt2012:bgBs2MuMuGamma:loSideband1:err}   {\ensuremath{{0.063 } } }
\vdef{bdt2012:bgBs2MuMuGamma:bdRare1}   {\ensuremath{{0.060582 } } }
\vdef{bdt2012:bgBs2MuMuGamma:bdRare1E}  {\ensuremath{{0.012116 } } }
\vdef{bdt2012:bgBs2MuMuGamma:bsRare1}   {\ensuremath{{0.019762 } } }
\vdef{bdt2012:bgBs2MuMuGamma:bsRare1E}  {\ensuremath{{0.003952 } } }
\vdef{bdt2012:bgBs2MuMuGamma:hiSideband1:val}   {\ensuremath{{0.001 } } }
\vdef{bdt2012:bgBs2MuMuGamma:hiSideband1:err}   {\ensuremath{{0.000 } } }
\vdef{bdt2012:bgBs2PiPi:loSideband1:val}   {\ensuremath{{0.000 } } }
\vdef{bdt2012:bgBs2PiPi:loSideband1:err}   {\ensuremath{{0.000 } } }
\vdef{bdt2012:bgBs2PiPi:bdRare1}   {\ensuremath{{0.000186 } } }
\vdef{bdt2012:bgBs2PiPi:bdRare1E}  {\ensuremath{{0.000189 } } }
\vdef{bdt2012:bgBs2PiPi:bsRare1}   {\ensuremath{{0.000794 } } }
\vdef{bdt2012:bgBs2PiPi:bsRare1E}  {\ensuremath{{0.000808 } } }
\vdef{bdt2012:bgBs2PiPi:hiSideband1:val}   {\ensuremath{{0.000 } } }
\vdef{bdt2012:bgBs2PiPi:hiSideband1:err}   {\ensuremath{{0.000 } } }
\vdef{bdt2012:bgBu2KMuMu:loSideband1:val}   {\ensuremath{{0.000 } } }
\vdef{bdt2012:bgBu2KMuMu:loSideband1:err}   {\ensuremath{{0.000 } } }
\vdef{bdt2012:bgBu2KMuMu:bdRare1}   {\ensuremath{{0.000000 } } }
\vdef{bdt2012:bgBu2KMuMu:bdRare1E}  {\ensuremath{{0.000000 } } }
\vdef{bdt2012:bgBu2KMuMu:bsRare1}   {\ensuremath{{0.000000 } } }
\vdef{bdt2012:bgBu2KMuMu:bsRare1E}  {\ensuremath{{0.000000 } } }
\vdef{bdt2012:bgBu2KMuMu:hiSideband1:val}   {\ensuremath{{0.000 } } }
\vdef{bdt2012:bgBu2KMuMu:hiSideband1:err}   {\ensuremath{{0.000 } } }
\vdef{bdt2012:bgBu2PiMuMu:loSideband1:val}   {\ensuremath{{0.660 } } }
\vdef{bdt2012:bgBu2PiMuMu:loSideband1:err}   {\ensuremath{{0.172 } } }
\vdef{bdt2012:bgBu2PiMuMu:bdRare1}   {\ensuremath{{0.000653 } } }
\vdef{bdt2012:bgBu2PiMuMu:bdRare1E}  {\ensuremath{{0.000170 } } }
\vdef{bdt2012:bgBu2PiMuMu:bsRare1}   {\ensuremath{{0.000000 } } }
\vdef{bdt2012:bgBu2PiMuMu:bsRare1E}  {\ensuremath{{0.000000 } } }
\vdef{bdt2012:bgBu2PiMuMu:hiSideband1:val}   {\ensuremath{{0.000 } } }
\vdef{bdt2012:bgBu2PiMuMu:hiSideband1:err}   {\ensuremath{{0.000 } } }
\vdef{bdt2012:bgLb2KP:loSideband1:val}   {\ensuremath{{0.000 } } }
\vdef{bdt2012:bgLb2KP:loSideband1:err}   {\ensuremath{{0.000 } } }
\vdef{bdt2012:bgLb2KP:bdRare1}   {\ensuremath{{0.000968 } } }
\vdef{bdt2012:bgLb2KP:bdRare1E}  {\ensuremath{{0.000728 } } }
\vdef{bdt2012:bgLb2KP:bsRare1}   {\ensuremath{{0.004899 } } }
\vdef{bdt2012:bgLb2KP:bsRare1E}  {\ensuremath{{0.003682 } } }
\vdef{bdt2012:bgLb2KP:hiSideband1:val}   {\ensuremath{{0.003 } } }
\vdef{bdt2012:bgLb2KP:hiSideband1:err}   {\ensuremath{{0.002 } } }
\vdef{bdt2012:bgLb2PMuNu:loSideband1:val}   {\ensuremath{{7.669 } } }
\vdef{bdt2012:bgLb2PMuNu:loSideband1:err}   {\ensuremath{{8.574 } } }
\vdef{bdt2012:bgLb2PMuNu:bdRare1}   {\ensuremath{{1.909107 } } }
\vdef{bdt2012:bgLb2PMuNu:bdRare1E}  {\ensuremath{{2.134446 } } }
\vdef{bdt2012:bgLb2PMuNu:bsRare1}   {\ensuremath{{1.489847 } } }
\vdef{bdt2012:bgLb2PMuNu:bsRare1E}  {\ensuremath{{1.665699 } } }
\vdef{bdt2012:bgLb2PMuNu:hiSideband1:val}   {\ensuremath{{0.314 } } }
\vdef{bdt2012:bgLb2PMuNu:hiSideband1:err}   {\ensuremath{{0.351 } } }
\vdef{bdt2012:bgLb2PiP:loSideband1:val}   {\ensuremath{{0.000 } } }
\vdef{bdt2012:bgLb2PiP:loSideband1:err}   {\ensuremath{{0.000 } } }
\vdef{bdt2012:bgLb2PiP:bdRare1}   {\ensuremath{{0.000225 } } }
\vdef{bdt2012:bgLb2PiP:bdRare1E}  {\ensuremath{{0.000172 } } }
\vdef{bdt2012:bgLb2PiP:bsRare1}   {\ensuremath{{0.001273 } } }
\vdef{bdt2012:bgLb2PiP:bsRare1E}  {\ensuremath{{0.000971 } } }
\vdef{bdt2012:bgLb2PiP:hiSideband1:val}   {\ensuremath{{0.002 } } }
\vdef{bdt2012:bgLb2PiP:hiSideband1:err}   {\ensuremath{{0.001 } } }
\vdef{bdt2012:bsRare1}   {\ensuremath{{0.000 } } }
\vdef{bdt2012:bsRare1E}  {\ensuremath{{0.000 } } }
\vdef{bdt2012:bdRare1}   {\ensuremath{{0.000 } } }
\vdef{bdt2012:bdRare1E}  {\ensuremath{{0.000 } } }
\vdef{bdt2012:N-EFF-TOT-BPLUS0:val}   {\ensuremath{{0.00082 } } }
\vdef{bdt2012:N-EFF-TOT-BPLUS0:err}   {\ensuremath{{0.000001 } } }
\vdef{bdt2012:N-EFF-TOT-BPLUS0:tot}   {\ensuremath{{0.00007 } } }
\vdef{bdt2012:N-EFF-TOT-BPLUS0:all}   {\ensuremath{{(0.82 \pm 0.07)\times 10^{-3}} } }
\vdef{bdt2012:N-EFF-PRODMC-BPLUS0:val}   {\ensuremath{{0.00083 } } }
\vdef{bdt2012:N-EFF-PRODMC-BPLUS0:err}   {\ensuremath{{0.000003 } } }
\vdef{bdt2012:N-EFF-PRODMC-BPLUS0:tot}   {\ensuremath{{0.00000 } } }
\vdef{bdt2012:N-EFF-PRODMC-BPLUS0:all}   {\ensuremath{{(0.83 \pm 0.00)\times 10^{-3}} } }
\vdef{bdt2012:N-EFF-PRODTNP-BPLUS0:val}   {\ensuremath{{0.00120 } } }
\vdef{bdt2012:N-EFF-PRODTNP-BPLUS0:err}   {\ensuremath{{1.589703 } } }
\vdef{bdt2012:N-EFF-PRODTNP-BPLUS0:tot}   {\ensuremath{{1.58970 } } }
\vdef{bdt2012:N-EFF-PRODTNP-BPLUS0:all}   {\ensuremath{{(1.20 \pm 1589.70)\times 10^{-3}} } }
\vdef{bdt2012:N-EFF-PRODTNPMC-BPLUS0:val}   {\ensuremath{{0.00083 } } }
\vdef{bdt2012:N-EFF-PRODTNPMC-BPLUS0:err}   {\ensuremath{{0.000003 } } }
\vdef{bdt2012:N-EFF-PRODTNPMC-BPLUS0:tot}   {\ensuremath{{0.00000 } } }
\vdef{bdt2012:N-EFF-PRODTNPMC-BPLUS0:all}   {\ensuremath{{(0.83 \pm 0.00)\times 10^{-3}} } }
\vdef{bdt2012:N-ACC-BPLUS0:val}   {\ensuremath{{0.011 } } }
\vdef{bdt2012:N-ACC-BPLUS0:err}   {\ensuremath{{0.000 } } }
\vdef{bdt2012:N-ACC-BPLUS0:tot}   {\ensuremath{{0.000 } } }
\vdef{bdt2012:N-ACC-BPLUS0:all}   {\ensuremath{{(1.12 \pm 0.04)\times 10^{-2}} } }
\vdef{bdt2012:N-EFF-MU-PID-BPLUS0:val}   {\ensuremath{{0.673 } } }
\vdef{bdt2012:N-EFF-MU-PID-BPLUS0:err}   {\ensuremath{{0.000 } } }
\vdef{bdt2012:N-EFF-MU-PID-BPLUS0:tot}   {\ensuremath{{0.027 } } }
\vdef{bdt2012:N-EFF-MU-PID-BPLUS0:all}   {\ensuremath{{(67.28 \pm 2.69)\times 10^{-2}} } }
\vdef{bdt2012:N-EFFRHO-MU-PID-BPLUS0:val}   {\ensuremath{{0.636 } } }
\vdef{bdt2012:N-EFFRHO-MU-PID-BPLUS0:err}   {\ensuremath{{0.000 } } }
\vdef{bdt2012:N-EFFRHO-MU-PID-BPLUS0:tot}   {\ensuremath{{0.000 } } }
\vdef{bdt2012:N-EFFRHO-MU-PID-BPLUS0:all}   {\ensuremath{{(63.60 \pm 0.00)\times 10^{-2}} } }
\vdef{bdt2012:N-EFF-MU-PIDMC-BPLUS0:val}   {\ensuremath{{0.635 } } }
\vdef{bdt2012:N-EFF-MU-PIDMC-BPLUS0:err}   {\ensuremath{{0.000 } } }
\vdef{bdt2012:N-EFF-MU-PIDMC-BPLUS0:tot}   {\ensuremath{{0.025 } } }
\vdef{bdt2012:N-EFF-MU-PIDMC-BPLUS0:all}   {\ensuremath{{(63.52 \pm 2.54)\times 10^{-2}} } }
\vdef{bdt2012:N-EFFRHO-MU-PIDMC-BPLUS0:val}   {\ensuremath{{0.600 } } }
\vdef{bdt2012:N-EFFRHO-MU-PIDMC-BPLUS0:err}   {\ensuremath{{0.000 } } }
\vdef{bdt2012:N-EFFRHO-MU-PIDMC-BPLUS0:tot}   {\ensuremath{{0.025 } } }
\vdef{bdt2012:N-EFFRHO-MU-PIDMC-BPLUS0:all}   {\ensuremath{{(60.05 \pm 2.54)\times 10^{-2}} } }
\vdef{bdt2012:N-EFF-MU-MC-BPLUS0:val}   {\ensuremath{{0.600 } } }
\vdef{bdt2012:N-EFF-MU-MC-BPLUS0:err}   {\ensuremath{{0.000 } } }
\vdef{bdt2012:N-EFF-MU-MC-BPLUS0:tot}   {\ensuremath{{0.024 } } }
\vdef{bdt2012:N-EFF-MU-MC-BPLUS0:all}   {\ensuremath{{(60.05 \pm 2.40)\times 10^{-2}} } }
\vdef{bdt2012:N-EFF-TRIG-PID-BPLUS0:val}   {\ensuremath{{0.707 } } }
\vdef{bdt2012:N-EFF-TRIG-PID-BPLUS0:err}   {\ensuremath{{0.000 } } }
\vdef{bdt2012:N-EFF-TRIG-PID-BPLUS0:tot}   {\ensuremath{{0.021 } } }
\vdef{bdt2012:N-EFF-TRIG-PID-BPLUS0:all}   {\ensuremath{{(70.72 \pm 2.12)\times 10^{-2}} } }
\vdef{bdt2012:N-EFFRHO-TRIG-PID-BPLUS0:val}   {\ensuremath{{0.585 } } }
\vdef{bdt2012:N-EFFRHO-TRIG-PID-BPLUS0:err}   {\ensuremath{{0.000 } } }
\vdef{bdt2012:N-EFFRHO-TRIG-PID-BPLUS0:tot}   {\ensuremath{{0.018 } } }
\vdef{bdt2012:N-EFFRHO-TRIG-PID-BPLUS0:all}   {\ensuremath{{(58.52 \pm 1.76)\times 10^{-2}} } }
\vdef{bdt2012:N-EFF-TRIG-PIDMC-BPLUS0:val}   {\ensuremath{{0.660 } } }
\vdef{bdt2012:N-EFF-TRIG-PIDMC-BPLUS0:err}   {\ensuremath{{0.000 } } }
\vdef{bdt2012:N-EFF-TRIG-PIDMC-BPLUS0:tot}   {\ensuremath{{0.020 } } }
\vdef{bdt2012:N-EFF-TRIG-PIDMC-BPLUS0:all}   {\ensuremath{{(65.99 \pm 1.98)\times 10^{-2}} } }
\vdef{bdt2012:N-EFFRHO-TRIG-PIDMC-BPLUS0:val}   {\ensuremath{{0.546 } } }
\vdef{bdt2012:N-EFFRHO-TRIG-PIDMC-BPLUS0:err}   {\ensuremath{{0.000 } } }
\vdef{bdt2012:N-EFFRHO-TRIG-PIDMC-BPLUS0:tot}   {\ensuremath{{0.016 } } }
\vdef{bdt2012:N-EFFRHO-TRIG-PIDMC-BPLUS0:all}   {\ensuremath{{(54.61 \pm 1.64)\times 10^{-2}} } }
\vdef{bdt2012:N-EFF-TRIG-MC-BPLUS0:val}   {\ensuremath{{0.546 } } }
\vdef{bdt2012:N-EFF-TRIG-MC-BPLUS0:err}   {\ensuremath{{0.001 } } }
\vdef{bdt2012:N-EFF-TRIG-MC-BPLUS0:tot}   {\ensuremath{{0.016 } } }
\vdef{bdt2012:N-EFF-TRIG-MC-BPLUS0:all}   {\ensuremath{{(54.61 \pm 1.64)\times 10^{-2}} } }
\vdef{bdt2012:N-EFF-CAND-BPLUS0:val}   {\ensuremath{{1.000 } } }
\vdef{bdt2012:N-EFF-CAND-BPLUS0:err}   {\ensuremath{{0.000 } } }
\vdef{bdt2012:N-EFF-CAND-BPLUS0:tot}   {\ensuremath{{0.010 } } }
\vdef{bdt2012:N-EFF-CAND-BPLUS0:all}   {\ensuremath{{(99.99 \pm 1.00)\times 10^{-2}} } }
\vdef{bdt2012:N-EFF-ANA-BPLUS0:val}   {\ensuremath{{0.2258 } } }
\vdef{bdt2012:N-EFF-ANA-BPLUS0:err}   {\ensuremath{{0.0002 } } }
\vdef{bdt2012:N-EFF-ANA-BPLUS0:tot}   {\ensuremath{{0.0128 } } }
\vdef{bdt2012:N-EFF-ANA-BPLUS0:all}   {\ensuremath{{(22.58 \pm 1.28)\times 10^{-2}} } }
\vdef{bdt2012:N-OBS-BPLUS0:val}   {\ensuremath{{308877 } } }
\vdef{bdt2012:N-OBS-BPLUS0:err}   {\ensuremath{{729 } } }
\vdef{bdt2012:N-OBS-BPLUS0:tot}   {\ensuremath{{15461 } } }
\vdef{bdt2012:N-OBS-BPLUS0:all}   {\ensuremath{{308877 } } }
\vdef{bdt2012:N-OBS-CBPLUS0:val}   {\ensuremath{{321086 } } }
\vdef{bdt2012:N-OBS-CBPLUS0:err}   {\ensuremath{{592 } } }
\vdef{bdt2012:N-EFF-TOT-BPLUS1:val}   {\ensuremath{{0.00021 } } }
\vdef{bdt2012:N-EFF-TOT-BPLUS1:err}   {\ensuremath{{0.000001 } } }
\vdef{bdt2012:N-EFF-TOT-BPLUS1:tot}   {\ensuremath{{0.00003 } } }
\vdef{bdt2012:N-EFF-TOT-BPLUS1:all}   {\ensuremath{{(0.21 \pm 0.03)\times 10^{-3}} } }
\vdef{bdt2012:N-EFF-PRODMC-BPLUS1:val}   {\ensuremath{{0.00021 } } }
\vdef{bdt2012:N-EFF-PRODMC-BPLUS1:err}   {\ensuremath{{0.000001 } } }
\vdef{bdt2012:N-EFF-PRODMC-BPLUS1:tot}   {\ensuremath{{0.00000 } } }
\vdef{bdt2012:N-EFF-PRODMC-BPLUS1:all}   {\ensuremath{{(0.21 \pm 0.00)\times 10^{-3}} } }
\vdef{bdt2012:N-EFF-PRODTNP-BPLUS1:val}   {\ensuremath{{0.00034 } } }
\vdef{bdt2012:N-EFF-PRODTNP-BPLUS1:err}   {\ensuremath{{0.237522 } } }
\vdef{bdt2012:N-EFF-PRODTNP-BPLUS1:tot}   {\ensuremath{{0.23752 } } }
\vdef{bdt2012:N-EFF-PRODTNP-BPLUS1:all}   {\ensuremath{{(0.34 \pm 237.52)\times 10^{-3}} } }
\vdef{bdt2012:N-EFF-PRODTNPMC-BPLUS1:val}   {\ensuremath{{0.00021 } } }
\vdef{bdt2012:N-EFF-PRODTNPMC-BPLUS1:err}   {\ensuremath{{0.000001 } } }
\vdef{bdt2012:N-EFF-PRODTNPMC-BPLUS1:tot}   {\ensuremath{{0.00000 } } }
\vdef{bdt2012:N-EFF-PRODTNPMC-BPLUS1:all}   {\ensuremath{{(0.21 \pm 0.00)\times 10^{-3}} } }
\vdef{bdt2012:N-ACC-BPLUS1:val}   {\ensuremath{{0.006 } } }
\vdef{bdt2012:N-ACC-BPLUS1:err}   {\ensuremath{{0.000 } } }
\vdef{bdt2012:N-ACC-BPLUS1:tot}   {\ensuremath{{0.000 } } }
\vdef{bdt2012:N-ACC-BPLUS1:all}   {\ensuremath{{(0.59 \pm 0.03)\times 10^{-2}} } }
\vdef{bdt2012:N-EFF-MU-PID-BPLUS1:val}   {\ensuremath{{0.584 } } }
\vdef{bdt2012:N-EFF-MU-PID-BPLUS1:err}   {\ensuremath{{0.000 } } }
\vdef{bdt2012:N-EFF-MU-PID-BPLUS1:tot}   {\ensuremath{{0.047 } } }
\vdef{bdt2012:N-EFF-MU-PID-BPLUS1:all}   {\ensuremath{{(58.40 \pm 4.67)\times 10^{-2}} } }
\vdef{bdt2012:N-EFFRHO-MU-PID-BPLUS1:val}   {\ensuremath{{0.569 } } }
\vdef{bdt2012:N-EFFRHO-MU-PID-BPLUS1:err}   {\ensuremath{{0.000 } } }
\vdef{bdt2012:N-EFFRHO-MU-PID-BPLUS1:tot}   {\ensuremath{{0.000 } } }
\vdef{bdt2012:N-EFFRHO-MU-PID-BPLUS1:all}   {\ensuremath{{(56.93 \pm 0.00)\times 10^{-2}} } }
\vdef{bdt2012:N-EFF-MU-PIDMC-BPLUS1:val}   {\ensuremath{{0.577 } } }
\vdef{bdt2012:N-EFF-MU-PIDMC-BPLUS1:err}   {\ensuremath{{0.000 } } }
\vdef{bdt2012:N-EFF-MU-PIDMC-BPLUS1:tot}   {\ensuremath{{0.046 } } }
\vdef{bdt2012:N-EFF-MU-PIDMC-BPLUS1:all}   {\ensuremath{{(57.67 \pm 4.61)\times 10^{-2}} } }
\vdef{bdt2012:N-EFFRHO-MU-PIDMC-BPLUS1:val}   {\ensuremath{{0.562 } } }
\vdef{bdt2012:N-EFFRHO-MU-PIDMC-BPLUS1:err}   {\ensuremath{{0.000 } } }
\vdef{bdt2012:N-EFFRHO-MU-PIDMC-BPLUS1:tot}   {\ensuremath{{0.046 } } }
\vdef{bdt2012:N-EFFRHO-MU-PIDMC-BPLUS1:all}   {\ensuremath{{(56.22 \pm 4.55)\times 10^{-2}} } }
\vdef{bdt2012:N-EFF-MU-MC-BPLUS1:val}   {\ensuremath{{0.562 } } }
\vdef{bdt2012:N-EFF-MU-MC-BPLUS1:err}   {\ensuremath{{0.001 } } }
\vdef{bdt2012:N-EFF-MU-MC-BPLUS1:tot}   {\ensuremath{{0.045 } } }
\vdef{bdt2012:N-EFF-MU-MC-BPLUS1:all}   {\ensuremath{{(56.22 \pm 4.50)\times 10^{-2}} } }
\vdef{bdt2012:N-EFF-TRIG-PID-BPLUS1:val}   {\ensuremath{{0.765 } } }
\vdef{bdt2012:N-EFF-TRIG-PID-BPLUS1:err}   {\ensuremath{{0.000 } } }
\vdef{bdt2012:N-EFF-TRIG-PID-BPLUS1:tot}   {\ensuremath{{0.046 } } }
\vdef{bdt2012:N-EFF-TRIG-PID-BPLUS1:all}   {\ensuremath{{(76.48 \pm 4.59)\times 10^{-2}} } }
\vdef{bdt2012:N-EFFRHO-TRIG-PID-BPLUS1:val}   {\ensuremath{{0.498 } } }
\vdef{bdt2012:N-EFFRHO-TRIG-PID-BPLUS1:err}   {\ensuremath{{0.000 } } }
\vdef{bdt2012:N-EFFRHO-TRIG-PID-BPLUS1:tot}   {\ensuremath{{0.030 } } }
\vdef{bdt2012:N-EFFRHO-TRIG-PID-BPLUS1:all}   {\ensuremath{{(49.82 \pm 2.99)\times 10^{-2}} } }
\vdef{bdt2012:N-EFF-TRIG-PIDMC-BPLUS1:val}   {\ensuremath{{0.749 } } }
\vdef{bdt2012:N-EFF-TRIG-PIDMC-BPLUS1:err}   {\ensuremath{{0.000 } } }
\vdef{bdt2012:N-EFF-TRIG-PIDMC-BPLUS1:tot}   {\ensuremath{{0.045 } } }
\vdef{bdt2012:N-EFF-TRIG-PIDMC-BPLUS1:all}   {\ensuremath{{(74.89 \pm 4.49)\times 10^{-2}} } }
\vdef{bdt2012:N-EFFRHO-TRIG-PIDMC-BPLUS1:val}   {\ensuremath{{0.488 } } }
\vdef{bdt2012:N-EFFRHO-TRIG-PIDMC-BPLUS1:err}   {\ensuremath{{0.000 } } }
\vdef{bdt2012:N-EFFRHO-TRIG-PIDMC-BPLUS1:tot}   {\ensuremath{{0.029 } } }
\vdef{bdt2012:N-EFFRHO-TRIG-PIDMC-BPLUS1:all}   {\ensuremath{{(48.79 \pm 2.93)\times 10^{-2}} } }
\vdef{bdt2012:N-EFF-TRIG-MC-BPLUS1:val}   {\ensuremath{{0.488 } } }
\vdef{bdt2012:N-EFF-TRIG-MC-BPLUS1:err}   {\ensuremath{{0.001 } } }
\vdef{bdt2012:N-EFF-TRIG-MC-BPLUS1:tot}   {\ensuremath{{0.029 } } }
\vdef{bdt2012:N-EFF-TRIG-MC-BPLUS1:all}   {\ensuremath{{(48.79 \pm 2.93)\times 10^{-2}} } }
\vdef{bdt2012:N-EFF-CAND-BPLUS1:val}   {\ensuremath{{1.000 } } }
\vdef{bdt2012:N-EFF-CAND-BPLUS1:err}   {\ensuremath{{0.000 } } }
\vdef{bdt2012:N-EFF-CAND-BPLUS1:tot}   {\ensuremath{{0.010 } } }
\vdef{bdt2012:N-EFF-CAND-BPLUS1:all}   {\ensuremath{{(99.98 \pm 1.00)\times 10^{-2}} } }
\vdef{bdt2012:N-EFF-ANA-BPLUS1:val}   {\ensuremath{{0.1316 } } }
\vdef{bdt2012:N-EFF-ANA-BPLUS1:err}   {\ensuremath{{0.0002 } } }
\vdef{bdt2012:N-EFF-ANA-BPLUS1:tot}   {\ensuremath{{0.0074 } } }
\vdef{bdt2012:N-EFF-ANA-BPLUS1:all}   {\ensuremath{{(13.16 \pm 0.74)\times 10^{-2}} } }
\vdef{bdt2012:N-OBS-BPLUS1:val}   {\ensuremath{{69260 } } }
\vdef{bdt2012:N-OBS-BPLUS1:err}   {\ensuremath{{291 } } }
\vdef{bdt2012:N-OBS-BPLUS1:tot}   {\ensuremath{{3475 } } }
\vdef{bdt2012:N-OBS-BPLUS1:all}   {\ensuremath{{69260 } } }
\vdef{bdt2012:N-OBS-CBPLUS1:val}   {\ensuremath{{65988 } } }
\vdef{bdt2012:N-OBS-CBPLUS1:err}   {\ensuremath{{276 } } }
\vdef{bdt2012:N-EXP2-SIG-BSMM0:val}   {\ensuremath{{10.72 } } }
\vdef{bdt2012:N-EXP2-SIG-BSMM0:err}   {\ensuremath{{ 1.61 } } }
\vdef{bdt2012:N-EXP2-SIG-BDMM0:val}   {\ensuremath{{1.001 } } }
\vdef{bdt2012:N-EXP2-SIG-BDMM0:err}   {\ensuremath{{0.100 } } }
\vdef{bdt2012:N-OBS-BKG0:val}   {\ensuremath{{34 } } }
\vdef{bdt2012:N-EXP-BSMM0:val}   {\ensuremath{{ 5.82 } } }
\vdef{bdt2012:N-EXP-BSMM0:err}   {\ensuremath{{ 2.91 } } }
\vdef{bdt2012:N-EXP-BDMM0:val}   {\ensuremath{{ 5.20 } } }
\vdef{bdt2012:N-EXP-BDMM0:err}   {\ensuremath{{ 2.60 } } }
\vdef{bdt2012:N-LOW-BD0:val}   {\ensuremath{{5.200 } } }
\vdef{bdt2012:N-HIGH-BD0:val}   {\ensuremath{{5.300 } } }
\vdef{bdt2012:N-LOW-BS0:val}   {\ensuremath{{5.300 } } }
\vdef{bdt2012:N-HIGH-BS0:val}   {\ensuremath{{5.450 } } }
\vdef{bdt2012:N-PSS0:val}   {\ensuremath{{0.881 } } }
\vdef{bdt2012:N-PSS0:err}   {\ensuremath{{0.000 } } }
\vdef{bdt2012:N-PSS0:tot}   {\ensuremath{{0.044 } } }
\vdef{bdt2012:N-PSD0:val}   {\ensuremath{{0.280 } } }
\vdef{bdt2012:N-PSD0:err}   {\ensuremath{{0.001 } } }
\vdef{bdt2012:N-PSD0:tot}   {\ensuremath{{0.014 } } }
\vdef{bdt2012:N-PDS0:val}   {\ensuremath{{0.070 } } }
\vdef{bdt2012:N-PDS0:err}   {\ensuremath{{0.000 } } }
\vdef{bdt2012:N-PDS0:tot}   {\ensuremath{{0.003 } } }
\vdef{bdt2012:N-PDD0:val}   {\ensuremath{{0.661 } } }
\vdef{bdt2012:N-PDD0:err}   {\ensuremath{{0.001 } } }
\vdef{bdt2012:N-PDD0:tot}   {\ensuremath{{0.033 } } }
\vdef{bdt2012:N-EFF-TOT-BSMM0:val}   {\ensuremath{{0.0023 } } }
\vdef{bdt2012:N-EFF-TOT-BSMM0:err}   {\ensuremath{{0.0000 } } }
\vdef{bdt2012:N-EFF-TOT-BSMM0:tot}   {\ensuremath{{0.0002 } } }
\vdef{bdt2012:N-EFF-TOT-BSMM0:all}   {\ensuremath{{(0.23 \pm 0.03)} } }
\vdef{bdt2012:N-EFF-PRODMC-BSMM0:val}   {\ensuremath{{0.0023 } } }
\vdef{bdt2012:N-EFF-PRODMC-BSMM0:err}   {\ensuremath{{0.0000 } } }
\vdef{bdt2012:N-EFF-PRODMC-BSMM0:tot}   {\ensuremath{{0.0000 } } }
\vdef{bdt2012:N-EFF-PRODMC-BSMM0:all}   {\ensuremath{{(2.32 \pm 0.01)\times 10^{-3}} } }
\vdef{bdt2012:N-EFFRATIO-TOT-BSMM0:val}   {\ensuremath{{0.362 } } }
\vdef{bdt2012:N-EFFRATIO-TOT-BSMM0:err}   {\ensuremath{{0.001 } } }
\vdef{bdt2012:N-EFFRATIO-PRODMC-BSMM0:val}   {\ensuremath{{0.357 } } }
\vdef{bdt2012:N-EFFRATIO-PRODMC-BSMM0:err}   {\ensuremath{{0.002 } } }
\vdef{bdt2012:N-EFFRATIO-PRODTNP-BSMM0:val}   {\ensuremath{{0.378 } } }
\vdef{bdt2012:N-EFFRATIO-PRODTNP-BSMM0:err}   {\ensuremath{{749.948 } } }
\vdef{bdt2012:N-EFFRATIO-PRODTNPMC-BSMM0:val}   {\ensuremath{{0.357 } } }
\vdef{bdt2012:N-EFFRATIO-PRODTNPMC-BSMM0:err}   {\ensuremath{{0.002 } } }
\vdef{bdt2012:N-EFF-PRODTNP-BSMM0:val}   {\ensuremath{{0.0032 } } }
\vdef{bdt2012:N-EFF-PRODTNP-BSMM0:err}   {\ensuremath{{4.7297 } } }
\vdef{bdt2012:N-EFF-PRODTNP-BSMM0:tot}   {\ensuremath{{4.7297 } } }
\vdef{bdt2012:N-EFF-PRODTNP-BSMM0:all}   {\ensuremath{{(3.19 \pm 4729.66)\times 10^{-3}} } }
\vdef{bdt2012:N-EFF-PRODTNPMC-BSMM0:val}   {\ensuremath{{0.0023 } } }
\vdef{bdt2012:N-EFF-PRODTNPMC-BSMM0:err}   {\ensuremath{{0.0000 } } }
\vdef{bdt2012:N-EFF-PRODTNPMC-BSMM0:tot}   {\ensuremath{{0.0000 } } }
\vdef{bdt2012:N-EFF-PRODTNPMC-BSMM0:all}   {\ensuremath{{(2.32 \pm 0.01)\times 10^{-3}} } }
\vdef{bdt2012:N-ACC-BSMM0:val}   {\ensuremath{{0.033 } } }
\vdef{bdt2012:N-ACC-BSMM0:err}   {\ensuremath{{0.000 } } }
\vdef{bdt2012:N-ACC-BSMM0:tot}   {\ensuremath{{0.001 } } }
\vdef{bdt2012:N-ACC-BSMM0:all}   {\ensuremath{{(3.33 \pm 0.12)\times 10^{-2}} } }
\vdef{bdt2012:N-EFF-MU-PID-BSMM0:val}   {\ensuremath{{0.680 } } }
\vdef{bdt2012:N-EFF-MU-PID-BSMM0:err}   {\ensuremath{{0.000 } } }
\vdef{bdt2012:N-EFF-MU-PID-BSMM0:tot}   {\ensuremath{{0.027 } } }
\vdef{bdt2012:N-EFF-MU-PID-BSMM0:all}   {\ensuremath{{(68.02 \pm 2.72)\times 10^{-2}} } }
\vdef{bdt2012:N-EFFRHO-MU-PID-BSMM0:val}   {\ensuremath{{0.599 } } }
\vdef{bdt2012:N-EFFRHO-MU-PID-BSMM0:err}   {\ensuremath{{0.000 } } }
\vdef{bdt2012:N-EFFRHO-MU-PID-BSMM0:tot}   {\ensuremath{{0.000 } } }
\vdef{bdt2012:N-EFFRHO-MU-PID-BSMM0:all}   {\ensuremath{{(59.93 \pm 0.00)\times 10^{-2}} } }
\vdef{bdt2012:N-EFF-MU-PIDMC-BSMM0:val}   {\ensuremath{{0.645 } } }
\vdef{bdt2012:N-EFF-MU-PIDMC-BSMM0:err}   {\ensuremath{{0.000 } } }
\vdef{bdt2012:N-EFF-MU-PIDMC-BSMM0:tot}   {\ensuremath{{0.026 } } }
\vdef{bdt2012:N-EFF-MU-PIDMC-BSMM0:all}   {\ensuremath{{(64.45 \pm 2.58)\times 10^{-2}} } }
\vdef{bdt2012:N-EFFRHO-MU-PIDMC-BSMM0:val}   {\ensuremath{{0.568 } } }
\vdef{bdt2012:N-EFFRHO-MU-PIDMC-BSMM0:err}   {\ensuremath{{0.000 } } }
\vdef{bdt2012:N-EFFRHO-MU-PIDMC-BSMM0:tot}   {\ensuremath{{0.024 } } }
\vdef{bdt2012:N-EFFRHO-MU-PIDMC-BSMM0:all}   {\ensuremath{{(56.79 \pm 2.40)\times 10^{-2}} } }
\vdef{bdt2012:N-EFF-MU-MC-BSMM0:val}   {\ensuremath{{0.568 } } }
\vdef{bdt2012:N-EFF-MU-MC-BSMM0:err}   {\ensuremath{{0.000 } } }
\vdef{bdt2012:N-EFF-MU-MC-BSMM0:tot}   {\ensuremath{{0.023 } } }
\vdef{bdt2012:N-EFF-MU-MC-BSMM0:all}   {\ensuremath{{(56.79 \pm 2.27)\times 10^{-2}} } }
\vdef{bdt2012:N-EFF-TRIG-PID-BSMM0:val}   {\ensuremath{{0.719 } } }
\vdef{bdt2012:N-EFF-TRIG-PID-BSMM0:err}   {\ensuremath{{0.000 } } }
\vdef{bdt2012:N-EFF-TRIG-PID-BSMM0:tot}   {\ensuremath{{0.022 } } }
\vdef{bdt2012:N-EFF-TRIG-PID-BSMM0:all}   {\ensuremath{{(71.95 \pm 2.16)\times 10^{-2}} } }
\vdef{bdt2012:N-EFFRHO-TRIG-PID-BSMM0:val}   {\ensuremath{{0.668 } } }
\vdef{bdt2012:N-EFFRHO-TRIG-PID-BSMM0:err}   {\ensuremath{{0.000 } } }
\vdef{bdt2012:N-EFFRHO-TRIG-PID-BSMM0:tot}   {\ensuremath{{0.020 } } }
\vdef{bdt2012:N-EFFRHO-TRIG-PID-BSMM0:all}   {\ensuremath{{(66.81 \pm 2.00)\times 10^{-2}} } }
\vdef{bdt2012:N-EFF-TRIG-PIDMC-BSMM0:val}   {\ensuremath{{0.676 } } }
\vdef{bdt2012:N-EFF-TRIG-PIDMC-BSMM0:err}   {\ensuremath{{0.000 } } }
\vdef{bdt2012:N-EFF-TRIG-PIDMC-BSMM0:tot}   {\ensuremath{{0.020 } } }
\vdef{bdt2012:N-EFF-TRIG-PIDMC-BSMM0:all}   {\ensuremath{{(67.61 \pm 2.03)\times 10^{-2}} } }
\vdef{bdt2012:N-EFFRHO-TRIG-PIDMC-BSMM0:val}   {\ensuremath{{0.628 } } }
\vdef{bdt2012:N-EFFRHO-TRIG-PIDMC-BSMM0:err}   {\ensuremath{{0.000 } } }
\vdef{bdt2012:N-EFFRHO-TRIG-PIDMC-BSMM0:tot}   {\ensuremath{{0.019 } } }
\vdef{bdt2012:N-EFFRHO-TRIG-PIDMC-BSMM0:all}   {\ensuremath{{(62.78 \pm 1.88)\times 10^{-2}} } }
\vdef{bdt2012:N-EFF-TRIG-MC-BSMM0:val}   {\ensuremath{{0.628 } } }
\vdef{bdt2012:N-EFF-TRIG-MC-BSMM0:err}   {\ensuremath{{0.000 } } }
\vdef{bdt2012:N-EFF-TRIG-MC-BSMM0:tot}   {\ensuremath{{0.019 } } }
\vdef{bdt2012:N-EFF-TRIG-MC-BSMM0:all}   {\ensuremath{{(62.78 \pm 1.88)\times 10^{-2}} } }
\vdef{bdt2012:N-EFF-CAND-BSMM0:val}   {\ensuremath{{1.000 } } }
\vdef{bdt2012:N-EFF-CAND-BSMM0:err}   {\ensuremath{{0.000 } } }
\vdef{bdt2012:N-EFF-CAND-BSMM0:tot}   {\ensuremath{{0.010 } } }
\vdef{bdt2012:N-EFF-CAND-BSMM0:all}   {\ensuremath{{(99.99 \pm 1.00)\times 10^{-2}} } }
\vdef{bdt2012:N-EFF-ANA-BSMM0:val}   {\ensuremath{{0.196 } } }
\vdef{bdt2012:N-EFF-ANA-BSMM0:err}   {\ensuremath{{0.000 } } }
\vdef{bdt2012:N-EFF-ANA-BSMM0:tot}   {\ensuremath{{0.006 } } }
\vdef{bdt2012:N-EFF-ANA-BSMM0:all}   {\ensuremath{{(19.59 \pm 0.59)\times 10^{-2}} } }
\vdef{bdt2012:N-EFF-TOT-BDMM0:val}   {\ensuremath{{0.0024 } } }
\vdef{bdt2012:N-EFF-TOT-BDMM0:err}   {\ensuremath{{0.0000 } } }
\vdef{bdt2012:N-EFF-TOT-BDMM0:tot}   {\ensuremath{{0.0002 } } }
\vdef{bdt2012:N-EFF-TOT-BDMM0:all}   {\ensuremath{{(0.24 \pm 0.02)} } }
\vdef{bdt2012:N-EFF-PRODMC-BDMM0:val}   {\ensuremath{{0.0024 } } }
\vdef{bdt2012:N-EFF-PRODMC-BDMM0:err}   {\ensuremath{{0.0000 } } }
\vdef{bdt2012:N-EFF-PRODMC-BDMM0:tot}   {\ensuremath{{0.0000 } } }
\vdef{bdt2012:N-EFF-PRODMC-BDMM0:all}   {\ensuremath{{(2.36 \pm 0.02)\times 10^{-3}} } }
\vdef{bdt2012:N-EFF-PRODTNP-BDMM0:val}   {\ensuremath{{0.0032 } } }
\vdef{bdt2012:N-EFF-PRODTNP-BDMM0:err}   {\ensuremath{{5.3791 } } }
\vdef{bdt2012:N-EFF-PRODTNP-BDMM0:tot}   {\ensuremath{{5.3791 } } }
\vdef{bdt2012:N-EFF-PRODTNP-BDMM0:all}   {\ensuremath{{(3.25 \pm 5379.13)\times 10^{-3}} } }
\vdef{bdt2012:N-EFF-PRODTNPMC-BDMM0:val}   {\ensuremath{{0.00236 } } }
\vdef{bdt2012:N-EFF-PRODTNPMC-BDMM0:err}   {\ensuremath{{0.000018 } } }
\vdef{bdt2012:N-EFF-PRODTNPMC-BDMM0:tot}   {\ensuremath{{0.00002 } } }
\vdef{bdt2012:N-EFF-PRODTNPMC-BDMM0:all}   {\ensuremath{{(2.36 \pm 0.02)\times 10^{-3}} } }
\vdef{bdt2012:N-ACC-BDMM0:val}   {\ensuremath{{0.032 } } }
\vdef{bdt2012:N-ACC-BDMM0:err}   {\ensuremath{{0.000 } } }
\vdef{bdt2012:N-ACC-BDMM0:tot}   {\ensuremath{{0.001 } } }
\vdef{bdt2012:N-ACC-BDMM0:all}   {\ensuremath{{(3.24 \pm 0.12)\times 10^{-2}} } }
\vdef{bdt2012:N-EFF-MU-PID-BDMM0:val}   {\ensuremath{{0.680 } } }
\vdef{bdt2012:N-EFF-MU-PID-BDMM0:err}   {\ensuremath{{0.000 } } }
\vdef{bdt2012:N-EFF-MU-PID-BDMM0:tot}   {\ensuremath{{0.027 } } }
\vdef{bdt2012:N-EFF-MU-PID-BDMM0:all}   {\ensuremath{{(67.99 \pm 2.72)\times 10^{-2}} } }
\vdef{bdt2012:N-EFFRHO-MU-PID-BDMM0:val}   {\ensuremath{{0.599 } } }
\vdef{bdt2012:N-EFFRHO-MU-PID-BDMM0:err}   {\ensuremath{{0.000 } } }
\vdef{bdt2012:N-EFFRHO-MU-PID-BDMM0:tot}   {\ensuremath{{0.000 } } }
\vdef{bdt2012:N-EFFRHO-MU-PID-BDMM0:all}   {\ensuremath{{(59.95 \pm 0.00)\times 10^{-2}} } }
\vdef{bdt2012:N-EFF-MU-PIDMC-BDMM0:val}   {\ensuremath{{0.644 } } }
\vdef{bdt2012:N-EFF-MU-PIDMC-BDMM0:err}   {\ensuremath{{0.000 } } }
\vdef{bdt2012:N-EFF-MU-PIDMC-BDMM0:tot}   {\ensuremath{{0.026 } } }
\vdef{bdt2012:N-EFF-MU-PIDMC-BDMM0:all}   {\ensuremath{{(64.42 \pm 2.58)\times 10^{-2}} } }
\vdef{bdt2012:N-EFFRHO-MU-PIDMC-BDMM0:val}   {\ensuremath{{0.568 } } }
\vdef{bdt2012:N-EFFRHO-MU-PIDMC-BDMM0:err}   {\ensuremath{{0.000 } } }
\vdef{bdt2012:N-EFFRHO-MU-PIDMC-BDMM0:tot}   {\ensuremath{{0.024 } } }
\vdef{bdt2012:N-EFFRHO-MU-PIDMC-BDMM0:all}   {\ensuremath{{(56.80 \pm 2.40)\times 10^{-2}} } }
\vdef{bdt2012:N-EFF-MU-MC-BDMM0:val}   {\ensuremath{{0.568 } } }
\vdef{bdt2012:N-EFF-MU-MC-BDMM0:err}   {\ensuremath{{0.000 } } }
\vdef{bdt2012:N-EFF-MU-MC-BDMM0:tot}   {\ensuremath{{0.023 } } }
\vdef{bdt2012:N-EFF-MU-MC-BDMM0:all}   {\ensuremath{{(56.80 \pm 2.27)\times 10^{-2}} } }
\vdef{bdt2012:N-EFF-TRIG-PID-BDMM0:val}   {\ensuremath{{0.719 } } }
\vdef{bdt2012:N-EFF-TRIG-PID-BDMM0:err}   {\ensuremath{{0.000 } } }
\vdef{bdt2012:N-EFF-TRIG-PID-BDMM0:tot}   {\ensuremath{{0.022 } } }
\vdef{bdt2012:N-EFF-TRIG-PID-BDMM0:all}   {\ensuremath{{(71.92 \pm 2.16)\times 10^{-2}} } }
\vdef{bdt2012:N-EFFRHO-TRIG-PID-BDMM0:val}   {\ensuremath{{0.665 } } }
\vdef{bdt2012:N-EFFRHO-TRIG-PID-BDMM0:err}   {\ensuremath{{0.000 } } }
\vdef{bdt2012:N-EFFRHO-TRIG-PID-BDMM0:tot}   {\ensuremath{{0.020 } } }
\vdef{bdt2012:N-EFFRHO-TRIG-PID-BDMM0:all}   {\ensuremath{{(66.54 \pm 2.00)\times 10^{-2}} } }
\vdef{bdt2012:N-EFF-TRIG-PIDMC-BDMM0:val}   {\ensuremath{{0.675 } } }
\vdef{bdt2012:N-EFF-TRIG-PIDMC-BDMM0:err}   {\ensuremath{{0.000 } } }
\vdef{bdt2012:N-EFF-TRIG-PIDMC-BDMM0:tot}   {\ensuremath{{0.020 } } }
\vdef{bdt2012:N-EFF-TRIG-PIDMC-BDMM0:all}   {\ensuremath{{(67.55 \pm 2.03)\times 10^{-2}} } }
\vdef{bdt2012:N-EFFRHO-TRIG-PIDMC-BDMM0:val}   {\ensuremath{{0.625 } } }
\vdef{bdt2012:N-EFFRHO-TRIG-PIDMC-BDMM0:err}   {\ensuremath{{0.000 } } }
\vdef{bdt2012:N-EFFRHO-TRIG-PIDMC-BDMM0:tot}   {\ensuremath{{0.019 } } }
\vdef{bdt2012:N-EFFRHO-TRIG-PIDMC-BDMM0:all}   {\ensuremath{{(62.50 \pm 1.88)\times 10^{-2}} } }
\vdef{bdt2012:N-EFF-TRIG-MC-BDMM0:val}   {\ensuremath{{0.625 } } }
\vdef{bdt2012:N-EFF-TRIG-MC-BDMM0:err}   {\ensuremath{{0.000 } } }
\vdef{bdt2012:N-EFF-TRIG-MC-BDMM0:tot}   {\ensuremath{{0.019 } } }
\vdef{bdt2012:N-EFF-TRIG-MC-BDMM0:all}   {\ensuremath{{(62.50 \pm 1.88)\times 10^{-2}} } }
\vdef{bdt2012:N-EFF-CAND-BDMM0:val}   {\ensuremath{{1.000 } } }
\vdef{bdt2012:N-EFF-CAND-BDMM0:err}   {\ensuremath{{0.000 } } }
\vdef{bdt2012:N-EFF-CAND-BDMM0:tot}   {\ensuremath{{0.010 } } }
\vdef{bdt2012:N-EFF-CAND-BDMM0:all}   {\ensuremath{{(100.00 \pm 1.00)\times 10^{-2}} } }
\vdef{bdt2012:N-EFF-ANA-BDMM0:val}   {\ensuremath{{0.205 } } }
\vdef{bdt2012:N-EFF-ANA-BDMM0:err}   {\ensuremath{{0.000 } } }
\vdef{bdt2012:N-EFF-ANA-BDMM0:tot}   {\ensuremath{{0.006 } } }
\vdef{bdt2012:N-EFF-ANA-BDMM0:all}   {\ensuremath{{(20.53 \pm 0.62)\times 10^{-2}} } }
\vdef{bdt2012:N-EXP-OBS-BS0:val}   {\ensuremath{{17.63 } } }
\vdef{bdt2012:N-EXP-OBS-BS0:err}   {\ensuremath{{ 3.32 } } }
\vdef{bdt2012:N-EXP-OBS-BD0:val}   {\ensuremath{{ 7.86 } } }
\vdef{bdt2012:N-EXP-OBS-BD0:err}   {\ensuremath{{ 2.60 } } }
\vdef{bdt2012:N-OBS-BSMM0:val}   {\ensuremath{{16 } } }
\vdef{bdt2012:N-OBS-BDMM0:val}   {\ensuremath{{11 } } }
\vdef{bdt2012:N-OFFLO-RARE0:val}   {\ensuremath{{ 0.22 } } }
\vdef{bdt2012:N-OFFLO-RARE0:err}   {\ensuremath{{ 0.13 } } }
\vdef{bdt2012:N-OFFHI-RARE0:val}   {\ensuremath{{ 0.02 } } }
\vdef{bdt2012:N-OFFHI-RARE0:err}   {\ensuremath{{ 0.01 } } }
\vdef{bdt2012:N-PEAK-BKG-BS0:val}   {\ensuremath{{ 0.24 } } }
\vdef{bdt2012:N-PEAK-BKG-BS0:err}   {\ensuremath{{ 0.18 } } }
\vdef{bdt2012:N-PEAK-BKG-BD0:val}   {\ensuremath{{ 0.80 } } }
\vdef{bdt2012:N-PEAK-BKG-BD0:err}   {\ensuremath{{ 0.59 } } }
\vdef{bdt2012:N-TAU-BS0:val}   {\ensuremath{{ 0.17 } } }
\vdef{bdt2012:N-TAU-BS0:err}   {\ensuremath{{ 0.01 } } }
\vdef{bdt2012:N-TAU-BD0:val}   {\ensuremath{{ 0.15 } } }
\vdef{bdt2012:N-TAU-BD0:err}   {\ensuremath{{ 0.01 } } }
\vdef{bdt2012:N-OBS-OFFHI0:val}   {\ensuremath{{12 } } }
\vdef{bdt2012:N-OBS-OFFLO0:val}   {\ensuremath{{22 } } }
\vdef{bdt2012:N-EXP-SoverB0:val}   {\ensuremath{{ 1.84 } } }
\vdef{bdt2012:N-EXP-SoverSplusB0:val}   {\ensuremath{{ 2.64 } } }
\vdef{bdt2012:N-EXP2-SIG-BSMM1:val}   {\ensuremath{{ 3.33 } } }
\vdef{bdt2012:N-EXP2-SIG-BSMM1:err}   {\ensuremath{{ 0.50 } } }
\vdef{bdt2012:N-EXP2-SIG-BDMM1:val}   {\ensuremath{{0.301 } } }
\vdef{bdt2012:N-EXP2-SIG-BDMM1:err}   {\ensuremath{{0.030 } } }
\vdef{bdt2012:N-OBS-BKG1:val}   {\ensuremath{{7 } } }
\vdef{bdt2012:N-EXP-BSMM1:val}   {\ensuremath{{ 1.26 } } }
\vdef{bdt2012:N-EXP-BSMM1:err}   {\ensuremath{{ 0.63 } } }
\vdef{bdt2012:N-EXP-BDMM1:val}   {\ensuremath{{ 1.03 } } }
\vdef{bdt2012:N-EXP-BDMM1:err}   {\ensuremath{{ 0.51 } } }
\vdef{bdt2012:N-LOW-BD1:val}   {\ensuremath{{5.200 } } }
\vdef{bdt2012:N-HIGH-BD1:val}   {\ensuremath{{5.300 } } }
\vdef{bdt2012:N-LOW-BS1:val}   {\ensuremath{{5.300 } } }
\vdef{bdt2012:N-HIGH-BS1:val}   {\ensuremath{{5.450 } } }
\vdef{bdt2012:N-PSS1:val}   {\ensuremath{{0.754 } } }
\vdef{bdt2012:N-PSS1:err}   {\ensuremath{{0.001 } } }
\vdef{bdt2012:N-PSS1:tot}   {\ensuremath{{0.038 } } }
\vdef{bdt2012:N-PSD1:val}   {\ensuremath{{0.327 } } }
\vdef{bdt2012:N-PSD1:err}   {\ensuremath{{0.001 } } }
\vdef{bdt2012:N-PSD1:tot}   {\ensuremath{{0.016 } } }
\vdef{bdt2012:N-PDS1:val}   {\ensuremath{{0.141 } } }
\vdef{bdt2012:N-PDS1:err}   {\ensuremath{{0.001 } } }
\vdef{bdt2012:N-PDS1:tot}   {\ensuremath{{0.007 } } }
\vdef{bdt2012:N-PDD1:val}   {\ensuremath{{0.550 } } }
\vdef{bdt2012:N-PDD1:err}   {\ensuremath{{0.001 } } }
\vdef{bdt2012:N-PDD1:tot}   {\ensuremath{{0.028 } } }
\vdef{bdt2012:N-EFF-TOT-BSMM1:val}   {\ensuremath{{0.0009 } } }
\vdef{bdt2012:N-EFF-TOT-BSMM1:err}   {\ensuremath{{0.0000 } } }
\vdef{bdt2012:N-EFF-TOT-BSMM1:tot}   {\ensuremath{{0.0001 } } }
\vdef{bdt2012:N-EFF-TOT-BSMM1:all}   {\ensuremath{{(0.09 \pm 0.01)} } }
\vdef{bdt2012:N-EFF-PRODMC-BSMM1:val}   {\ensuremath{{0.0010 } } }
\vdef{bdt2012:N-EFF-PRODMC-BSMM1:err}   {\ensuremath{{0.0000 } } }
\vdef{bdt2012:N-EFF-PRODMC-BSMM1:tot}   {\ensuremath{{0.0000 } } }
\vdef{bdt2012:N-EFF-PRODMC-BSMM1:all}   {\ensuremath{{(0.96 \pm 0.00)\times 10^{-3}} } }
\vdef{bdt2012:N-EFFRATIO-TOT-BSMM1:val}   {\ensuremath{{0.223 } } }
\vdef{bdt2012:N-EFFRATIO-TOT-BSMM1:err}   {\ensuremath{{0.001 } } }
\vdef{bdt2012:N-EFFRATIO-PRODMC-BSMM1:val}   {\ensuremath{{0.219 } } }
\vdef{bdt2012:N-EFFRATIO-PRODMC-BSMM1:err}   {\ensuremath{{0.002 } } }
\vdef{bdt2012:N-EFFRATIO-PRODTNP-BSMM1:val}   {\ensuremath{{0.341 } } }
\vdef{bdt2012:N-EFFRATIO-PRODTNP-BSMM1:err}   {\ensuremath{{375.909 } } }
\vdef{bdt2012:N-EFFRATIO-PRODTNPMC-BSMM1:val}   {\ensuremath{{0.219 } } }
\vdef{bdt2012:N-EFFRATIO-PRODTNPMC-BSMM1:err}   {\ensuremath{{0.001 } } }
\vdef{bdt2012:N-EFF-PRODTNP-BSMM1:val}   {\ensuremath{{0.0010 } } }
\vdef{bdt2012:N-EFF-PRODTNP-BSMM1:err}   {\ensuremath{{0.8691 } } }
\vdef{bdt2012:N-EFF-PRODTNP-BSMM1:tot}   {\ensuremath{{0.8691 } } }
\vdef{bdt2012:N-EFF-PRODTNP-BSMM1:all}   {\ensuremath{{(1.01 \pm 869.11)\times 10^{-3}} } }
\vdef{bdt2012:N-EFF-PRODTNPMC-BSMM1:val}   {\ensuremath{{0.0010 } } }
\vdef{bdt2012:N-EFF-PRODTNPMC-BSMM1:err}   {\ensuremath{{0.0000 } } }
\vdef{bdt2012:N-EFF-PRODTNPMC-BSMM1:tot}   {\ensuremath{{0.0000 } } }
\vdef{bdt2012:N-EFF-PRODTNPMC-BSMM1:all}   {\ensuremath{{(0.96 \pm 0.00)\times 10^{-3}} } }
\vdef{bdt2012:N-ACC-BSMM1:val}   {\ensuremath{{0.024 } } }
\vdef{bdt2012:N-ACC-BSMM1:err}   {\ensuremath{{0.000 } } }
\vdef{bdt2012:N-ACC-BSMM1:tot}   {\ensuremath{{0.001 } } }
\vdef{bdt2012:N-ACC-BSMM1:all}   {\ensuremath{{(2.38 \pm 0.12)\times 10^{-2}} } }
\vdef{bdt2012:N-EFF-MU-PID-BSMM1:val}   {\ensuremath{{0.555 } } }
\vdef{bdt2012:N-EFF-MU-PID-BSMM1:err}   {\ensuremath{{0.000 } } }
\vdef{bdt2012:N-EFF-MU-PID-BSMM1:tot}   {\ensuremath{{0.044 } } }
\vdef{bdt2012:N-EFF-MU-PID-BSMM1:all}   {\ensuremath{{(55.54 \pm 4.44)\times 10^{-2}} } }
\vdef{bdt2012:N-EFFRHO-MU-PID-BSMM1:val}   {\ensuremath{{0.646 } } }
\vdef{bdt2012:N-EFFRHO-MU-PID-BSMM1:err}   {\ensuremath{{0.000 } } }
\vdef{bdt2012:N-EFFRHO-MU-PID-BSMM1:tot}   {\ensuremath{{0.000 } } }
\vdef{bdt2012:N-EFFRHO-MU-PID-BSMM1:all}   {\ensuremath{{(64.64 \pm 0.00)\times 10^{-2}} } }
\vdef{bdt2012:N-EFF-MU-PIDMC-BSMM1:val}   {\ensuremath{{0.557 } } }
\vdef{bdt2012:N-EFF-MU-PIDMC-BSMM1:err}   {\ensuremath{{0.000 } } }
\vdef{bdt2012:N-EFF-MU-PIDMC-BSMM1:tot}   {\ensuremath{{0.045 } } }
\vdef{bdt2012:N-EFF-MU-PIDMC-BSMM1:all}   {\ensuremath{{(55.74 \pm 4.46)\times 10^{-2}} } }
\vdef{bdt2012:N-EFFRHO-MU-PIDMC-BSMM1:val}   {\ensuremath{{0.649 } } }
\vdef{bdt2012:N-EFFRHO-MU-PIDMC-BSMM1:err}   {\ensuremath{{0.000 } } }
\vdef{bdt2012:N-EFFRHO-MU-PIDMC-BSMM1:tot}   {\ensuremath{{0.052 } } }
\vdef{bdt2012:N-EFFRHO-MU-PIDMC-BSMM1:all}   {\ensuremath{{(64.87 \pm 5.17)\times 10^{-2}} } }
\vdef{bdt2012:N-EFF-MU-MC-BSMM1:val}   {\ensuremath{{0.649 } } }
\vdef{bdt2012:N-EFF-MU-MC-BSMM1:err}   {\ensuremath{{0.001 } } }
\vdef{bdt2012:N-EFF-MU-MC-BSMM1:tot}   {\ensuremath{{0.052 } } }
\vdef{bdt2012:N-EFF-MU-MC-BSMM1:all}   {\ensuremath{{(64.87 \pm 5.19)\times 10^{-2}} } }
\vdef{bdt2012:N-EFF-TRIG-PID-BSMM1:val}   {\ensuremath{{0.743 } } }
\vdef{bdt2012:N-EFF-TRIG-PID-BSMM1:err}   {\ensuremath{{0.000 } } }
\vdef{bdt2012:N-EFF-TRIG-PID-BSMM1:tot}   {\ensuremath{{0.045 } } }
\vdef{bdt2012:N-EFF-TRIG-PID-BSMM1:all}   {\ensuremath{{(74.32 \pm 4.46)\times 10^{-2}} } }
\vdef{bdt2012:N-EFFRHO-TRIG-PID-BSMM1:val}   {\ensuremath{{0.622 } } }
\vdef{bdt2012:N-EFFRHO-TRIG-PID-BSMM1:err}   {\ensuremath{{0.000 } } }
\vdef{bdt2012:N-EFFRHO-TRIG-PID-BSMM1:tot}   {\ensuremath{{0.037 } } }
\vdef{bdt2012:N-EFFRHO-TRIG-PID-BSMM1:all}   {\ensuremath{{(62.20 \pm 3.73)\times 10^{-2}} } }
\vdef{bdt2012:N-EFF-TRIG-PIDMC-BSMM1:val}   {\ensuremath{{0.726 } } }
\vdef{bdt2012:N-EFF-TRIG-PIDMC-BSMM1:err}   {\ensuremath{{0.000 } } }
\vdef{bdt2012:N-EFF-TRIG-PIDMC-BSMM1:tot}   {\ensuremath{{0.044 } } }
\vdef{bdt2012:N-EFF-TRIG-PIDMC-BSMM1:all}   {\ensuremath{{(72.60 \pm 4.36)\times 10^{-2}} } }
\vdef{bdt2012:N-EFFRHO-TRIG-PIDMC-BSMM1:val}   {\ensuremath{{0.608 } } }
\vdef{bdt2012:N-EFFRHO-TRIG-PIDMC-BSMM1:err}   {\ensuremath{{0.000 } } }
\vdef{bdt2012:N-EFFRHO-TRIG-PIDMC-BSMM1:tot}   {\ensuremath{{0.036 } } }
\vdef{bdt2012:N-EFFRHO-TRIG-PIDMC-BSMM1:all}   {\ensuremath{{(60.77 \pm 3.65)\times 10^{-2}} } }
\vdef{bdt2012:N-EFF-TRIG-MC-BSMM1:val}   {\ensuremath{{0.608 } } }
\vdef{bdt2012:N-EFF-TRIG-MC-BSMM1:err}   {\ensuremath{{0.001 } } }
\vdef{bdt2012:N-EFF-TRIG-MC-BSMM1:tot}   {\ensuremath{{0.036 } } }
\vdef{bdt2012:N-EFF-TRIG-MC-BSMM1:all}   {\ensuremath{{(60.77 \pm 3.65)\times 10^{-2}} } }
\vdef{bdt2012:N-EFF-CAND-BSMM1:val}   {\ensuremath{{1.000 } } }
\vdef{bdt2012:N-EFF-CAND-BSMM1:err}   {\ensuremath{{0.000 } } }
\vdef{bdt2012:N-EFF-CAND-BSMM1:tot}   {\ensuremath{{0.010 } } }
\vdef{bdt2012:N-EFF-CAND-BSMM1:all}   {\ensuremath{{(99.99 \pm 1.00)\times 10^{-2}} } }
\vdef{bdt2012:N-EFF-ANA-BSMM1:val}   {\ensuremath{{0.103 } } }
\vdef{bdt2012:N-EFF-ANA-BSMM1:err}   {\ensuremath{{0.000 } } }
\vdef{bdt2012:N-EFF-ANA-BSMM1:tot}   {\ensuremath{{0.003 } } }
\vdef{bdt2012:N-EFF-ANA-BSMM1:all}   {\ensuremath{{(10.28 \pm 0.31)\times 10^{-2}} } }
\vdef{bdt2012:N-EFF-TOT-BDMM1:val}   {\ensuremath{{0.0010 } } }
\vdef{bdt2012:N-EFF-TOT-BDMM1:err}   {\ensuremath{{0.0000 } } }
\vdef{bdt2012:N-EFF-TOT-BDMM1:tot}   {\ensuremath{{0.0001 } } }
\vdef{bdt2012:N-EFF-TOT-BDMM1:all}   {\ensuremath{{(0.10 \pm 0.01)} } }
\vdef{bdt2012:N-EFF-PRODMC-BDMM1:val}   {\ensuremath{{0.0010 } } }
\vdef{bdt2012:N-EFF-PRODMC-BDMM1:err}   {\ensuremath{{0.0000 } } }
\vdef{bdt2012:N-EFF-PRODMC-BDMM1:tot}   {\ensuremath{{0.0000 } } }
\vdef{bdt2012:N-EFF-PRODMC-BDMM1:all}   {\ensuremath{{(0.97 \pm 0.01)\times 10^{-3}} } }
\vdef{bdt2012:N-EFF-PRODTNP-BDMM1:val}   {\ensuremath{{0.0010 } } }
\vdef{bdt2012:N-EFF-PRODTNP-BDMM1:err}   {\ensuremath{{0.9751 } } }
\vdef{bdt2012:N-EFF-PRODTNP-BDMM1:tot}   {\ensuremath{{0.9751 } } }
\vdef{bdt2012:N-EFF-PRODTNP-BDMM1:all}   {\ensuremath{{(1.02 \pm 975.08)\times 10^{-3}} } }
\vdef{bdt2012:N-EFF-PRODTNPMC-BDMM1:val}   {\ensuremath{{0.00097 } } }
\vdef{bdt2012:N-EFF-PRODTNPMC-BDMM1:err}   {\ensuremath{{0.000009 } } }
\vdef{bdt2012:N-EFF-PRODTNPMC-BDMM1:tot}   {\ensuremath{{0.00001 } } }
\vdef{bdt2012:N-EFF-PRODTNPMC-BDMM1:all}   {\ensuremath{{(0.97 \pm 0.01)\times 10^{-3}} } }
\vdef{bdt2012:N-ACC-BDMM1:val}   {\ensuremath{{0.023 } } }
\vdef{bdt2012:N-ACC-BDMM1:err}   {\ensuremath{{0.000 } } }
\vdef{bdt2012:N-ACC-BDMM1:tot}   {\ensuremath{{0.001 } } }
\vdef{bdt2012:N-ACC-BDMM1:all}   {\ensuremath{{(2.29 \pm 0.12)\times 10^{-2}} } }
\vdef{bdt2012:N-EFF-MU-PID-BDMM1:val}   {\ensuremath{{0.557 } } }
\vdef{bdt2012:N-EFF-MU-PID-BDMM1:err}   {\ensuremath{{0.000 } } }
\vdef{bdt2012:N-EFF-MU-PID-BDMM1:tot}   {\ensuremath{{0.045 } } }
\vdef{bdt2012:N-EFF-MU-PID-BDMM1:all}   {\ensuremath{{(55.66 \pm 4.45)\times 10^{-2}} } }
\vdef{bdt2012:N-EFFRHO-MU-PID-BDMM1:val}   {\ensuremath{{0.644 } } }
\vdef{bdt2012:N-EFFRHO-MU-PID-BDMM1:err}   {\ensuremath{{0.000 } } }
\vdef{bdt2012:N-EFFRHO-MU-PID-BDMM1:tot}   {\ensuremath{{0.000 } } }
\vdef{bdt2012:N-EFFRHO-MU-PID-BDMM1:all}   {\ensuremath{{(64.44 \pm 0.00)\times 10^{-2}} } }
\vdef{bdt2012:N-EFF-MU-PIDMC-BDMM1:val}   {\ensuremath{{0.558 } } }
\vdef{bdt2012:N-EFF-MU-PIDMC-BDMM1:err}   {\ensuremath{{0.000 } } }
\vdef{bdt2012:N-EFF-MU-PIDMC-BDMM1:tot}   {\ensuremath{{0.045 } } }
\vdef{bdt2012:N-EFF-MU-PIDMC-BDMM1:all}   {\ensuremath{{(55.79 \pm 4.46)\times 10^{-2}} } }
\vdef{bdt2012:N-EFFRHO-MU-PIDMC-BDMM1:val}   {\ensuremath{{0.646 } } }
\vdef{bdt2012:N-EFFRHO-MU-PIDMC-BDMM1:err}   {\ensuremath{{0.000 } } }
\vdef{bdt2012:N-EFFRHO-MU-PIDMC-BDMM1:tot}   {\ensuremath{{0.052 } } }
\vdef{bdt2012:N-EFFRHO-MU-PIDMC-BDMM1:all}   {\ensuremath{{(64.59 \pm 5.16)\times 10^{-2}} } }
\vdef{bdt2012:N-EFF-MU-MC-BDMM1:val}   {\ensuremath{{0.646 } } }
\vdef{bdt2012:N-EFF-MU-MC-BDMM1:err}   {\ensuremath{{0.001 } } }
\vdef{bdt2012:N-EFF-MU-MC-BDMM1:tot}   {\ensuremath{{0.052 } } }
\vdef{bdt2012:N-EFF-MU-MC-BDMM1:all}   {\ensuremath{{(64.59 \pm 5.17)\times 10^{-2}} } }
\vdef{bdt2012:N-EFF-TRIG-PID-BDMM1:val}   {\ensuremath{{0.743 } } }
\vdef{bdt2012:N-EFF-TRIG-PID-BDMM1:err}   {\ensuremath{{0.000 } } }
\vdef{bdt2012:N-EFF-TRIG-PID-BDMM1:tot}   {\ensuremath{{0.045 } } }
\vdef{bdt2012:N-EFF-TRIG-PID-BDMM1:all}   {\ensuremath{{(74.28 \pm 4.46)\times 10^{-2}} } }
\vdef{bdt2012:N-EFFRHO-TRIG-PID-BDMM1:val}   {\ensuremath{{0.620 } } }
\vdef{bdt2012:N-EFFRHO-TRIG-PID-BDMM1:err}   {\ensuremath{{0.000 } } }
\vdef{bdt2012:N-EFFRHO-TRIG-PID-BDMM1:tot}   {\ensuremath{{0.037 } } }
\vdef{bdt2012:N-EFFRHO-TRIG-PID-BDMM1:all}   {\ensuremath{{(62.02 \pm 3.72)\times 10^{-2}} } }
\vdef{bdt2012:N-EFF-TRIG-PIDMC-BDMM1:val}   {\ensuremath{{0.725 } } }
\vdef{bdt2012:N-EFF-TRIG-PIDMC-BDMM1:err}   {\ensuremath{{0.000 } } }
\vdef{bdt2012:N-EFF-TRIG-PIDMC-BDMM1:tot}   {\ensuremath{{0.044 } } }
\vdef{bdt2012:N-EFF-TRIG-PIDMC-BDMM1:all}   {\ensuremath{{(72.52 \pm 4.35)\times 10^{-2}} } }
\vdef{bdt2012:N-EFFRHO-TRIG-PIDMC-BDMM1:val}   {\ensuremath{{0.605 } } }
\vdef{bdt2012:N-EFFRHO-TRIG-PIDMC-BDMM1:err}   {\ensuremath{{0.000 } } }
\vdef{bdt2012:N-EFFRHO-TRIG-PIDMC-BDMM1:tot}   {\ensuremath{{0.036 } } }
\vdef{bdt2012:N-EFFRHO-TRIG-PIDMC-BDMM1:all}   {\ensuremath{{(60.55 \pm 3.63)\times 10^{-2}} } }
\vdef{bdt2012:N-EFF-TRIG-MC-BDMM1:val}   {\ensuremath{{0.605 } } }
\vdef{bdt2012:N-EFF-TRIG-MC-BDMM1:err}   {\ensuremath{{0.001 } } }
\vdef{bdt2012:N-EFF-TRIG-MC-BDMM1:tot}   {\ensuremath{{0.036 } } }
\vdef{bdt2012:N-EFF-TRIG-MC-BDMM1:all}   {\ensuremath{{(60.55 \pm 3.63)\times 10^{-2}} } }
\vdef{bdt2012:N-EFF-CAND-BDMM1:val}   {\ensuremath{{1.000 } } }
\vdef{bdt2012:N-EFF-CAND-BDMM1:err}   {\ensuremath{{0.000 } } }
\vdef{bdt2012:N-EFF-CAND-BDMM1:tot}   {\ensuremath{{0.010 } } }
\vdef{bdt2012:N-EFF-CAND-BDMM1:all}   {\ensuremath{{(99.99 \pm 1.00)\times 10^{-2}} } }
\vdef{bdt2012:N-EFF-ANA-BDMM1:val}   {\ensuremath{{0.108 } } }
\vdef{bdt2012:N-EFF-ANA-BDMM1:err}   {\ensuremath{{0.000 } } }
\vdef{bdt2012:N-EFF-ANA-BDMM1:tot}   {\ensuremath{{0.003 } } }
\vdef{bdt2012:N-EFF-ANA-BDMM1:all}   {\ensuremath{{(10.79 \pm 0.32)\times 10^{-2}} } }
\vdef{bdt2012:N-EXP-OBS-BS1:val}   {\ensuremath{{ 5.31 } } }
\vdef{bdt2012:N-EXP-OBS-BS1:err}   {\ensuremath{{ 0.80 } } }
\vdef{bdt2012:N-EXP-OBS-BD1:val}   {\ensuremath{{ 1.85 } } }
\vdef{bdt2012:N-EXP-OBS-BD1:err}   {\ensuremath{{ 0.52 } } }
\vdef{bdt2012:N-OBS-BSMM1:val}   {\ensuremath{{4 } } }
\vdef{bdt2012:N-OBS-BDMM1:val}   {\ensuremath{{3 } } }
\vdef{bdt2012:N-OFFLO-RARE1:val}   {\ensuremath{{ 0.11 } } }
\vdef{bdt2012:N-OFFLO-RARE1:err}   {\ensuremath{{ 0.06 } } }
\vdef{bdt2012:N-OFFHI-RARE1:val}   {\ensuremath{{ 0.01 } } }
\vdef{bdt2012:N-OFFHI-RARE1:err}   {\ensuremath{{ 0.00 } } }
\vdef{bdt2012:N-PEAK-BKG-BS1:val}   {\ensuremath{{ 0.09 } } }
\vdef{bdt2012:N-PEAK-BKG-BS1:err}   {\ensuremath{{ 0.07 } } }
\vdef{bdt2012:N-PEAK-BKG-BD1:val}   {\ensuremath{{ 0.19 } } }
\vdef{bdt2012:N-PEAK-BKG-BD1:err}   {\ensuremath{{ 0.14 } } }
\vdef{bdt2012:N-TAU-BS1:val}   {\ensuremath{{ 0.18 } } }
\vdef{bdt2012:N-TAU-BS1:err}   {\ensuremath{{ 0.01 } } }
\vdef{bdt2012:N-TAU-BD1:val}   {\ensuremath{{ 0.15 } } }
\vdef{bdt2012:N-TAU-BD1:err}   {\ensuremath{{ 0.01 } } }
\vdef{bdt2012:N-OBS-OFFHI1:val}   {\ensuremath{{3 } } }
\vdef{bdt2012:N-OBS-OFFLO1:val}   {\ensuremath{{4 } } }
\vdef{bdt2012:N-EXP-SoverB1:val}   {\ensuremath{{ 2.64 } } }
\vdef{bdt2012:N-EXP-SoverSplusB1:val}   {\ensuremath{{ 1.56 } } }
\vdef{bdt2012:SgBd0:val}  {\ensuremath{{0.847 } } }
\vdef{bdt2012:SgBd0:e1}   {\ensuremath{{0.920 } } }
\vdef{bdt2012:SgBd0:e2}   {\ensuremath{{0.127 } } }
\vdef{bdt2012:SgBs0:val}  {\ensuremath{{10.717 } } }
\vdef{bdt2012:SgBs0:e1}   {\ensuremath{{3.274 } } }
\vdef{bdt2012:SgBs0:e2}   {\ensuremath{{1.608 } } }
\vdef{bdt2012:SgLo0:val}  {\ensuremath{{0.245 } } }
\vdef{bdt2012:SgLo0:e1}   {\ensuremath{{0.495 } } }
\vdef{bdt2012:SgLo0:e2}   {\ensuremath{{0.037 } } }
\vdef{bdt2012:SgHi0:val}  {\ensuremath{{0.369 } } }
\vdef{bdt2012:SgHi0:e1}   {\ensuremath{{0.607 } } }
\vdef{bdt2012:SgHi0:e2}   {\ensuremath{{0.055 } } }
\vdef{bdt2012:BdBd0:val}  {\ensuremath{{1.001 } } }
\vdef{bdt2012:BdBd0:e1}   {\ensuremath{{1.000 } } }
\vdef{bdt2012:BdBd0:e2}   {\ensuremath{{0.100 } } }
\vdef{bdt2012:BdBs0:val}  {\ensuremath{{0.424 } } }
\vdef{bdt2012:BdBs0:e1}   {\ensuremath{{0.651 } } }
\vdef{bdt2012:BdBs0:e2}   {\ensuremath{{0.042 } } }
\vdef{bdt2012:BdLo0:val}  {\ensuremath{{0.099 } } }
\vdef{bdt2012:BdLo0:e1}   {\ensuremath{{0.314 } } }
\vdef{bdt2012:BdLo0:e2}   {\ensuremath{{0.010 } } }
\vdef{bdt2012:BdHi0:val}  {\ensuremath{{0.001 } } }
\vdef{bdt2012:BdHi0:e1}   {\ensuremath{{0.028 } } }
\vdef{bdt2012:BdHi0:e2}   {\ensuremath{{0.000 } } }
\vdef{bdt2012:BgPeakLo0:val}   {\ensuremath{{0.223 } } }
\vdef{bdt2012:BgPeakLo0:e1}   {\ensuremath{{0.011 } } }
\vdef{bdt2012:BgPeakLo0:e2}   {\ensuremath{{0.159 } } }
\vdef{bdt2012:BgPeakBd0:val}   {\ensuremath{{0.804 } } }
\vdef{bdt2012:BgPeakBd0:e1}   {\ensuremath{{0.040 } } }
\vdef{bdt2012:BgPeakBd0:e2}   {\ensuremath{{0.587 } } }
\vdef{bdt2012:BgPeakBs0:val}   {\ensuremath{{0.242 } } }
\vdef{bdt2012:BgPeakBs0:e1}   {\ensuremath{{0.012 } } }
\vdef{bdt2012:BgPeakBs0:e2}   {\ensuremath{{0.179 } } }
\vdef{bdt2012:BgPeakHi0:val}   {\ensuremath{{0.019 } } }
\vdef{bdt2012:BgPeakHi0:e1}   {\ensuremath{{0.001 } } }
\vdef{bdt2012:BgPeakHi0:e2}   {\ensuremath{{0.014 } } }
\vdef{bdt2012:BgRslLo0:val}   {\ensuremath{{50.385 } } }
\vdef{bdt2012:BgRslLo0:e1}   {\ensuremath{{2.519 } } }
\vdef{bdt2012:BgRslLo0:e2}   {\ensuremath{{37.957 } } }
\vdef{bdt2012:BgRslBd0:val}   {\ensuremath{{9.132 } } }
\vdef{bdt2012:BgRslBd0:e1}   {\ensuremath{{0.457 } } }
\vdef{bdt2012:BgRslBd0:e2}   {\ensuremath{{9.525 } } }
\vdef{bdt2012:BgRslBs0:val}   {\ensuremath{{6.553 } } }
\vdef{bdt2012:BgRslBs0:e1}   {\ensuremath{{0.328 } } }
\vdef{bdt2012:BgRslBs0:e2}   {\ensuremath{{7.198 } } }
\vdef{bdt2012:BgRslHi0:val}   {\ensuremath{{0.706 } } }
\vdef{bdt2012:BgRslHi0:e1}   {\ensuremath{{0.035 } } }
\vdef{bdt2012:BgRslHi0:e2}   {\ensuremath{{0.778 } } }
\vdef{bdt2012:BgRareLo0:val}   {\ensuremath{{50.608 } } }
\vdef{bdt2012:BgRareLo0:e1}   {\ensuremath{{2.519 } } }
\vdef{bdt2012:BgRareLo0:e2}   {\ensuremath{{37.958 } } }
\vdef{bdt2012:BgRareBd0:val}   {\ensuremath{{9.936 } } }
\vdef{bdt2012:BgRareBd0:e1}   {\ensuremath{{0.458 } } }
\vdef{bdt2012:BgRareBd0:e2}   {\ensuremath{{9.543 } } }
\vdef{bdt2012:BgRareBs0:val}   {\ensuremath{{6.795 } } }
\vdef{bdt2012:BgRareBs0:e1}   {\ensuremath{{0.328 } } }
\vdef{bdt2012:BgRareBs0:e2}   {\ensuremath{{7.201 } } }
\vdef{bdt2012:BgRareHi0:val}   {\ensuremath{{0.725 } } }
\vdef{bdt2012:BgRareHi0:e1}   {\ensuremath{{0.035 } } }
\vdef{bdt2012:BgRareHi0:e2}   {\ensuremath{{0.778 } } }
\vdef{bdt2012:BgRslsLo0:val}   {\ensuremath{{14.000 } } }
\vdef{bdt2012:BgRslsLo0:e1}   {\ensuremath{{0.700 } } }
\vdef{bdt2012:BgRslsLo0:e2}   {\ensuremath{{10.547 } } }
\vdef{bdt2012:BgRslsBd0:val}   {\ensuremath{{2.537 } } }
\vdef{bdt2012:BgRslsBd0:e1}   {\ensuremath{{0.127 } } }
\vdef{bdt2012:BgRslsBd0:e2}   {\ensuremath{{2.647 } } }
\vdef{bdt2012:BgRslsBs0:val}   {\ensuremath{{1.821 } } }
\vdef{bdt2012:BgRslsBs0:e1}   {\ensuremath{{0.091 } } }
\vdef{bdt2012:BgRslsBs0:e2}   {\ensuremath{{2.000 } } }
\vdef{bdt2012:BgRslsHi0:val}   {\ensuremath{{0.196 } } }
\vdef{bdt2012:BgRslsHi0:e1}   {\ensuremath{{0.010 } } }
\vdef{bdt2012:BgRslsHi0:e2}   {\ensuremath{{0.216 } } }
\vdef{bdt2012:BgCombLo0:val}   {\ensuremath{{8.000 } } }
\vdef{bdt2012:BgCombLo0:e1}   {\ensuremath{{2.309 } } }
\vdef{bdt2012:BgCombLo0:e2}   {\ensuremath{{2.444 } } }
\vdef{bdt2012:BgCombBd0:val}   {\ensuremath{{2.667 } } }
\vdef{bdt2012:BgCombBd0:e1}   {\ensuremath{{0.770 } } }
\vdef{bdt2012:BgCombBd0:e2}   {\ensuremath{{0.815 } } }
\vdef{bdt2012:BgCombBs0:val}   {\ensuremath{{4.000 } } }
\vdef{bdt2012:BgCombBs0:e1}   {\ensuremath{{1.155 } } }
\vdef{bdt2012:BgCombBs0:e2}   {\ensuremath{{1.222 } } }
\vdef{bdt2012:BgCombHi0:val}   {\ensuremath{{12.000 } } }
\vdef{bdt2012:BgCombHi0:e1}   {\ensuremath{{3.464 } } }
\vdef{bdt2012:BgCombHi0:e2}   {\ensuremath{{3.666 } } }
\vdef{bdt2012:BgNonpLo0:val}   {\ensuremath{{22.000 } } }
\vdef{bdt2012:BgNonpLo0:e1}   {\ensuremath{{2.413 } } }
\vdef{bdt2012:BgNonpLo0:e2}   {\ensuremath{{10.826 } } }
\vdef{bdt2012:BgNonpBd0:val}   {\ensuremath{{5.204 } } }
\vdef{bdt2012:BgNonpBd0:e1}   {\ensuremath{{0.780 } } }
\vdef{bdt2012:BgNonpBd0:e2}   {\ensuremath{{2.769 } } }
\vdef{bdt2012:BgNonpBs0:val}   {\ensuremath{{5.821 } } }
\vdef{bdt2012:BgNonpBs0:e1}   {\ensuremath{{1.158 } } }
\vdef{bdt2012:BgNonpBs0:e2}   {\ensuremath{{2.160 } } }
\vdef{bdt2012:BgNonpHi0:val}   {\ensuremath{{12.196 } } }
\vdef{bdt2012:BgNonpHi0:e1}   {\ensuremath{{3.464 } } }
\vdef{bdt2012:BgNonpHi0:e2}   {\ensuremath{{3.672 } } }
\vdef{bdt2012:BgTotLo0:val}   {\ensuremath{{22.223 } } }
\vdef{bdt2012:BgTotLo0:e1}   {\ensuremath{{0.000 } } }
\vdef{bdt2012:BgTotLo0:e2}   {\ensuremath{{10.827 } } }
\vdef{bdt2012:BgTotBd0:val}   {\ensuremath{{6.008 } } }
\vdef{bdt2012:BgTotBd0:e1}   {\ensuremath{{0.000 } } }
\vdef{bdt2012:BgTotBd0:e2}   {\ensuremath{{2.831 } } }
\vdef{bdt2012:BgTotBs0:val}   {\ensuremath{{6.062 } } }
\vdef{bdt2012:BgTotBs0:e1}   {\ensuremath{{0.000 } } }
\vdef{bdt2012:BgTotBs0:e2}   {\ensuremath{{2.167 } } }
\vdef{bdt2012:BgTotHi0:val}   {\ensuremath{{12.215 } } }
\vdef{bdt2012:BgTotHi0:e1}   {\ensuremath{{0.000 } } }
\vdef{bdt2012:BgTotHi0:e2}   {\ensuremath{{3.672 } } }
\vdef{bdt2012:SgAndBgLo0:val}   {\ensuremath{{22.568 } } }
\vdef{bdt2012:SgAndBgLo0:e1}   {\ensuremath{{0.000 } } }
\vdef{bdt2012:SgAndBgLo0:e2}   {\ensuremath{{10.828 } } }
\vdef{bdt2012:SgAndBgBd0:val}   {\ensuremath{{7.856 } } }
\vdef{bdt2012:SgAndBgBd0:e1}   {\ensuremath{{0.000 } } }
\vdef{bdt2012:SgAndBgBd0:e2}   {\ensuremath{{3.005 } } }
\vdef{bdt2012:SgAndBgBs0:val}   {\ensuremath{{17.203 } } }
\vdef{bdt2012:SgAndBgBs0:e1}   {\ensuremath{{0.000 } } }
\vdef{bdt2012:SgAndBgBs0:e2}   {\ensuremath{{2.699 } } }
\vdef{bdt2012:SgAndBgHi0:val}   {\ensuremath{{12.585 } } }
\vdef{bdt2012:SgAndBgHi0:e1}   {\ensuremath{{0.000 } } }
\vdef{bdt2012:SgAndBgHi0:e2}   {\ensuremath{{3.673 } } }
\vdef{bdt2012:SgBd1:val}  {\ensuremath{{0.625 } } }
\vdef{bdt2012:SgBd1:e1}   {\ensuremath{{0.791 } } }
\vdef{bdt2012:SgBd1:e2}   {\ensuremath{{0.094 } } }
\vdef{bdt2012:SgBs1:val}  {\ensuremath{{3.332 } } }
\vdef{bdt2012:SgBs1:e1}   {\ensuremath{{1.825 } } }
\vdef{bdt2012:SgBs1:e2}   {\ensuremath{{0.500 } } }
\vdef{bdt2012:SgLo1:val}  {\ensuremath{{0.107 } } }
\vdef{bdt2012:SgLo1:e1}   {\ensuremath{{0.327 } } }
\vdef{bdt2012:SgLo1:e2}   {\ensuremath{{0.016 } } }
\vdef{bdt2012:SgHi1:val}  {\ensuremath{{0.376 } } }
\vdef{bdt2012:SgHi1:e1}   {\ensuremath{{0.613 } } }
\vdef{bdt2012:SgHi1:e2}   {\ensuremath{{0.056 } } }
\vdef{bdt2012:BdBd1:val}  {\ensuremath{{0.301 } } }
\vdef{bdt2012:BdBd1:e1}   {\ensuremath{{0.549 } } }
\vdef{bdt2012:BdBd1:e2}   {\ensuremath{{0.030 } } }
\vdef{bdt2012:BdBs1:val}  {\ensuremath{{0.179 } } }
\vdef{bdt2012:BdBs1:e1}   {\ensuremath{{0.423 } } }
\vdef{bdt2012:BdBs1:e2}   {\ensuremath{{0.018 } } }
\vdef{bdt2012:BdLo1:val}  {\ensuremath{{0.068 } } }
\vdef{bdt2012:BdLo1:e1}   {\ensuremath{{0.261 } } }
\vdef{bdt2012:BdLo1:e2}   {\ensuremath{{0.007 } } }
\vdef{bdt2012:BdHi1:val}  {\ensuremath{{0.002 } } }
\vdef{bdt2012:BdHi1:e1}   {\ensuremath{{0.045 } } }
\vdef{bdt2012:BdHi1:e2}   {\ensuremath{{0.000 } } }
\vdef{bdt2012:BgPeakLo1:val}   {\ensuremath{{0.107 } } }
\vdef{bdt2012:BgPeakLo1:e1}   {\ensuremath{{0.005 } } }
\vdef{bdt2012:BgPeakLo1:e2}   {\ensuremath{{0.076 } } }
\vdef{bdt2012:BgPeakBd1:val}   {\ensuremath{{0.194 } } }
\vdef{bdt2012:BgPeakBd1:e1}   {\ensuremath{{0.010 } } }
\vdef{bdt2012:BgPeakBd1:e2}   {\ensuremath{{0.142 } } }
\vdef{bdt2012:BgPeakBs1:val}   {\ensuremath{{0.092 } } }
\vdef{bdt2012:BgPeakBs1:e1}   {\ensuremath{{0.005 } } }
\vdef{bdt2012:BgPeakBs1:e2}   {\ensuremath{{0.067 } } }
\vdef{bdt2012:BgPeakHi1:val}   {\ensuremath{{0.009 } } }
\vdef{bdt2012:BgPeakHi1:e1}   {\ensuremath{{0.000 } } }
\vdef{bdt2012:BgPeakHi1:e2}   {\ensuremath{{0.006 } } }
\vdef{bdt2012:BgRslLo1:val}   {\ensuremath{{12.268 } } }
\vdef{bdt2012:BgRslLo1:e1}   {\ensuremath{{0.613 } } }
\vdef{bdt2012:BgRslLo1:e2}   {\ensuremath{{8.660 } } }
\vdef{bdt2012:BgRslBd1:val}   {\ensuremath{{2.145 } } }
\vdef{bdt2012:BgRslBd1:e1}   {\ensuremath{{0.107 } } }
\vdef{bdt2012:BgRslBd1:e2}   {\ensuremath{{2.135 } } }
\vdef{bdt2012:BgRslBs1:val}   {\ensuremath{{1.535 } } }
\vdef{bdt2012:BgRslBs1:e1}   {\ensuremath{{0.077 } } }
\vdef{bdt2012:BgRslBs1:e2}   {\ensuremath{{1.666 } } }
\vdef{bdt2012:BgRslHi1:val}   {\ensuremath{{0.316 } } }
\vdef{bdt2012:BgRslHi1:e1}   {\ensuremath{{0.016 } } }
\vdef{bdt2012:BgRslHi1:e2}   {\ensuremath{{0.349 } } }
\vdef{bdt2012:BgRareLo1:val}   {\ensuremath{{12.374 } } }
\vdef{bdt2012:BgRareLo1:e1}   {\ensuremath{{0.613 } } }
\vdef{bdt2012:BgRareLo1:e2}   {\ensuremath{{8.660 } } }
\vdef{bdt2012:BgRareBd1:val}   {\ensuremath{{2.339 } } }
\vdef{bdt2012:BgRareBd1:e1}   {\ensuremath{{0.108 } } }
\vdef{bdt2012:BgRareBd1:e2}   {\ensuremath{{2.140 } } }
\vdef{bdt2012:BgRareBs1:val}   {\ensuremath{{1.627 } } }
\vdef{bdt2012:BgRareBs1:e1}   {\ensuremath{{0.077 } } }
\vdef{bdt2012:BgRareBs1:e2}   {\ensuremath{{1.667 } } }
\vdef{bdt2012:BgRareHi1:val}   {\ensuremath{{0.326 } } }
\vdef{bdt2012:BgRareHi1:e1}   {\ensuremath{{0.016 } } }
\vdef{bdt2012:BgRareHi1:e2}   {\ensuremath{{0.349 } } }
\vdef{bdt2012:BgRslsLo1:val}   {\ensuremath{{2.000 } } }
\vdef{bdt2012:BgRslsLo1:e1}   {\ensuremath{{0.100 } } }
\vdef{bdt2012:BgRslsLo1:e2}   {\ensuremath{{1.412 } } }
\vdef{bdt2012:BgRslsBd1:val}   {\ensuremath{{0.362 } } }
\vdef{bdt2012:BgRslsBd1:e1}   {\ensuremath{{0.018 } } }
\vdef{bdt2012:BgRslsBd1:e2}   {\ensuremath{{0.361 } } }
\vdef{bdt2012:BgRslsBs1:val}   {\ensuremath{{0.260 } } }
\vdef{bdt2012:BgRslsBs1:e1}   {\ensuremath{{0.013 } } }
\vdef{bdt2012:BgRslsBs1:e2}   {\ensuremath{{0.282 } } }
\vdef{bdt2012:BgRslsHi1:val}   {\ensuremath{{0.028 } } }
\vdef{bdt2012:BgRslsHi1:e1}   {\ensuremath{{0.001 } } }
\vdef{bdt2012:BgRslsHi1:e2}   {\ensuremath{{0.031 } } }
\vdef{bdt2012:BgCombLo1:val}   {\ensuremath{{2.000 } } }
\vdef{bdt2012:BgCombLo1:e1}   {\ensuremath{{1.155 } } }
\vdef{bdt2012:BgCombLo1:e2}   {\ensuremath{{1.172 } } }
\vdef{bdt2012:BgCombBd1:val}   {\ensuremath{{0.667 } } }
\vdef{bdt2012:BgCombBd1:e1}   {\ensuremath{{0.385 } } }
\vdef{bdt2012:BgCombBd1:e2}   {\ensuremath{{0.391 } } }
\vdef{bdt2012:BgCombBs1:val}   {\ensuremath{{1.000 } } }
\vdef{bdt2012:BgCombBs1:e1}   {\ensuremath{{0.577 } } }
\vdef{bdt2012:BgCombBs1:e2}   {\ensuremath{{0.586 } } }
\vdef{bdt2012:BgCombHi1:val}   {\ensuremath{{3.000 } } }
\vdef{bdt2012:BgCombHi1:e1}   {\ensuremath{{1.732 } } }
\vdef{bdt2012:BgCombHi1:e2}   {\ensuremath{{1.758 } } }
\vdef{bdt2012:BgNonpLo1:val}   {\ensuremath{{4.000 } } }
\vdef{bdt2012:BgNonpLo1:e1}   {\ensuremath{{1.159 } } }
\vdef{bdt2012:BgNonpLo1:e2}   {\ensuremath{{1.835 } } }
\vdef{bdt2012:BgNonpBd1:val}   {\ensuremath{{1.029 } } }
\vdef{bdt2012:BgNonpBd1:e1}   {\ensuremath{{0.385 } } }
\vdef{bdt2012:BgNonpBd1:e2}   {\ensuremath{{0.532 } } }
\vdef{bdt2012:BgNonpBs1:val}   {\ensuremath{{1.260 } } }
\vdef{bdt2012:BgNonpBs1:e1}   {\ensuremath{{0.577 } } }
\vdef{bdt2012:BgNonpBs1:e2}   {\ensuremath{{0.482 } } }
\vdef{bdt2012:BgNonpHi1:val}   {\ensuremath{{3.028 } } }
\vdef{bdt2012:BgNonpHi1:e1}   {\ensuremath{{1.732 } } }
\vdef{bdt2012:BgNonpHi1:e2}   {\ensuremath{{1.758 } } }
\vdef{bdt2012:BgTotLo1:val}   {\ensuremath{{4.107 } } }
\vdef{bdt2012:BgTotLo1:e1}   {\ensuremath{{0.000 } } }
\vdef{bdt2012:BgTotLo1:e2}   {\ensuremath{{1.836 } } }
\vdef{bdt2012:BgTotBd1:val}   {\ensuremath{{1.224 } } }
\vdef{bdt2012:BgTotBd1:e1}   {\ensuremath{{0.000 } } }
\vdef{bdt2012:BgTotBd1:e2}   {\ensuremath{{0.550 } } }
\vdef{bdt2012:BgTotBs1:val}   {\ensuremath{{1.352 } } }
\vdef{bdt2012:BgTotBs1:e1}   {\ensuremath{{0.000 } } }
\vdef{bdt2012:BgTotBs1:e2}   {\ensuremath{{0.487 } } }
\vdef{bdt2012:BgTotHi1:val}   {\ensuremath{{3.037 } } }
\vdef{bdt2012:BgTotHi1:e1}   {\ensuremath{{0.000 } } }
\vdef{bdt2012:BgTotHi1:e2}   {\ensuremath{{1.758 } } }
\vdef{bdt2012:SgAndBgLo1:val}   {\ensuremath{{4.282 } } }
\vdef{bdt2012:SgAndBgLo1:e1}   {\ensuremath{{0.000 } } }
\vdef{bdt2012:SgAndBgLo1:e2}   {\ensuremath{{1.836 } } }
\vdef{bdt2012:SgAndBgBd1:val}   {\ensuremath{{2.150 } } }
\vdef{bdt2012:SgAndBgBd1:e1}   {\ensuremath{{0.000 } } }
\vdef{bdt2012:SgAndBgBd1:e2}   {\ensuremath{{0.783 } } }
\vdef{bdt2012:SgAndBgBs1:val}   {\ensuremath{{4.864 } } }
\vdef{bdt2012:SgAndBgBs1:e1}   {\ensuremath{{0.000 } } }
\vdef{bdt2012:SgAndBgBs1:e2}   {\ensuremath{{0.698 } } }
\vdef{bdt2012:SgAndBgHi1:val}   {\ensuremath{{3.415 } } }
\vdef{bdt2012:SgAndBgHi1:e1}   {\ensuremath{{0.000 } } }
\vdef{bdt2012:SgAndBgHi1:e2}   {\ensuremath{{1.759 } } }
\vdef{bdt2012:N-EFF-TOT-BS0:val}   {\ensuremath{{0.000498 } } }
\vdef{bdt2012:N-EFF-TOT-BS0:err}   {\ensuremath{{0.000001 } } }
\vdef{bdt2012:N-ACC-BS0:val}   {\ensuremath{{0.0082 } } }
\vdef{bdt2012:N-ACC-BS0:err}   {\ensuremath{{0.0000 } } }
\vdef{bdt2012:N-EFF-MU-PID-BS0:val}   {\ensuremath{{0.6749 } } }
\vdef{bdt2012:N-EFF-MU-PID-BS0:err}   {\ensuremath{{0.0002 } } }
\vdef{bdt2012:N-EFF-MU-PIDMC-BS0:val}   {\ensuremath{{0.6379 } } }
\vdef{bdt2012:N-EFF-MU-PIDMC-BS0:err}   {\ensuremath{{0.0002 } } }
\vdef{bdt2012:N-EFF-MU-MC-BS0:val}   {\ensuremath{{0.6000 } } }
\vdef{bdt2012:N-EFF-MU-MC-BS0:err}   {\ensuremath{{0.0007 } } }
\vdef{bdt2012:N-EFF-TRIG-PID-BS0:val}   {\ensuremath{{0.7135 } } }
\vdef{bdt2012:N-EFF-TRIG-PID-BS0:err}   {\ensuremath{{0.0007 } } }
\vdef{bdt2012:N-EFF-TRIG-PIDMC-BS0:val}   {\ensuremath{{0.6678 } } }
\vdef{bdt2012:N-EFF-TRIG-PIDMC-BS0:err}   {\ensuremath{{0.0008 } } }
\vdef{bdt2012:N-EFF-TRIG-MC-BS0:val}   {\ensuremath{{0.5500 } } }
\vdef{bdt2012:N-EFF-TRIG-MC-BS0:err}   {\ensuremath{{0.0009 } } }
\vdef{bdt2012:N-EFF-CAND-BS0:val}   {\ensuremath{{0.9952 } } }
\vdef{bdt2012:N-EFF-CAND-BS0:err}   {\ensuremath{{0.0004 } } }
\vdef{bdt2012:N-EFF-ANA-BS0:val}   {\ensuremath{{0.1914 } } }
\vdef{bdt2012:N-EFF-ANA-BS0:err}   {\ensuremath{{0.0003 } } }
\vdef{bdt2012:N-OBS-BS0:val}   {\ensuremath{{20673 } } }
\vdef{bdt2012:N-OBS-BS0:err}   {\ensuremath{{535 } } }
\vdef{bdt2012:N-EFF-TOT-BS1:val}   {\ensuremath{{0.000120 } } }
\vdef{bdt2012:N-EFF-TOT-BS1:err}   {\ensuremath{{0.000001 } } }
\vdef{bdt2012:N-ACC-BS1:val}   {\ensuremath{{0.0043 } } }
\vdef{bdt2012:N-ACC-BS1:err}   {\ensuremath{{0.0000 } } }
\vdef{bdt2012:N-EFF-MU-PID-BS1:val}   {\ensuremath{{0.5806 } } }
\vdef{bdt2012:N-EFF-MU-PID-BS1:err}   {\ensuremath{{0.0006 } } }
\vdef{bdt2012:N-EFF-MU-PIDMC-BS1:val}   {\ensuremath{{0.5753 } } }
\vdef{bdt2012:N-EFF-MU-PIDMC-BS1:err}   {\ensuremath{{0.0005 } } }
\vdef{bdt2012:N-EFF-MU-MC-BS1:val}   {\ensuremath{{0.5606 } } }
\vdef{bdt2012:N-EFF-MU-MC-BS1:err}   {\ensuremath{{0.0013 } } }
\vdef{bdt2012:N-EFF-TRIG-PID-BS1:val}   {\ensuremath{{0.7661 } } }
\vdef{bdt2012:N-EFF-TRIG-PID-BS1:err}   {\ensuremath{{0.0005 } } }
\vdef{bdt2012:N-EFF-TRIG-PIDMC-BS1:val}   {\ensuremath{{0.7503 } } }
\vdef{bdt2012:N-EFF-TRIG-PIDMC-BS1:err}   {\ensuremath{{0.0006 } } }
\vdef{bdt2012:N-EFF-TRIG-MC-BS1:val}   {\ensuremath{{0.4818 } } }
\vdef{bdt2012:N-EFF-TRIG-MC-BS1:err}   {\ensuremath{{0.0018 } } }
\vdef{bdt2012:N-EFF-CAND-BS1:val}   {\ensuremath{{0.9932 } } }
\vdef{bdt2012:N-EFF-CAND-BS1:err}   {\ensuremath{{0.0007 } } }
\vdef{bdt2012:N-EFF-ANA-BS1:val}   {\ensuremath{{0.1091 } } }
\vdef{bdt2012:N-EFF-ANA-BS1:err}   {\ensuremath{{0.0003 } } }
\vdef{bdt2012:N-OBS-BS1:val}   {\ensuremath{{5311 } } }
\vdef{bdt2012:N-OBS-BS1:err}   {\ensuremath{{86 } } }
\vdef{2012:bdt:0}     {\ensuremath{{0.360 } } }
\vdef{2012:bdtMax:0}  {\ensuremath{{10.000 } } }
\vdef{2012:mBdLo:0}   {\ensuremath{{5.200 } } }
\vdef{2012:mBdHi:0}   {\ensuremath{{5.300 } } }
\vdef{2012:mBsLo:0}   {\ensuremath{{5.300 } } }
\vdef{2012:mBsHi:0}   {\ensuremath{{5.450 } } }
\vdef{2012:etaMin:0}   {\ensuremath{{0.0 } } }
\vdef{2012:etaMax:0}   {\ensuremath{{1.4 } } }
\vdef{2012:pt:0}   {\ensuremath{{6.5 } } }
\vdef{2012:m1pt:0}   {\ensuremath{{4.5 } } }
\vdef{2012:m2pt:0}   {\ensuremath{{4.0 } } }
\vdef{2012:m1eta:0}   {\ensuremath{{1.4 } } }
\vdef{2012:m2eta:0}   {\ensuremath{{1.4 } } }
\vdef{2012:iso:0}   {\ensuremath{{0.80 } } }
\vdef{2012:chi2dof:0}   {\ensuremath{{2.2 } } }
\vdef{2012:alpha:0}   {\ensuremath{{0.050 } } }
\vdef{2012:fls3d:0}   {\ensuremath{{13.0 } } }
\vdef{2012:docatrk:0}   {\ensuremath{{0.015 } } }
\vdef{2012:closetrk:0}   {\ensuremath{{2 } } }
\vdef{2012:pvlip:0}   {\ensuremath{{100.000 } } }
\vdef{2012:pvlips:0}   {\ensuremath{{100.000 } } }
\vdef{2012:pvlip2:0}   {\ensuremath{{-100.000 } } }
\vdef{2012:pvlips2:0}   {\ensuremath{{-100.000 } } }
\vdef{2012:maxdoca:0}   {\ensuremath{{100.000 } } }
\vdef{2012:pvip:0}   {\ensuremath{{0.008 } } }
\vdef{2012:pvips:0}   {\ensuremath{{2.000 } } }
\vdef{2012:doApplyCowboyVeto:0}   {no }
\vdef{2012:fDoApplyCowboyVetoAlsoInSignal:0}   {no }
\vdef{2012:bdt:1}     {\ensuremath{{0.380 } } }
\vdef{2012:bdtMax:1}  {\ensuremath{{10.000 } } }
\vdef{2012:mBdLo:1}   {\ensuremath{{5.200 } } }
\vdef{2012:mBdHi:1}   {\ensuremath{{5.300 } } }
\vdef{2012:mBsLo:1}   {\ensuremath{{5.300 } } }
\vdef{2012:mBsHi:1}   {\ensuremath{{5.450 } } }
\vdef{2012:etaMin:1}   {\ensuremath{{1.4 } } }
\vdef{2012:etaMax:1}   {\ensuremath{{2.4 } } }
\vdef{2012:pt:1}   {\ensuremath{{8.5 } } }
\vdef{2012:m1pt:1}   {\ensuremath{{4.5 } } }
\vdef{2012:m2pt:1}   {\ensuremath{{4.2 } } }
\vdef{2012:m1eta:1}   {\ensuremath{{2.0 } } }
\vdef{2012:m2eta:1}   {\ensuremath{{2.0 } } }
\vdef{2012:iso:1}   {\ensuremath{{0.80 } } }
\vdef{2012:chi2dof:1}   {\ensuremath{{1.8 } } }
\vdef{2012:alpha:1}   {\ensuremath{{0.030 } } }
\vdef{2012:fls3d:1}   {\ensuremath{{15.0 } } }
\vdef{2012:docatrk:1}   {\ensuremath{{0.015 } } }
\vdef{2012:closetrk:1}   {\ensuremath{{2 } } }
\vdef{2012:pvlip:1}   {\ensuremath{{100.000 } } }
\vdef{2012:pvlips:1}   {\ensuremath{{100.000 } } }
\vdef{2012:pvlip2:1}   {\ensuremath{{-100.000 } } }
\vdef{2012:pvlips2:1}   {\ensuremath{{-100.000 } } }
\vdef{2012:maxdoca:1}   {\ensuremath{{100.000 } } }
\vdef{2012:pvip:1}   {\ensuremath{{0.008 } } }
\vdef{2012:pvips:1}   {\ensuremath{{2.000 } } }
\vdef{2012:doApplyCowboyVeto:1}   {no }
\vdef{2012:fDoApplyCowboyVetoAlsoInSignal:1}   {no }
\vdef{bdt2012:N-CSBF-TNP-BS0:val}   {\ensuremath{{0.000024 } } }
\vdef{bdt2012:N-CSBF-TNP-BS0:err}   {\ensuremath{{0.000001 } } }
\vdef{bdt2012:N-CSBF-MC-BS0:val}   {\ensuremath{{0.000025 } } }
\vdef{bdt2012:N-CSBF-MC-BS0:err}   {\ensuremath{{0.000001 } } }
\vdef{bdt2012:N-CSBF-MC-BS0:syst}   {\ensuremath{{0.000002 } } }
\vdef{bdt2012:N-CSBF-BS0:val}   {\ensuremath{{0.000025 } } }
\vdef{bdt2012:N-CSBF-BS0:err}   {\ensuremath{{0.000001 } } }
\vdef{bdt2012:N-CSBF-BS0:syst}   {\ensuremath{{0.000002 } } }
\vdef{bdt2012:bgBd2K0MuMu:loSideband0:val}   {\ensuremath{{0.000 } } }
\vdef{bdt2012:bgBd2K0MuMu:loSideband0:err}   {\ensuremath{{0.000 } } }
\vdef{bdt2012:bgBd2K0MuMu:bdRare0}   {\ensuremath{{0.000000 } } }
\vdef{bdt2012:bgBd2K0MuMu:bdRare0E}  {\ensuremath{{0.000000 } } }
\vdef{bdt2012:bgBd2K0MuMu:bsRare0}   {\ensuremath{{0.000000 } } }
\vdef{bdt2012:bgBd2K0MuMu:bsRare0E}  {\ensuremath{{0.000000 } } }
\vdef{bdt2012:bgBd2K0MuMu:hiSideband0:val}   {\ensuremath{{0.000 } } }
\vdef{bdt2012:bgBd2K0MuMu:hiSideband0:err}   {\ensuremath{{0.000 } } }
\vdef{bdt2012:bgBd2KK:loSideband0:val}   {\ensuremath{{0.004 } } }
\vdef{bdt2012:bgBd2KK:loSideband0:err}   {\ensuremath{{0.005 } } }
\vdef{bdt2012:bgBd2KK:bdRare0}   {\ensuremath{{0.001967 } } }
\vdef{bdt2012:bgBd2KK:bdRare0E}  {\ensuremath{{0.002482 } } }
\vdef{bdt2012:bgBd2KK:bsRare0}   {\ensuremath{{0.000072 } } }
\vdef{bdt2012:bgBd2KK:bsRare0E}  {\ensuremath{{0.000091 } } }
\vdef{bdt2012:bgBd2KK:hiSideband0:val}   {\ensuremath{{0.000 } } }
\vdef{bdt2012:bgBd2KK:hiSideband0:err}   {\ensuremath{{0.000 } } }
\vdef{bdt2012:bgBd2KPi:loSideband0:val}   {\ensuremath{{0.177 } } }
\vdef{bdt2012:bgBd2KPi:loSideband0:err}   {\ensuremath{{0.125 } } }
\vdef{bdt2012:bgBd2KPi:bdRare0}   {\ensuremath{{0.455567 } } }
\vdef{bdt2012:bgBd2KPi:bdRare0E}  {\ensuremath{{0.322424 } } }
\vdef{bdt2012:bgBd2KPi:bsRare0}   {\ensuremath{{0.044416 } } }
\vdef{bdt2012:bgBd2KPi:bsRare0E}  {\ensuremath{{0.031435 } } }
\vdef{bdt2012:bgBd2KPi:hiSideband0:val}   {\ensuremath{{0.002 } } }
\vdef{bdt2012:bgBd2KPi:hiSideband0:err}   {\ensuremath{{0.001 } } }
\vdef{bdt2012:bgBd2MuMuGamma:loSideband0:val}   {\ensuremath{{0.028 } } }
\vdef{bdt2012:bgBd2MuMuGamma:loSideband0:err}   {\ensuremath{{0.006 } } }
\vdef{bdt2012:bgBd2MuMuGamma:bdRare0}   {\ensuremath{{0.002021 } } }
\vdef{bdt2012:bgBd2MuMuGamma:bdRare0E}  {\ensuremath{{0.000404 } } }
\vdef{bdt2012:bgBd2MuMuGamma:bsRare0}   {\ensuremath{{0.000204 } } }
\vdef{bdt2012:bgBd2MuMuGamma:bsRare0E}  {\ensuremath{{0.000041 } } }
\vdef{bdt2012:bgBd2MuMuGamma:hiSideband0:val}   {\ensuremath{{0.000 } } }
\vdef{bdt2012:bgBd2MuMuGamma:hiSideband0:err}   {\ensuremath{{0.000 } } }
\vdef{bdt2012:bgBd2Pi0MuMu:loSideband0:val}   {\ensuremath{{1.464 } } }
\vdef{bdt2012:bgBd2Pi0MuMu:loSideband0:err}   {\ensuremath{{0.732 } } }
\vdef{bdt2012:bgBd2Pi0MuMu:bdRare0}   {\ensuremath{{0.001312 } } }
\vdef{bdt2012:bgBd2Pi0MuMu:bdRare0E}  {\ensuremath{{0.000656 } } }
\vdef{bdt2012:bgBd2Pi0MuMu:bsRare0}   {\ensuremath{{0.000000 } } }
\vdef{bdt2012:bgBd2Pi0MuMu:bsRare0E}  {\ensuremath{{0.000000 } } }
\vdef{bdt2012:bgBd2Pi0MuMu:hiSideband0:val}   {\ensuremath{{0.001 } } }
\vdef{bdt2012:bgBd2Pi0MuMu:hiSideband0:err}   {\ensuremath{{0.000 } } }
\vdef{bdt2012:bgBd2PiMuNu:loSideband0:val}   {\ensuremath{{7.039 } } }
\vdef{bdt2012:bgBd2PiMuNu:loSideband0:err}   {\ensuremath{{3.537 } } }
\vdef{bdt2012:bgBd2PiMuNu:bdRare0}   {\ensuremath{{0.175771 } } }
\vdef{bdt2012:bgBd2PiMuNu:bdRare0E}  {\ensuremath{{0.088324 } } }
\vdef{bdt2012:bgBd2PiMuNu:bsRare0}   {\ensuremath{{0.016416 } } }
\vdef{bdt2012:bgBd2PiMuNu:bsRare0E}  {\ensuremath{{0.008249 } } }
\vdef{bdt2012:bgBd2PiMuNu:hiSideband0:val}   {\ensuremath{{0.004 } } }
\vdef{bdt2012:bgBd2PiMuNu:hiSideband0:err}   {\ensuremath{{0.002 } } }
\vdef{bdt2012:bgBd2PiPi:loSideband0:val}   {\ensuremath{{0.008 } } }
\vdef{bdt2012:bgBd2PiPi:loSideband0:err}   {\ensuremath{{0.008 } } }
\vdef{bdt2012:bgBd2PiPi:bdRare0}   {\ensuremath{{0.061578 } } }
\vdef{bdt2012:bgBd2PiPi:bdRare0E}  {\ensuremath{{0.061635 } } }
\vdef{bdt2012:bgBd2PiPi:bsRare0}   {\ensuremath{{0.025867 } } }
\vdef{bdt2012:bgBd2PiPi:bsRare0E}  {\ensuremath{{0.025891 } } }
\vdef{bdt2012:bgBd2PiPi:hiSideband0:val}   {\ensuremath{{0.001 } } }
\vdef{bdt2012:bgBd2PiPi:hiSideband0:err}   {\ensuremath{{0.001 } } }
\vdef{bdt2012:bgBs2KK:loSideband0:val}   {\ensuremath{{0.031 } } }
\vdef{bdt2012:bgBs2KK:loSideband0:err}   {\ensuremath{{0.031 } } }
\vdef{bdt2012:bgBs2KK:bdRare0}   {\ensuremath{{0.257948 } } }
\vdef{bdt2012:bgBs2KK:bdRare0E}  {\ensuremath{{0.260834 } } }
\vdef{bdt2012:bgBs2KK:bsRare0}   {\ensuremath{{0.108502 } } }
\vdef{bdt2012:bgBs2KK:bsRare0E}  {\ensuremath{{0.109716 } } }
\vdef{bdt2012:bgBs2KK:hiSideband0:val}   {\ensuremath{{0.002 } } }
\vdef{bdt2012:bgBs2KK:hiSideband0:err}   {\ensuremath{{0.002 } } }
\vdef{bdt2012:bgBs2KMuNu:loSideband0:val}   {\ensuremath{{4.331 } } }
\vdef{bdt2012:bgBs2KMuNu:loSideband0:err}   {\ensuremath{{2.176 } } }
\vdef{bdt2012:bgBs2KMuNu:bdRare0}   {\ensuremath{{0.200169 } } }
\vdef{bdt2012:bgBs2KMuNu:bdRare0E}  {\ensuremath{{0.100584 } } }
\vdef{bdt2012:bgBs2KMuNu:bsRare0}   {\ensuremath{{0.022405 } } }
\vdef{bdt2012:bgBs2KMuNu:bsRare0E}  {\ensuremath{{0.011258 } } }
\vdef{bdt2012:bgBs2KMuNu:hiSideband0:val}   {\ensuremath{{0.004 } } }
\vdef{bdt2012:bgBs2KMuNu:hiSideband0:err}   {\ensuremath{{0.002 } } }
\vdef{bdt2012:bgBs2KPi:loSideband0:val}   {\ensuremath{{0.002 } } }
\vdef{bdt2012:bgBs2KPi:loSideband0:err}   {\ensuremath{{0.001 } } }
\vdef{bdt2012:bgBs2KPi:bdRare0}   {\ensuremath{{0.012350 } } }
\vdef{bdt2012:bgBs2KPi:bdRare0E}  {\ensuremath{{0.009146 } } }
\vdef{bdt2012:bgBs2KPi:bsRare0}   {\ensuremath{{0.027979 } } }
\vdef{bdt2012:bgBs2KPi:bsRare0E}  {\ensuremath{{0.020719 } } }
\vdef{bdt2012:bgBs2KPi:hiSideband0:val}   {\ensuremath{{0.001 } } }
\vdef{bdt2012:bgBs2KPi:hiSideband0:err}   {\ensuremath{{0.000 } } }
\vdef{bdt2012:bgBs2MuMuGamma:loSideband0:val}   {\ensuremath{{1.118 } } }
\vdef{bdt2012:bgBs2MuMuGamma:loSideband0:err}   {\ensuremath{{0.224 } } }
\vdef{bdt2012:bgBs2MuMuGamma:bdRare0}   {\ensuremath{{0.212783 } } }
\vdef{bdt2012:bgBs2MuMuGamma:bdRare0E}  {\ensuremath{{0.042557 } } }
\vdef{bdt2012:bgBs2MuMuGamma:bsRare0}   {\ensuremath{{0.070117 } } }
\vdef{bdt2012:bgBs2MuMuGamma:bsRare0E}  {\ensuremath{{0.014023 } } }
\vdef{bdt2012:bgBs2MuMuGamma:hiSideband0:val}   {\ensuremath{{0.001 } } }
\vdef{bdt2012:bgBs2MuMuGamma:hiSideband0:err}   {\ensuremath{{0.000 } } }
\vdef{bdt2012:bgBs2PiPi:loSideband0:val}   {\ensuremath{{0.000 } } }
\vdef{bdt2012:bgBs2PiPi:loSideband0:err}   {\ensuremath{{0.000 } } }
\vdef{bdt2012:bgBs2PiPi:bdRare0}   {\ensuremath{{0.000280 } } }
\vdef{bdt2012:bgBs2PiPi:bdRare0E}  {\ensuremath{{0.000285 } } }
\vdef{bdt2012:bgBs2PiPi:bsRare0}   {\ensuremath{{0.002828 } } }
\vdef{bdt2012:bgBs2PiPi:bsRare0E}  {\ensuremath{{0.002880 } } }
\vdef{bdt2012:bgBs2PiPi:hiSideband0:val}   {\ensuremath{{0.000 } } }
\vdef{bdt2012:bgBs2PiPi:hiSideband0:err}   {\ensuremath{{0.000 } } }
\vdef{bdt2012:bgBu2KMuMu:loSideband0:val}   {\ensuremath{{0.000 } } }
\vdef{bdt2012:bgBu2KMuMu:loSideband0:err}   {\ensuremath{{0.000 } } }
\vdef{bdt2012:bgBu2KMuMu:bdRare0}   {\ensuremath{{0.000000 } } }
\vdef{bdt2012:bgBu2KMuMu:bdRare0E}  {\ensuremath{{0.000000 } } }
\vdef{bdt2012:bgBu2KMuMu:bsRare0}   {\ensuremath{{0.000000 } } }
\vdef{bdt2012:bgBu2KMuMu:bsRare0E}  {\ensuremath{{0.000000 } } }
\vdef{bdt2012:bgBu2KMuMu:hiSideband0:val}   {\ensuremath{{0.000 } } }
\vdef{bdt2012:bgBu2KMuMu:hiSideband0:err}   {\ensuremath{{0.000 } } }
\vdef{bdt2012:bgBu2PiMuMu:loSideband0:val}   {\ensuremath{{2.440 } } }
\vdef{bdt2012:bgBu2PiMuMu:loSideband0:err}   {\ensuremath{{0.634 } } }
\vdef{bdt2012:bgBu2PiMuMu:bdRare0}   {\ensuremath{{0.003499 } } }
\vdef{bdt2012:bgBu2PiMuMu:bdRare0E}  {\ensuremath{{0.000910 } } }
\vdef{bdt2012:bgBu2PiMuMu:bsRare0}   {\ensuremath{{0.001759 } } }
\vdef{bdt2012:bgBu2PiMuMu:bsRare0E}  {\ensuremath{{0.000457 } } }
\vdef{bdt2012:bgBu2PiMuMu:hiSideband0:val}   {\ensuremath{{0.000 } } }
\vdef{bdt2012:bgBu2PiMuMu:hiSideband0:err}   {\ensuremath{{0.000 } } }
\vdef{bdt2012:bgLb2KP:loSideband0:val}   {\ensuremath{{0.001 } } }
\vdef{bdt2012:bgLb2KP:loSideband0:err}   {\ensuremath{{0.001 } } }
\vdef{bdt2012:bgLb2KP:bdRare0}   {\ensuremath{{0.001883 } } }
\vdef{bdt2012:bgLb2KP:bdRare0E}  {\ensuremath{{0.001415 } } }
\vdef{bdt2012:bgLb2KP:bsRare0}   {\ensuremath{{0.020697 } } }
\vdef{bdt2012:bgLb2KP:bsRare0E}  {\ensuremath{{0.015554 } } }
\vdef{bdt2012:bgLb2KP:hiSideband0:val}   {\ensuremath{{0.008 } } }
\vdef{bdt2012:bgLb2KP:hiSideband0:err}   {\ensuremath{{0.006 } } }
\vdef{bdt2012:bgLb2PMuNu:loSideband0:val}   {\ensuremath{{32.340 } } }
\vdef{bdt2012:bgLb2PMuNu:loSideband0:err}   {\ensuremath{{36.157 } } }
\vdef{bdt2012:bgLb2PMuNu:bdRare0}   {\ensuremath{{8.167563 } } }
\vdef{bdt2012:bgLb2PMuNu:bdRare0E}  {\ensuremath{{9.131613 } } }
\vdef{bdt2012:bgLb2PMuNu:bsRare0}   {\ensuremath{{6.173069 } } }
\vdef{bdt2012:bgLb2PMuNu:bsRare0E}  {\ensuremath{{6.901701 } } }
\vdef{bdt2012:bgLb2PMuNu:hiSideband0:val}   {\ensuremath{{0.667 } } }
\vdef{bdt2012:bgLb2PMuNu:hiSideband0:err}   {\ensuremath{{0.746 } } }
\vdef{bdt2012:bgLb2PiP:loSideband0:val}   {\ensuremath{{0.000 } } }
\vdef{bdt2012:bgLb2PiP:loSideband0:err}   {\ensuremath{{0.000 } } }
\vdef{bdt2012:bgLb2PiP:bdRare0}   {\ensuremath{{0.000444 } } }
\vdef{bdt2012:bgLb2PiP:bdRare0E}  {\ensuremath{{0.000339 } } }
\vdef{bdt2012:bgLb2PiP:bsRare0}   {\ensuremath{{0.004119 } } }
\vdef{bdt2012:bgLb2PiP:bsRare0E}  {\ensuremath{{0.003142 } } }
\vdef{bdt2012:bgLb2PiP:hiSideband0:val}   {\ensuremath{{0.006 } } }
\vdef{bdt2012:bgLb2PiP:hiSideband0:err}   {\ensuremath{{0.005 } } }
\vdef{bdt2012:bsRare0}   {\ensuremath{{0.000 } } }
\vdef{bdt2012:bsRare0E}  {\ensuremath{{0.000 } } }
\vdef{bdt2012:bdRare0}   {\ensuremath{{0.000 } } }
\vdef{bdt2012:bdRare0E}  {\ensuremath{{0.000 } } }
\vdef{bdt2012:N-CSBF-TNP-BS1:val}   {\ensuremath{{0.000030 } } }
\vdef{bdt2012:N-CSBF-TNP-BS1:err}   {\ensuremath{{0.000000 } } }
\vdef{bdt2012:N-CSBF-MC-BS1:val}   {\ensuremath{{0.000030 } } }
\vdef{bdt2012:N-CSBF-MC-BS1:err}   {\ensuremath{{0.000001 } } }
\vdef{bdt2012:N-CSBF-MC-BS1:syst}   {\ensuremath{{0.000002 } } }
\vdef{bdt2012:N-CSBF-BS1:val}   {\ensuremath{{0.000031 } } }
\vdef{bdt2012:N-CSBF-BS1:err}   {\ensuremath{{0.000001 } } }
\vdef{bdt2012:N-CSBF-BS1:syst}   {\ensuremath{{0.000002 } } }
\vdef{bdt2012:bgBd2K0MuMu:loSideband1:val}   {\ensuremath{{0.000 } } }
\vdef{bdt2012:bgBd2K0MuMu:loSideband1:err}   {\ensuremath{{0.000 } } }
\vdef{bdt2012:bgBd2K0MuMu:bdRare1}   {\ensuremath{{0.020378 } } }
\vdef{bdt2012:bgBd2K0MuMu:bdRare1E}  {\ensuremath{{0.000000 } } }
\vdef{bdt2012:bgBd2K0MuMu:bsRare1}   {\ensuremath{{0.000000 } } }
\vdef{bdt2012:bgBd2K0MuMu:bsRare1E}  {\ensuremath{{0.000000 } } }
\vdef{bdt2012:bgBd2K0MuMu:hiSideband1:val}   {\ensuremath{{0.000 } } }
\vdef{bdt2012:bgBd2K0MuMu:hiSideband1:err}   {\ensuremath{{0.000 } } }
\vdef{bdt2012:bgBd2KK:loSideband1:val}   {\ensuremath{{0.001 } } }
\vdef{bdt2012:bgBd2KK:loSideband1:err}   {\ensuremath{{0.002 } } }
\vdef{bdt2012:bgBd2KK:bdRare1}   {\ensuremath{{0.000599 } } }
\vdef{bdt2012:bgBd2KK:bdRare1E}  {\ensuremath{{0.000756 } } }
\vdef{bdt2012:bgBd2KK:bsRare1}   {\ensuremath{{0.000082 } } }
\vdef{bdt2012:bgBd2KK:bsRare1E}  {\ensuremath{{0.000103 } } }
\vdef{bdt2012:bgBd2KK:hiSideband1:val}   {\ensuremath{{0.000 } } }
\vdef{bdt2012:bgBd2KK:hiSideband1:err}   {\ensuremath{{0.000 } } }
\vdef{bdt2012:bgBd2KPi:loSideband1:val}   {\ensuremath{{0.078 } } }
\vdef{bdt2012:bgBd2KPi:loSideband1:err}   {\ensuremath{{0.055 } } }
\vdef{bdt2012:bgBd2KPi:bdRare1}   {\ensuremath{{0.106973 } } }
\vdef{bdt2012:bgBd2KPi:bdRare1E}  {\ensuremath{{0.075710 } } }
\vdef{bdt2012:bgBd2KPi:bsRare1}   {\ensuremath{{0.029467 } } }
\vdef{bdt2012:bgBd2KPi:bsRare1E}  {\ensuremath{{0.020855 } } }
\vdef{bdt2012:bgBd2KPi:hiSideband1:val}   {\ensuremath{{0.002 } } }
\vdef{bdt2012:bgBd2KPi:hiSideband1:err}   {\ensuremath{{0.001 } } }
\vdef{bdt2012:bgBd2MuMuGamma:loSideband1:val}   {\ensuremath{{0.008 } } }
\vdef{bdt2012:bgBd2MuMuGamma:loSideband1:err}   {\ensuremath{{0.002 } } }
\vdef{bdt2012:bgBd2MuMuGamma:bdRare1}   {\ensuremath{{0.000788 } } }
\vdef{bdt2012:bgBd2MuMuGamma:bdRare1E}  {\ensuremath{{0.000158 } } }
\vdef{bdt2012:bgBd2MuMuGamma:bsRare1}   {\ensuremath{{0.000104 } } }
\vdef{bdt2012:bgBd2MuMuGamma:bsRare1E}  {\ensuremath{{0.000021 } } }
\vdef{bdt2012:bgBd2MuMuGamma:hiSideband1:val}   {\ensuremath{{0.000 } } }
\vdef{bdt2012:bgBd2MuMuGamma:hiSideband1:err}   {\ensuremath{{0.000 } } }
\vdef{bdt2012:bgBd2Pi0MuMu:loSideband1:val}   {\ensuremath{{0.382 } } }
\vdef{bdt2012:bgBd2Pi0MuMu:loSideband1:err}   {\ensuremath{{0.191 } } }
\vdef{bdt2012:bgBd2Pi0MuMu:bdRare1}   {\ensuremath{{0.003567 } } }
\vdef{bdt2012:bgBd2Pi0MuMu:bdRare1E}  {\ensuremath{{0.001783 } } }
\vdef{bdt2012:bgBd2Pi0MuMu:bsRare1}   {\ensuremath{{0.000474 } } }
\vdef{bdt2012:bgBd2Pi0MuMu:bsRare1E}  {\ensuremath{{0.000237 } } }
\vdef{bdt2012:bgBd2Pi0MuMu:hiSideband1:val}   {\ensuremath{{0.000 } } }
\vdef{bdt2012:bgBd2Pi0MuMu:hiSideband1:err}   {\ensuremath{{0.000 } } }
\vdef{bdt2012:bgBd2PiMuNu:loSideband1:val}   {\ensuremath{{2.029 } } }
\vdef{bdt2012:bgBd2PiMuNu:loSideband1:err}   {\ensuremath{{1.019 } } }
\vdef{bdt2012:bgBd2PiMuNu:bdRare1}   {\ensuremath{{0.080086 } } }
\vdef{bdt2012:bgBd2PiMuNu:bdRare1E}  {\ensuremath{{0.040243 } } }
\vdef{bdt2012:bgBd2PiMuNu:bsRare1}   {\ensuremath{{0.012728 } } }
\vdef{bdt2012:bgBd2PiMuNu:bsRare1E}  {\ensuremath{{0.006396 } } }
\vdef{bdt2012:bgBd2PiMuNu:hiSideband1:val}   {\ensuremath{{0.003 } } }
\vdef{bdt2012:bgBd2PiMuNu:hiSideband1:err}   {\ensuremath{{0.001 } } }
\vdef{bdt2012:bgBd2PiPi:loSideband1:val}   {\ensuremath{{0.005 } } }
\vdef{bdt2012:bgBd2PiPi:loSideband1:err}   {\ensuremath{{0.005 } } }
\vdef{bdt2012:bgBd2PiPi:bdRare1}   {\ensuremath{{0.016006 } } }
\vdef{bdt2012:bgBd2PiPi:bdRare1E}  {\ensuremath{{0.016021 } } }
\vdef{bdt2012:bgBd2PiPi:bsRare1}   {\ensuremath{{0.010246 } } }
\vdef{bdt2012:bgBd2PiPi:bsRare1E}  {\ensuremath{{0.010255 } } }
\vdef{bdt2012:bgBd2PiPi:hiSideband1:val}   {\ensuremath{{0.001 } } }
\vdef{bdt2012:bgBd2PiPi:hiSideband1:err}   {\ensuremath{{0.001 } } }
\vdef{bdt2012:bgBs2KK:loSideband1:val}   {\ensuremath{{0.020 } } }
\vdef{bdt2012:bgBs2KK:loSideband1:err}   {\ensuremath{{0.020 } } }
\vdef{bdt2012:bgBs2KK:bdRare1}   {\ensuremath{{0.061978 } } }
\vdef{bdt2012:bgBs2KK:bdRare1E}  {\ensuremath{{0.062671 } } }
\vdef{bdt2012:bgBs2KK:bsRare1}   {\ensuremath{{0.036131 } } }
\vdef{bdt2012:bgBs2KK:bsRare1E}  {\ensuremath{{0.036535 } } }
\vdef{bdt2012:bgBs2KK:hiSideband1:val}   {\ensuremath{{0.002 } } }
\vdef{bdt2012:bgBs2KK:hiSideband1:err}   {\ensuremath{{0.002 } } }
\vdef{bdt2012:bgBs2KMuNu:loSideband1:val}   {\ensuremath{{1.154 } } }
\vdef{bdt2012:bgBs2KMuNu:loSideband1:err}   {\ensuremath{{0.580 } } }
\vdef{bdt2012:bgBs2KMuNu:bdRare1}   {\ensuremath{{0.066940 } } }
\vdef{bdt2012:bgBs2KMuNu:bdRare1E}  {\ensuremath{{0.033637 } } }
\vdef{bdt2012:bgBs2KMuNu:bsRare1}   {\ensuremath{{0.011943 } } }
\vdef{bdt2012:bgBs2KMuNu:bsRare1E}  {\ensuremath{{0.006001 } } }
\vdef{bdt2012:bgBs2KMuNu:hiSideband1:val}   {\ensuremath{{0.001 } } }
\vdef{bdt2012:bgBs2KMuNu:hiSideband1:err}   {\ensuremath{{0.000 } } }
\vdef{bdt2012:bgBs2KPi:loSideband1:val}   {\ensuremath{{0.001 } } }
\vdef{bdt2012:bgBs2KPi:loSideband1:err}   {\ensuremath{{0.001 } } }
\vdef{bdt2012:bgBs2KPi:bdRare1}   {\ensuremath{{0.004583 } } }
\vdef{bdt2012:bgBs2KPi:bdRare1E}  {\ensuremath{{0.003394 } } }
\vdef{bdt2012:bgBs2KPi:bsRare1}   {\ensuremath{{0.007297 } } }
\vdef{bdt2012:bgBs2KPi:bsRare1E}  {\ensuremath{{0.005404 } } }
\vdef{bdt2012:bgBs2KPi:hiSideband1:val}   {\ensuremath{{0.001 } } }
\vdef{bdt2012:bgBs2KPi:hiSideband1:err}   {\ensuremath{{0.000 } } }
\vdef{bdt2012:bgBs2MuMuGamma:loSideband1:val}   {\ensuremath{{0.303 } } }
\vdef{bdt2012:bgBs2MuMuGamma:loSideband1:err}   {\ensuremath{{0.061 } } }
\vdef{bdt2012:bgBs2MuMuGamma:bdRare1}   {\ensuremath{{0.058086 } } }
\vdef{bdt2012:bgBs2MuMuGamma:bdRare1E}  {\ensuremath{{0.011617 } } }
\vdef{bdt2012:bgBs2MuMuGamma:bsRare1}   {\ensuremath{{0.018948 } } }
\vdef{bdt2012:bgBs2MuMuGamma:bsRare1E}  {\ensuremath{{0.003790 } } }
\vdef{bdt2012:bgBs2MuMuGamma:hiSideband1:val}   {\ensuremath{{0.001 } } }
\vdef{bdt2012:bgBs2MuMuGamma:hiSideband1:err}   {\ensuremath{{0.000 } } }
\vdef{bdt2012:bgBs2PiPi:loSideband1:val}   {\ensuremath{{0.000 } } }
\vdef{bdt2012:bgBs2PiPi:loSideband1:err}   {\ensuremath{{0.000 } } }
\vdef{bdt2012:bgBs2PiPi:bdRare1}   {\ensuremath{{0.000178 } } }
\vdef{bdt2012:bgBs2PiPi:bdRare1E}  {\ensuremath{{0.000182 } } }
\vdef{bdt2012:bgBs2PiPi:bsRare1}   {\ensuremath{{0.000761 } } }
\vdef{bdt2012:bgBs2PiPi:bsRare1E}  {\ensuremath{{0.000775 } } }
\vdef{bdt2012:bgBs2PiPi:hiSideband1:val}   {\ensuremath{{0.000 } } }
\vdef{bdt2012:bgBs2PiPi:hiSideband1:err}   {\ensuremath{{0.000 } } }
\vdef{bdt2012:bgBu2KMuMu:loSideband1:val}   {\ensuremath{{0.000 } } }
\vdef{bdt2012:bgBu2KMuMu:loSideband1:err}   {\ensuremath{{0.000 } } }
\vdef{bdt2012:bgBu2KMuMu:bdRare1}   {\ensuremath{{0.000000 } } }
\vdef{bdt2012:bgBu2KMuMu:bdRare1E}  {\ensuremath{{0.000000 } } }
\vdef{bdt2012:bgBu2KMuMu:bsRare1}   {\ensuremath{{0.000000 } } }
\vdef{bdt2012:bgBu2KMuMu:bsRare1E}  {\ensuremath{{0.000000 } } }
\vdef{bdt2012:bgBu2KMuMu:hiSideband1:val}   {\ensuremath{{0.000 } } }
\vdef{bdt2012:bgBu2KMuMu:hiSideband1:err}   {\ensuremath{{0.000 } } }
\vdef{bdt2012:bgBu2PiMuMu:loSideband1:val}   {\ensuremath{{0.660 } } }
\vdef{bdt2012:bgBu2PiMuMu:loSideband1:err}   {\ensuremath{{0.172 } } }
\vdef{bdt2012:bgBu2PiMuMu:bdRare1}   {\ensuremath{{0.000653 } } }
\vdef{bdt2012:bgBu2PiMuMu:bdRare1E}  {\ensuremath{{0.000170 } } }
\vdef{bdt2012:bgBu2PiMuMu:bsRare1}   {\ensuremath{{0.000000 } } }
\vdef{bdt2012:bgBu2PiMuMu:bsRare1E}  {\ensuremath{{0.000000 } } }
\vdef{bdt2012:bgBu2PiMuMu:hiSideband1:val}   {\ensuremath{{0.000 } } }
\vdef{bdt2012:bgBu2PiMuMu:hiSideband1:err}   {\ensuremath{{0.000 } } }
\vdef{bdt2012:bgLb2KP:loSideband1:val}   {\ensuremath{{0.000 } } }
\vdef{bdt2012:bgLb2KP:loSideband1:err}   {\ensuremath{{0.000 } } }
\vdef{bdt2012:bgLb2KP:bdRare1}   {\ensuremath{{0.000928 } } }
\vdef{bdt2012:bgLb2KP:bdRare1E}  {\ensuremath{{0.000698 } } }
\vdef{bdt2012:bgLb2KP:bsRare1}   {\ensuremath{{0.004697 } } }
\vdef{bdt2012:bgLb2KP:bsRare1E}  {\ensuremath{{0.003530 } } }
\vdef{bdt2012:bgLb2KP:hiSideband1:val}   {\ensuremath{{0.002 } } }
\vdef{bdt2012:bgLb2KP:hiSideband1:err}   {\ensuremath{{0.002 } } }
\vdef{bdt2012:bgLb2PMuNu:loSideband1:val}   {\ensuremath{{7.353 } } }
\vdef{bdt2012:bgLb2PMuNu:loSideband1:err}   {\ensuremath{{8.221 } } }
\vdef{bdt2012:bgLb2PMuNu:bdRare1}   {\ensuremath{{1.830454 } } }
\vdef{bdt2012:bgLb2PMuNu:bdRare1E}  {\ensuremath{{2.046510 } } }
\vdef{bdt2012:bgLb2PMuNu:bsRare1}   {\ensuremath{{1.428467 } } }
\vdef{bdt2012:bgLb2PMuNu:bsRare1E}  {\ensuremath{{1.597075 } } }
\vdef{bdt2012:bgLb2PMuNu:hiSideband1:val}   {\ensuremath{{0.30 } } }
\vdef{bdt2012:bgLb2PMuNu:hiSideband1:err}   {\ensuremath{{0.34 } } }
\vdef{bdt2012:bgLb2PiP:loSideband1:val}   {\ensuremath{{0.000 } } }
\vdef{bdt2012:bgLb2PiP:loSideband1:err}   {\ensuremath{{0.000 } } }
\vdef{bdt2012:bgLb2PiP:bdRare1}   {\ensuremath{{0.000216 } } }
\vdef{bdt2012:bgLb2PiP:bdRare1E}  {\ensuremath{{0.000165 } } }
\vdef{bdt2012:bgLb2PiP:bsRare1}   {\ensuremath{{0.001221 } } }
\vdef{bdt2012:bgLb2PiP:bsRare1E}  {\ensuremath{{0.000931 } } }
\vdef{bdt2012:bgLb2PiP:hiSideband1:val}   {\ensuremath{{0.002 } } }
\vdef{bdt2012:bgLb2PiP:hiSideband1:err}   {\ensuremath{{0.001 } } }
\vdef{bdt2012:bsRare1}   {\ensuremath{{0.000 } } }
\vdef{bdt2012:bsRare1E}  {\ensuremath{{0.000 } } }
\vdef{bdt2012:bdRare1}   {\ensuremath{{0.000 } } }
\vdef{bdt2012:bdRare1E}  {\ensuremath{{0.000 } } }
\vdef{bdt2012:N-EFF-TOT-BPLUS0:val}   {\ensuremath{{0.00082 } } }
\vdef{bdt2012:N-EFF-TOT-BPLUS0:err}   {\ensuremath{{0.000001 } } }
\vdef{bdt2012:N-EFF-TOT-BPLUS0:tot}   {\ensuremath{{0.00007 } } }
\vdef{bdt2012:N-EFF-TOT-BPLUS0:all}   {\ensuremath{{(0.82 \pm 0.07)\times 10^{-3}} } }
\vdef{bdt2012:N-EFF-PRODMC-BPLUS0:val}   {\ensuremath{{0.00083 } } }
\vdef{bdt2012:N-EFF-PRODMC-BPLUS0:err}   {\ensuremath{{0.000003 } } }
\vdef{bdt2012:N-EFF-PRODMC-BPLUS0:tot}   {\ensuremath{{0.00000 } } }
\vdef{bdt2012:N-EFF-PRODMC-BPLUS0:all}   {\ensuremath{{(0.83 \pm 0.00)\times 10^{-3}} } }
\vdef{bdt2012:N-EFF-PRODTNP-BPLUS0:val}   {\ensuremath{{0.00120 } } }
\vdef{bdt2012:N-EFF-PRODTNP-BPLUS0:err}   {\ensuremath{{1.589703 } } }
\vdef{bdt2012:N-EFF-PRODTNP-BPLUS0:tot}   {\ensuremath{{1.58970 } } }
\vdef{bdt2012:N-EFF-PRODTNP-BPLUS0:all}   {\ensuremath{{(1.20 \pm 1589.70)\times 10^{-3}} } }
\vdef{bdt2012:N-EFF-PRODTNPMC-BPLUS0:val}   {\ensuremath{{0.00083 } } }
\vdef{bdt2012:N-EFF-PRODTNPMC-BPLUS0:err}   {\ensuremath{{0.000003 } } }
\vdef{bdt2012:N-EFF-PRODTNPMC-BPLUS0:tot}   {\ensuremath{{0.00000 } } }
\vdef{bdt2012:N-EFF-PRODTNPMC-BPLUS0:all}   {\ensuremath{{(0.83 \pm 0.00)\times 10^{-3}} } }
\vdef{bdt2012:N-ACC-BPLUS0:val}   {\ensuremath{{0.011 } } }
\vdef{bdt2012:N-ACC-BPLUS0:err}   {\ensuremath{{0.000 } } }
\vdef{bdt2012:N-ACC-BPLUS0:tot}   {\ensuremath{{0.000 } } }
\vdef{bdt2012:N-ACC-BPLUS0:all}   {\ensuremath{{(1.12 \pm 0.04)\times 10^{-2}} } }
\vdef{bdt2012:N-EFF-MU-PID-BPLUS0:val}   {\ensuremath{{0.673 } } }
\vdef{bdt2012:N-EFF-MU-PID-BPLUS0:err}   {\ensuremath{{0.000 } } }
\vdef{bdt2012:N-EFF-MU-PID-BPLUS0:tot}   {\ensuremath{{0.027 } } }
\vdef{bdt2012:N-EFF-MU-PID-BPLUS0:all}   {\ensuremath{{(67.28 \pm 2.69)\times 10^{-2}} } }
\vdef{bdt2012:N-EFFRHO-MU-PID-BPLUS0:val}   {\ensuremath{{0.636 } } }
\vdef{bdt2012:N-EFFRHO-MU-PID-BPLUS0:err}   {\ensuremath{{0.000 } } }
\vdef{bdt2012:N-EFFRHO-MU-PID-BPLUS0:tot}   {\ensuremath{{0.000 } } }
\vdef{bdt2012:N-EFFRHO-MU-PID-BPLUS0:all}   {\ensuremath{{(63.60 \pm 0.00)\times 10^{-2}} } }
\vdef{bdt2012:N-EFF-MU-PIDMC-BPLUS0:val}   {\ensuremath{{0.635 } } }
\vdef{bdt2012:N-EFF-MU-PIDMC-BPLUS0:err}   {\ensuremath{{0.000 } } }
\vdef{bdt2012:N-EFF-MU-PIDMC-BPLUS0:tot}   {\ensuremath{{0.025 } } }
\vdef{bdt2012:N-EFF-MU-PIDMC-BPLUS0:all}   {\ensuremath{{(63.52 \pm 2.54)\times 10^{-2}} } }
\vdef{bdt2012:N-EFFRHO-MU-PIDMC-BPLUS0:val}   {\ensuremath{{0.600 } } }
\vdef{bdt2012:N-EFFRHO-MU-PIDMC-BPLUS0:err}   {\ensuremath{{0.000 } } }
\vdef{bdt2012:N-EFFRHO-MU-PIDMC-BPLUS0:tot}   {\ensuremath{{0.025 } } }
\vdef{bdt2012:N-EFFRHO-MU-PIDMC-BPLUS0:all}   {\ensuremath{{(60.05 \pm 2.54)\times 10^{-2}} } }
\vdef{bdt2012:N-EFF-MU-MC-BPLUS0:val}   {\ensuremath{{0.600 } } }
\vdef{bdt2012:N-EFF-MU-MC-BPLUS0:err}   {\ensuremath{{0.000 } } }
\vdef{bdt2012:N-EFF-MU-MC-BPLUS0:tot}   {\ensuremath{{0.024 } } }
\vdef{bdt2012:N-EFF-MU-MC-BPLUS0:all}   {\ensuremath{{(60.05 \pm 2.40)\times 10^{-2}} } }
\vdef{bdt2012:N-EFF-TRIG-PID-BPLUS0:val}   {\ensuremath{{0.707 } } }
\vdef{bdt2012:N-EFF-TRIG-PID-BPLUS0:err}   {\ensuremath{{0.000 } } }
\vdef{bdt2012:N-EFF-TRIG-PID-BPLUS0:tot}   {\ensuremath{{0.021 } } }
\vdef{bdt2012:N-EFF-TRIG-PID-BPLUS0:all}   {\ensuremath{{(70.72 \pm 2.12)\times 10^{-2}} } }
\vdef{bdt2012:N-EFFRHO-TRIG-PID-BPLUS0:val}   {\ensuremath{{0.585 } } }
\vdef{bdt2012:N-EFFRHO-TRIG-PID-BPLUS0:err}   {\ensuremath{{0.000 } } }
\vdef{bdt2012:N-EFFRHO-TRIG-PID-BPLUS0:tot}   {\ensuremath{{0.018 } } }
\vdef{bdt2012:N-EFFRHO-TRIG-PID-BPLUS0:all}   {\ensuremath{{(58.52 \pm 1.76)\times 10^{-2}} } }
\vdef{bdt2012:N-EFF-TRIG-PIDMC-BPLUS0:val}   {\ensuremath{{0.660 } } }
\vdef{bdt2012:N-EFF-TRIG-PIDMC-BPLUS0:err}   {\ensuremath{{0.000 } } }
\vdef{bdt2012:N-EFF-TRIG-PIDMC-BPLUS0:tot}   {\ensuremath{{0.020 } } }
\vdef{bdt2012:N-EFF-TRIG-PIDMC-BPLUS0:all}   {\ensuremath{{(65.99 \pm 1.98)\times 10^{-2}} } }
\vdef{bdt2012:N-EFFRHO-TRIG-PIDMC-BPLUS0:val}   {\ensuremath{{0.546 } } }
\vdef{bdt2012:N-EFFRHO-TRIG-PIDMC-BPLUS0:err}   {\ensuremath{{0.000 } } }
\vdef{bdt2012:N-EFFRHO-TRIG-PIDMC-BPLUS0:tot}   {\ensuremath{{0.016 } } }
\vdef{bdt2012:N-EFFRHO-TRIG-PIDMC-BPLUS0:all}   {\ensuremath{{(54.61 \pm 1.64)\times 10^{-2}} } }
\vdef{bdt2012:N-EFF-TRIG-MC-BPLUS0:val}   {\ensuremath{{0.546 } } }
\vdef{bdt2012:N-EFF-TRIG-MC-BPLUS0:err}   {\ensuremath{{0.001 } } }
\vdef{bdt2012:N-EFF-TRIG-MC-BPLUS0:tot}   {\ensuremath{{0.016 } } }
\vdef{bdt2012:N-EFF-TRIG-MC-BPLUS0:all}   {\ensuremath{{(54.61 \pm 1.64)\times 10^{-2}} } }
\vdef{bdt2012:N-EFF-CAND-BPLUS0:val}   {\ensuremath{{1.000 } } }
\vdef{bdt2012:N-EFF-CAND-BPLUS0:err}   {\ensuremath{{0.000 } } }
\vdef{bdt2012:N-EFF-CAND-BPLUS0:tot}   {\ensuremath{{0.010 } } }
\vdef{bdt2012:N-EFF-CAND-BPLUS0:all}   {\ensuremath{{(99.99 \pm 1.00)\times 10^{-2}} } }
\vdef{bdt2012:N-EFF-ANA-BPLUS0:val}   {\ensuremath{{0.2258 } } }
\vdef{bdt2012:N-EFF-ANA-BPLUS0:err}   {\ensuremath{{0.0002 } } }
\vdef{bdt2012:N-EFF-ANA-BPLUS0:tot}   {\ensuremath{{0.0128 } } }
\vdef{bdt2012:N-EFF-ANA-BPLUS0:all}   {\ensuremath{{(22.58 \pm 1.28)\times 10^{-2}} } }
\vdef{bdt2012:N-OBS-BPLUS0:val}   {\ensuremath{{308877 } } }
\vdef{bdt2012:N-OBS-BPLUS0:err}   {\ensuremath{{729 } } }
\vdef{bdt2012:N-OBS-BPLUS0:tot}   {\ensuremath{{15461 } } }
\vdef{bdt2012:N-OBS-BPLUS0:all}   {\ensuremath{{308877 } } }
\vdef{bdt2012:N-OBS-CBPLUS0:val}   {\ensuremath{{321086 } } }
\vdef{bdt2012:N-OBS-CBPLUS0:err}   {\ensuremath{{592 } } }
\vdef{bdt2012:N-EFF-TOT-BPLUS1:val}   {\ensuremath{{0.00021 } } }
\vdef{bdt2012:N-EFF-TOT-BPLUS1:err}   {\ensuremath{{0.000001 } } }
\vdef{bdt2012:N-EFF-TOT-BPLUS1:tot}   {\ensuremath{{0.00003 } } }
\vdef{bdt2012:N-EFF-TOT-BPLUS1:all}   {\ensuremath{{(0.21 \pm 0.03)\times 10^{-3}} } }
\vdef{bdt2012:N-EFF-PRODMC-BPLUS1:val}   {\ensuremath{{0.00021 } } }
\vdef{bdt2012:N-EFF-PRODMC-BPLUS1:err}   {\ensuremath{{0.000001 } } }
\vdef{bdt2012:N-EFF-PRODMC-BPLUS1:tot}   {\ensuremath{{0.00000 } } }
\vdef{bdt2012:N-EFF-PRODMC-BPLUS1:all}   {\ensuremath{{(0.21 \pm 0.00)\times 10^{-3}} } }
\vdef{bdt2012:N-EFF-PRODTNP-BPLUS1:val}   {\ensuremath{{0.00034 } } }
\vdef{bdt2012:N-EFF-PRODTNP-BPLUS1:err}   {\ensuremath{{0.237522 } } }
\vdef{bdt2012:N-EFF-PRODTNP-BPLUS1:tot}   {\ensuremath{{0.23752 } } }
\vdef{bdt2012:N-EFF-PRODTNP-BPLUS1:all}   {\ensuremath{{(0.34 \pm 237.52)\times 10^{-3}} } }
\vdef{bdt2012:N-EFF-PRODTNPMC-BPLUS1:val}   {\ensuremath{{0.00021 } } }
\vdef{bdt2012:N-EFF-PRODTNPMC-BPLUS1:err}   {\ensuremath{{0.000001 } } }
\vdef{bdt2012:N-EFF-PRODTNPMC-BPLUS1:tot}   {\ensuremath{{0.00000 } } }
\vdef{bdt2012:N-EFF-PRODTNPMC-BPLUS1:all}   {\ensuremath{{(0.21 \pm 0.00)\times 10^{-3}} } }
\vdef{bdt2012:N-ACC-BPLUS1:val}   {\ensuremath{{0.006 } } }
\vdef{bdt2012:N-ACC-BPLUS1:err}   {\ensuremath{{0.000 } } }
\vdef{bdt2012:N-ACC-BPLUS1:tot}   {\ensuremath{{0.000 } } }
\vdef{bdt2012:N-ACC-BPLUS1:all}   {\ensuremath{{(0.59 \pm 0.03)\times 10^{-2}} } }
\vdef{bdt2012:N-EFF-MU-PID-BPLUS1:val}   {\ensuremath{{0.584 } } }
\vdef{bdt2012:N-EFF-MU-PID-BPLUS1:err}   {\ensuremath{{0.000 } } }
\vdef{bdt2012:N-EFF-MU-PID-BPLUS1:tot}   {\ensuremath{{0.047 } } }
\vdef{bdt2012:N-EFF-MU-PID-BPLUS1:all}   {\ensuremath{{(58.40 \pm 4.67)\times 10^{-2}} } }
\vdef{bdt2012:N-EFFRHO-MU-PID-BPLUS1:val}   {\ensuremath{{0.569 } } }
\vdef{bdt2012:N-EFFRHO-MU-PID-BPLUS1:err}   {\ensuremath{{0.000 } } }
\vdef{bdt2012:N-EFFRHO-MU-PID-BPLUS1:tot}   {\ensuremath{{0.000 } } }
\vdef{bdt2012:N-EFFRHO-MU-PID-BPLUS1:all}   {\ensuremath{{(56.93 \pm 0.00)\times 10^{-2}} } }
\vdef{bdt2012:N-EFF-MU-PIDMC-BPLUS1:val}   {\ensuremath{{0.577 } } }
\vdef{bdt2012:N-EFF-MU-PIDMC-BPLUS1:err}   {\ensuremath{{0.000 } } }
\vdef{bdt2012:N-EFF-MU-PIDMC-BPLUS1:tot}   {\ensuremath{{0.046 } } }
\vdef{bdt2012:N-EFF-MU-PIDMC-BPLUS1:all}   {\ensuremath{{(57.67 \pm 4.61)\times 10^{-2}} } }
\vdef{bdt2012:N-EFFRHO-MU-PIDMC-BPLUS1:val}   {\ensuremath{{0.562 } } }
\vdef{bdt2012:N-EFFRHO-MU-PIDMC-BPLUS1:err}   {\ensuremath{{0.000 } } }
\vdef{bdt2012:N-EFFRHO-MU-PIDMC-BPLUS1:tot}   {\ensuremath{{0.046 } } }
\vdef{bdt2012:N-EFFRHO-MU-PIDMC-BPLUS1:all}   {\ensuremath{{(56.22 \pm 4.55)\times 10^{-2}} } }
\vdef{bdt2012:N-EFF-MU-MC-BPLUS1:val}   {\ensuremath{{0.562 } } }
\vdef{bdt2012:N-EFF-MU-MC-BPLUS1:err}   {\ensuremath{{0.001 } } }
\vdef{bdt2012:N-EFF-MU-MC-BPLUS1:tot}   {\ensuremath{{0.045 } } }
\vdef{bdt2012:N-EFF-MU-MC-BPLUS1:all}   {\ensuremath{{(56.22 \pm 4.50)\times 10^{-2}} } }
\vdef{bdt2012:N-EFF-TRIG-PID-BPLUS1:val}   {\ensuremath{{0.765 } } }
\vdef{bdt2012:N-EFF-TRIG-PID-BPLUS1:err}   {\ensuremath{{0.000 } } }
\vdef{bdt2012:N-EFF-TRIG-PID-BPLUS1:tot}   {\ensuremath{{0.046 } } }
\vdef{bdt2012:N-EFF-TRIG-PID-BPLUS1:all}   {\ensuremath{{(76.48 \pm 4.59)\times 10^{-2}} } }
\vdef{bdt2012:N-EFFRHO-TRIG-PID-BPLUS1:val}   {\ensuremath{{0.498 } } }
\vdef{bdt2012:N-EFFRHO-TRIG-PID-BPLUS1:err}   {\ensuremath{{0.000 } } }
\vdef{bdt2012:N-EFFRHO-TRIG-PID-BPLUS1:tot}   {\ensuremath{{0.030 } } }
\vdef{bdt2012:N-EFFRHO-TRIG-PID-BPLUS1:all}   {\ensuremath{{(49.82 \pm 2.99)\times 10^{-2}} } }
\vdef{bdt2012:N-EFF-TRIG-PIDMC-BPLUS1:val}   {\ensuremath{{0.749 } } }
\vdef{bdt2012:N-EFF-TRIG-PIDMC-BPLUS1:err}   {\ensuremath{{0.000 } } }
\vdef{bdt2012:N-EFF-TRIG-PIDMC-BPLUS1:tot}   {\ensuremath{{0.045 } } }
\vdef{bdt2012:N-EFF-TRIG-PIDMC-BPLUS1:all}   {\ensuremath{{(74.89 \pm 4.49)\times 10^{-2}} } }
\vdef{bdt2012:N-EFFRHO-TRIG-PIDMC-BPLUS1:val}   {\ensuremath{{0.488 } } }
\vdef{bdt2012:N-EFFRHO-TRIG-PIDMC-BPLUS1:err}   {\ensuremath{{0.000 } } }
\vdef{bdt2012:N-EFFRHO-TRIG-PIDMC-BPLUS1:tot}   {\ensuremath{{0.029 } } }
\vdef{bdt2012:N-EFFRHO-TRIG-PIDMC-BPLUS1:all}   {\ensuremath{{(48.79 \pm 2.93)\times 10^{-2}} } }
\vdef{bdt2012:N-EFF-TRIG-MC-BPLUS1:val}   {\ensuremath{{0.488 } } }
\vdef{bdt2012:N-EFF-TRIG-MC-BPLUS1:err}   {\ensuremath{{0.001 } } }
\vdef{bdt2012:N-EFF-TRIG-MC-BPLUS1:tot}   {\ensuremath{{0.029 } } }
\vdef{bdt2012:N-EFF-TRIG-MC-BPLUS1:all}   {\ensuremath{{(48.79 \pm 2.93)\times 10^{-2}} } }
\vdef{bdt2012:N-EFF-CAND-BPLUS1:val}   {\ensuremath{{1.000 } } }
\vdef{bdt2012:N-EFF-CAND-BPLUS1:err}   {\ensuremath{{0.000 } } }
\vdef{bdt2012:N-EFF-CAND-BPLUS1:tot}   {\ensuremath{{0.010 } } }
\vdef{bdt2012:N-EFF-CAND-BPLUS1:all}   {\ensuremath{{(99.98 \pm 1.00)\times 10^{-2}} } }
\vdef{bdt2012:N-EFF-ANA-BPLUS1:val}   {\ensuremath{{0.1316 } } }
\vdef{bdt2012:N-EFF-ANA-BPLUS1:err}   {\ensuremath{{0.0002 } } }
\vdef{bdt2012:N-EFF-ANA-BPLUS1:tot}   {\ensuremath{{0.0074 } } }
\vdef{bdt2012:N-EFF-ANA-BPLUS1:all}   {\ensuremath{{(13.16 \pm 0.74)\times 10^{-2}} } }
\vdef{bdt2012:N-OBS-BPLUS1:val}   {\ensuremath{{69260 } } }
\vdef{bdt2012:N-OBS-BPLUS1:err}   {\ensuremath{{291 } } }
\vdef{bdt2012:N-OBS-BPLUS1:tot}   {\ensuremath{{3475 } } }
\vdef{bdt2012:N-OBS-BPLUS1:all}   {\ensuremath{{69260 } } }
\vdef{bdt2012:N-OBS-CBPLUS1:val}   {\ensuremath{{65988 } } }
\vdef{bdt2012:N-OBS-CBPLUS1:err}   {\ensuremath{{276 } } }
\vdef{bdt2012:N-EXP2-SIG-BSMM0:val}   {\ensuremath{{11.46 } } }
\vdef{bdt2012:N-EXP2-SIG-BSMM0:err}   {\ensuremath{{ 1.72 } } }
\vdef{bdt2012:N-EXP2-SIG-BDMM0:val}   {\ensuremath{{1.00 } } }
\vdef{bdt2012:N-EXP2-SIG-BDMM0:err}   {\ensuremath{{0.10 } } }
\vdef{bdt2012:N-OBS-BKG0:val}   {\ensuremath{{34 } } }
\vdef{bdt2012:N-EXP-BSMM0:val}   {\ensuremath{{ 5.80 } } }
\vdef{bdt2012:N-EXP-BSMM0:err}   {\ensuremath{{ 2.90 } } }
\vdef{bdt2012:N-EXP-BDMM0:val}   {\ensuremath{{ 5.18 } } }
\vdef{bdt2012:N-EXP-BDMM0:err}   {\ensuremath{{ 2.59 } } }
\vdef{bdt2012:N-LOW-BD0:val}   {\ensuremath{{5.200 } } }
\vdef{bdt2012:N-HIGH-BD0:val}   {\ensuremath{{5.300 } } }
\vdef{bdt2012:N-LOW-BS0:val}   {\ensuremath{{5.300 } } }
\vdef{bdt2012:N-HIGH-BS0:val}   {\ensuremath{{5.450 } } }
\vdef{bdt2012:N-PSS0:val}   {\ensuremath{{0.881 } } }
\vdef{bdt2012:N-PSS0:err}   {\ensuremath{{0.000 } } }
\vdef{bdt2012:N-PSS0:tot}   {\ensuremath{{0.044 } } }
\vdef{bdt2012:N-PSD0:val}   {\ensuremath{{0.280 } } }
\vdef{bdt2012:N-PSD0:err}   {\ensuremath{{0.001 } } }
\vdef{bdt2012:N-PSD0:tot}   {\ensuremath{{0.014 } } }
\vdef{bdt2012:N-PDS0:val}   {\ensuremath{{0.070 } } }
\vdef{bdt2012:N-PDS0:err}   {\ensuremath{{0.000 } } }
\vdef{bdt2012:N-PDS0:tot}   {\ensuremath{{0.003 } } }
\vdef{bdt2012:N-PDD0:val}   {\ensuremath{{0.661 } } }
\vdef{bdt2012:N-PDD0:err}   {\ensuremath{{0.001 } } }
\vdef{bdt2012:N-PDD0:tot}   {\ensuremath{{0.033 } } }
\vdef{bdt2012:N-EFF-TOT-BSMM0:val}   {\ensuremath{{0.0023 } } }
\vdef{bdt2012:N-EFF-TOT-BSMM0:err}   {\ensuremath{{0.0000 } } }
\vdef{bdt2012:N-EFF-TOT-BSMM0:tot}   {\ensuremath{{0.0002 } } }
\vdef{bdt2012:N-EFF-TOT-BSMM0:all}   {\ensuremath{{(0.23 \pm 0.03)} } }
\vdef{bdt2012:N-EFF-PRODMC-BSMM0:val}   {\ensuremath{{0.0023 } } }
\vdef{bdt2012:N-EFF-PRODMC-BSMM0:err}   {\ensuremath{{0.0000 } } }
\vdef{bdt2012:N-EFF-PRODMC-BSMM0:tot}   {\ensuremath{{0.0000 } } }
\vdef{bdt2012:N-EFF-PRODMC-BSMM0:all}   {\ensuremath{{(2.32 \pm 0.01)\times 10^{-3}} } }
\vdef{bdt2012:N-EFFRATIO-TOT-BSMM0:val}   {\ensuremath{{0.362 } } }
\vdef{bdt2012:N-EFFRATIO-TOT-BSMM0:err}   {\ensuremath{{0.001 } } }
\vdef{bdt2012:N-EFFRATIO-PRODMC-BSMM0:val}   {\ensuremath{{0.357 } } }
\vdef{bdt2012:N-EFFRATIO-PRODMC-BSMM0:err}   {\ensuremath{{0.002 } } }
\vdef{bdt2012:N-EFFRATIO-PRODTNP-BSMM0:val}   {\ensuremath{{0.378 } } }
\vdef{bdt2012:N-EFFRATIO-PRODTNP-BSMM0:err}   {\ensuremath{{749.948 } } }
\vdef{bdt2012:N-EFFRATIO-PRODTNPMC-BSMM0:val}   {\ensuremath{{0.357 } } }
\vdef{bdt2012:N-EFFRATIO-PRODTNPMC-BSMM0:err}   {\ensuremath{{0.002 } } }
\vdef{bdt2012:N-EFF-PRODTNP-BSMM0:val}   {\ensuremath{{0.0032 } } }
\vdef{bdt2012:N-EFF-PRODTNP-BSMM0:err}   {\ensuremath{{4.7297 } } }
\vdef{bdt2012:N-EFF-PRODTNP-BSMM0:tot}   {\ensuremath{{4.7297 } } }
\vdef{bdt2012:N-EFF-PRODTNP-BSMM0:all}   {\ensuremath{{(3.19 \pm 4729.66)\times 10^{-3}} } }
\vdef{bdt2012:N-EFF-PRODTNPMC-BSMM0:val}   {\ensuremath{{0.0023 } } }
\vdef{bdt2012:N-EFF-PRODTNPMC-BSMM0:err}   {\ensuremath{{0.0000 } } }
\vdef{bdt2012:N-EFF-PRODTNPMC-BSMM0:tot}   {\ensuremath{{0.0000 } } }
\vdef{bdt2012:N-EFF-PRODTNPMC-BSMM0:all}   {\ensuremath{{(2.32 \pm 0.01)\times 10^{-3}} } }
\vdef{bdt2012:N-ACC-BSMM0:val}   {\ensuremath{{0.033 } } }
\vdef{bdt2012:N-ACC-BSMM0:err}   {\ensuremath{{0.000 } } }
\vdef{bdt2012:N-ACC-BSMM0:tot}   {\ensuremath{{0.001 } } }
\vdef{bdt2012:N-ACC-BSMM0:all}   {\ensuremath{{(3.33 \pm 0.12)\times 10^{-2}} } }
\vdef{bdt2012:N-EFF-MU-PID-BSMM0:val}   {\ensuremath{{0.680 } } }
\vdef{bdt2012:N-EFF-MU-PID-BSMM0:err}   {\ensuremath{{0.000 } } }
\vdef{bdt2012:N-EFF-MU-PID-BSMM0:tot}   {\ensuremath{{0.027 } } }
\vdef{bdt2012:N-EFF-MU-PID-BSMM0:all}   {\ensuremath{{(68.02 \pm 2.72)\times 10^{-2}} } }
\vdef{bdt2012:N-EFFRHO-MU-PID-BSMM0:val}   {\ensuremath{{0.599 } } }
\vdef{bdt2012:N-EFFRHO-MU-PID-BSMM0:err}   {\ensuremath{{0.000 } } }
\vdef{bdt2012:N-EFFRHO-MU-PID-BSMM0:tot}   {\ensuremath{{0.000 } } }
\vdef{bdt2012:N-EFFRHO-MU-PID-BSMM0:all}   {\ensuremath{{(59.93 \pm 0.00)\times 10^{-2}} } }
\vdef{bdt2012:N-EFF-MU-PIDMC-BSMM0:val}   {\ensuremath{{0.645 } } }
\vdef{bdt2012:N-EFF-MU-PIDMC-BSMM0:err}   {\ensuremath{{0.000 } } }
\vdef{bdt2012:N-EFF-MU-PIDMC-BSMM0:tot}   {\ensuremath{{0.026 } } }
\vdef{bdt2012:N-EFF-MU-PIDMC-BSMM0:all}   {\ensuremath{{(64.45 \pm 2.58)\times 10^{-2}} } }
\vdef{bdt2012:N-EFFRHO-MU-PIDMC-BSMM0:val}   {\ensuremath{{0.568 } } }
\vdef{bdt2012:N-EFFRHO-MU-PIDMC-BSMM0:err}   {\ensuremath{{0.000 } } }
\vdef{bdt2012:N-EFFRHO-MU-PIDMC-BSMM0:tot}   {\ensuremath{{0.024 } } }
\vdef{bdt2012:N-EFFRHO-MU-PIDMC-BSMM0:all}   {\ensuremath{{(56.79 \pm 2.40)\times 10^{-2}} } }
\vdef{bdt2012:N-EFF-MU-MC-BSMM0:val}   {\ensuremath{{0.568 } } }
\vdef{bdt2012:N-EFF-MU-MC-BSMM0:err}   {\ensuremath{{0.000 } } }
\vdef{bdt2012:N-EFF-MU-MC-BSMM0:tot}   {\ensuremath{{0.023 } } }
\vdef{bdt2012:N-EFF-MU-MC-BSMM0:all}   {\ensuremath{{(56.79 \pm 2.27)\times 10^{-2}} } }
\vdef{bdt2012:N-EFF-TRIG-PID-BSMM0:val}   {\ensuremath{{0.719 } } }
\vdef{bdt2012:N-EFF-TRIG-PID-BSMM0:err}   {\ensuremath{{0.000 } } }
\vdef{bdt2012:N-EFF-TRIG-PID-BSMM0:tot}   {\ensuremath{{0.022 } } }
\vdef{bdt2012:N-EFF-TRIG-PID-BSMM0:all}   {\ensuremath{{(71.95 \pm 2.16)\times 10^{-2}} } }
\vdef{bdt2012:N-EFFRHO-TRIG-PID-BSMM0:val}   {\ensuremath{{0.668 } } }
\vdef{bdt2012:N-EFFRHO-TRIG-PID-BSMM0:err}   {\ensuremath{{0.000 } } }
\vdef{bdt2012:N-EFFRHO-TRIG-PID-BSMM0:tot}   {\ensuremath{{0.020 } } }
\vdef{bdt2012:N-EFFRHO-TRIG-PID-BSMM0:all}   {\ensuremath{{(66.81 \pm 2.00)\times 10^{-2}} } }
\vdef{bdt2012:N-EFF-TRIG-PIDMC-BSMM0:val}   {\ensuremath{{0.676 } } }
\vdef{bdt2012:N-EFF-TRIG-PIDMC-BSMM0:err}   {\ensuremath{{0.000 } } }
\vdef{bdt2012:N-EFF-TRIG-PIDMC-BSMM0:tot}   {\ensuremath{{0.020 } } }
\vdef{bdt2012:N-EFF-TRIG-PIDMC-BSMM0:all}   {\ensuremath{{(67.61 \pm 2.03)\times 10^{-2}} } }
\vdef{bdt2012:N-EFFRHO-TRIG-PIDMC-BSMM0:val}   {\ensuremath{{0.628 } } }
\vdef{bdt2012:N-EFFRHO-TRIG-PIDMC-BSMM0:err}   {\ensuremath{{0.000 } } }
\vdef{bdt2012:N-EFFRHO-TRIG-PIDMC-BSMM0:tot}   {\ensuremath{{0.019 } } }
\vdef{bdt2012:N-EFFRHO-TRIG-PIDMC-BSMM0:all}   {\ensuremath{{(62.78 \pm 1.88)\times 10^{-2}} } }
\vdef{bdt2012:N-EFF-TRIG-MC-BSMM0:val}   {\ensuremath{{0.628 } } }
\vdef{bdt2012:N-EFF-TRIG-MC-BSMM0:err}   {\ensuremath{{0.000 } } }
\vdef{bdt2012:N-EFF-TRIG-MC-BSMM0:tot}   {\ensuremath{{0.019 } } }
\vdef{bdt2012:N-EFF-TRIG-MC-BSMM0:all}   {\ensuremath{{(62.78 \pm 1.88)\times 10^{-2}} } }
\vdef{bdt2012:N-EFF-CAND-BSMM0:val}   {\ensuremath{{1.000 } } }
\vdef{bdt2012:N-EFF-CAND-BSMM0:err}   {\ensuremath{{0.000 } } }
\vdef{bdt2012:N-EFF-CAND-BSMM0:tot}   {\ensuremath{{0.010 } } }
\vdef{bdt2012:N-EFF-CAND-BSMM0:all}   {\ensuremath{{(99.99 \pm 1.00)\times 10^{-2}} } }
\vdef{bdt2012:N-EFF-ANA-BSMM0:val}   {\ensuremath{{0.196 } } }
\vdef{bdt2012:N-EFF-ANA-BSMM0:err}   {\ensuremath{{0.000 } } }
\vdef{bdt2012:N-EFF-ANA-BSMM0:tot}   {\ensuremath{{0.006 } } }
\vdef{bdt2012:N-EFF-ANA-BSMM0:all}   {\ensuremath{{(19.59 \pm 0.59)\times 10^{-2}} } }
\vdef{bdt2012:N-EFF-TOT-BDMM0:val}   {\ensuremath{{0.0024 } } }
\vdef{bdt2012:N-EFF-TOT-BDMM0:err}   {\ensuremath{{0.0000 } } }
\vdef{bdt2012:N-EFF-TOT-BDMM0:tot}   {\ensuremath{{0.0002 } } }
\vdef{bdt2012:N-EFF-TOT-BDMM0:all}   {\ensuremath{{(0.24 \pm 0.02)} } }
\vdef{bdt2012:N-EFF-PRODMC-BDMM0:val}   {\ensuremath{{0.0024 } } }
\vdef{bdt2012:N-EFF-PRODMC-BDMM0:err}   {\ensuremath{{0.0000 } } }
\vdef{bdt2012:N-EFF-PRODMC-BDMM0:tot}   {\ensuremath{{0.0000 } } }
\vdef{bdt2012:N-EFF-PRODMC-BDMM0:all}   {\ensuremath{{(2.36 \pm 0.02)\times 10^{-3}} } }
\vdef{bdt2012:N-EFF-PRODTNP-BDMM0:val}   {\ensuremath{{0.0032 } } }
\vdef{bdt2012:N-EFF-PRODTNP-BDMM0:err}   {\ensuremath{{5.3791 } } }
\vdef{bdt2012:N-EFF-PRODTNP-BDMM0:tot}   {\ensuremath{{5.3791 } } }
\vdef{bdt2012:N-EFF-PRODTNP-BDMM0:all}   {\ensuremath{{(3.25 \pm 5379.13)\times 10^{-3}} } }
\vdef{bdt2012:N-EFF-PRODTNPMC-BDMM0:val}   {\ensuremath{{0.00236 } } }
\vdef{bdt2012:N-EFF-PRODTNPMC-BDMM0:err}   {\ensuremath{{0.000018 } } }
\vdef{bdt2012:N-EFF-PRODTNPMC-BDMM0:tot}   {\ensuremath{{0.00002 } } }
\vdef{bdt2012:N-EFF-PRODTNPMC-BDMM0:all}   {\ensuremath{{(2.36 \pm 0.02)\times 10^{-3}} } }
\vdef{bdt2012:N-ACC-BDMM0:val}   {\ensuremath{{0.032 } } }
\vdef{bdt2012:N-ACC-BDMM0:err}   {\ensuremath{{0.000 } } }
\vdef{bdt2012:N-ACC-BDMM0:tot}   {\ensuremath{{0.001 } } }
\vdef{bdt2012:N-ACC-BDMM0:all}   {\ensuremath{{(3.24 \pm 0.12)\times 10^{-2}} } }
\vdef{bdt2012:N-EFF-MU-PID-BDMM0:val}   {\ensuremath{{0.680 } } }
\vdef{bdt2012:N-EFF-MU-PID-BDMM0:err}   {\ensuremath{{0.000 } } }
\vdef{bdt2012:N-EFF-MU-PID-BDMM0:tot}   {\ensuremath{{0.027 } } }
\vdef{bdt2012:N-EFF-MU-PID-BDMM0:all}   {\ensuremath{{(67.99 \pm 2.72)\times 10^{-2}} } }
\vdef{bdt2012:N-EFFRHO-MU-PID-BDMM0:val}   {\ensuremath{{0.599 } } }
\vdef{bdt2012:N-EFFRHO-MU-PID-BDMM0:err}   {\ensuremath{{0.000 } } }
\vdef{bdt2012:N-EFFRHO-MU-PID-BDMM0:tot}   {\ensuremath{{0.000 } } }
\vdef{bdt2012:N-EFFRHO-MU-PID-BDMM0:all}   {\ensuremath{{(59.95 \pm 0.00)\times 10^{-2}} } }
\vdef{bdt2012:N-EFF-MU-PIDMC-BDMM0:val}   {\ensuremath{{0.644 } } }
\vdef{bdt2012:N-EFF-MU-PIDMC-BDMM0:err}   {\ensuremath{{0.000 } } }
\vdef{bdt2012:N-EFF-MU-PIDMC-BDMM0:tot}   {\ensuremath{{0.026 } } }
\vdef{bdt2012:N-EFF-MU-PIDMC-BDMM0:all}   {\ensuremath{{(64.42 \pm 2.58)\times 10^{-2}} } }
\vdef{bdt2012:N-EFFRHO-MU-PIDMC-BDMM0:val}   {\ensuremath{{0.568 } } }
\vdef{bdt2012:N-EFFRHO-MU-PIDMC-BDMM0:err}   {\ensuremath{{0.000 } } }
\vdef{bdt2012:N-EFFRHO-MU-PIDMC-BDMM0:tot}   {\ensuremath{{0.024 } } }
\vdef{bdt2012:N-EFFRHO-MU-PIDMC-BDMM0:all}   {\ensuremath{{(56.80 \pm 2.40)\times 10^{-2}} } }
\vdef{bdt2012:N-EFF-MU-MC-BDMM0:val}   {\ensuremath{{0.568 } } }
\vdef{bdt2012:N-EFF-MU-MC-BDMM0:err}   {\ensuremath{{0.000 } } }
\vdef{bdt2012:N-EFF-MU-MC-BDMM0:tot}   {\ensuremath{{0.023 } } }
\vdef{bdt2012:N-EFF-MU-MC-BDMM0:all}   {\ensuremath{{(56.80 \pm 2.27)\times 10^{-2}} } }
\vdef{bdt2012:N-EFF-TRIG-PID-BDMM0:val}   {\ensuremath{{0.719 } } }
\vdef{bdt2012:N-EFF-TRIG-PID-BDMM0:err}   {\ensuremath{{0.000 } } }
\vdef{bdt2012:N-EFF-TRIG-PID-BDMM0:tot}   {\ensuremath{{0.022 } } }
\vdef{bdt2012:N-EFF-TRIG-PID-BDMM0:all}   {\ensuremath{{(71.92 \pm 2.16)\times 10^{-2}} } }
\vdef{bdt2012:N-EFFRHO-TRIG-PID-BDMM0:val}   {\ensuremath{{0.665 } } }
\vdef{bdt2012:N-EFFRHO-TRIG-PID-BDMM0:err}   {\ensuremath{{0.000 } } }
\vdef{bdt2012:N-EFFRHO-TRIG-PID-BDMM0:tot}   {\ensuremath{{0.020 } } }
\vdef{bdt2012:N-EFFRHO-TRIG-PID-BDMM0:all}   {\ensuremath{{(66.54 \pm 2.00)\times 10^{-2}} } }
\vdef{bdt2012:N-EFF-TRIG-PIDMC-BDMM0:val}   {\ensuremath{{0.675 } } }
\vdef{bdt2012:N-EFF-TRIG-PIDMC-BDMM0:err}   {\ensuremath{{0.000 } } }
\vdef{bdt2012:N-EFF-TRIG-PIDMC-BDMM0:tot}   {\ensuremath{{0.020 } } }
\vdef{bdt2012:N-EFF-TRIG-PIDMC-BDMM0:all}   {\ensuremath{{(67.55 \pm 2.03)\times 10^{-2}} } }
\vdef{bdt2012:N-EFFRHO-TRIG-PIDMC-BDMM0:val}   {\ensuremath{{0.625 } } }
\vdef{bdt2012:N-EFFRHO-TRIG-PIDMC-BDMM0:err}   {\ensuremath{{0.000 } } }
\vdef{bdt2012:N-EFFRHO-TRIG-PIDMC-BDMM0:tot}   {\ensuremath{{0.019 } } }
\vdef{bdt2012:N-EFFRHO-TRIG-PIDMC-BDMM0:all}   {\ensuremath{{(62.50 \pm 1.88)\times 10^{-2}} } }
\vdef{bdt2012:N-EFF-TRIG-MC-BDMM0:val}   {\ensuremath{{0.625 } } }
\vdef{bdt2012:N-EFF-TRIG-MC-BDMM0:err}   {\ensuremath{{0.000 } } }
\vdef{bdt2012:N-EFF-TRIG-MC-BDMM0:tot}   {\ensuremath{{0.019 } } }
\vdef{bdt2012:N-EFF-TRIG-MC-BDMM0:all}   {\ensuremath{{(62.50 \pm 1.88)\times 10^{-2}} } }
\vdef{bdt2012:N-EFF-CAND-BDMM0:val}   {\ensuremath{{1.000 } } }
\vdef{bdt2012:N-EFF-CAND-BDMM0:err}   {\ensuremath{{0.000 } } }
\vdef{bdt2012:N-EFF-CAND-BDMM0:tot}   {\ensuremath{{0.010 } } }
\vdef{bdt2012:N-EFF-CAND-BDMM0:all}   {\ensuremath{{(100.00 \pm 1.00)\times 10^{-2}} } }
\vdef{bdt2012:N-EFF-ANA-BDMM0:val}   {\ensuremath{{0.205 } } }
\vdef{bdt2012:N-EFF-ANA-BDMM0:err}   {\ensuremath{{0.000 } } }
\vdef{bdt2012:N-EFF-ANA-BDMM0:tot}   {\ensuremath{{0.006 } } }
\vdef{bdt2012:N-EFF-ANA-BDMM0:all}   {\ensuremath{{(20.53 \pm 0.62)\times 10^{-2}} } }
\vdef{bdt2012:N-EXP-OBS-BS0:val}   {\ensuremath{{18.41 } } }
\vdef{bdt2012:N-EXP-OBS-BS0:err}   {\ensuremath{{ 3.37 } } }
\vdef{bdt2012:N-EXP-OBS-BD0:val}   {\ensuremath{{ 7.88 } } }
\vdef{bdt2012:N-EXP-OBS-BD0:err}   {\ensuremath{{ 2.59 } } }
\vdef{bdt2012:N-OBS-BSMM0:val}   {\ensuremath{{16 } } }
\vdef{bdt2012:N-OBS-BDMM0:val}   {\ensuremath{{11 } } }
\vdef{bdt2012:N-OFFLO-RARE0:val}   {\ensuremath{{ 0.22 } } }
\vdef{bdt2012:N-OFFLO-RARE0:err}   {\ensuremath{{ 0.13 } } }
\vdef{bdt2012:N-OFFHI-RARE0:val}   {\ensuremath{{ 0.02 } } }
\vdef{bdt2012:N-OFFHI-RARE0:err}   {\ensuremath{{ 0.01 } } }
\vdef{bdt2012:N-PEAK-BKG-BS0:val}   {\ensuremath{{ 0.23 } } }
\vdef{bdt2012:N-PEAK-BKG-BS0:err}   {\ensuremath{{ 0.17 } } }
\vdef{bdt2012:N-PEAK-BKG-BD0:val}   {\ensuremath{{ 0.79 } } }
\vdef{bdt2012:N-PEAK-BKG-BD0:err}   {\ensuremath{{ 0.58 } } }
\vdef{bdt2012:N-TAU-BS0:val}   {\ensuremath{{ 0.17 } } }
\vdef{bdt2012:N-TAU-BS0:err}   {\ensuremath{{ 0.01 } } }
\vdef{bdt2012:N-TAU-BD0:val}   {\ensuremath{{ 0.15 } } }
\vdef{bdt2012:N-TAU-BD0:err}   {\ensuremath{{ 0.01 } } }
\vdef{bdt2012:N-OBS-OFFHI0:val}   {\ensuremath{{12 } } }
\vdef{bdt2012:N-OBS-OFFLO0:val}   {\ensuremath{{22 } } }
\vdef{bdt2012:N-EXP-SoverB0:val}   {\ensuremath{{ 1.98 } } }
\vdef{bdt2012:N-EXP-SoverSplusB0:val}   {\ensuremath{{ 2.76 } } }
\vdef{bdt2012:N-EXP2-SIG-BSMM1:val}   {\ensuremath{{ 3.56 } } }
\vdef{bdt2012:N-EXP2-SIG-BSMM1:err}   {\ensuremath{{ 0.53 } } }
\vdef{bdt2012:N-EXP2-SIG-BDMM1:val}   {\ensuremath{{0.30 } } }
\vdef{bdt2012:N-EXP2-SIG-BDMM1:err}   {\ensuremath{{0.03 } } }
\vdef{bdt2012:N-OBS-BKG1:val}   {\ensuremath{{7 } } }
\vdef{bdt2012:N-EXP-BSMM1:val}   {\ensuremath{{ 1.26 } } }
\vdef{bdt2012:N-EXP-BSMM1:err}   {\ensuremath{{ 0.63 } } }
\vdef{bdt2012:N-EXP-BDMM1:val}   {\ensuremath{{ 1.03 } } }
\vdef{bdt2012:N-EXP-BDMM1:err}   {\ensuremath{{ 0.51 } } }
\vdef{bdt2012:N-LOW-BD1:val}   {\ensuremath{{5.200 } } }
\vdef{bdt2012:N-HIGH-BD1:val}   {\ensuremath{{5.300 } } }
\vdef{bdt2012:N-LOW-BS1:val}   {\ensuremath{{5.300 } } }
\vdef{bdt2012:N-HIGH-BS1:val}   {\ensuremath{{5.450 } } }
\vdef{bdt2012:N-PSS1:val}   {\ensuremath{{0.754 } } }
\vdef{bdt2012:N-PSS1:err}   {\ensuremath{{0.001 } } }
\vdef{bdt2012:N-PSS1:tot}   {\ensuremath{{0.038 } } }
\vdef{bdt2012:N-PSD1:val}   {\ensuremath{{0.327 } } }
\vdef{bdt2012:N-PSD1:err}   {\ensuremath{{0.001 } } }
\vdef{bdt2012:N-PSD1:tot}   {\ensuremath{{0.016 } } }
\vdef{bdt2012:N-PDS1:val}   {\ensuremath{{0.141 } } }
\vdef{bdt2012:N-PDS1:err}   {\ensuremath{{0.001 } } }
\vdef{bdt2012:N-PDS1:tot}   {\ensuremath{{0.007 } } }
\vdef{bdt2012:N-PDD1:val}   {\ensuremath{{0.550 } } }
\vdef{bdt2012:N-PDD1:err}   {\ensuremath{{0.001 } } }
\vdef{bdt2012:N-PDD1:tot}   {\ensuremath{{0.028 } } }
\vdef{bdt2012:N-EFF-TOT-BSMM1:val}   {\ensuremath{{0.0009 } } }
\vdef{bdt2012:N-EFF-TOT-BSMM1:err}   {\ensuremath{{0.0000 } } }
\vdef{bdt2012:N-EFF-TOT-BSMM1:tot}   {\ensuremath{{0.0001 } } }
\vdef{bdt2012:N-EFF-TOT-BSMM1:all}   {\ensuremath{{(0.09 \pm 0.01)} } }
\vdef{bdt2012:N-EFF-PRODMC-BSMM1:val}   {\ensuremath{{0.0010 } } }
\vdef{bdt2012:N-EFF-PRODMC-BSMM1:err}   {\ensuremath{{0.0000 } } }
\vdef{bdt2012:N-EFF-PRODMC-BSMM1:tot}   {\ensuremath{{0.0000 } } }
\vdef{bdt2012:N-EFF-PRODMC-BSMM1:all}   {\ensuremath{{(0.96 \pm 0.00)\times 10^{-3}} } }
\vdef{bdt2012:N-EFFRATIO-TOT-BSMM1:val}   {\ensuremath{{0.223 } } }
\vdef{bdt2012:N-EFFRATIO-TOT-BSMM1:err}   {\ensuremath{{0.001 } } }
\vdef{bdt2012:N-EFFRATIO-PRODMC-BSMM1:val}   {\ensuremath{{0.219 } } }
\vdef{bdt2012:N-EFFRATIO-PRODMC-BSMM1:err}   {\ensuremath{{0.002 } } }
\vdef{bdt2012:N-EFFRATIO-PRODTNP-BSMM1:val}   {\ensuremath{{0.341 } } }
\vdef{bdt2012:N-EFFRATIO-PRODTNP-BSMM1:err}   {\ensuremath{{375.909 } } }
\vdef{bdt2012:N-EFFRATIO-PRODTNPMC-BSMM1:val}   {\ensuremath{{0.219 } } }
\vdef{bdt2012:N-EFFRATIO-PRODTNPMC-BSMM1:err}   {\ensuremath{{0.001 } } }
\vdef{bdt2012:N-EFF-PRODTNP-BSMM1:val}   {\ensuremath{{0.0010 } } }
\vdef{bdt2012:N-EFF-PRODTNP-BSMM1:err}   {\ensuremath{{0.8691 } } }
\vdef{bdt2012:N-EFF-PRODTNP-BSMM1:tot}   {\ensuremath{{0.8691 } } }
\vdef{bdt2012:N-EFF-PRODTNP-BSMM1:all}   {\ensuremath{{(1.01 \pm 869.11)\times 10^{-3}} } }
\vdef{bdt2012:N-EFF-PRODTNPMC-BSMM1:val}   {\ensuremath{{0.0010 } } }
\vdef{bdt2012:N-EFF-PRODTNPMC-BSMM1:err}   {\ensuremath{{0.0000 } } }
\vdef{bdt2012:N-EFF-PRODTNPMC-BSMM1:tot}   {\ensuremath{{0.0000 } } }
\vdef{bdt2012:N-EFF-PRODTNPMC-BSMM1:all}   {\ensuremath{{(0.96 \pm 0.00)\times 10^{-3}} } }
\vdef{bdt2012:N-ACC-BSMM1:val}   {\ensuremath{{0.024 } } }
\vdef{bdt2012:N-ACC-BSMM1:err}   {\ensuremath{{0.000 } } }
\vdef{bdt2012:N-ACC-BSMM1:tot}   {\ensuremath{{0.001 } } }
\vdef{bdt2012:N-ACC-BSMM1:all}   {\ensuremath{{(2.38 \pm 0.12)\times 10^{-2}} } }
\vdef{bdt2012:N-EFF-MU-PID-BSMM1:val}   {\ensuremath{{0.555 } } }
\vdef{bdt2012:N-EFF-MU-PID-BSMM1:err}   {\ensuremath{{0.000 } } }
\vdef{bdt2012:N-EFF-MU-PID-BSMM1:tot}   {\ensuremath{{0.044 } } }
\vdef{bdt2012:N-EFF-MU-PID-BSMM1:all}   {\ensuremath{{(55.54 \pm 4.44)\times 10^{-2}} } }
\vdef{bdt2012:N-EFFRHO-MU-PID-BSMM1:val}   {\ensuremath{{0.646 } } }
\vdef{bdt2012:N-EFFRHO-MU-PID-BSMM1:err}   {\ensuremath{{0.000 } } }
\vdef{bdt2012:N-EFFRHO-MU-PID-BSMM1:tot}   {\ensuremath{{0.000 } } }
\vdef{bdt2012:N-EFFRHO-MU-PID-BSMM1:all}   {\ensuremath{{(64.64 \pm 0.00)\times 10^{-2}} } }
\vdef{bdt2012:N-EFF-MU-PIDMC-BSMM1:val}   {\ensuremath{{0.557 } } }
\vdef{bdt2012:N-EFF-MU-PIDMC-BSMM1:err}   {\ensuremath{{0.000 } } }
\vdef{bdt2012:N-EFF-MU-PIDMC-BSMM1:tot}   {\ensuremath{{0.045 } } }
\vdef{bdt2012:N-EFF-MU-PIDMC-BSMM1:all}   {\ensuremath{{(55.74 \pm 4.46)\times 10^{-2}} } }
\vdef{bdt2012:N-EFFRHO-MU-PIDMC-BSMM1:val}   {\ensuremath{{0.649 } } }
\vdef{bdt2012:N-EFFRHO-MU-PIDMC-BSMM1:err}   {\ensuremath{{0.000 } } }
\vdef{bdt2012:N-EFFRHO-MU-PIDMC-BSMM1:tot}   {\ensuremath{{0.052 } } }
\vdef{bdt2012:N-EFFRHO-MU-PIDMC-BSMM1:all}   {\ensuremath{{(64.87 \pm 5.17)\times 10^{-2}} } }
\vdef{bdt2012:N-EFF-MU-MC-BSMM1:val}   {\ensuremath{{0.649 } } }
\vdef{bdt2012:N-EFF-MU-MC-BSMM1:err}   {\ensuremath{{0.001 } } }
\vdef{bdt2012:N-EFF-MU-MC-BSMM1:tot}   {\ensuremath{{0.052 } } }
\vdef{bdt2012:N-EFF-MU-MC-BSMM1:all}   {\ensuremath{{(64.87 \pm 5.19)\times 10^{-2}} } }
\vdef{bdt2012:N-EFF-TRIG-PID-BSMM1:val}   {\ensuremath{{0.743 } } }
\vdef{bdt2012:N-EFF-TRIG-PID-BSMM1:err}   {\ensuremath{{0.000 } } }
\vdef{bdt2012:N-EFF-TRIG-PID-BSMM1:tot}   {\ensuremath{{0.045 } } }
\vdef{bdt2012:N-EFF-TRIG-PID-BSMM1:all}   {\ensuremath{{(74.32 \pm 4.46)\times 10^{-2}} } }
\vdef{bdt2012:N-EFFRHO-TRIG-PID-BSMM1:val}   {\ensuremath{{0.622 } } }
\vdef{bdt2012:N-EFFRHO-TRIG-PID-BSMM1:err}   {\ensuremath{{0.000 } } }
\vdef{bdt2012:N-EFFRHO-TRIG-PID-BSMM1:tot}   {\ensuremath{{0.037 } } }
\vdef{bdt2012:N-EFFRHO-TRIG-PID-BSMM1:all}   {\ensuremath{{(62.20 \pm 3.73)\times 10^{-2}} } }
\vdef{bdt2012:N-EFF-TRIG-PIDMC-BSMM1:val}   {\ensuremath{{0.726 } } }
\vdef{bdt2012:N-EFF-TRIG-PIDMC-BSMM1:err}   {\ensuremath{{0.000 } } }
\vdef{bdt2012:N-EFF-TRIG-PIDMC-BSMM1:tot}   {\ensuremath{{0.044 } } }
\vdef{bdt2012:N-EFF-TRIG-PIDMC-BSMM1:all}   {\ensuremath{{(72.60 \pm 4.36)\times 10^{-2}} } }
\vdef{bdt2012:N-EFFRHO-TRIG-PIDMC-BSMM1:val}   {\ensuremath{{0.608 } } }
\vdef{bdt2012:N-EFFRHO-TRIG-PIDMC-BSMM1:err}   {\ensuremath{{0.000 } } }
\vdef{bdt2012:N-EFFRHO-TRIG-PIDMC-BSMM1:tot}   {\ensuremath{{0.036 } } }
\vdef{bdt2012:N-EFFRHO-TRIG-PIDMC-BSMM1:all}   {\ensuremath{{(60.77 \pm 3.65)\times 10^{-2}} } }
\vdef{bdt2012:N-EFF-TRIG-MC-BSMM1:val}   {\ensuremath{{0.608 } } }
\vdef{bdt2012:N-EFF-TRIG-MC-BSMM1:err}   {\ensuremath{{0.001 } } }
\vdef{bdt2012:N-EFF-TRIG-MC-BSMM1:tot}   {\ensuremath{{0.036 } } }
\vdef{bdt2012:N-EFF-TRIG-MC-BSMM1:all}   {\ensuremath{{(60.77 \pm 3.65)\times 10^{-2}} } }
\vdef{bdt2012:N-EFF-CAND-BSMM1:val}   {\ensuremath{{1.000 } } }
\vdef{bdt2012:N-EFF-CAND-BSMM1:err}   {\ensuremath{{0.000 } } }
\vdef{bdt2012:N-EFF-CAND-BSMM1:tot}   {\ensuremath{{0.010 } } }
\vdef{bdt2012:N-EFF-CAND-BSMM1:all}   {\ensuremath{{(99.99 \pm 1.00)\times 10^{-2}} } }
\vdef{bdt2012:N-EFF-ANA-BSMM1:val}   {\ensuremath{{0.103 } } }
\vdef{bdt2012:N-EFF-ANA-BSMM1:err}   {\ensuremath{{0.000 } } }
\vdef{bdt2012:N-EFF-ANA-BSMM1:tot}   {\ensuremath{{0.003 } } }
\vdef{bdt2012:N-EFF-ANA-BSMM1:all}   {\ensuremath{{(10.28 \pm 0.31)\times 10^{-2}} } }
\vdef{bdt2012:N-EFF-TOT-BDMM1:val}   {\ensuremath{{0.0010 } } }
\vdef{bdt2012:N-EFF-TOT-BDMM1:err}   {\ensuremath{{0.0000 } } }
\vdef{bdt2012:N-EFF-TOT-BDMM1:tot}   {\ensuremath{{0.0001 } } }
\vdef{bdt2012:N-EFF-TOT-BDMM1:all}   {\ensuremath{{(0.10 \pm 0.01)} } }
\vdef{bdt2012:N-EFF-PRODMC-BDMM1:val}   {\ensuremath{{0.0010 } } }
\vdef{bdt2012:N-EFF-PRODMC-BDMM1:err}   {\ensuremath{{0.0000 } } }
\vdef{bdt2012:N-EFF-PRODMC-BDMM1:tot}   {\ensuremath{{0.0000 } } }
\vdef{bdt2012:N-EFF-PRODMC-BDMM1:all}   {\ensuremath{{(0.97 \pm 0.01)\times 10^{-3}} } }
\vdef{bdt2012:N-EFF-PRODTNP-BDMM1:val}   {\ensuremath{{0.0010 } } }
\vdef{bdt2012:N-EFF-PRODTNP-BDMM1:err}   {\ensuremath{{0.9751 } } }
\vdef{bdt2012:N-EFF-PRODTNP-BDMM1:tot}   {\ensuremath{{0.9751 } } }
\vdef{bdt2012:N-EFF-PRODTNP-BDMM1:all}   {\ensuremath{{(1.02 \pm 975.08)\times 10^{-3}} } }
\vdef{bdt2012:N-EFF-PRODTNPMC-BDMM1:val}   {\ensuremath{{0.00097 } } }
\vdef{bdt2012:N-EFF-PRODTNPMC-BDMM1:err}   {\ensuremath{{0.000009 } } }
\vdef{bdt2012:N-EFF-PRODTNPMC-BDMM1:tot}   {\ensuremath{{0.00001 } } }
\vdef{bdt2012:N-EFF-PRODTNPMC-BDMM1:all}   {\ensuremath{{(0.97 \pm 0.01)\times 10^{-3}} } }
\vdef{bdt2012:N-ACC-BDMM1:val}   {\ensuremath{{0.023 } } }
\vdef{bdt2012:N-ACC-BDMM1:err}   {\ensuremath{{0.000 } } }
\vdef{bdt2012:N-ACC-BDMM1:tot}   {\ensuremath{{0.001 } } }
\vdef{bdt2012:N-ACC-BDMM1:all}   {\ensuremath{{(2.29 \pm 0.12)\times 10^{-2}} } }
\vdef{bdt2012:N-EFF-MU-PID-BDMM1:val}   {\ensuremath{{0.557 } } }
\vdef{bdt2012:N-EFF-MU-PID-BDMM1:err}   {\ensuremath{{0.000 } } }
\vdef{bdt2012:N-EFF-MU-PID-BDMM1:tot}   {\ensuremath{{0.045 } } }
\vdef{bdt2012:N-EFF-MU-PID-BDMM1:all}   {\ensuremath{{(55.66 \pm 4.45)\times 10^{-2}} } }
\vdef{bdt2012:N-EFFRHO-MU-PID-BDMM1:val}   {\ensuremath{{0.644 } } }
\vdef{bdt2012:N-EFFRHO-MU-PID-BDMM1:err}   {\ensuremath{{0.000 } } }
\vdef{bdt2012:N-EFFRHO-MU-PID-BDMM1:tot}   {\ensuremath{{0.000 } } }
\vdef{bdt2012:N-EFFRHO-MU-PID-BDMM1:all}   {\ensuremath{{(64.44 \pm 0.00)\times 10^{-2}} } }
\vdef{bdt2012:N-EFF-MU-PIDMC-BDMM1:val}   {\ensuremath{{0.558 } } }
\vdef{bdt2012:N-EFF-MU-PIDMC-BDMM1:err}   {\ensuremath{{0.000 } } }
\vdef{bdt2012:N-EFF-MU-PIDMC-BDMM1:tot}   {\ensuremath{{0.045 } } }
\vdef{bdt2012:N-EFF-MU-PIDMC-BDMM1:all}   {\ensuremath{{(55.79 \pm 4.46)\times 10^{-2}} } }
\vdef{bdt2012:N-EFFRHO-MU-PIDMC-BDMM1:val}   {\ensuremath{{0.646 } } }
\vdef{bdt2012:N-EFFRHO-MU-PIDMC-BDMM1:err}   {\ensuremath{{0.000 } } }
\vdef{bdt2012:N-EFFRHO-MU-PIDMC-BDMM1:tot}   {\ensuremath{{0.052 } } }
\vdef{bdt2012:N-EFFRHO-MU-PIDMC-BDMM1:all}   {\ensuremath{{(64.59 \pm 5.16)\times 10^{-2}} } }
\vdef{bdt2012:N-EFF-MU-MC-BDMM1:val}   {\ensuremath{{0.646 } } }
\vdef{bdt2012:N-EFF-MU-MC-BDMM1:err}   {\ensuremath{{0.001 } } }
\vdef{bdt2012:N-EFF-MU-MC-BDMM1:tot}   {\ensuremath{{0.052 } } }
\vdef{bdt2012:N-EFF-MU-MC-BDMM1:all}   {\ensuremath{{(64.59 \pm 5.17)\times 10^{-2}} } }
\vdef{bdt2012:N-EFF-TRIG-PID-BDMM1:val}   {\ensuremath{{0.743 } } }
\vdef{bdt2012:N-EFF-TRIG-PID-BDMM1:err}   {\ensuremath{{0.000 } } }
\vdef{bdt2012:N-EFF-TRIG-PID-BDMM1:tot}   {\ensuremath{{0.045 } } }
\vdef{bdt2012:N-EFF-TRIG-PID-BDMM1:all}   {\ensuremath{{(74.28 \pm 4.46)\times 10^{-2}} } }
\vdef{bdt2012:N-EFFRHO-TRIG-PID-BDMM1:val}   {\ensuremath{{0.620 } } }
\vdef{bdt2012:N-EFFRHO-TRIG-PID-BDMM1:err}   {\ensuremath{{0.000 } } }
\vdef{bdt2012:N-EFFRHO-TRIG-PID-BDMM1:tot}   {\ensuremath{{0.037 } } }
\vdef{bdt2012:N-EFFRHO-TRIG-PID-BDMM1:all}   {\ensuremath{{(62.02 \pm 3.72)\times 10^{-2}} } }
\vdef{bdt2012:N-EFF-TRIG-PIDMC-BDMM1:val}   {\ensuremath{{0.725 } } }
\vdef{bdt2012:N-EFF-TRIG-PIDMC-BDMM1:err}   {\ensuremath{{0.000 } } }
\vdef{bdt2012:N-EFF-TRIG-PIDMC-BDMM1:tot}   {\ensuremath{{0.044 } } }
\vdef{bdt2012:N-EFF-TRIG-PIDMC-BDMM1:all}   {\ensuremath{{(72.52 \pm 4.35)\times 10^{-2}} } }
\vdef{bdt2012:N-EFFRHO-TRIG-PIDMC-BDMM1:val}   {\ensuremath{{0.605 } } }
\vdef{bdt2012:N-EFFRHO-TRIG-PIDMC-BDMM1:err}   {\ensuremath{{0.000 } } }
\vdef{bdt2012:N-EFFRHO-TRIG-PIDMC-BDMM1:tot}   {\ensuremath{{0.036 } } }
\vdef{bdt2012:N-EFFRHO-TRIG-PIDMC-BDMM1:all}   {\ensuremath{{(60.55 \pm 3.63)\times 10^{-2}} } }
\vdef{bdt2012:N-EFF-TRIG-MC-BDMM1:val}   {\ensuremath{{0.605 } } }
\vdef{bdt2012:N-EFF-TRIG-MC-BDMM1:err}   {\ensuremath{{0.001 } } }
\vdef{bdt2012:N-EFF-TRIG-MC-BDMM1:tot}   {\ensuremath{{0.036 } } }
\vdef{bdt2012:N-EFF-TRIG-MC-BDMM1:all}   {\ensuremath{{(60.55 \pm 3.63)\times 10^{-2}} } }
\vdef{bdt2012:N-EFF-CAND-BDMM1:val}   {\ensuremath{{1.000 } } }
\vdef{bdt2012:N-EFF-CAND-BDMM1:err}   {\ensuremath{{0.000 } } }
\vdef{bdt2012:N-EFF-CAND-BDMM1:tot}   {\ensuremath{{0.010 } } }
\vdef{bdt2012:N-EFF-CAND-BDMM1:all}   {\ensuremath{{(99.99 \pm 1.00)\times 10^{-2}} } }
\vdef{bdt2012:N-EFF-ANA-BDMM1:val}   {\ensuremath{{0.108 } } }
\vdef{bdt2012:N-EFF-ANA-BDMM1:err}   {\ensuremath{{0.000 } } }
\vdef{bdt2012:N-EFF-ANA-BDMM1:tot}   {\ensuremath{{0.003 } } }
\vdef{bdt2012:N-EFF-ANA-BDMM1:all}   {\ensuremath{{(10.79 \pm 0.32)\times 10^{-2}} } }
\vdef{bdt2012:N-EXP-OBS-BS1:val}   {\ensuremath{{ 5.58 } } }
\vdef{bdt2012:N-EXP-OBS-BS1:err}   {\ensuremath{{ 0.83 } } }
\vdef{bdt2012:N-EXP-OBS-BD1:val}   {\ensuremath{{ 1.89 } } }
\vdef{bdt2012:N-EXP-OBS-BD1:err}   {\ensuremath{{ 0.51 } } }
\vdef{bdt2012:N-OBS-BSMM1:val}   {\ensuremath{{4 } } }
\vdef{bdt2012:N-OBS-BDMM1:val}   {\ensuremath{{3 } } }
\vdef{bdt2012:N-OFFLO-RARE1:val}   {\ensuremath{{ 0.11 } } }
\vdef{bdt2012:N-OFFLO-RARE1:err}   {\ensuremath{{ 0.06 } } }
\vdef{bdt2012:N-OFFHI-RARE1:val}   {\ensuremath{{ 0.01 } } }
\vdef{bdt2012:N-OFFHI-RARE1:err}   {\ensuremath{{ 0.00 } } }
\vdef{bdt2012:N-PEAK-BKG-BS1:val}   {\ensuremath{{ 0.09 } } }
\vdef{bdt2012:N-PEAK-BKG-BS1:err}   {\ensuremath{{ 0.07 } } }
\vdef{bdt2012:N-PEAK-BKG-BD1:val}   {\ensuremath{{ 0.19 } } }
\vdef{bdt2012:N-PEAK-BKG-BD1:err}   {\ensuremath{{ 0.14 } } }
\vdef{bdt2012:N-TAU-BS1:val}   {\ensuremath{{ 0.18 } } }
\vdef{bdt2012:N-TAU-BS1:err}   {\ensuremath{{ 0.01 } } }
\vdef{bdt2012:N-TAU-BD1:val}   {\ensuremath{{ 0.15 } } }
\vdef{bdt2012:N-TAU-BD1:err}   {\ensuremath{{ 0.01 } } }
\vdef{bdt2012:N-OBS-OFFHI1:val}   {\ensuremath{{3 } } }
\vdef{bdt2012:N-OBS-OFFLO1:val}   {\ensuremath{{4 } } }
\vdef{bdt2012:N-EXP-SoverB1:val}   {\ensuremath{{ 2.83 } } }
\vdef{bdt2012:N-EXP-SoverSplusB1:val}   {\ensuremath{{ 1.62 } } }
\vdef{bdt2012:SgBd0:val}  {\ensuremath{{0.906 } } }
\vdef{bdt2012:SgBd0:e1}   {\ensuremath{{0.952 } } }
\vdef{bdt2012:SgBd0:e2}   {\ensuremath{{0.136 } } }
\vdef{bdt2012:SgBs0:val}  {\ensuremath{{11.463 } } }
\vdef{bdt2012:SgBs0:e1}   {\ensuremath{{3.386 } } }
\vdef{bdt2012:SgBs0:e2}   {\ensuremath{{1.720 } } }
\vdef{bdt2012:SgLo0:val}  {\ensuremath{{0.263 } } }
\vdef{bdt2012:SgLo0:e1}   {\ensuremath{{0.512 } } }
\vdef{bdt2012:SgLo0:e2}   {\ensuremath{{0.039 } } }
\vdef{bdt2012:SgHi0:val}  {\ensuremath{{0.394 } } }
\vdef{bdt2012:SgHi0:e1}   {\ensuremath{{0.628 } } }
\vdef{bdt2012:SgHi0:e2}   {\ensuremath{{0.059 } } }
\vdef{bdt2012:BdBd0:val}  {\ensuremath{{1.001 } } }
\vdef{bdt2012:BdBd0:e1}   {\ensuremath{{1.000 } } }
\vdef{bdt2012:BdBd0:e2}   {\ensuremath{{0.100 } } }
\vdef{bdt2012:BdBs0:val}  {\ensuremath{{0.424 } } }
\vdef{bdt2012:BdBs0:e1}   {\ensuremath{{0.651 } } }
\vdef{bdt2012:BdBs0:e2}   {\ensuremath{{0.042 } } }
\vdef{bdt2012:BdLo0:val}  {\ensuremath{{0.099 } } }
\vdef{bdt2012:BdLo0:e1}   {\ensuremath{{0.314 } } }
\vdef{bdt2012:BdLo0:e2}   {\ensuremath{{0.010 } } }
\vdef{bdt2012:BdHi0:val}  {\ensuremath{{0.001 } } }
\vdef{bdt2012:BdHi0:e1}   {\ensuremath{{0.028 } } }
\vdef{bdt2012:BdHi0:e2}   {\ensuremath{{0.000 } } }
\vdef{bdt2012:BgPeakLo0:val}   {\ensuremath{{0.222 } } }
\vdef{bdt2012:BgPeakLo0:e1}   {\ensuremath{{0.011 } } }
\vdef{bdt2012:BgPeakLo0:e2}   {\ensuremath{{0.158 } } }
\vdef{bdt2012:BgPeakBd0:val}   {\ensuremath{{0.792 } } }
\vdef{bdt2012:BgPeakBd0:e1}   {\ensuremath{{0.040 } } }
\vdef{bdt2012:BgPeakBd0:e2}   {\ensuremath{{0.577 } } }
\vdef{bdt2012:BgPeakBs0:val}   {\ensuremath{{0.234 } } }
\vdef{bdt2012:BgPeakBs0:e1}   {\ensuremath{{0.012 } } }
\vdef{bdt2012:BgPeakBs0:e2}   {\ensuremath{{0.173 } } }
\vdef{bdt2012:BgPeakHi0:val}   {\ensuremath{{0.019 } } }
\vdef{bdt2012:BgPeakHi0:e1}   {\ensuremath{{0.001 } } }
\vdef{bdt2012:BgPeakHi0:e2}   {\ensuremath{{0.013 } } }
\vdef{bdt2012:BgRslLo0:val}   {\ensuremath{{48.761 } } }
\vdef{bdt2012:BgRslLo0:e1}   {\ensuremath{{2.438 } } }
\vdef{bdt2012:BgRslLo0:e2}   {\ensuremath{{36.409 } } }
\vdef{bdt2012:BgRslBd0:val}   {\ensuremath{{8.763 } } }
\vdef{bdt2012:BgRslBd0:e1}   {\ensuremath{{0.438 } } }
\vdef{bdt2012:BgRslBd0:e2}   {\ensuremath{{9.133 } } }
\vdef{bdt2012:BgRslBs0:val}   {\ensuremath{{6.284 } } }
\vdef{bdt2012:BgRslBs0:e1}   {\ensuremath{{0.314 } } }
\vdef{bdt2012:BgRslBs0:e2}   {\ensuremath{{6.902 } } }
\vdef{bdt2012:BgRslHi0:val}   {\ensuremath{{0.677 } } }
\vdef{bdt2012:BgRslHi0:e1}   {\ensuremath{{0.034 } } }
\vdef{bdt2012:BgRslHi0:e2}   {\ensuremath{{0.746 } } }
\vdef{bdt2012:BgRareLo0:val}   {\ensuremath{{48.983 } } }
\vdef{bdt2012:BgRareLo0:e1}   {\ensuremath{{2.438 } } }
\vdef{bdt2012:BgRareLo0:e2}   {\ensuremath{{36.409 } } }
\vdef{bdt2012:BgRareBd0:val}   {\ensuremath{{9.555 } } }
\vdef{bdt2012:BgRareBd0:e1}   {\ensuremath{{0.440 } } }
\vdef{bdt2012:BgRareBd0:e2}   {\ensuremath{{9.151 } } }
\vdef{bdt2012:BgRareBs0:val}   {\ensuremath{{6.518 } } }
\vdef{bdt2012:BgRareBs0:e1}   {\ensuremath{{0.314 } } }
\vdef{bdt2012:BgRareBs0:e2}   {\ensuremath{{6.904 } } }
\vdef{bdt2012:BgRareHi0:val}   {\ensuremath{{0.695 } } }
\vdef{bdt2012:BgRareHi0:e1}   {\ensuremath{{0.034 } } }
\vdef{bdt2012:BgRareHi0:e2}   {\ensuremath{{0.746 } } }
\vdef{bdt2012:BgRslsLo0:val}   {\ensuremath{{14.000 } } }
\vdef{bdt2012:BgRslsLo0:e1}   {\ensuremath{{0.700 } } }
\vdef{bdt2012:BgRslsLo0:e2}   {\ensuremath{{10.453 } } }
\vdef{bdt2012:BgRslsBd0:val}   {\ensuremath{{2.516 } } }
\vdef{bdt2012:BgRslsBd0:e1}   {\ensuremath{{0.126 } } }
\vdef{bdt2012:BgRslsBd0:e2}   {\ensuremath{{2.622 } } }
\vdef{bdt2012:BgRslsBs0:val}   {\ensuremath{{1.804 } } }
\vdef{bdt2012:BgRslsBs0:e1}   {\ensuremath{{0.090 } } }
\vdef{bdt2012:BgRslsBs0:e2}   {\ensuremath{{1.982 } } }
\vdef{bdt2012:BgRslsHi0:val}   {\ensuremath{{0.194 } } }
\vdef{bdt2012:BgRslsHi0:e1}   {\ensuremath{{0.010 } } }
\vdef{bdt2012:BgRslsHi0:e2}   {\ensuremath{{0.214 } } }
\vdef{bdt2012:BgCombLo0:val}   {\ensuremath{{8.000 } } }
\vdef{bdt2012:BgCombLo0:e1}   {\ensuremath{{2.309 } } }
\vdef{bdt2012:BgCombLo0:e2}   {\ensuremath{{2.444 } } }
\vdef{bdt2012:BgCombBd0:val}   {\ensuremath{{2.667 } } }
\vdef{bdt2012:BgCombBd0:e1}   {\ensuremath{{0.770 } } }
\vdef{bdt2012:BgCombBd0:e2}   {\ensuremath{{0.815 } } }
\vdef{bdt2012:BgCombBs0:val}   {\ensuremath{{4.000 } } }
\vdef{bdt2012:BgCombBs0:e1}   {\ensuremath{{1.155 } } }
\vdef{bdt2012:BgCombBs0:e2}   {\ensuremath{{1.222 } } }
\vdef{bdt2012:BgCombHi0:val}   {\ensuremath{{12.000 } } }
\vdef{bdt2012:BgCombHi0:e1}   {\ensuremath{{3.464 } } }
\vdef{bdt2012:BgCombHi0:e2}   {\ensuremath{{3.666 } } }
\vdef{bdt2012:BgNonpLo0:val}   {\ensuremath{{22.000 } } }
\vdef{bdt2012:BgNonpLo0:e1}   {\ensuremath{{2.413 } } }
\vdef{bdt2012:BgNonpLo0:e2}   {\ensuremath{{10.735 } } }
\vdef{bdt2012:BgNonpBd0:val}   {\ensuremath{{5.183 } } }
\vdef{bdt2012:BgNonpBd0:e1}   {\ensuremath{{0.780 } } }
\vdef{bdt2012:BgNonpBd0:e2}   {\ensuremath{{2.746 } } }
\vdef{bdt2012:BgNonpBs0:val}   {\ensuremath{{5.804 } } }
\vdef{bdt2012:BgNonpBs0:e1}   {\ensuremath{{1.158 } } }
\vdef{bdt2012:BgNonpBs0:e2}   {\ensuremath{{2.143 } } }
\vdef{bdt2012:BgNonpHi0:val}   {\ensuremath{{12.194 } } }
\vdef{bdt2012:BgNonpHi0:e1}   {\ensuremath{{3.464 } } }
\vdef{bdt2012:BgNonpHi0:e2}   {\ensuremath{{3.672 } } }
\vdef{bdt2012:BgTotLo0:val}   {\ensuremath{{22.222 } } }
\vdef{bdt2012:BgTotLo0:e1}   {\ensuremath{{0.000 } } }
\vdef{bdt2012:BgTotLo0:e2}   {\ensuremath{{10.736 } } }
\vdef{bdt2012:BgTotBd0:val}   {\ensuremath{{5.975 } } }
\vdef{bdt2012:BgTotBd0:e1}   {\ensuremath{{0.000 } } }
\vdef{bdt2012:BgTotBd0:e2}   {\ensuremath{{2.806 } } }
\vdef{bdt2012:BgTotBs0:val}   {\ensuremath{{6.039 } } }
\vdef{bdt2012:BgTotBs0:e1}   {\ensuremath{{0.000 } } }
\vdef{bdt2012:BgTotBs0:e2}   {\ensuremath{{2.149 } } }
\vdef{bdt2012:BgTotHi0:val}   {\ensuremath{{12.213 } } }
\vdef{bdt2012:BgTotHi0:e1}   {\ensuremath{{0.000 } } }
\vdef{bdt2012:BgTotHi0:e2}   {\ensuremath{{3.672 } } }
\vdef{bdt2012:SgAndBgLo0:val}   {\ensuremath{{22.583 } } }
\vdef{bdt2012:SgAndBgLo0:e1}   {\ensuremath{{0.000 } } }
\vdef{bdt2012:SgAndBgLo0:e2}   {\ensuremath{{10.737 } } }
\vdef{bdt2012:SgAndBgBd0:val}   {\ensuremath{{7.9 } } }
\vdef{bdt2012:SgAndBgBd0:e1}   {\ensuremath{{0.000 } } }
\vdef{bdt2012:SgAndBgBd0:e2}   {\ensuremath{{3.0 } } }
\vdef{bdt2012:SgAndBgBs0:val}   {\ensuremath{{17.9 } } }
\vdef{bdt2012:SgAndBgBs0:e1}   {\ensuremath{{0.000 } } }
\vdef{bdt2012:SgAndBgBs0:e2}   {\ensuremath{{2.8 } } }
\vdef{bdt2012:SgAndBgHi0:val}   {\ensuremath{{12.608 } } }
\vdef{bdt2012:SgAndBgHi0:e1}   {\ensuremath{{0.000 } } }
\vdef{bdt2012:SgAndBgHi0:e2}   {\ensuremath{{3.673 } } }
\vdef{bdt2012:SgBd1:val}  {\ensuremath{{0.669 } } }
\vdef{bdt2012:SgBd1:e1}   {\ensuremath{{0.818 } } }
\vdef{bdt2012:SgBd1:e2}   {\ensuremath{{0.100 } } }
\vdef{bdt2012:SgBs1:val}  {\ensuremath{{3.565 } } }
\vdef{bdt2012:SgBs1:e1}   {\ensuremath{{1.888 } } }
\vdef{bdt2012:SgBs1:e2}   {\ensuremath{{0.535 } } }
\vdef{bdt2012:SgLo1:val}  {\ensuremath{{0.114 } } }
\vdef{bdt2012:SgLo1:e1}   {\ensuremath{{0.338 } } }
\vdef{bdt2012:SgLo1:e2}   {\ensuremath{{0.017 } } }
\vdef{bdt2012:SgHi1:val}  {\ensuremath{{0.402 } } }
\vdef{bdt2012:SgHi1:e1}   {\ensuremath{{0.634 } } }
\vdef{bdt2012:SgHi1:e2}   {\ensuremath{{0.060 } } }
\vdef{bdt2012:BdBd1:val}  {\ensuremath{{0.301 } } }
\vdef{bdt2012:BdBd1:e1}   {\ensuremath{{0.549 } } }
\vdef{bdt2012:BdBd1:e2}   {\ensuremath{{0.030 } } }
\vdef{bdt2012:BdBs1:val}  {\ensuremath{{0.179 } } }
\vdef{bdt2012:BdBs1:e1}   {\ensuremath{{0.423 } } }
\vdef{bdt2012:BdBs1:e2}   {\ensuremath{{0.018 } } }
\vdef{bdt2012:BdLo1:val}  {\ensuremath{{0.068 } } }
\vdef{bdt2012:BdLo1:e1}   {\ensuremath{{0.261 } } }
\vdef{bdt2012:BdLo1:e2}   {\ensuremath{{0.007 } } }
\vdef{bdt2012:BdHi1:val}  {\ensuremath{{0.002 } } }
\vdef{bdt2012:BdHi1:e1}   {\ensuremath{{0.045 } } }
\vdef{bdt2012:BdHi1:e2}   {\ensuremath{{0.000 } } }
\vdef{bdt2012:BgPeakLo1:val}   {\ensuremath{{0.106 } } }
\vdef{bdt2012:BgPeakLo1:e1}   {\ensuremath{{0.005 } } }
\vdef{bdt2012:BgPeakLo1:e2}   {\ensuremath{{0.076 } } }
\vdef{bdt2012:BgPeakBd1:val}   {\ensuremath{{0.191 } } }
\vdef{bdt2012:BgPeakBd1:e1}   {\ensuremath{{0.010 } } }
\vdef{bdt2012:BgPeakBd1:e2}   {\ensuremath{{0.139 } } }
\vdef{bdt2012:BgPeakBs1:val}   {\ensuremath{{0.090 } } }
\vdef{bdt2012:BgPeakBs1:e1}   {\ensuremath{{0.004 } } }
\vdef{bdt2012:BgPeakBs1:e2}   {\ensuremath{{0.065 } } }
\vdef{bdt2012:BgPeakHi1:val}   {\ensuremath{{0.009 } } }
\vdef{bdt2012:BgPeakHi1:e1}   {\ensuremath{{0.000 } } }
\vdef{bdt2012:BgPeakHi1:e2}   {\ensuremath{{0.006 } } }
\vdef{bdt2012:BgRslLo1:val}   {\ensuremath{{11.889 } } }
\vdef{bdt2012:BgRslLo1:e1}   {\ensuremath{{0.594 } } }
\vdef{bdt2012:BgRslLo1:e2}   {\ensuremath{{8.308 } } }
\vdef{bdt2012:BgRslBd1:val}   {\ensuremath{{2.061 } } }
\vdef{bdt2012:BgRslBd1:e1}   {\ensuremath{{0.103 } } }
\vdef{bdt2012:BgRslBd1:e2}   {\ensuremath{{2.047 } } }
\vdef{bdt2012:BgRslBs1:val}   {\ensuremath{{1.473 } } }
\vdef{bdt2012:BgRslBs1:e1}   {\ensuremath{{0.074 } } }
\vdef{bdt2012:BgRslBs1:e2}   {\ensuremath{{1.597 } } }
\vdef{bdt2012:BgRslHi1:val}   {\ensuremath{{0.303 } } }
\vdef{bdt2012:BgRslHi1:e1}   {\ensuremath{{0.015 } } }
\vdef{bdt2012:BgRslHi1:e2}   {\ensuremath{{0.334 } } }
\vdef{bdt2012:BgRareLo1:val}   {\ensuremath{{11.995 } } }
\vdef{bdt2012:BgRareLo1:e1}   {\ensuremath{{0.594 } } }
\vdef{bdt2012:BgRareLo1:e2}   {\ensuremath{{8.309 } } }
\vdef{bdt2012:BgRareBd1:val}   {\ensuremath{{2.252 } } }
\vdef{bdt2012:BgRareBd1:e1}   {\ensuremath{{0.103 } } }
\vdef{bdt2012:BgRareBd1:e2}   {\ensuremath{{2.052 } } }
\vdef{bdt2012:BgRareBs1:val}   {\ensuremath{{1.563 } } }
\vdef{bdt2012:BgRareBs1:e1}   {\ensuremath{{0.074 } } }
\vdef{bdt2012:BgRareBs1:e2}   {\ensuremath{{1.598 } } }
\vdef{bdt2012:BgRareHi1:val}   {\ensuremath{{0.312 } } }
\vdef{bdt2012:BgRareHi1:e1}   {\ensuremath{{0.015 } } }
\vdef{bdt2012:BgRareHi1:e2}   {\ensuremath{{0.334 } } }
\vdef{bdt2012:BgRslsLo1:val}   {\ensuremath{{2.000 } } }
\vdef{bdt2012:BgRslsLo1:e1}   {\ensuremath{{0.100 } } }
\vdef{bdt2012:BgRslsLo1:e2}   {\ensuremath{{1.398 } } }
\vdef{bdt2012:BgRslsBd1:val}   {\ensuremath{{0.359 } } }
\vdef{bdt2012:BgRslsBd1:e1}   {\ensuremath{{0.018 } } }
\vdef{bdt2012:BgRslsBd1:e2}   {\ensuremath{{0.357 } } }
\vdef{bdt2012:BgRslsBs1:val}   {\ensuremath{{0.258 } } }
\vdef{bdt2012:BgRslsBs1:e1}   {\ensuremath{{0.013 } } }
\vdef{bdt2012:BgRslsBs1:e2}   {\ensuremath{{0.280 } } }
\vdef{bdt2012:BgRslsHi1:val}   {\ensuremath{{0.028 } } }
\vdef{bdt2012:BgRslsHi1:e1}   {\ensuremath{{0.001 } } }
\vdef{bdt2012:BgRslsHi1:e2}   {\ensuremath{{0.031 } } }
\vdef{bdt2012:BgCombLo1:val}   {\ensuremath{{2.000 } } }
\vdef{bdt2012:BgCombLo1:e1}   {\ensuremath{{1.155 } } }
\vdef{bdt2012:BgCombLo1:e2}   {\ensuremath{{1.172 } } }
\vdef{bdt2012:BgCombBd1:val}   {\ensuremath{{0.667 } } }
\vdef{bdt2012:BgCombBd1:e1}   {\ensuremath{{0.385 } } }
\vdef{bdt2012:BgCombBd1:e2}   {\ensuremath{{0.391 } } }
\vdef{bdt2012:BgCombBs1:val}   {\ensuremath{{1.000 } } }
\vdef{bdt2012:BgCombBs1:e1}   {\ensuremath{{0.577 } } }
\vdef{bdt2012:BgCombBs1:e2}   {\ensuremath{{0.586 } } }
\vdef{bdt2012:BgCombHi1:val}   {\ensuremath{{3.000 } } }
\vdef{bdt2012:BgCombHi1:e1}   {\ensuremath{{1.732 } } }
\vdef{bdt2012:BgCombHi1:e2}   {\ensuremath{{1.758 } } }
\vdef{bdt2012:BgNonpLo1:val}   {\ensuremath{{4.000 } } }
\vdef{bdt2012:BgNonpLo1:e1}   {\ensuremath{{1.159 } } }
\vdef{bdt2012:BgNonpLo1:e2}   {\ensuremath{{1.824 } } }
\vdef{bdt2012:BgNonpBd1:val}   {\ensuremath{{1.026 } } }
\vdef{bdt2012:BgNonpBd1:e1}   {\ensuremath{{0.385 } } }
\vdef{bdt2012:BgNonpBd1:e2}   {\ensuremath{{0.529 } } }
\vdef{bdt2012:BgNonpBs1:val}   {\ensuremath{{1.258 } } }
\vdef{bdt2012:BgNonpBs1:e1}   {\ensuremath{{0.577 } } }
\vdef{bdt2012:BgNonpBs1:e2}   {\ensuremath{{0.480 } } }
\vdef{bdt2012:BgNonpHi1:val}   {\ensuremath{{3.028 } } }
\vdef{bdt2012:BgNonpHi1:e1}   {\ensuremath{{1.732 } } }
\vdef{bdt2012:BgNonpHi1:e2}   {\ensuremath{{1.758 } } }
\vdef{bdt2012:BgTotLo1:val}   {\ensuremath{{4.106 } } }
\vdef{bdt2012:BgTotLo1:e1}   {\ensuremath{{0.000 } } }
\vdef{bdt2012:BgTotLo1:e2}   {\ensuremath{{1.826 } } }
\vdef{bdt2012:BgTotBd1:val}   {\ensuremath{{1.218 } } }
\vdef{bdt2012:BgTotBd1:e1}   {\ensuremath{{0.000 } } }
\vdef{bdt2012:BgTotBd1:e2}   {\ensuremath{{0.547 } } }
\vdef{bdt2012:BgTotBs1:val}   {\ensuremath{{1.348 } } }
\vdef{bdt2012:BgTotBs1:e1}   {\ensuremath{{0.000 } } }
\vdef{bdt2012:BgTotBs1:e2}   {\ensuremath{{0.485 } } }
\vdef{bdt2012:BgTotHi1:val}   {\ensuremath{{3.037 } } }
\vdef{bdt2012:BgTotHi1:e1}   {\ensuremath{{0.000 } } }
\vdef{bdt2012:BgTotHi1:e2}   {\ensuremath{{1.758 } } }
\vdef{bdt2012:SgAndBgLo1:val}   {\ensuremath{{4.288 } } }
\vdef{bdt2012:SgAndBgLo1:e1}   {\ensuremath{{0.000 } } }
\vdef{bdt2012:SgAndBgLo1:e2}   {\ensuremath{{1.826 } } }
\vdef{bdt2012:SgAndBgBd1:val}   {\ensuremath{{2.2 } } }
\vdef{bdt2012:SgAndBgBd1:e1}   {\ensuremath{{0.000 } } }
\vdef{bdt2012:SgAndBgBd1:e2}   {\ensuremath{{0.8 } } }
\vdef{bdt2012:SgAndBgBs1:val}   {\ensuremath{{5.1 } } }
\vdef{bdt2012:SgAndBgBs1:e1}   {\ensuremath{{0.000 } } }
\vdef{bdt2012:SgAndBgBs1:e2}   {\ensuremath{{0.7 } } }
\vdef{bdt2012:SgAndBgHi1:val}   {\ensuremath{{3.440 } } }
\vdef{bdt2012:SgAndBgHi1:e1}   {\ensuremath{{0.000 } } }
\vdef{bdt2012:SgAndBgHi1:e2}   {\ensuremath{{1.759 } } }
\vdef{bdt2012:N-EFF-TOT-BS0:val}   {\ensuremath{{0.000498 } } }
\vdef{bdt2012:N-EFF-TOT-BS0:err}   {\ensuremath{{0.000001 } } }
\vdef{bdt2012:N-ACC-BS0:val}   {\ensuremath{{0.0082 } } }
\vdef{bdt2012:N-ACC-BS0:err}   {\ensuremath{{0.0000 } } }
\vdef{bdt2012:N-EFF-MU-PID-BS0:val}   {\ensuremath{{0.6749 } } }
\vdef{bdt2012:N-EFF-MU-PID-BS0:err}   {\ensuremath{{0.0002 } } }
\vdef{bdt2012:N-EFF-MU-PIDMC-BS0:val}   {\ensuremath{{0.6379 } } }
\vdef{bdt2012:N-EFF-MU-PIDMC-BS0:err}   {\ensuremath{{0.0002 } } }
\vdef{bdt2012:N-EFF-MU-MC-BS0:val}   {\ensuremath{{0.6000 } } }
\vdef{bdt2012:N-EFF-MU-MC-BS0:err}   {\ensuremath{{0.0007 } } }
\vdef{bdt2012:N-EFF-TRIG-PID-BS0:val}   {\ensuremath{{0.7135 } } }
\vdef{bdt2012:N-EFF-TRIG-PID-BS0:err}   {\ensuremath{{0.0007 } } }
\vdef{bdt2012:N-EFF-TRIG-PIDMC-BS0:val}   {\ensuremath{{0.6678 } } }
\vdef{bdt2012:N-EFF-TRIG-PIDMC-BS0:err}   {\ensuremath{{0.0008 } } }
\vdef{bdt2012:N-EFF-TRIG-MC-BS0:val}   {\ensuremath{{0.5500 } } }
\vdef{bdt2012:N-EFF-TRIG-MC-BS0:err}   {\ensuremath{{0.0009 } } }
\vdef{bdt2012:N-EFF-CAND-BS0:val}   {\ensuremath{{0.9952 } } }
\vdef{bdt2012:N-EFF-CAND-BS0:err}   {\ensuremath{{0.0004 } } }
\vdef{bdt2012:N-EFF-ANA-BS0:val}   {\ensuremath{{0.1914 } } }
\vdef{bdt2012:N-EFF-ANA-BS0:err}   {\ensuremath{{0.0003 } } }
\vdef{bdt2012:N-OBS-BS0:val}   {\ensuremath{{20673 } } }
\vdef{bdt2012:N-OBS-BS0:err}   {\ensuremath{{535 } } }
\vdef{bdt2012:N-EFF-TOT-BS1:val}   {\ensuremath{{0.000120 } } }
\vdef{bdt2012:N-EFF-TOT-BS1:err}   {\ensuremath{{0.000001 } } }
\vdef{bdt2012:N-ACC-BS1:val}   {\ensuremath{{0.0043 } } }
\vdef{bdt2012:N-ACC-BS1:err}   {\ensuremath{{0.0000 } } }
\vdef{bdt2012:N-EFF-MU-PID-BS1:val}   {\ensuremath{{0.5806 } } }
\vdef{bdt2012:N-EFF-MU-PID-BS1:err}   {\ensuremath{{0.0006 } } }
\vdef{bdt2012:N-EFF-MU-PIDMC-BS1:val}   {\ensuremath{{0.5753 } } }
\vdef{bdt2012:N-EFF-MU-PIDMC-BS1:err}   {\ensuremath{{0.0005 } } }
\vdef{bdt2012:N-EFF-MU-MC-BS1:val}   {\ensuremath{{0.5606 } } }
\vdef{bdt2012:N-EFF-MU-MC-BS1:err}   {\ensuremath{{0.0013 } } }
\vdef{bdt2012:N-EFF-TRIG-PID-BS1:val}   {\ensuremath{{0.7661 } } }
\vdef{bdt2012:N-EFF-TRIG-PID-BS1:err}   {\ensuremath{{0.0005 } } }
\vdef{bdt2012:N-EFF-TRIG-PIDMC-BS1:val}   {\ensuremath{{0.7503 } } }
\vdef{bdt2012:N-EFF-TRIG-PIDMC-BS1:err}   {\ensuremath{{0.0006 } } }
\vdef{bdt2012:N-EFF-TRIG-MC-BS1:val}   {\ensuremath{{0.4818 } } }
\vdef{bdt2012:N-EFF-TRIG-MC-BS1:err}   {\ensuremath{{0.0018 } } }
\vdef{bdt2012:N-EFF-CAND-BS1:val}   {\ensuremath{{0.9932 } } }
\vdef{bdt2012:N-EFF-CAND-BS1:err}   {\ensuremath{{0.0007 } } }
\vdef{bdt2012:N-EFF-ANA-BS1:val}   {\ensuremath{{0.1091 } } }
\vdef{bdt2012:N-EFF-ANA-BS1:err}   {\ensuremath{{0.0003 } } }
\vdef{bdt2012:N-OBS-BS1:val}   {\ensuremath{{5311 } } }
\vdef{bdt2012:N-OBS-BS1:err}   {\ensuremath{{86 } } }
\vdef{2011:bdt:0}     {\ensuremath{{0.290 } } }
\vdef{2011:bdtMax:0}  {\ensuremath{{10.000 } } }
\vdef{2011:mBdLo:0}   {\ensuremath{{5.200 } } }
\vdef{2011:mBdHi:0}   {\ensuremath{{5.300 } } }
\vdef{2011:mBsLo:0}   {\ensuremath{{5.300 } } }
\vdef{2011:mBsHi:0}   {\ensuremath{{5.450 } } }
\vdef{2011:etaMin:0}   {\ensuremath{{0.0 } } }
\vdef{2011:etaMax:0}   {\ensuremath{{1.4 } } }
\vdef{2011:pt:0}   {\ensuremath{{6.5 } } }
\vdef{2011:m1pt:0}   {\ensuremath{{4.5 } } }
\vdef{2011:m2pt:0}   {\ensuremath{{4.0 } } }
\vdef{2011:m1eta:0}   {\ensuremath{{1.4 } } }
\vdef{2011:m2eta:0}   {\ensuremath{{1.4 } } }
\vdef{2011:iso:0}   {\ensuremath{{0.80 } } }
\vdef{2011:chi2dof:0}   {\ensuremath{{2.2 } } }
\vdef{2011:alpha:0}   {\ensuremath{{0.050 } } }
\vdef{2011:fls3d:0}   {\ensuremath{{13.0 } } }
\vdef{2011:docatrk:0}   {\ensuremath{{0.015 } } }
\vdef{2011:closetrk:0}   {\ensuremath{{2 } } }
\vdef{2011:pvlip:0}   {\ensuremath{{100.000 } } }
\vdef{2011:pvlips:0}   {\ensuremath{{100.000 } } }
\vdef{2011:pvlip2:0}   {\ensuremath{{-100.000 } } }
\vdef{2011:pvlips2:0}   {\ensuremath{{-100.000 } } }
\vdef{2011:maxdoca:0}   {\ensuremath{{100.000 } } }
\vdef{2011:pvip:0}   {\ensuremath{{0.008 } } }
\vdef{2011:pvips:0}   {\ensuremath{{2.000 } } }
\vdef{2011:doApplyCowboyVeto:0}   {no }
\vdef{2011:fDoApplyCowboyVetoAlsoInSignal:0}   {no }
\vdef{2011:bdt:1}     {\ensuremath{{0.290 } } }
\vdef{2011:bdtMax:1}  {\ensuremath{{10.000 } } }
\vdef{2011:mBdLo:1}   {\ensuremath{{5.200 } } }
\vdef{2011:mBdHi:1}   {\ensuremath{{5.300 } } }
\vdef{2011:mBsLo:1}   {\ensuremath{{5.300 } } }
\vdef{2011:mBsHi:1}   {\ensuremath{{5.450 } } }
\vdef{2011:etaMin:1}   {\ensuremath{{1.4 } } }
\vdef{2011:etaMax:1}   {\ensuremath{{2.4 } } }
\vdef{2011:pt:1}   {\ensuremath{{8.5 } } }
\vdef{2011:m1pt:1}   {\ensuremath{{4.5 } } }
\vdef{2011:m2pt:1}   {\ensuremath{{4.2 } } }
\vdef{2011:m1eta:1}   {\ensuremath{{2.4 } } }
\vdef{2011:m2eta:1}   {\ensuremath{{2.4 } } }
\vdef{2011:iso:1}   {\ensuremath{{0.80 } } }
\vdef{2011:chi2dof:1}   {\ensuremath{{1.8 } } }
\vdef{2011:alpha:1}   {\ensuremath{{0.030 } } }
\vdef{2011:fls3d:1}   {\ensuremath{{15.0 } } }
\vdef{2011:docatrk:1}   {\ensuremath{{0.015 } } }
\vdef{2011:closetrk:1}   {\ensuremath{{2 } } }
\vdef{2011:pvlip:1}   {\ensuremath{{100.000 } } }
\vdef{2011:pvlips:1}   {\ensuremath{{100.000 } } }
\vdef{2011:pvlip2:1}   {\ensuremath{{-100.000 } } }
\vdef{2011:pvlips2:1}   {\ensuremath{{-100.000 } } }
\vdef{2011:maxdoca:1}   {\ensuremath{{100.000 } } }
\vdef{2011:pvip:1}   {\ensuremath{{0.008 } } }
\vdef{2011:pvips:1}   {\ensuremath{{2.000 } } }
\vdef{2011:doApplyCowboyVeto:1}   {no }
\vdef{2011:fDoApplyCowboyVetoAlsoInSignal:1}   {no }
\vdef{2011:bdt:0}     {\ensuremath{{0.290 } } }
\vdef{2011:bdtMax:0}  {\ensuremath{{10.000 } } }
\vdef{2011:mBdLo:0}   {\ensuremath{{5.200 } } }
\vdef{2011:mBdHi:0}   {\ensuremath{{5.300 } } }
\vdef{2011:mBsLo:0}   {\ensuremath{{5.300 } } }
\vdef{2011:mBsHi:0}   {\ensuremath{{5.450 } } }
\vdef{2011:etaMin:0}   {\ensuremath{{0.0 } } }
\vdef{2011:etaMax:0}   {\ensuremath{{1.4 } } }
\vdef{2011:pt:0}   {\ensuremath{{6.5 } } }
\vdef{2011:m1pt:0}   {\ensuremath{{4.5 } } }
\vdef{2011:m2pt:0}   {\ensuremath{{4.0 } } }
\vdef{2011:m1eta:0}   {\ensuremath{{1.4 } } }
\vdef{2011:m2eta:0}   {\ensuremath{{1.4 } } }
\vdef{2011:iso:0}   {\ensuremath{{0.80 } } }
\vdef{2011:chi2dof:0}   {\ensuremath{{2.2 } } }
\vdef{2011:alpha:0}   {\ensuremath{{0.050 } } }
\vdef{2011:fls3d:0}   {\ensuremath{{13.0 } } }
\vdef{2011:docatrk:0}   {\ensuremath{{0.015 } } }
\vdef{2011:closetrk:0}   {\ensuremath{{2 } } }
\vdef{2011:pvlip:0}   {\ensuremath{{100.000 } } }
\vdef{2011:pvlips:0}   {\ensuremath{{100.000 } } }
\vdef{2011:pvlip2:0}   {\ensuremath{{-100.000 } } }
\vdef{2011:pvlips2:0}   {\ensuremath{{-100.000 } } }
\vdef{2011:maxdoca:0}   {\ensuremath{{100.000 } } }
\vdef{2011:pvip:0}   {\ensuremath{{0.008 } } }
\vdef{2011:pvips:0}   {\ensuremath{{2.000 } } }
\vdef{2011:doApplyCowboyVeto:0}   {no }
\vdef{2011:fDoApplyCowboyVetoAlsoInSignal:0}   {no }
\vdef{2011:bdt:1}     {\ensuremath{{0.290 } } }
\vdef{2011:bdtMax:1}  {\ensuremath{{10.000 } } }
\vdef{2011:mBdLo:1}   {\ensuremath{{5.200 } } }
\vdef{2011:mBdHi:1}   {\ensuremath{{5.300 } } }
\vdef{2011:mBsLo:1}   {\ensuremath{{5.300 } } }
\vdef{2011:mBsHi:1}   {\ensuremath{{5.450 } } }
\vdef{2011:etaMin:1}   {\ensuremath{{1.4 } } }
\vdef{2011:etaMax:1}   {\ensuremath{{2.4 } } }
\vdef{2011:pt:1}   {\ensuremath{{8.5 } } }
\vdef{2011:m1pt:1}   {\ensuremath{{4.5 } } }
\vdef{2011:m2pt:1}   {\ensuremath{{4.2 } } }
\vdef{2011:m1eta:1}   {\ensuremath{{2.4 } } }
\vdef{2011:m2eta:1}   {\ensuremath{{2.4 } } }
\vdef{2011:iso:1}   {\ensuremath{{0.80 } } }
\vdef{2011:chi2dof:1}   {\ensuremath{{1.8 } } }
\vdef{2011:alpha:1}   {\ensuremath{{0.030 } } }
\vdef{2011:fls3d:1}   {\ensuremath{{15.0 } } }
\vdef{2011:docatrk:1}   {\ensuremath{{0.015 } } }
\vdef{2011:closetrk:1}   {\ensuremath{{2 } } }
\vdef{2011:pvlip:1}   {\ensuremath{{100.000 } } }
\vdef{2011:pvlips:1}   {\ensuremath{{100.000 } } }
\vdef{2011:pvlip2:1}   {\ensuremath{{-100.000 } } }
\vdef{2011:pvlips2:1}   {\ensuremath{{-100.000 } } }
\vdef{2011:maxdoca:1}   {\ensuremath{{100.000 } } }
\vdef{2011:pvip:1}   {\ensuremath{{0.008 } } }
\vdef{2011:pvips:1}   {\ensuremath{{2.000 } } }
\vdef{2011:doApplyCowboyVeto:1}   {no }
\vdef{2011:fDoApplyCowboyVetoAlsoInSignal:1}   {no }
\vdef{bdt2011:N-CSBF-TNP-BS0:val}   {\ensuremath{{0.000024 } } }
\vdef{bdt2011:N-CSBF-TNP-BS0:err}   {\ensuremath{{0.000001 } } }
\vdef{bdt2011:N-CSBF-MC-BS0:val}   {\ensuremath{{0.000025 } } }
\vdef{bdt2011:N-CSBF-MC-BS0:err}   {\ensuremath{{0.000001 } } }
\vdef{bdt2011:N-CSBF-MC-BS0:syst}   {\ensuremath{{0.000002 } } }
\vdef{bdt2011:N-CSBF-BS0:val}   {\ensuremath{{0.000025 } } }
\vdef{bdt2011:N-CSBF-BS0:err}   {\ensuremath{{0.000001 } } }
\vdef{bdt2011:N-CSBF-BS0:syst}   {\ensuremath{{0.000002 } } }
\vdef{bdt2011:bgBd2K0MuMu:loSideband0:val}   {\ensuremath{{0.000 } } }
\vdef{bdt2011:bgBd2K0MuMu:loSideband0:err}   {\ensuremath{{0.000 } } }
\vdef{bdt2011:bgBd2K0MuMu:bdRare0}   {\ensuremath{{0.000000 } } }
\vdef{bdt2011:bgBd2K0MuMu:bdRare0E}  {\ensuremath{{0.000000 } } }
\vdef{bdt2011:bgBd2K0MuMu:bsRare0}   {\ensuremath{{0.000000 } } }
\vdef{bdt2011:bgBd2K0MuMu:bsRare0E}  {\ensuremath{{0.000000 } } }
\vdef{bdt2011:bgBd2K0MuMu:hiSideband0:val}   {\ensuremath{{0.000 } } }
\vdef{bdt2011:bgBd2K0MuMu:hiSideband0:err}   {\ensuremath{{0.000 } } }
\vdef{bdt2011:bgBd2KK:loSideband0:val}   {\ensuremath{{0.002 } } }
\vdef{bdt2011:bgBd2KK:loSideband0:err}   {\ensuremath{{0.003 } } }
\vdef{bdt2011:bgBd2KK:bdRare0}   {\ensuremath{{0.001114 } } }
\vdef{bdt2011:bgBd2KK:bdRare0E}  {\ensuremath{{0.001407 } } }
\vdef{bdt2011:bgBd2KK:bsRare0}   {\ensuremath{{0.000051 } } }
\vdef{bdt2011:bgBd2KK:bsRare0E}  {\ensuremath{{0.000064 } } }
\vdef{bdt2011:bgBd2KK:hiSideband0:val}   {\ensuremath{{0.000 } } }
\vdef{bdt2011:bgBd2KK:hiSideband0:err}   {\ensuremath{{0.000 } } }
\vdef{bdt2011:bgBd2KPi:loSideband0:val}   {\ensuremath{{0.085 } } }
\vdef{bdt2011:bgBd2KPi:loSideband0:err}   {\ensuremath{{0.060 } } }
\vdef{bdt2011:bgBd2KPi:bdRare0}   {\ensuremath{{0.217428 } } }
\vdef{bdt2011:bgBd2KPi:bdRare0E}  {\ensuremath{{0.153883 } } }
\vdef{bdt2011:bgBd2KPi:bsRare0}   {\ensuremath{{0.022318 } } }
\vdef{bdt2011:bgBd2KPi:bsRare0E}  {\ensuremath{{0.015795 } } }
\vdef{bdt2011:bgBd2KPi:hiSideband0:val}   {\ensuremath{{0.001 } } }
\vdef{bdt2011:bgBd2KPi:hiSideband0:err}   {\ensuremath{{0.001 } } }
\vdef{bdt2011:bgBd2MuMuGamma:loSideband0:val}   {\ensuremath{{0.010 } } }
\vdef{bdt2011:bgBd2MuMuGamma:loSideband0:err}   {\ensuremath{{0.002 } } }
\vdef{bdt2011:bgBd2MuMuGamma:bdRare0}   {\ensuremath{{0.000683 } } }
\vdef{bdt2011:bgBd2MuMuGamma:bdRare0E}  {\ensuremath{{0.000137 } } }
\vdef{bdt2011:bgBd2MuMuGamma:bsRare0}   {\ensuremath{{0.000060 } } }
\vdef{bdt2011:bgBd2MuMuGamma:bsRare0E}  {\ensuremath{{0.000012 } } }
\vdef{bdt2011:bgBd2MuMuGamma:hiSideband0:val}   {\ensuremath{{0.000 } } }
\vdef{bdt2011:bgBd2MuMuGamma:hiSideband0:err}   {\ensuremath{{0.000 } } }
\vdef{bdt2011:bgBd2Pi0MuMu:loSideband0:val}   {\ensuremath{{0.505 } } }
\vdef{bdt2011:bgBd2Pi0MuMu:loSideband0:err}   {\ensuremath{{0.252 } } }
\vdef{bdt2011:bgBd2Pi0MuMu:bdRare0}   {\ensuremath{{0.000362 } } }
\vdef{bdt2011:bgBd2Pi0MuMu:bdRare0E}  {\ensuremath{{0.000181 } } }
\vdef{bdt2011:bgBd2Pi0MuMu:bsRare0}   {\ensuremath{{0.000000 } } }
\vdef{bdt2011:bgBd2Pi0MuMu:bsRare0E}  {\ensuremath{{0.000000 } } }
\vdef{bdt2011:bgBd2Pi0MuMu:hiSideband0:val}   {\ensuremath{{0.000 } } }
\vdef{bdt2011:bgBd2Pi0MuMu:hiSideband0:err}   {\ensuremath{{0.000 } } }
\vdef{bdt2011:bgBd2PiMuNu:loSideband0:val}   {\ensuremath{{3.584 } } }
\vdef{bdt2011:bgBd2PiMuNu:loSideband0:err}   {\ensuremath{{1.801 } } }
\vdef{bdt2011:bgBd2PiMuNu:bdRare0}   {\ensuremath{{0.099800 } } }
\vdef{bdt2011:bgBd2PiMuNu:bdRare0E}  {\ensuremath{{0.050149 } } }
\vdef{bdt2011:bgBd2PiMuNu:bsRare0}   {\ensuremath{{0.012668 } } }
\vdef{bdt2011:bgBd2PiMuNu:bsRare0E}  {\ensuremath{{0.006366 } } }
\vdef{bdt2011:bgBd2PiMuNu:hiSideband0:val}   {\ensuremath{{0.004 } } }
\vdef{bdt2011:bgBd2PiMuNu:hiSideband0:err}   {\ensuremath{{0.002 } } }
\vdef{bdt2011:bgBd2PiPi:loSideband0:val}   {\ensuremath{{0.005 } } }
\vdef{bdt2011:bgBd2PiPi:loSideband0:err}   {\ensuremath{{0.005 } } }
\vdef{bdt2011:bgBd2PiPi:bdRare0}   {\ensuremath{{0.033956 } } }
\vdef{bdt2011:bgBd2PiPi:bdRare0E}  {\ensuremath{{0.033988 } } }
\vdef{bdt2011:bgBd2PiPi:bsRare0}   {\ensuremath{{0.014331 } } }
\vdef{bdt2011:bgBd2PiPi:bsRare0E}  {\ensuremath{{0.014344 } } }
\vdef{bdt2011:bgBd2PiPi:hiSideband0:val}   {\ensuremath{{0.001 } } }
\vdef{bdt2011:bgBd2PiPi:hiSideband0:err}   {\ensuremath{{0.001 } } }
\vdef{bdt2011:bgBs2KK:loSideband0:val}   {\ensuremath{{0.013 } } }
\vdef{bdt2011:bgBs2KK:loSideband0:err}   {\ensuremath{{0.013 } } }
\vdef{bdt2011:bgBs2KK:bdRare0}   {\ensuremath{{0.108586 } } }
\vdef{bdt2011:bgBs2KK:bdRare0E}  {\ensuremath{{0.109800 } } }
\vdef{bdt2011:bgBs2KK:bsRare0}   {\ensuremath{{0.045901 } } }
\vdef{bdt2011:bgBs2KK:bsRare0E}  {\ensuremath{{0.046414 } } }
\vdef{bdt2011:bgBs2KK:hiSideband0:val}   {\ensuremath{{0.001 } } }
\vdef{bdt2011:bgBs2KK:hiSideband0:err}   {\ensuremath{{0.001 } } }
\vdef{bdt2011:bgBs2KMuNu:loSideband0:val}   {\ensuremath{{1.781 } } }
\vdef{bdt2011:bgBs2KMuNu:loSideband0:err}   {\ensuremath{{0.895 } } }
\vdef{bdt2011:bgBs2KMuNu:bdRare0}   {\ensuremath{{0.053153 } } }
\vdef{bdt2011:bgBs2KMuNu:bdRare0E}  {\ensuremath{{0.026709 } } }
\vdef{bdt2011:bgBs2KMuNu:bsRare0}   {\ensuremath{{0.007908 } } }
\vdef{bdt2011:bgBs2KMuNu:bsRare0E}  {\ensuremath{{0.003974 } } }
\vdef{bdt2011:bgBs2KMuNu:hiSideband0:val}   {\ensuremath{{0.000 } } }
\vdef{bdt2011:bgBs2KMuNu:hiSideband0:err}   {\ensuremath{{0.000 } } }
\vdef{bdt2011:bgBs2KPi:loSideband0:val}   {\ensuremath{{0.001 } } }
\vdef{bdt2011:bgBs2KPi:loSideband0:err}   {\ensuremath{{0.001 } } }
\vdef{bdt2011:bgBs2KPi:bdRare0}   {\ensuremath{{0.006143 } } }
\vdef{bdt2011:bgBs2KPi:bdRare0E}  {\ensuremath{{0.004549 } } }
\vdef{bdt2011:bgBs2KPi:bsRare0}   {\ensuremath{{0.013459 } } }
\vdef{bdt2011:bgBs2KPi:bsRare0E}  {\ensuremath{{0.009967 } } }
\vdef{bdt2011:bgBs2KPi:hiSideband0:val}   {\ensuremath{{0.000 } } }
\vdef{bdt2011:bgBs2KPi:hiSideband0:err}   {\ensuremath{{0.000 } } }
\vdef{bdt2011:bgBs2MuMuGamma:loSideband0:val}   {\ensuremath{{0.418 } } }
\vdef{bdt2011:bgBs2MuMuGamma:loSideband0:err}   {\ensuremath{{0.084 } } }
\vdef{bdt2011:bgBs2MuMuGamma:bdRare0}   {\ensuremath{{0.071812 } } }
\vdef{bdt2011:bgBs2MuMuGamma:bdRare0E}  {\ensuremath{{0.014362 } } }
\vdef{bdt2011:bgBs2MuMuGamma:bsRare0}   {\ensuremath{{0.020626 } } }
\vdef{bdt2011:bgBs2MuMuGamma:bsRare0E}  {\ensuremath{{0.004125 } } }
\vdef{bdt2011:bgBs2MuMuGamma:hiSideband0:val}   {\ensuremath{{0.000 } } }
\vdef{bdt2011:bgBs2MuMuGamma:hiSideband0:err}   {\ensuremath{{0.000 } } }
\vdef{bdt2011:bgBs2PiPi:loSideband0:val}   {\ensuremath{{0.000 } } }
\vdef{bdt2011:bgBs2PiPi:loSideband0:err}   {\ensuremath{{0.000 } } }
\vdef{bdt2011:bgBs2PiPi:bdRare0}   {\ensuremath{{0.000175 } } }
\vdef{bdt2011:bgBs2PiPi:bdRare0E}  {\ensuremath{{0.000178 } } }
\vdef{bdt2011:bgBs2PiPi:bsRare0}   {\ensuremath{{0.001632 } } }
\vdef{bdt2011:bgBs2PiPi:bsRare0E}  {\ensuremath{{0.001662 } } }
\vdef{bdt2011:bgBs2PiPi:hiSideband0:val}   {\ensuremath{{0.000 } } }
\vdef{bdt2011:bgBs2PiPi:hiSideband0:err}   {\ensuremath{{0.000 } } }
\vdef{bdt2011:bgBu2KMuMu:loSideband0:val}   {\ensuremath{{0.015 } } }
\vdef{bdt2011:bgBu2KMuMu:loSideband0:err}   {\ensuremath{{0.001 } } }
\vdef{bdt2011:bgBu2KMuMu:bdRare0}   {\ensuremath{{0.000000 } } }
\vdef{bdt2011:bgBu2KMuMu:bdRare0E}  {\ensuremath{{0.000000 } } }
\vdef{bdt2011:bgBu2KMuMu:bsRare0}   {\ensuremath{{0.000000 } } }
\vdef{bdt2011:bgBu2KMuMu:bsRare0E}  {\ensuremath{{0.000000 } } }
\vdef{bdt2011:bgBu2KMuMu:hiSideband0:val}   {\ensuremath{{0.000 } } }
\vdef{bdt2011:bgBu2KMuMu:hiSideband0:err}   {\ensuremath{{0.000 } } }
\vdef{bdt2011:bgBu2PiMuMu:loSideband0:val}   {\ensuremath{{0.828 } } }
\vdef{bdt2011:bgBu2PiMuMu:loSideband0:err}   {\ensuremath{{0.215 } } }
\vdef{bdt2011:bgBu2PiMuMu:bdRare0}   {\ensuremath{{0.001298 } } }
\vdef{bdt2011:bgBu2PiMuMu:bdRare0E}  {\ensuremath{{0.000338 } } }
\vdef{bdt2011:bgBu2PiMuMu:bsRare0}   {\ensuremath{{0.000604 } } }
\vdef{bdt2011:bgBu2PiMuMu:bsRare0E}  {\ensuremath{{0.000157 } } }
\vdef{bdt2011:bgBu2PiMuMu:hiSideband0:val}   {\ensuremath{{0.000 } } }
\vdef{bdt2011:bgBu2PiMuMu:hiSideband0:err}   {\ensuremath{{0.000 } } }
\vdef{bdt2011:bgLb2KP:loSideband0:val}   {\ensuremath{{0.001 } } }
\vdef{bdt2011:bgLb2KP:loSideband0:err}   {\ensuremath{{0.000 } } }
\vdef{bdt2011:bgLb2KP:bdRare0}   {\ensuremath{{0.001151 } } }
\vdef{bdt2011:bgLb2KP:bdRare0E}  {\ensuremath{{0.000865 } } }
\vdef{bdt2011:bgLb2KP:bsRare0}   {\ensuremath{{0.011220 } } }
\vdef{bdt2011:bgLb2KP:bsRare0E}  {\ensuremath{{0.008432 } } }
\vdef{bdt2011:bgLb2KP:hiSideband0:val}   {\ensuremath{{0.004 } } }
\vdef{bdt2011:bgLb2KP:hiSideband0:err}   {\ensuremath{{0.003 } } }
\vdef{bdt2011:bgLb2PMuNu:loSideband0:val}   {\ensuremath{{16.634 } } }
\vdef{bdt2011:bgLb2PMuNu:loSideband0:err}   {\ensuremath{{18.598 } } }
\vdef{bdt2011:bgLb2PMuNu:bdRare0}   {\ensuremath{{3.477172 } } }
\vdef{bdt2011:bgLb2PMuNu:bdRare0E}  {\ensuremath{{3.887597 } } }
\vdef{bdt2011:bgLb2PMuNu:bsRare0}   {\ensuremath{{2.876504 } } }
\vdef{bdt2011:bgLb2PMuNu:bsRare0E}  {\ensuremath{{3.216029 } } }
\vdef{bdt2011:bgLb2PMuNu:hiSideband0:val}   {\ensuremath{{0.278 } } }
\vdef{bdt2011:bgLb2PMuNu:hiSideband0:err}   {\ensuremath{{0.310 } } }
\vdef{bdt2011:bgLb2PiP:loSideband0:val}   {\ensuremath{{0.000 } } }
\vdef{bdt2011:bgLb2PiP:loSideband0:err}   {\ensuremath{{0.000 } } }
\vdef{bdt2011:bgLb2PiP:bdRare0}   {\ensuremath{{0.000347 } } }
\vdef{bdt2011:bgLb2PiP:bdRare0E}  {\ensuremath{{0.000264 } } }
\vdef{bdt2011:bgLb2PiP:bsRare0}   {\ensuremath{{0.002732 } } }
\vdef{bdt2011:bgLb2PiP:bsRare0E}  {\ensuremath{{0.002084 } } }
\vdef{bdt2011:bgLb2PiP:hiSideband0:val}   {\ensuremath{{0.004 } } }
\vdef{bdt2011:bgLb2PiP:hiSideband0:err}   {\ensuremath{{0.003 } } }
\vdef{bdt2011:bsRare0}   {\ensuremath{{0.000 } } }
\vdef{bdt2011:bsRare0E}  {\ensuremath{{0.000 } } }
\vdef{bdt2011:bdRare0}   {\ensuremath{{0.000 } } }
\vdef{bdt2011:bdRare0E}  {\ensuremath{{0.000 } } }
\vdef{bdt2011:N-CSBF-TNP-BS1:val}   {\ensuremath{{0.000030 } } }
\vdef{bdt2011:N-CSBF-TNP-BS1:err}   {\ensuremath{{0.000001 } } }
\vdef{bdt2011:N-CSBF-MC-BS1:val}   {\ensuremath{{0.000032 } } }
\vdef{bdt2011:N-CSBF-MC-BS1:err}   {\ensuremath{{0.000001 } } }
\vdef{bdt2011:N-CSBF-MC-BS1:syst}   {\ensuremath{{0.000002 } } }
\vdef{bdt2011:N-CSBF-BS1:val}   {\ensuremath{{0.000031 } } }
\vdef{bdt2011:N-CSBF-BS1:err}   {\ensuremath{{0.000001 } } }
\vdef{bdt2011:N-CSBF-BS1:syst}   {\ensuremath{{0.000002 } } }
\vdef{bdt2011:bgBd2K0MuMu:loSideband1:val}   {\ensuremath{{0.000 } } }
\vdef{bdt2011:bgBd2K0MuMu:loSideband1:err}   {\ensuremath{{0.000 } } }
\vdef{bdt2011:bgBd2K0MuMu:bdRare1}   {\ensuremath{{0.004221 } } }
\vdef{bdt2011:bgBd2K0MuMu:bdRare1E}  {\ensuremath{{0.000000 } } }
\vdef{bdt2011:bgBd2K0MuMu:bsRare1}   {\ensuremath{{0.000000 } } }
\vdef{bdt2011:bgBd2K0MuMu:bsRare1E}  {\ensuremath{{0.000000 } } }
\vdef{bdt2011:bgBd2K0MuMu:hiSideband1:val}   {\ensuremath{{0.000 } } }
\vdef{bdt2011:bgBd2K0MuMu:hiSideband1:err}   {\ensuremath{{0.000 } } }
\vdef{bdt2011:bgBd2KK:loSideband1:val}   {\ensuremath{{0.001 } } }
\vdef{bdt2011:bgBd2KK:loSideband1:err}   {\ensuremath{{0.001 } } }
\vdef{bdt2011:bgBd2KK:bdRare1}   {\ensuremath{{0.000420 } } }
\vdef{bdt2011:bgBd2KK:bdRare1E}  {\ensuremath{{0.000530 } } }
\vdef{bdt2011:bgBd2KK:bsRare1}   {\ensuremath{{0.000076 } } }
\vdef{bdt2011:bgBd2KK:bsRare1E}  {\ensuremath{{0.000095 } } }
\vdef{bdt2011:bgBd2KK:hiSideband1:val}   {\ensuremath{{0.000 } } }
\vdef{bdt2011:bgBd2KK:hiSideband1:err}   {\ensuremath{{0.000 } } }
\vdef{bdt2011:bgBd2KPi:loSideband1:val}   {\ensuremath{{0.048 } } }
\vdef{bdt2011:bgBd2KPi:loSideband1:err}   {\ensuremath{{0.034 } } }
\vdef{bdt2011:bgBd2KPi:bdRare1}   {\ensuremath{{0.058786 } } }
\vdef{bdt2011:bgBd2KPi:bdRare1E}  {\ensuremath{{0.041605 } } }
\vdef{bdt2011:bgBd2KPi:bsRare1}   {\ensuremath{{0.019412 } } }
\vdef{bdt2011:bgBd2KPi:bsRare1E}  {\ensuremath{{0.013739 } } }
\vdef{bdt2011:bgBd2KPi:hiSideband1:val}   {\ensuremath{{0.002 } } }
\vdef{bdt2011:bgBd2KPi:hiSideband1:err}   {\ensuremath{{0.001 } } }
\vdef{bdt2011:bgBd2MuMuGamma:loSideband1:val}   {\ensuremath{{0.003 } } }
\vdef{bdt2011:bgBd2MuMuGamma:loSideband1:err}   {\ensuremath{{0.001 } } }
\vdef{bdt2011:bgBd2MuMuGamma:bdRare1}   {\ensuremath{{0.000299 } } }
\vdef{bdt2011:bgBd2MuMuGamma:bdRare1E}  {\ensuremath{{0.000060 } } }
\vdef{bdt2011:bgBd2MuMuGamma:bsRare1}   {\ensuremath{{0.000045 } } }
\vdef{bdt2011:bgBd2MuMuGamma:bsRare1E}  {\ensuremath{{0.000009 } } }
\vdef{bdt2011:bgBd2MuMuGamma:hiSideband1:val}   {\ensuremath{{0.000 } } }
\vdef{bdt2011:bgBd2MuMuGamma:hiSideband1:err}   {\ensuremath{{0.000 } } }
\vdef{bdt2011:bgBd2Pi0MuMu:loSideband1:val}   {\ensuremath{{0.164 } } }
\vdef{bdt2011:bgBd2Pi0MuMu:loSideband1:err}   {\ensuremath{{0.082 } } }
\vdef{bdt2011:bgBd2Pi0MuMu:bdRare1}   {\ensuremath{{0.001651 } } }
\vdef{bdt2011:bgBd2Pi0MuMu:bdRare1E}  {\ensuremath{{0.000825 } } }
\vdef{bdt2011:bgBd2Pi0MuMu:bsRare1}   {\ensuremath{{0.000000 } } }
\vdef{bdt2011:bgBd2Pi0MuMu:bsRare1E}  {\ensuremath{{0.000000 } } }
\vdef{bdt2011:bgBd2Pi0MuMu:hiSideband1:val}   {\ensuremath{{0.000 } } }
\vdef{bdt2011:bgBd2Pi0MuMu:hiSideband1:err}   {\ensuremath{{0.000 } } }
\vdef{bdt2011:bgBd2PiMuNu:loSideband1:val}   {\ensuremath{{1.301 } } }
\vdef{bdt2011:bgBd2PiMuNu:loSideband1:err}   {\ensuremath{{0.654 } } }
\vdef{bdt2011:bgBd2PiMuNu:bdRare1}   {\ensuremath{{0.052250 } } }
\vdef{bdt2011:bgBd2PiMuNu:bdRare1E}  {\ensuremath{{0.026255 } } }
\vdef{bdt2011:bgBd2PiMuNu:bsRare1}   {\ensuremath{{0.003781 } } }
\vdef{bdt2011:bgBd2PiMuNu:bsRare1E}  {\ensuremath{{0.001900 } } }
\vdef{bdt2011:bgBd2PiMuNu:hiSideband1:val}   {\ensuremath{{0.003 } } }
\vdef{bdt2011:bgBd2PiMuNu:hiSideband1:err}   {\ensuremath{{0.001 } } }
\vdef{bdt2011:bgBd2PiPi:loSideband1:val}   {\ensuremath{{0.004 } } }
\vdef{bdt2011:bgBd2PiPi:loSideband1:err}   {\ensuremath{{0.004 } } }
\vdef{bdt2011:bgBd2PiPi:bdRare1}   {\ensuremath{{0.009850 } } }
\vdef{bdt2011:bgBd2PiPi:bdRare1E}  {\ensuremath{{0.009859 } } }
\vdef{bdt2011:bgBd2PiPi:bsRare1}   {\ensuremath{{0.006885 } } }
\vdef{bdt2011:bgBd2PiPi:bsRare1E}  {\ensuremath{{0.006891 } } }
\vdef{bdt2011:bgBd2PiPi:hiSideband1:val}   {\ensuremath{{0.001 } } }
\vdef{bdt2011:bgBd2PiPi:hiSideband1:err}   {\ensuremath{{0.001 } } }
\vdef{bdt2011:bgBs2KK:loSideband1:val}   {\ensuremath{{0.013 } } }
\vdef{bdt2011:bgBs2KK:loSideband1:err}   {\ensuremath{{0.013 } } }
\vdef{bdt2011:bgBs2KK:bdRare1}   {\ensuremath{{0.033026 } } }
\vdef{bdt2011:bgBs2KK:bdRare1E}  {\ensuremath{{0.033396 } } }
\vdef{bdt2011:bgBs2KK:bsRare1}   {\ensuremath{{0.021042 } } }
\vdef{bdt2011:bgBs2KK:bsRare1E}  {\ensuremath{{0.021277 } } }
\vdef{bdt2011:bgBs2KK:hiSideband1:val}   {\ensuremath{{0.002 } } }
\vdef{bdt2011:bgBs2KK:hiSideband1:err}   {\ensuremath{{0.002 } } }
\vdef{bdt2011:bgBs2KMuNu:loSideband1:val}   {\ensuremath{{0.591 } } }
\vdef{bdt2011:bgBs2KMuNu:loSideband1:err}   {\ensuremath{{0.297 } } }
\vdef{bdt2011:bgBs2KMuNu:bdRare1}   {\ensuremath{{0.033433 } } }
\vdef{bdt2011:bgBs2KMuNu:bdRare1E}  {\ensuremath{{0.016800 } } }
\vdef{bdt2011:bgBs2KMuNu:bsRare1}   {\ensuremath{{0.018882 } } }
\vdef{bdt2011:bgBs2KMuNu:bsRare1E}  {\ensuremath{{0.009488 } } }
\vdef{bdt2011:bgBs2KMuNu:hiSideband1:val}   {\ensuremath{{0.000 } } }
\vdef{bdt2011:bgBs2KMuNu:hiSideband1:err}   {\ensuremath{{0.000 } } }
\vdef{bdt2011:bgBs2KPi:loSideband1:val}   {\ensuremath{{0.001 } } }
\vdef{bdt2011:bgBs2KPi:loSideband1:err}   {\ensuremath{{0.001 } } }
\vdef{bdt2011:bgBs2KPi:bdRare1}   {\ensuremath{{0.002933 } } }
\vdef{bdt2011:bgBs2KPi:bdRare1E}  {\ensuremath{{0.002172 } } }
\vdef{bdt2011:bgBs2KPi:bsRare1}   {\ensuremath{{0.004596 } } }
\vdef{bdt2011:bgBs2KPi:bsRare1E}  {\ensuremath{{0.003404 } } }
\vdef{bdt2011:bgBs2KPi:hiSideband1:val}   {\ensuremath{{0.001 } } }
\vdef{bdt2011:bgBs2KPi:hiSideband1:err}   {\ensuremath{{0.000 } } }
\vdef{bdt2011:bgBs2MuMuGamma:loSideband1:val}   {\ensuremath{{0.142 } } }
\vdef{bdt2011:bgBs2MuMuGamma:loSideband1:err}   {\ensuremath{{0.028 } } }
\vdef{bdt2011:bgBs2MuMuGamma:bdRare1}   {\ensuremath{{0.023949 } } }
\vdef{bdt2011:bgBs2MuMuGamma:bdRare1E}  {\ensuremath{{0.004790 } } }
\vdef{bdt2011:bgBs2MuMuGamma:bsRare1}   {\ensuremath{{0.008756 } } }
\vdef{bdt2011:bgBs2MuMuGamma:bsRare1E}  {\ensuremath{{0.001751 } } }
\vdef{bdt2011:bgBs2MuMuGamma:hiSideband1:val}   {\ensuremath{{0.000 } } }
\vdef{bdt2011:bgBs2MuMuGamma:hiSideband1:err}   {\ensuremath{{0.000 } } }
\vdef{bdt2011:bgBs2PiPi:loSideband1:val}   {\ensuremath{{0.000 } } }
\vdef{bdt2011:bgBs2PiPi:loSideband1:err}   {\ensuremath{{0.000 } } }
\vdef{bdt2011:bgBs2PiPi:bdRare1}   {\ensuremath{{0.000154 } } }
\vdef{bdt2011:bgBs2PiPi:bdRare1E}  {\ensuremath{{0.000156 } } }
\vdef{bdt2011:bgBs2PiPi:bsRare1}   {\ensuremath{{0.000493 } } }
\vdef{bdt2011:bgBs2PiPi:bsRare1E}  {\ensuremath{{0.000502 } } }
\vdef{bdt2011:bgBs2PiPi:hiSideband1:val}   {\ensuremath{{0.000 } } }
\vdef{bdt2011:bgBs2PiPi:hiSideband1:err}   {\ensuremath{{0.000 } } }
\vdef{bdt2011:bgBu2KMuMu:loSideband1:val}   {\ensuremath{{0.016 } } }
\vdef{bdt2011:bgBu2KMuMu:loSideband1:err}   {\ensuremath{{0.001 } } }
\vdef{bdt2011:bgBu2KMuMu:bdRare1}   {\ensuremath{{0.000000 } } }
\vdef{bdt2011:bgBu2KMuMu:bdRare1E}  {\ensuremath{{0.000000 } } }
\vdef{bdt2011:bgBu2KMuMu:bsRare1}   {\ensuremath{{0.000000 } } }
\vdef{bdt2011:bgBu2KMuMu:bsRare1E}  {\ensuremath{{0.000000 } } }
\vdef{bdt2011:bgBu2KMuMu:hiSideband1:val}   {\ensuremath{{0.000 } } }
\vdef{bdt2011:bgBu2KMuMu:hiSideband1:err}   {\ensuremath{{0.000 } } }
\vdef{bdt2011:bgBu2PiMuMu:loSideband1:val}   {\ensuremath{{0.295 } } }
\vdef{bdt2011:bgBu2PiMuMu:loSideband1:err}   {\ensuremath{{0.077 } } }
\vdef{bdt2011:bgBu2PiMuMu:bdRare1}   {\ensuremath{{0.001153 } } }
\vdef{bdt2011:bgBu2PiMuMu:bdRare1E}  {\ensuremath{{0.000300 } } }
\vdef{bdt2011:bgBu2PiMuMu:bsRare1}   {\ensuremath{{0.000000 } } }
\vdef{bdt2011:bgBu2PiMuMu:bsRare1E}  {\ensuremath{{0.000000 } } }
\vdef{bdt2011:bgBu2PiMuMu:hiSideband1:val}   {\ensuremath{{0.000 } } }
\vdef{bdt2011:bgBu2PiMuMu:hiSideband1:err}   {\ensuremath{{0.000 } } }
\vdef{bdt2011:bgLb2KP:loSideband1:val}   {\ensuremath{{0.000 } } }
\vdef{bdt2011:bgLb2KP:loSideband1:err}   {\ensuremath{{0.000 } } }
\vdef{bdt2011:bgLb2KP:bdRare1}   {\ensuremath{{0.000816 } } }
\vdef{bdt2011:bgLb2KP:bdRare1E}  {\ensuremath{{0.000613 } } }
\vdef{bdt2011:bgLb2KP:bsRare1}   {\ensuremath{{0.003436 } } }
\vdef{bdt2011:bgLb2KP:bsRare1E}  {\ensuremath{{0.002582 } } }
\vdef{bdt2011:bgLb2KP:hiSideband1:val}   {\ensuremath{{0.002 } } }
\vdef{bdt2011:bgLb2KP:hiSideband1:err}   {\ensuremath{{0.001 } } }
\vdef{bdt2011:bgLb2PMuNu:loSideband1:val}   {\ensuremath{{4.751 } } }
\vdef{bdt2011:bgLb2PMuNu:loSideband1:err}   {\ensuremath{{5.312 } } }
\vdef{bdt2011:bgLb2PMuNu:bdRare1}   {\ensuremath{{1.284575 } } }
\vdef{bdt2011:bgLb2PMuNu:bdRare1E}  {\ensuremath{{1.436198 } } }
\vdef{bdt2011:bgLb2PMuNu:bsRare1}   {\ensuremath{{0.803126 } } }
\vdef{bdt2011:bgLb2PMuNu:bsRare1E}  {\ensuremath{{0.897922 } } }
\vdef{bdt2011:bgLb2PMuNu:hiSideband1:val}   {\ensuremath{{0.215 } } }
\vdef{bdt2011:bgLb2PMuNu:hiSideband1:err}   {\ensuremath{{0.241 } } }
\vdef{bdt2011:bgLb2PiP:loSideband1:val}   {\ensuremath{{0.000 } } }
\vdef{bdt2011:bgLb2PiP:loSideband1:err}   {\ensuremath{{0.000 } } }
\vdef{bdt2011:bgLb2PiP:bdRare1}   {\ensuremath{{0.000210 } } }
\vdef{bdt2011:bgLb2PiP:bdRare1E}  {\ensuremath{{0.000160 } } }
\vdef{bdt2011:bgLb2PiP:bsRare1}   {\ensuremath{{0.000949 } } }
\vdef{bdt2011:bgLb2PiP:bsRare1E}  {\ensuremath{{0.000724 } } }
\vdef{bdt2011:bgLb2PiP:hiSideband1:val}   {\ensuremath{{0.001 } } }
\vdef{bdt2011:bgLb2PiP:hiSideband1:err}   {\ensuremath{{0.001 } } }
\vdef{bdt2011:bsRare1}   {\ensuremath{{0.000 } } }
\vdef{bdt2011:bsRare1E}  {\ensuremath{{0.000 } } }
\vdef{bdt2011:bdRare1}   {\ensuremath{{0.000 } } }
\vdef{bdt2011:bdRare1E}  {\ensuremath{{0.000 } } }
\vdef{bdt2011:N-EFF-TOT-BPLUS0:val}   {\ensuremath{{0.00098 } } }
\vdef{bdt2011:N-EFF-TOT-BPLUS0:err}   {\ensuremath{{0.000003 } } }
\vdef{bdt2011:N-EFF-TOT-BPLUS0:tot}   {\ensuremath{{0.00008 } } }
\vdef{bdt2011:N-EFF-TOT-BPLUS0:all}   {\ensuremath{{(0.98 \pm 0.08)\times 10^{-3}} } }
\vdef{bdt2011:N-EFF-PRODMC-BPLUS0:val}   {\ensuremath{{0.00099 } } }
\vdef{bdt2011:N-EFF-PRODMC-BPLUS0:err}   {\ensuremath{{0.000012 } } }
\vdef{bdt2011:N-EFF-PRODMC-BPLUS0:tot}   {\ensuremath{{0.00001 } } }
\vdef{bdt2011:N-EFF-PRODMC-BPLUS0:all}   {\ensuremath{{(0.99 \pm 0.01)\times 10^{-3}} } }
\vdef{bdt2011:N-EFF-PRODTNP-BPLUS0:val}   {\ensuremath{{0.00141 } } }
\vdef{bdt2011:N-EFF-PRODTNP-BPLUS0:err}   {\ensuremath{{0.759123 } } }
\vdef{bdt2011:N-EFF-PRODTNP-BPLUS0:tot}   {\ensuremath{{0.75912 } } }
\vdef{bdt2011:N-EFF-PRODTNP-BPLUS0:all}   {\ensuremath{{(1.41 \pm 759.12)\times 10^{-3}} } }
\vdef{bdt2011:N-EFF-PRODTNPMC-BPLUS0:val}   {\ensuremath{{0.00099 } } }
\vdef{bdt2011:N-EFF-PRODTNPMC-BPLUS0:err}   {\ensuremath{{0.000012 } } }
\vdef{bdt2011:N-EFF-PRODTNPMC-BPLUS0:tot}   {\ensuremath{{0.00001 } } }
\vdef{bdt2011:N-EFF-PRODTNPMC-BPLUS0:all}   {\ensuremath{{(0.99 \pm 0.01)\times 10^{-3}} } }
\vdef{bdt2011:N-ACC-BPLUS0:val}   {\ensuremath{{0.010 } } }
\vdef{bdt2011:N-ACC-BPLUS0:err}   {\ensuremath{{0.000 } } }
\vdef{bdt2011:N-ACC-BPLUS0:tot}   {\ensuremath{{0.000 } } }
\vdef{bdt2011:N-ACC-BPLUS0:all}   {\ensuremath{{(1.03 \pm 0.04)\times 10^{-2}} } }
\vdef{bdt2011:N-EFF-MU-PID-BPLUS0:val}   {\ensuremath{{0.714 } } }
\vdef{bdt2011:N-EFF-MU-PID-BPLUS0:err}   {\ensuremath{{0.000 } } }
\vdef{bdt2011:N-EFF-MU-PID-BPLUS0:tot}   {\ensuremath{{0.029 } } }
\vdef{bdt2011:N-EFF-MU-PID-BPLUS0:all}   {\ensuremath{{(71.42 \pm 2.86)\times 10^{-2}} } }
\vdef{bdt2011:N-EFFRHO-MU-PID-BPLUS0:val}   {\ensuremath{{0.515 } } }
\vdef{bdt2011:N-EFFRHO-MU-PID-BPLUS0:err}   {\ensuremath{{0.000 } } }
\vdef{bdt2011:N-EFFRHO-MU-PID-BPLUS0:tot}   {\ensuremath{{0.000 } } }
\vdef{bdt2011:N-EFFRHO-MU-PID-BPLUS0:all}   {\ensuremath{{(51.55 \pm 0.00)\times 10^{-2}} } }
\vdef{bdt2011:N-EFF-MU-PIDMC-BPLUS0:val}   {\ensuremath{{0.690 } } }
\vdef{bdt2011:N-EFF-MU-PIDMC-BPLUS0:err}   {\ensuremath{{0.000 } } }
\vdef{bdt2011:N-EFF-MU-PIDMC-BPLUS0:tot}   {\ensuremath{{0.028 } } }
\vdef{bdt2011:N-EFF-MU-PIDMC-BPLUS0:all}   {\ensuremath{{(68.95 \pm 2.76)\times 10^{-2}} } }
\vdef{bdt2011:N-EFFRHO-MU-PIDMC-BPLUS0:val}   {\ensuremath{{0.498 } } }
\vdef{bdt2011:N-EFFRHO-MU-PIDMC-BPLUS0:err}   {\ensuremath{{0.000 } } }
\vdef{bdt2011:N-EFFRHO-MU-PIDMC-BPLUS0:tot}   {\ensuremath{{0.021 } } }
\vdef{bdt2011:N-EFFRHO-MU-PIDMC-BPLUS0:all}   {\ensuremath{{(49.77 \pm 2.06)\times 10^{-2}} } }
\vdef{bdt2011:N-EFF-MU-MC-BPLUS0:val}   {\ensuremath{{0.498 } } }
\vdef{bdt2011:N-EFF-MU-MC-BPLUS0:err}   {\ensuremath{{0.001 } } }
\vdef{bdt2011:N-EFF-MU-MC-BPLUS0:tot}   {\ensuremath{{0.020 } } }
\vdef{bdt2011:N-EFF-MU-MC-BPLUS0:all}   {\ensuremath{{(49.77 \pm 1.99)\times 10^{-2}} } }
\vdef{bdt2011:N-EFF-TRIG-PID-BPLUS0:val}   {\ensuremath{{0.771 } } }
\vdef{bdt2011:N-EFF-TRIG-PID-BPLUS0:err}   {\ensuremath{{0.000 } } }
\vdef{bdt2011:N-EFF-TRIG-PID-BPLUS0:tot}   {\ensuremath{{0.023 } } }
\vdef{bdt2011:N-EFF-TRIG-PID-BPLUS0:all}   {\ensuremath{{(77.11 \pm 2.31)\times 10^{-2}} } }
\vdef{bdt2011:N-EFFRHO-TRIG-PID-BPLUS0:val}   {\ensuremath{{0.821 } } }
\vdef{bdt2011:N-EFFRHO-TRIG-PID-BPLUS0:err}   {\ensuremath{{0.000 } } }
\vdef{bdt2011:N-EFFRHO-TRIG-PID-BPLUS0:tot}   {\ensuremath{{0.025 } } }
\vdef{bdt2011:N-EFFRHO-TRIG-PID-BPLUS0:all}   {\ensuremath{{(82.10 \pm 2.46)\times 10^{-2}} } }
\vdef{bdt2011:N-EFF-TRIG-PIDMC-BPLUS0:val}   {\ensuremath{{0.726 } } }
\vdef{bdt2011:N-EFF-TRIG-PIDMC-BPLUS0:err}   {\ensuremath{{0.001 } } }
\vdef{bdt2011:N-EFF-TRIG-PIDMC-BPLUS0:tot}   {\ensuremath{{0.022 } } }
\vdef{bdt2011:N-EFF-TRIG-PIDMC-BPLUS0:all}   {\ensuremath{{(72.61 \pm 2.18)\times 10^{-2}} } }
\vdef{bdt2011:N-EFFRHO-TRIG-PIDMC-BPLUS0:val}   {\ensuremath{{0.773 } } }
\vdef{bdt2011:N-EFFRHO-TRIG-PIDMC-BPLUS0:err}   {\ensuremath{{0.001 } } }
\vdef{bdt2011:N-EFFRHO-TRIG-PIDMC-BPLUS0:tot}   {\ensuremath{{0.023 } } }
\vdef{bdt2011:N-EFFRHO-TRIG-PIDMC-BPLUS0:all}   {\ensuremath{{(77.30 \pm 2.32)\times 10^{-2}} } }
\vdef{bdt2011:N-EFF-TRIG-MC-BPLUS0:val}   {\ensuremath{{0.773 } } }
\vdef{bdt2011:N-EFF-TRIG-MC-BPLUS0:err}   {\ensuremath{{0.001 } } }
\vdef{bdt2011:N-EFF-TRIG-MC-BPLUS0:tot}   {\ensuremath{{0.023 } } }
\vdef{bdt2011:N-EFF-TRIG-MC-BPLUS0:all}   {\ensuremath{{(77.30 \pm 2.32)\times 10^{-2}} } }
\vdef{bdt2011:N-EFF-CAND-BPLUS0:val}   {\ensuremath{{1.000 } } }
\vdef{bdt2011:N-EFF-CAND-BPLUS0:err}   {\ensuremath{{0.000 } } }
\vdef{bdt2011:N-EFF-CAND-BPLUS0:tot}   {\ensuremath{{0.010 } } }
\vdef{bdt2011:N-EFF-CAND-BPLUS0:all}   {\ensuremath{{(99.99 \pm 1.00)\times 10^{-2}} } }
\vdef{bdt2011:N-EFF-ANA-BPLUS0:val}   {\ensuremath{{0.2497 } } }
\vdef{bdt2011:N-EFF-ANA-BPLUS0:err}   {\ensuremath{{0.0005 } } }
\vdef{bdt2011:N-EFF-ANA-BPLUS0:tot}   {\ensuremath{{0.0141 } } }
\vdef{bdt2011:N-EFF-ANA-BPLUS0:all}   {\ensuremath{{(24.97 \pm 1.41)\times 10^{-2}} } }
\vdef{bdt2011:N-OBS-BPLUS0:val}   {\ensuremath{{71\,191 } } }
\vdef{bdt2011:N-OBS-BPLUS0:err}   {\ensuremath{{2050 } } }
\vdef{bdt2011:N-OBS-BPLUS0:tot}   {\ensuremath{{4107 } } }
\vdef{bdt2011:N-OBS-BPLUS0:all}   {\ensuremath{{71191 } } }
\vdef{bdt2011:N-OBS-CBPLUS0:val}   {\ensuremath{{72810 } } }
\vdef{bdt2011:N-OBS-CBPLUS0:err}   {\ensuremath{{277 } } }
\vdef{bdt2011:N-EFF-TOT-BPLUS1:val}   {\ensuremath{{0.00036 } } }
\vdef{bdt2011:N-EFF-TOT-BPLUS1:err}   {\ensuremath{{0.000002 } } }
\vdef{bdt2011:N-EFF-TOT-BPLUS1:tot}   {\ensuremath{{0.00004 } } }
\vdef{bdt2011:N-EFF-TOT-BPLUS1:all}   {\ensuremath{{(0.36 \pm 0.04)\times 10^{-3}} } }
\vdef{bdt2011:N-EFF-PRODMC-BPLUS1:val}   {\ensuremath{{0.00035 } } }
\vdef{bdt2011:N-EFF-PRODMC-BPLUS1:err}   {\ensuremath{{0.000006 } } }
\vdef{bdt2011:N-EFF-PRODMC-BPLUS1:tot}   {\ensuremath{{0.00001 } } }
\vdef{bdt2011:N-EFF-PRODMC-BPLUS1:all}   {\ensuremath{{(0.35 \pm 0.01)\times 10^{-3}} } }
\vdef{bdt2011:N-EFF-PRODTNP-BPLUS1:val}   {\ensuremath{{0.00044 } } }
\vdef{bdt2011:N-EFF-PRODTNP-BPLUS1:err}   {\ensuremath{{0.143284 } } }
\vdef{bdt2011:N-EFF-PRODTNP-BPLUS1:tot}   {\ensuremath{{0.14328 } } }
\vdef{bdt2011:N-EFF-PRODTNP-BPLUS1:all}   {\ensuremath{{(0.44 \pm 143.28)\times 10^{-3}} } }
\vdef{bdt2011:N-EFF-PRODTNPMC-BPLUS1:val}   {\ensuremath{{0.00035 } } }
\vdef{bdt2011:N-EFF-PRODTNPMC-BPLUS1:err}   {\ensuremath{{0.000006 } } }
\vdef{bdt2011:N-EFF-PRODTNPMC-BPLUS1:tot}   {\ensuremath{{0.00001 } } }
\vdef{bdt2011:N-EFF-PRODTNPMC-BPLUS1:all}   {\ensuremath{{(0.35 \pm 0.01)\times 10^{-3}} } }
\vdef{bdt2011:N-ACC-BPLUS1:val}   {\ensuremath{{0.005 } } }
\vdef{bdt2011:N-ACC-BPLUS1:err}   {\ensuremath{{0.000 } } }
\vdef{bdt2011:N-ACC-BPLUS1:tot}   {\ensuremath{{0.000 } } }
\vdef{bdt2011:N-ACC-BPLUS1:all}   {\ensuremath{{(0.53 \pm 0.03)\times 10^{-2}} } }
\vdef{bdt2011:N-EFF-MU-PID-BPLUS1:val}   {\ensuremath{{0.648 } } }
\vdef{bdt2011:N-EFF-MU-PID-BPLUS1:err}   {\ensuremath{{0.001 } } }
\vdef{bdt2011:N-EFF-MU-PID-BPLUS1:tot}   {\ensuremath{{0.052 } } }
\vdef{bdt2011:N-EFF-MU-PID-BPLUS1:all}   {\ensuremath{{(64.78 \pm 5.18)\times 10^{-2}} } }
\vdef{bdt2011:N-EFFRHO-MU-PID-BPLUS1:val}   {\ensuremath{{0.566 } } }
\vdef{bdt2011:N-EFFRHO-MU-PID-BPLUS1:err}   {\ensuremath{{0.001 } } }
\vdef{bdt2011:N-EFFRHO-MU-PID-BPLUS1:tot}   {\ensuremath{{0.000 } } }
\vdef{bdt2011:N-EFFRHO-MU-PID-BPLUS1:all}   {\ensuremath{{(56.65 \pm 0.00)\times 10^{-2}} } }
\vdef{bdt2011:N-EFF-MU-PIDMC-BPLUS1:val}   {\ensuremath{{0.682 } } }
\vdef{bdt2011:N-EFF-MU-PIDMC-BPLUS1:err}   {\ensuremath{{0.001 } } }
\vdef{bdt2011:N-EFF-MU-PIDMC-BPLUS1:tot}   {\ensuremath{{0.055 } } }
\vdef{bdt2011:N-EFF-MU-PIDMC-BPLUS1:all}   {\ensuremath{{(68.17 \pm 5.45)\times 10^{-2}} } }
\vdef{bdt2011:N-EFFRHO-MU-PIDMC-BPLUS1:val}   {\ensuremath{{0.596 } } }
\vdef{bdt2011:N-EFFRHO-MU-PIDMC-BPLUS1:err}   {\ensuremath{{0.001 } } }
\vdef{bdt2011:N-EFFRHO-MU-PIDMC-BPLUS1:tot}   {\ensuremath{{0.045 } } }
\vdef{bdt2011:N-EFFRHO-MU-PIDMC-BPLUS1:all}   {\ensuremath{{(59.62 \pm 4.53)\times 10^{-2}} } }
\vdef{bdt2011:N-EFF-MU-MC-BPLUS1:val}   {\ensuremath{{0.596 } } }
\vdef{bdt2011:N-EFF-MU-MC-BPLUS1:err}   {\ensuremath{{0.002 } } }
\vdef{bdt2011:N-EFF-MU-MC-BPLUS1:tot}   {\ensuremath{{0.048 } } }
\vdef{bdt2011:N-EFF-MU-MC-BPLUS1:all}   {\ensuremath{{(59.62 \pm 4.77)\times 10^{-2}} } }
\vdef{bdt2011:N-EFF-TRIG-PID-BPLUS1:val}   {\ensuremath{{0.689 } } }
\vdef{bdt2011:N-EFF-TRIG-PID-BPLUS1:err}   {\ensuremath{{0.001 } } }
\vdef{bdt2011:N-EFF-TRIG-PID-BPLUS1:tot}   {\ensuremath{{0.041 } } }
\vdef{bdt2011:N-EFF-TRIG-PID-BPLUS1:all}   {\ensuremath{{(68.87 \pm 4.13)\times 10^{-2}} } }
\vdef{bdt2011:N-EFFRHO-TRIG-PID-BPLUS1:val}   {\ensuremath{{0.764 } } }
\vdef{bdt2011:N-EFFRHO-TRIG-PID-BPLUS1:err}   {\ensuremath{{0.002 } } }
\vdef{bdt2011:N-EFFRHO-TRIG-PID-BPLUS1:tot}   {\ensuremath{{0.046 } } }
\vdef{bdt2011:N-EFFRHO-TRIG-PID-BPLUS1:all}   {\ensuremath{{(76.40 \pm 4.59)\times 10^{-2}} } }
\vdef{bdt2011:N-EFF-TRIG-PIDMC-BPLUS1:val}   {\ensuremath{{0.541 } } }
\vdef{bdt2011:N-EFF-TRIG-PIDMC-BPLUS1:err}   {\ensuremath{{0.002 } } }
\vdef{bdt2011:N-EFF-TRIG-PIDMC-BPLUS1:tot}   {\ensuremath{{0.033 } } }
\vdef{bdt2011:N-EFF-TRIG-PIDMC-BPLUS1:all}   {\ensuremath{{(54.15 \pm 3.25)\times 10^{-2}} } }
\vdef{bdt2011:N-EFFRHO-TRIG-PIDMC-BPLUS1:val}   {\ensuremath{{0.601 } } }
\vdef{bdt2011:N-EFFRHO-TRIG-PIDMC-BPLUS1:err}   {\ensuremath{{0.002 } } }
\vdef{bdt2011:N-EFFRHO-TRIG-PIDMC-BPLUS1:tot}   {\ensuremath{{0.036 } } }
\vdef{bdt2011:N-EFFRHO-TRIG-PIDMC-BPLUS1:all}   {\ensuremath{{(60.06 \pm 3.61)\times 10^{-2}} } }
\vdef{bdt2011:N-EFF-TRIG-MC-BPLUS1:val}   {\ensuremath{{0.601 } } }
\vdef{bdt2011:N-EFF-TRIG-MC-BPLUS1:err}   {\ensuremath{{0.002 } } }
\vdef{bdt2011:N-EFF-TRIG-MC-BPLUS1:tot}   {\ensuremath{{0.036 } } }
\vdef{bdt2011:N-EFF-TRIG-MC-BPLUS1:all}   {\ensuremath{{(60.06 \pm 3.61)\times 10^{-2}} } }
\vdef{bdt2011:N-EFF-CAND-BPLUS1:val}   {\ensuremath{{1.000 } } }
\vdef{bdt2011:N-EFF-CAND-BPLUS1:err}   {\ensuremath{{0.000 } } }
\vdef{bdt2011:N-EFF-CAND-BPLUS1:tot}   {\ensuremath{{0.010 } } }
\vdef{bdt2011:N-EFF-CAND-BPLUS1:all}   {\ensuremath{{(99.97 \pm 1.00)\times 10^{-2}} } }
\vdef{bdt2011:N-EFF-ANA-BPLUS1:val}   {\ensuremath{{0.1872 } } }
\vdef{bdt2011:N-EFF-ANA-BPLUS1:err}   {\ensuremath{{0.0006 } } }
\vdef{bdt2011:N-EFF-ANA-BPLUS1:tot}   {\ensuremath{{0.0106 } } }
\vdef{bdt2011:N-EFF-ANA-BPLUS1:all}   {\ensuremath{{(18.72 \pm 1.06)\times 10^{-2}} } }
\vdef{bdt2011:N-OBS-BPLUS1:val}   {\ensuremath{{21373 } } }
\vdef{bdt2011:N-OBS-BPLUS1:err}   {\ensuremath{{161 } } }
\vdef{bdt2011:N-OBS-BPLUS1:tot}   {\ensuremath{{1080 } } }
\vdef{bdt2011:N-OBS-BPLUS1:all}   {\ensuremath{{21373 } } }
\vdef{bdt2011:N-OBS-CBPLUS1:val}   {\ensuremath{{20295 } } }
\vdef{bdt2011:N-OBS-CBPLUS1:err}   {\ensuremath{{152 } } }
\vdef{bdt2011:N-EXP2-SIG-BSMM0:val}   {\ensuremath{{ 2.77 } } }
\vdef{bdt2011:N-EXP2-SIG-BSMM0:err}   {\ensuremath{{ 0.42 } } }
\vdef{bdt2011:N-EXP2-SIG-BDMM0:val}   {\ensuremath{{0.27 } } }
\vdef{bdt2011:N-EXP2-SIG-BDMM0:err}   {\ensuremath{{0.03 } } }
\vdef{bdt2011:N-OBS-BKG0:val}   {\ensuremath{{3 } } }
\vdef{bdt2011:N-EXP-BSMM0:val}   {\ensuremath{{ 0.37 } } }
\vdef{bdt2011:N-EXP-BSMM0:err}   {\ensuremath{{ 0.18 } } }
\vdef{bdt2011:N-EXP-BDMM0:val}   {\ensuremath{{ 0.47 } } }
\vdef{bdt2011:N-EXP-BDMM0:err}   {\ensuremath{{ 0.23 } } }
\vdef{bdt2011:N-LOW-BD0:val}   {\ensuremath{{5.200 } } }
\vdef{bdt2011:N-HIGH-BD0:val}   {\ensuremath{{5.300 } } }
\vdef{bdt2011:N-LOW-BS0:val}   {\ensuremath{{5.300 } } }
\vdef{bdt2011:N-HIGH-BS0:val}   {\ensuremath{{5.450 } } }
\vdef{bdt2011:N-PSS0:val}   {\ensuremath{{0.889 } } }
\vdef{bdt2011:N-PSS0:err}   {\ensuremath{{0.003 } } }
\vdef{bdt2011:N-PSS0:tot}   {\ensuremath{{0.045 } } }
\vdef{bdt2011:N-PSD0:val}   {\ensuremath{{0.273 } } }
\vdef{bdt2011:N-PSD0:err}   {\ensuremath{{0.016 } } }
\vdef{bdt2011:N-PSD0:tot}   {\ensuremath{{0.021 } } }
\vdef{bdt2011:N-PDS0:val}   {\ensuremath{{0.061 } } }
\vdef{bdt2011:N-PDS0:err}   {\ensuremath{{0.002 } } }
\vdef{bdt2011:N-PDS0:tot}   {\ensuremath{{0.004 } } }
\vdef{bdt2011:N-PDD0:val}   {\ensuremath{{0.675 } } }
\vdef{bdt2011:N-PDD0:err}   {\ensuremath{{0.017 } } }
\vdef{bdt2011:N-PDD0:tot}   {\ensuremath{{0.038 } } }
\vdef{bdt2011:N-EFF-TOT-BSMM0:val}   {\ensuremath{{0.0030 } } }
\vdef{bdt2011:N-EFF-TOT-BSMM0:err}   {\ensuremath{{0.0000 } } }
\vdef{bdt2011:N-EFF-TOT-BSMM0:tot}   {\ensuremath{{0.0003 } } }
\vdef{bdt2011:N-EFF-TOT-BSMM0:all}   {\ensuremath{{(0.30 \pm 0.04)} } }
\vdef{bdt2011:N-EFF-PRODMC-BSMM0:val}   {\ensuremath{{0.0030 } } }
\vdef{bdt2011:N-EFF-PRODMC-BSMM0:err}   {\ensuremath{{0.0000 } } }
\vdef{bdt2011:N-EFF-PRODMC-BSMM0:tot}   {\ensuremath{{0.0000 } } }
\vdef{bdt2011:N-EFF-PRODMC-BSMM0:all}   {\ensuremath{{(3.00 \pm 0.03)\times 10^{-3}} } }
\vdef{bdt2011:N-EFFRATIO-TOT-BSMM0:val}   {\ensuremath{{0.325 } } }
\vdef{bdt2011:N-EFFRATIO-TOT-BSMM0:err}   {\ensuremath{{0.003 } } }
\vdef{bdt2011:N-EFFRATIO-PRODMC-BSMM0:val}   {\ensuremath{{0.328 } } }
\vdef{bdt2011:N-EFFRATIO-PRODMC-BSMM0:err}   {\ensuremath{{0.005 } } }
\vdef{bdt2011:N-EFFRATIO-PRODTNP-BSMM0:val}   {\ensuremath{{0.324 } } }
\vdef{bdt2011:N-EFFRATIO-PRODTNP-BSMM0:err}   {\ensuremath{{185.253 } } }
\vdef{bdt2011:N-EFFRATIO-PRODTNPMC-BSMM0:val}   {\ensuremath{{0.328 } } }
\vdef{bdt2011:N-EFFRATIO-PRODTNPMC-BSMM0:err}   {\ensuremath{{0.005 } } }
\vdef{bdt2011:N-EFF-PRODTNP-BSMM0:val}   {\ensuremath{{0.0044 } } }
\vdef{bdt2011:N-EFF-PRODTNP-BSMM0:err}   {\ensuremath{{0.8467 } } }
\vdef{bdt2011:N-EFF-PRODTNP-BSMM0:tot}   {\ensuremath{{0.8467 } } }
\vdef{bdt2011:N-EFF-PRODTNP-BSMM0:all}   {\ensuremath{{(4.36 \pm 846.67)\times 10^{-3}} } }
\vdef{bdt2011:N-EFF-PRODTNPMC-BSMM0:val}   {\ensuremath{{0.0030 } } }
\vdef{bdt2011:N-EFF-PRODTNPMC-BSMM0:err}   {\ensuremath{{0.0000 } } }
\vdef{bdt2011:N-EFF-PRODTNPMC-BSMM0:tot}   {\ensuremath{{0.0000 } } }
\vdef{bdt2011:N-EFF-PRODTNPMC-BSMM0:all}   {\ensuremath{{(3.00 \pm 0.03)\times 10^{-3}} } }
\vdef{bdt2011:N-ACC-BSMM0:val}   {\ensuremath{{0.034 } } }
\vdef{bdt2011:N-ACC-BSMM0:err}   {\ensuremath{{0.000 } } }
\vdef{bdt2011:N-ACC-BSMM0:tot}   {\ensuremath{{0.001 } } }
\vdef{bdt2011:N-ACC-BSMM0:all}   {\ensuremath{{(3.36 \pm 0.12)\times 10^{-2}} } }
\vdef{bdt2011:N-EFF-MU-PID-BSMM0:val}   {\ensuremath{{0.722 } } }
\vdef{bdt2011:N-EFF-MU-PID-BSMM0:err}   {\ensuremath{{0.001 } } }
\vdef{bdt2011:N-EFF-MU-PID-BSMM0:tot}   {\ensuremath{{0.029 } } }
\vdef{bdt2011:N-EFF-MU-PID-BSMM0:all}   {\ensuremath{{(72.23 \pm 2.89)\times 10^{-2}} } }
\vdef{bdt2011:N-EFFRHO-MU-PID-BSMM0:val}   {\ensuremath{{0.486 } } }
\vdef{bdt2011:N-EFFRHO-MU-PID-BSMM0:err}   {\ensuremath{{0.001 } } }
\vdef{bdt2011:N-EFFRHO-MU-PID-BSMM0:tot}   {\ensuremath{{0.000 } } }
\vdef{bdt2011:N-EFFRHO-MU-PID-BSMM0:all}   {\ensuremath{{(48.57 \pm 0.00)\times 10^{-2}} } }
\vdef{bdt2011:N-EFF-MU-PIDMC-BSMM0:val}   {\ensuremath{{0.701 } } }
\vdef{bdt2011:N-EFF-MU-PIDMC-BSMM0:err}   {\ensuremath{{0.001 } } }
\vdef{bdt2011:N-EFF-MU-PIDMC-BSMM0:tot}   {\ensuremath{{0.028 } } }
\vdef{bdt2011:N-EFF-MU-PIDMC-BSMM0:all}   {\ensuremath{{(70.11 \pm 2.81)\times 10^{-2}} } }
\vdef{bdt2011:N-EFFRHO-MU-PIDMC-BSMM0:val}   {\ensuremath{{0.471 } } }
\vdef{bdt2011:N-EFFRHO-MU-PIDMC-BSMM0:err}   {\ensuremath{{0.001 } } }
\vdef{bdt2011:N-EFFRHO-MU-PIDMC-BSMM0:tot}   {\ensuremath{{0.019 } } }
\vdef{bdt2011:N-EFFRHO-MU-PIDMC-BSMM0:all}   {\ensuremath{{(47.14 \pm 1.94)\times 10^{-2}} } }
\vdef{bdt2011:N-EFF-MU-MC-BSMM0:val}   {\ensuremath{{0.471 } } }
\vdef{bdt2011:N-EFF-MU-MC-BSMM0:err}   {\ensuremath{{0.003 } } }
\vdef{bdt2011:N-EFF-MU-MC-BSMM0:tot}   {\ensuremath{{0.019 } } }
\vdef{bdt2011:N-EFF-MU-MC-BSMM0:all}   {\ensuremath{{(47.14 \pm 1.91)\times 10^{-2}} } }
\vdef{bdt2011:N-EFF-TRIG-PID-BSMM0:val}   {\ensuremath{{0.782 } } }
\vdef{bdt2011:N-EFF-TRIG-PID-BSMM0:err}   {\ensuremath{{0.001 } } }
\vdef{bdt2011:N-EFF-TRIG-PID-BSMM0:tot}   {\ensuremath{{0.023 } } }
\vdef{bdt2011:N-EFF-TRIG-PID-BSMM0:all}   {\ensuremath{{(78.21 \pm 2.35)\times 10^{-2}} } }
\vdef{bdt2011:N-EFFRHO-TRIG-PID-BSMM0:val}   {\ensuremath{{0.873 } } }
\vdef{bdt2011:N-EFFRHO-TRIG-PID-BSMM0:err}   {\ensuremath{{0.001 } } }
\vdef{bdt2011:N-EFFRHO-TRIG-PID-BSMM0:tot}   {\ensuremath{{0.026 } } }
\vdef{bdt2011:N-EFFRHO-TRIG-PID-BSMM0:all}   {\ensuremath{{(87.27 \pm 2.62)\times 10^{-2}} } }
\vdef{bdt2011:N-EFF-TRIG-PIDMC-BSMM0:val}   {\ensuremath{{0.740 } } }
\vdef{bdt2011:N-EFF-TRIG-PIDMC-BSMM0:err}   {\ensuremath{{0.001 } } }
\vdef{bdt2011:N-EFF-TRIG-PIDMC-BSMM0:tot}   {\ensuremath{{0.022 } } }
\vdef{bdt2011:N-EFF-TRIG-PIDMC-BSMM0:all}   {\ensuremath{{(74.05 \pm 2.23)\times 10^{-2}} } }
\vdef{bdt2011:N-EFFRHO-TRIG-PIDMC-BSMM0:val}   {\ensuremath{{0.826 } } }
\vdef{bdt2011:N-EFFRHO-TRIG-PIDMC-BSMM0:err}   {\ensuremath{{0.002 } } }
\vdef{bdt2011:N-EFFRHO-TRIG-PIDMC-BSMM0:tot}   {\ensuremath{{0.025 } } }
\vdef{bdt2011:N-EFFRHO-TRIG-PIDMC-BSMM0:all}   {\ensuremath{{(82.63 \pm 2.48)\times 10^{-2}} } }
\vdef{bdt2011:N-EFF-TRIG-MC-BSMM0:val}   {\ensuremath{{0.826 } } }
\vdef{bdt2011:N-EFF-TRIG-MC-BSMM0:err}   {\ensuremath{{0.003 } } }
\vdef{bdt2011:N-EFF-TRIG-MC-BSMM0:tot}   {\ensuremath{{0.025 } } }
\vdef{bdt2011:N-EFF-TRIG-MC-BSMM0:all}   {\ensuremath{{(82.63 \pm 2.50)\times 10^{-2}} } }
\vdef{bdt2011:N-EFF-CAND-BSMM0:val}   {\ensuremath{{1.000 } } }
\vdef{bdt2011:N-EFF-CAND-BSMM0:err}   {\ensuremath{{0.000 } } }
\vdef{bdt2011:N-EFF-CAND-BSMM0:tot}   {\ensuremath{{0.010 } } }
\vdef{bdt2011:N-EFF-CAND-BSMM0:all}   {\ensuremath{{(99.99 \pm 1.00)\times 10^{-2}} } }
\vdef{bdt2011:N-EFF-ANA-BSMM0:val}   {\ensuremath{{0.229 } } }
\vdef{bdt2011:N-EFF-ANA-BSMM0:err}   {\ensuremath{{0.001 } } }
\vdef{bdt2011:N-EFF-ANA-BSMM0:tot}   {\ensuremath{{0.007 } } }
\vdef{bdt2011:N-EFF-ANA-BSMM0:all}   {\ensuremath{{(22.92 \pm 0.70)\times 10^{-2}} } }
\vdef{bdt2011:N-EFF-TOT-BDMM0:val}   {\ensuremath{{0.0033 } } }
\vdef{bdt2011:N-EFF-TOT-BDMM0:err}   {\ensuremath{{0.0001 } } }
\vdef{bdt2011:N-EFF-TOT-BDMM0:tot}   {\ensuremath{{0.0003 } } }
\vdef{bdt2011:N-EFF-TOT-BDMM0:all}   {\ensuremath{{(0.33 \pm 0.03)} } }
\vdef{bdt2011:N-EFF-PRODMC-BDMM0:val}   {\ensuremath{{0.0033 } } }
\vdef{bdt2011:N-EFF-PRODMC-BDMM0:err}   {\ensuremath{{0.0001 } } }
\vdef{bdt2011:N-EFF-PRODMC-BDMM0:tot}   {\ensuremath{{0.0001 } } }
\vdef{bdt2011:N-EFF-PRODMC-BDMM0:all}   {\ensuremath{{(3.30 \pm 0.12)\times 10^{-3}} } }
\vdef{bdt2011:N-EFF-PRODTNP-BDMM0:val}   {\ensuremath{{0.0047 } } }
\vdef{bdt2011:N-EFF-PRODTNP-BDMM0:err}   {\ensuremath{{0.2439 } } }
\vdef{bdt2011:N-EFF-PRODTNP-BDMM0:tot}   {\ensuremath{{0.2439 } } }
\vdef{bdt2011:N-EFF-PRODTNP-BDMM0:all}   {\ensuremath{{(4.70 \pm 243.95)\times 10^{-3}} } }
\vdef{bdt2011:N-EFF-PRODTNPMC-BDMM0:val}   {\ensuremath{{0.00330 } } }
\vdef{bdt2011:N-EFF-PRODTNPMC-BDMM0:err}   {\ensuremath{{0.000110 } } }
\vdef{bdt2011:N-EFF-PRODTNPMC-BDMM0:tot}   {\ensuremath{{0.00011 } } }
\vdef{bdt2011:N-EFF-PRODTNPMC-BDMM0:all}   {\ensuremath{{(3.30 \pm 0.11)\times 10^{-3}} } }
\vdef{bdt2011:N-ACC-BDMM0:val}   {\ensuremath{{0.034 } } }
\vdef{bdt2011:N-ACC-BDMM0:err}   {\ensuremath{{0.000 } } }
\vdef{bdt2011:N-ACC-BDMM0:tot}   {\ensuremath{{0.001 } } }
\vdef{bdt2011:N-ACC-BDMM0:all}   {\ensuremath{{(3.37 \pm 0.12)\times 10^{-2}} } }
\vdef{bdt2011:N-EFF-MU-PID-BDMM0:val}   {\ensuremath{{0.718 } } }
\vdef{bdt2011:N-EFF-MU-PID-BDMM0:err}   {\ensuremath{{0.004 } } }
\vdef{bdt2011:N-EFF-MU-PID-BDMM0:tot}   {\ensuremath{{0.029 } } }
\vdef{bdt2011:N-EFF-MU-PID-BDMM0:all}   {\ensuremath{{(71.82 \pm 2.90)\times 10^{-2}} } }
\vdef{bdt2011:N-EFFRHO-MU-PID-BDMM0:val}   {\ensuremath{{0.476 } } }
\vdef{bdt2011:N-EFFRHO-MU-PID-BDMM0:err}   {\ensuremath{{0.003 } } }
\vdef{bdt2011:N-EFFRHO-MU-PID-BDMM0:tot}   {\ensuremath{{0.000 } } }
\vdef{bdt2011:N-EFFRHO-MU-PID-BDMM0:all}   {\ensuremath{{(47.57 \pm 0.00)\times 10^{-2}} } }
\vdef{bdt2011:N-EFF-MU-PIDMC-BDMM0:val}   {\ensuremath{{0.701 } } }
\vdef{bdt2011:N-EFF-MU-PIDMC-BDMM0:err}   {\ensuremath{{0.004 } } }
\vdef{bdt2011:N-EFF-MU-PIDMC-BDMM0:tot}   {\ensuremath{{0.028 } } }
\vdef{bdt2011:N-EFF-MU-PIDMC-BDMM0:all}   {\ensuremath{{(70.07 \pm 2.84)\times 10^{-2}} } }
\vdef{bdt2011:N-EFFRHO-MU-PIDMC-BDMM0:val}   {\ensuremath{{0.464 } } }
\vdef{bdt2011:N-EFFRHO-MU-PIDMC-BDMM0:err}   {\ensuremath{{0.003 } } }
\vdef{bdt2011:N-EFFRHO-MU-PIDMC-BDMM0:tot}   {\ensuremath{{0.019 } } }
\vdef{bdt2011:N-EFFRHO-MU-PIDMC-BDMM0:all}   {\ensuremath{{(46.41 \pm 1.92)\times 10^{-2}} } }
\vdef{bdt2011:N-EFF-MU-MC-BDMM0:val}   {\ensuremath{{0.464 } } }
\vdef{bdt2011:N-EFF-MU-MC-BDMM0:err}   {\ensuremath{{0.011 } } }
\vdef{bdt2011:N-EFF-MU-MC-BDMM0:tot}   {\ensuremath{{0.022 } } }
\vdef{bdt2011:N-EFF-MU-MC-BDMM0:all}   {\ensuremath{{(46.41 \pm 2.16)\times 10^{-2}} } }
\vdef{bdt2011:N-EFF-TRIG-PID-BDMM0:val}   {\ensuremath{{0.781 } } }
\vdef{bdt2011:N-EFF-TRIG-PID-BDMM0:err}   {\ensuremath{{0.004 } } }
\vdef{bdt2011:N-EFF-TRIG-PID-BDMM0:tot}   {\ensuremath{{0.024 } } }
\vdef{bdt2011:N-EFF-TRIG-PID-BDMM0:all}   {\ensuremath{{(78.12 \pm 2.38)\times 10^{-2}} } }
\vdef{bdt2011:N-EFFRHO-TRIG-PID-BDMM0:val}   {\ensuremath{{0.896 } } }
\vdef{bdt2011:N-EFFRHO-TRIG-PID-BDMM0:err}   {\ensuremath{{0.005 } } }
\vdef{bdt2011:N-EFFRHO-TRIG-PID-BDMM0:tot}   {\ensuremath{{0.027 } } }
\vdef{bdt2011:N-EFFRHO-TRIG-PID-BDMM0:all}   {\ensuremath{{(89.59 \pm 2.74)\times 10^{-2}} } }
\vdef{bdt2011:N-EFF-TRIG-PIDMC-BDMM0:val}   {\ensuremath{{0.739 } } }
\vdef{bdt2011:N-EFF-TRIG-PIDMC-BDMM0:err}   {\ensuremath{{0.005 } } }
\vdef{bdt2011:N-EFF-TRIG-PIDMC-BDMM0:tot}   {\ensuremath{{0.023 } } }
\vdef{bdt2011:N-EFF-TRIG-PIDMC-BDMM0:all}   {\ensuremath{{(73.90 \pm 2.28)\times 10^{-2}} } }
\vdef{bdt2011:N-EFFRHO-TRIG-PIDMC-BDMM0:val}   {\ensuremath{{0.848 } } }
\vdef{bdt2011:N-EFFRHO-TRIG-PIDMC-BDMM0:err}   {\ensuremath{{0.006 } } }
\vdef{bdt2011:N-EFFRHO-TRIG-PIDMC-BDMM0:tot}   {\ensuremath{{0.026 } } }
\vdef{bdt2011:N-EFFRHO-TRIG-PIDMC-BDMM0:all}   {\ensuremath{{(84.75 \pm 2.61)\times 10^{-2}} } }
\vdef{bdt2011:N-EFF-TRIG-MC-BDMM0:val}   {\ensuremath{{0.848 } } }
\vdef{bdt2011:N-EFF-TRIG-MC-BDMM0:err}   {\ensuremath{{0.012 } } }
\vdef{bdt2011:N-EFF-TRIG-MC-BDMM0:tot}   {\ensuremath{{0.028 } } }
\vdef{bdt2011:N-EFF-TRIG-MC-BDMM0:all}   {\ensuremath{{(84.75 \pm 2.80)\times 10^{-2}} } }
\vdef{bdt2011:N-EFF-CAND-BDMM0:val}   {\ensuremath{{1.000 } } }
\vdef{bdt2011:N-EFF-CAND-BDMM0:err}   {\ensuremath{{0.000 } } }
\vdef{bdt2011:N-EFF-CAND-BDMM0:tot}   {\ensuremath{{0.010 } } }
\vdef{bdt2011:N-EFF-CAND-BDMM0:all}   {\ensuremath{{(100.00 \pm 1.00)\times 10^{-2}} } }
\vdef{bdt2011:N-EFF-ANA-BDMM0:val}   {\ensuremath{{0.249 } } }
\vdef{bdt2011:N-EFF-ANA-BDMM0:err}   {\ensuremath{{0.005 } } }
\vdef{bdt2011:N-EFF-ANA-BDMM0:tot}   {\ensuremath{{0.009 } } }
\vdef{bdt2011:N-EFF-ANA-BDMM0:all}   {\ensuremath{{(24.87 \pm 0.89)\times 10^{-2}} } }
\vdef{bdt2011:N-EXP-OBS-BS0:val}   {\ensuremath{{ 3.44 } } }
\vdef{bdt2011:N-EXP-OBS-BS0:err}   {\ensuremath{{ 0.45 } } }
\vdef{bdt2011:N-EXP-OBS-BD0:val}   {\ensuremath{{ 1.3 } } }
\vdef{bdt2011:N-EXP-OBS-BD0:err}   {\ensuremath{{ 0.2 } } }
\vdef{bdt2011:N-OBS-BSMM0:val}   {\ensuremath{{4 } } }
\vdef{bdt2011:N-OBS-BDMM0:val}   {\ensuremath{{3 } } }
\vdef{bdt2011:N-OFFLO-RARE0:val}   {\ensuremath{{ 0.11 } } }
\vdef{bdt2011:N-OFFLO-RARE0:err}   {\ensuremath{{ 0.06 } } }
\vdef{bdt2011:N-OFFHI-RARE0:val}   {\ensuremath{{ 0.01 } } }
\vdef{bdt2011:N-OFFHI-RARE0:err}   {\ensuremath{{ 0.00 } } }
\vdef{bdt2011:N-PEAK-BKG-BS0:val}   {\ensuremath{{ 0.11 } } }
\vdef{bdt2011:N-PEAK-BKG-BS0:err}   {\ensuremath{{ 0.08 } } }
\vdef{bdt2011:N-PEAK-BKG-BD0:val}   {\ensuremath{{ 0.37 } } }
\vdef{bdt2011:N-PEAK-BKG-BD0:err}   {\ensuremath{{ 0.27 } } }
\vdef{bdt2011:N-TAU-BS0:val}   {\ensuremath{{ 0.12 } } }
\vdef{bdt2011:N-TAU-BS0:err}   {\ensuremath{{ 0.00 } } }
\vdef{bdt2011:N-TAU-BD0:val}   {\ensuremath{{ 0.15 } } }
\vdef{bdt2011:N-TAU-BD0:err}   {\ensuremath{{ 0.01 } } }
\vdef{bdt2011:N-OBS-OFFHI0:val}   {\ensuremath{{0 } } }
\vdef{bdt2011:N-OBS-OFFLO0:val}   {\ensuremath{{3 } } }
\vdef{bdt2011:N-EXP-SoverB0:val}   {\ensuremath{{ 7.53 } } }
\vdef{bdt2011:N-EXP-SoverSplusB0:val}   {\ensuremath{{ 1.56 } } }
\vdef{bdt2011:N-EXP2-SIG-BSMM1:val}   {\ensuremath{{ 1.19 } } }
\vdef{bdt2011:N-EXP2-SIG-BSMM1:err}   {\ensuremath{{ 0.18 } } }
\vdef{bdt2011:N-EXP2-SIG-BDMM1:val}   {\ensuremath{{0.11 } } }
\vdef{bdt2011:N-EXP2-SIG-BDMM1:err}   {\ensuremath{{0.01 } } }
\vdef{bdt2011:N-OBS-BKG1:val}   {\ensuremath{{7 } } }
\vdef{bdt2011:N-EXP-BSMM1:val}   {\ensuremath{{ 1.25 } } }
\vdef{bdt2011:N-EXP-BSMM1:err}   {\ensuremath{{ 0.62 } } }
\vdef{bdt2011:N-EXP-BDMM1:val}   {\ensuremath{{ 0.98 } } }
\vdef{bdt2011:N-EXP-BDMM1:err}   {\ensuremath{{ 0.49 } } }
\vdef{bdt2011:N-LOW-BD1:val}   {\ensuremath{{5.200 } } }
\vdef{bdt2011:N-HIGH-BD1:val}   {\ensuremath{{5.300 } } }
\vdef{bdt2011:N-LOW-BS1:val}   {\ensuremath{{5.300 } } }
\vdef{bdt2011:N-HIGH-BS1:val}   {\ensuremath{{5.450 } } }
\vdef{bdt2011:N-PSS1:val}   {\ensuremath{{0.716 } } }
\vdef{bdt2011:N-PSS1:err}   {\ensuremath{{0.005 } } }
\vdef{bdt2011:N-PSS1:tot}   {\ensuremath{{0.036 } } }
\vdef{bdt2011:N-PSD1:val}   {\ensuremath{{0.315 } } }
\vdef{bdt2011:N-PSD1:err}   {\ensuremath{{0.021 } } }
\vdef{bdt2011:N-PSD1:tot}   {\ensuremath{{0.026 } } }
\vdef{bdt2011:N-PDS1:val}   {\ensuremath{{0.158 } } }
\vdef{bdt2011:N-PDS1:err}   {\ensuremath{{0.004 } } }
\vdef{bdt2011:N-PDS1:tot}   {\ensuremath{{0.009 } } }
\vdef{bdt2011:N-PDD1:val}   {\ensuremath{{0.537 } } }
\vdef{bdt2011:N-PDD1:err}   {\ensuremath{{0.023 } } }
\vdef{bdt2011:N-PDD1:tot}   {\ensuremath{{0.035 } } }
\vdef{bdt2011:N-EFF-TOT-BSMM1:val}   {\ensuremath{{0.0020 } } }
\vdef{bdt2011:N-EFF-TOT-BSMM1:err}   {\ensuremath{{0.0000 } } }
\vdef{bdt2011:N-EFF-TOT-BSMM1:tot}   {\ensuremath{{0.0002 } } }
\vdef{bdt2011:N-EFF-TOT-BSMM1:all}   {\ensuremath{{(0.20 \pm 0.02)} } }
\vdef{bdt2011:N-EFF-PRODMC-BSMM1:val}   {\ensuremath{{0.0020 } } }
\vdef{bdt2011:N-EFF-PRODMC-BSMM1:err}   {\ensuremath{{0.0000 } } }
\vdef{bdt2011:N-EFF-PRODMC-BSMM1:tot}   {\ensuremath{{0.0000 } } }
\vdef{bdt2011:N-EFF-PRODMC-BSMM1:all}   {\ensuremath{{(1.96 \pm 0.02)\times 10^{-3}} } }
\vdef{bdt2011:N-EFFRATIO-TOT-BSMM1:val}   {\ensuremath{{0.183 } } }
\vdef{bdt2011:N-EFFRATIO-TOT-BSMM1:err}   {\ensuremath{{0.002 } } }
\vdef{bdt2011:N-EFFRATIO-PRODMC-BSMM1:val}   {\ensuremath{{0.181 } } }
\vdef{bdt2011:N-EFFRATIO-PRODMC-BSMM1:err}   {\ensuremath{{0.004 } } }
\vdef{bdt2011:N-EFFRATIO-PRODTNP-BSMM1:val}   {\ensuremath{{0.339 } } }
\vdef{bdt2011:N-EFFRATIO-PRODTNP-BSMM1:err}   {\ensuremath{{119.311 } } }
\vdef{bdt2011:N-EFFRATIO-PRODTNPMC-BSMM1:val}   {\ensuremath{{0.181 } } }
\vdef{bdt2011:N-EFFRATIO-PRODTNPMC-BSMM1:err}   {\ensuremath{{0.004 } } }
\vdef{bdt2011:N-EFF-PRODTNP-BSMM1:val}   {\ensuremath{{0.0013 } } }
\vdef{bdt2011:N-EFF-PRODTNP-BSMM1:err}   {\ensuremath{{0.1770 } } }
\vdef{bdt2011:N-EFF-PRODTNP-BSMM1:tot}   {\ensuremath{{0.1770 } } }
\vdef{bdt2011:N-EFF-PRODTNP-BSMM1:all}   {\ensuremath{{(1.30 \pm 176.96)\times 10^{-3}} } }
\vdef{bdt2011:N-EFF-PRODTNPMC-BSMM1:val}   {\ensuremath{{0.0020 } } }
\vdef{bdt2011:N-EFF-PRODTNPMC-BSMM1:err}   {\ensuremath{{0.0000 } } }
\vdef{bdt2011:N-EFF-PRODTNPMC-BSMM1:tot}   {\ensuremath{{0.0000 } } }
\vdef{bdt2011:N-EFF-PRODTNPMC-BSMM1:all}   {\ensuremath{{(1.96 \pm 0.03)\times 10^{-3}} } }
\vdef{bdt2011:N-ACC-BSMM1:val}   {\ensuremath{{0.024 } } }
\vdef{bdt2011:N-ACC-BSMM1:err}   {\ensuremath{{0.000 } } }
\vdef{bdt2011:N-ACC-BSMM1:tot}   {\ensuremath{{0.001 } } }
\vdef{bdt2011:N-ACC-BSMM1:all}   {\ensuremath{{(2.39 \pm 0.12)\times 10^{-2}} } }
\vdef{bdt2011:N-EFF-MU-PID-BSMM1:val}   {\ensuremath{{0.626 } } }
\vdef{bdt2011:N-EFF-MU-PID-BSMM1:err}   {\ensuremath{{0.002 } } }
\vdef{bdt2011:N-EFF-MU-PID-BSMM1:tot}   {\ensuremath{{0.050 } } }
\vdef{bdt2011:N-EFF-MU-PID-BSMM1:all}   {\ensuremath{{(62.63 \pm 5.01)\times 10^{-2}} } }
\vdef{bdt2011:N-EFFRHO-MU-PID-BSMM1:val}   {\ensuremath{{0.631 } } }
\vdef{bdt2011:N-EFFRHO-MU-PID-BSMM1:err}   {\ensuremath{{0.002 } } }
\vdef{bdt2011:N-EFFRHO-MU-PID-BSMM1:tot}   {\ensuremath{{0.000 } } }
\vdef{bdt2011:N-EFFRHO-MU-PID-BSMM1:all}   {\ensuremath{{(63.11 \pm 0.00)\times 10^{-2}} } }
\vdef{bdt2011:N-EFF-MU-PIDMC-BSMM1:val}   {\ensuremath{{0.658 } } }
\vdef{bdt2011:N-EFF-MU-PIDMC-BSMM1:err}   {\ensuremath{{0.002 } } }
\vdef{bdt2011:N-EFF-MU-PIDMC-BSMM1:tot}   {\ensuremath{{0.053 } } }
\vdef{bdt2011:N-EFF-MU-PIDMC-BSMM1:all}   {\ensuremath{{(65.77 \pm 5.27)\times 10^{-2}} } }
\vdef{bdt2011:N-EFFRHO-MU-PIDMC-BSMM1:val}   {\ensuremath{{0.663 } } }
\vdef{bdt2011:N-EFFRHO-MU-PIDMC-BSMM1:err}   {\ensuremath{{0.002 } } }
\vdef{bdt2011:N-EFFRHO-MU-PIDMC-BSMM1:tot}   {\ensuremath{{0.051 } } }
\vdef{bdt2011:N-EFFRHO-MU-PIDMC-BSMM1:all}   {\ensuremath{{(66.27 \pm 5.05)\times 10^{-2}} } }
\vdef{bdt2011:N-EFF-MU-MC-BSMM1:val}   {\ensuremath{{0.663 } } }
\vdef{bdt2011:N-EFF-MU-MC-BSMM1:err}   {\ensuremath{{0.004 } } }
\vdef{bdt2011:N-EFF-MU-MC-BSMM1:tot}   {\ensuremath{{0.053 } } }
\vdef{bdt2011:N-EFF-MU-MC-BSMM1:all}   {\ensuremath{{(66.27 \pm 5.32)\times 10^{-2}} } }
\vdef{bdt2011:N-EFF-TRIG-PID-BSMM1:val}   {\ensuremath{{0.512 } } }
\vdef{bdt2011:N-EFF-TRIG-PID-BSMM1:err}   {\ensuremath{{0.006 } } }
\vdef{bdt2011:N-EFF-TRIG-PID-BSMM1:tot}   {\ensuremath{{0.031 } } }
\vdef{bdt2011:N-EFF-TRIG-PID-BSMM1:all}   {\ensuremath{{(51.18 \pm 3.13)\times 10^{-2}} } }
\vdef{bdt2011:N-EFFRHO-TRIG-PID-BSMM1:val}   {\ensuremath{{0.901 } } }
\vdef{bdt2011:N-EFFRHO-TRIG-PID-BSMM1:err}   {\ensuremath{{0.010 } } }
\vdef{bdt2011:N-EFFRHO-TRIG-PID-BSMM1:tot}   {\ensuremath{{0.055 } } }
\vdef{bdt2011:N-EFFRHO-TRIG-PID-BSMM1:all}   {\ensuremath{{(90.10 \pm 5.50)\times 10^{-2}} } }
\vdef{bdt2011:N-EFF-TRIG-PIDMC-BSMM1:val}   {\ensuremath{{0.413 } } }
\vdef{bdt2011:N-EFF-TRIG-PIDMC-BSMM1:err}   {\ensuremath{{0.004 } } }
\vdef{bdt2011:N-EFF-TRIG-PIDMC-BSMM1:tot}   {\ensuremath{{0.025 } } }
\vdef{bdt2011:N-EFF-TRIG-PIDMC-BSMM1:all}   {\ensuremath{{(41.33 \pm 2.52)\times 10^{-2}} } }
\vdef{bdt2011:N-EFFRHO-TRIG-PIDMC-BSMM1:val}   {\ensuremath{{0.727 } } }
\vdef{bdt2011:N-EFFRHO-TRIG-PIDMC-BSMM1:err}   {\ensuremath{{0.008 } } }
\vdef{bdt2011:N-EFFRHO-TRIG-PIDMC-BSMM1:tot}   {\ensuremath{{0.044 } } }
\vdef{bdt2011:N-EFFRHO-TRIG-PIDMC-BSMM1:all}   {\ensuremath{{(72.74 \pm 4.43)\times 10^{-2}} } }
\vdef{bdt2011:N-EFF-TRIG-MC-BSMM1:val}   {\ensuremath{{0.727 } } }
\vdef{bdt2011:N-EFF-TRIG-MC-BSMM1:err}   {\ensuremath{{0.004 } } }
\vdef{bdt2011:N-EFF-TRIG-MC-BSMM1:tot}   {\ensuremath{{0.044 } } }
\vdef{bdt2011:N-EFF-TRIG-MC-BSMM1:all}   {\ensuremath{{(72.74 \pm 4.39)\times 10^{-2}} } }
\vdef{bdt2011:N-EFF-CAND-BSMM1:val}   {\ensuremath{{1.000 } } }
\vdef{bdt2011:N-EFF-CAND-BSMM1:err}   {\ensuremath{{0.000 } } }
\vdef{bdt2011:N-EFF-CAND-BSMM1:tot}   {\ensuremath{{0.010 } } }
\vdef{bdt2011:N-EFF-CAND-BSMM1:all}   {\ensuremath{{(99.99 \pm 1.00)\times 10^{-2}} } }
\vdef{bdt2011:N-EFF-ANA-BSMM1:val}   {\ensuremath{{0.170 } } }
\vdef{bdt2011:N-EFF-ANA-BSMM1:err}   {\ensuremath{{0.001 } } }
\vdef{bdt2011:N-EFF-ANA-BSMM1:tot}   {\ensuremath{{0.005 } } }
\vdef{bdt2011:N-EFF-ANA-BSMM1:all}   {\ensuremath{{(16.99 \pm 0.52)\times 10^{-2}} } }
\vdef{bdt2011:N-EFF-TOT-BDMM1:val}   {\ensuremath{{0.0020 } } }
\vdef{bdt2011:N-EFF-TOT-BDMM1:err}   {\ensuremath{{0.0001 } } }
\vdef{bdt2011:N-EFF-TOT-BDMM1:tot}   {\ensuremath{{0.0002 } } }
\vdef{bdt2011:N-EFF-TOT-BDMM1:all}   {\ensuremath{{(0.20 \pm 0.02)} } }
\vdef{bdt2011:N-EFF-PRODMC-BDMM1:val}   {\ensuremath{{0.0020 } } }
\vdef{bdt2011:N-EFF-PRODMC-BDMM1:err}   {\ensuremath{{0.0001 } } }
\vdef{bdt2011:N-EFF-PRODMC-BDMM1:tot}   {\ensuremath{{0.0001 } } }
\vdef{bdt2011:N-EFF-PRODMC-BDMM1:all}   {\ensuremath{{(1.96 \pm 0.09)\times 10^{-3}} } }
\vdef{bdt2011:N-EFF-PRODTNP-BDMM1:val}   {\ensuremath{{0.0013 } } }
\vdef{bdt2011:N-EFF-PRODTNP-BDMM1:err}   {\ensuremath{{0.0450 } } }
\vdef{bdt2011:N-EFF-PRODTNP-BDMM1:tot}   {\ensuremath{{0.0450 } } }
\vdef{bdt2011:N-EFF-PRODTNP-BDMM1:all}   {\ensuremath{{(1.29 \pm 44.95)\times 10^{-3}} } }
\vdef{bdt2011:N-EFF-PRODTNPMC-BDMM1:val}   {\ensuremath{{0.00196 } } }
\vdef{bdt2011:N-EFF-PRODTNPMC-BDMM1:err}   {\ensuremath{{0.000103 } } }
\vdef{bdt2011:N-EFF-PRODTNPMC-BDMM1:tot}   {\ensuremath{{0.00010 } } }
\vdef{bdt2011:N-EFF-PRODTNPMC-BDMM1:all}   {\ensuremath{{(1.96 \pm 0.10)\times 10^{-3}} } }
\vdef{bdt2011:N-ACC-BDMM1:val}   {\ensuremath{{0.023 } } }
\vdef{bdt2011:N-ACC-BDMM1:err}   {\ensuremath{{0.000 } } }
\vdef{bdt2011:N-ACC-BDMM1:tot}   {\ensuremath{{0.001 } } }
\vdef{bdt2011:N-ACC-BDMM1:all}   {\ensuremath{{(2.30 \pm 0.12)\times 10^{-2}} } }
\vdef{bdt2011:N-EFF-MU-PID-BDMM1:val}   {\ensuremath{{0.627 } } }
\vdef{bdt2011:N-EFF-MU-PID-BDMM1:err}   {\ensuremath{{0.006 } } }
\vdef{bdt2011:N-EFF-MU-PID-BDMM1:tot}   {\ensuremath{{0.051 } } }
\vdef{bdt2011:N-EFF-MU-PID-BDMM1:all}   {\ensuremath{{(62.75 \pm 5.06)\times 10^{-2}} } }
\vdef{bdt2011:N-EFFRHO-MU-PID-BDMM1:val}   {\ensuremath{{0.613 } } }
\vdef{bdt2011:N-EFFRHO-MU-PID-BDMM1:err}   {\ensuremath{{0.006 } } }
\vdef{bdt2011:N-EFFRHO-MU-PID-BDMM1:tot}   {\ensuremath{{0.000 } } }
\vdef{bdt2011:N-EFFRHO-MU-PID-BDMM1:all}   {\ensuremath{{(61.35 \pm 0.00)\times 10^{-2}} } }
\vdef{bdt2011:N-EFF-MU-PIDMC-BDMM1:val}   {\ensuremath{{0.670 } } }
\vdef{bdt2011:N-EFF-MU-PIDMC-BDMM1:err}   {\ensuremath{{0.007 } } }
\vdef{bdt2011:N-EFF-MU-PIDMC-BDMM1:tot}   {\ensuremath{{0.054 } } }
\vdef{bdt2011:N-EFF-MU-PIDMC-BDMM1:all}   {\ensuremath{{(66.95 \pm 5.40)\times 10^{-2}} } }
\vdef{bdt2011:N-EFFRHO-MU-PIDMC-BDMM1:val}   {\ensuremath{{0.655 } } }
\vdef{bdt2011:N-EFFRHO-MU-PIDMC-BDMM1:err}   {\ensuremath{{0.007 } } }
\vdef{bdt2011:N-EFFRHO-MU-PIDMC-BDMM1:tot}   {\ensuremath{{0.049 } } }
\vdef{bdt2011:N-EFFRHO-MU-PIDMC-BDMM1:all}   {\ensuremath{{(65.46 \pm 4.95)\times 10^{-2}} } }
\vdef{bdt2011:N-EFF-MU-MC-BDMM1:val}   {\ensuremath{{0.655 } } }
\vdef{bdt2011:N-EFF-MU-MC-BDMM1:err}   {\ensuremath{{0.015 } } }
\vdef{bdt2011:N-EFF-MU-MC-BDMM1:tot}   {\ensuremath{{0.054 } } }
\vdef{bdt2011:N-EFF-MU-MC-BDMM1:all}   {\ensuremath{{(65.46 \pm 5.45)\times 10^{-2}} } }
\vdef{bdt2011:N-EFF-TRIG-PID-BDMM1:val}   {\ensuremath{{0.501 } } }
\vdef{bdt2011:N-EFF-TRIG-PID-BDMM1:err}   {\ensuremath{{0.024 } } }
\vdef{bdt2011:N-EFF-TRIG-PID-BDMM1:tot}   {\ensuremath{{0.038 } } }
\vdef{bdt2011:N-EFF-TRIG-PID-BDMM1:all}   {\ensuremath{{(50.07 \pm 3.82)\times 10^{-2}} } }
\vdef{bdt2011:N-EFFRHO-TRIG-PID-BDMM1:val}   {\ensuremath{{0.838 } } }
\vdef{bdt2011:N-EFFRHO-TRIG-PID-BDMM1:err}   {\ensuremath{{0.040 } } }
\vdef{bdt2011:N-EFFRHO-TRIG-PID-BDMM1:tot}   {\ensuremath{{0.064 } } }
\vdef{bdt2011:N-EFFRHO-TRIG-PID-BDMM1:all}   {\ensuremath{{(83.84 \pm 6.40)\times 10^{-2}} } }
\vdef{bdt2011:N-EFF-TRIG-PIDMC-BDMM1:val}   {\ensuremath{{0.435 } } }
\vdef{bdt2011:N-EFF-TRIG-PIDMC-BDMM1:err}   {\ensuremath{{0.015 } } }
\vdef{bdt2011:N-EFF-TRIG-PIDMC-BDMM1:tot}   {\ensuremath{{0.030 } } }
\vdef{bdt2011:N-EFF-TRIG-PIDMC-BDMM1:all}   {\ensuremath{{(43.46 \pm 3.03)\times 10^{-2}} } }
\vdef{bdt2011:N-EFFRHO-TRIG-PIDMC-BDMM1:val}   {\ensuremath{{0.728 } } }
\vdef{bdt2011:N-EFFRHO-TRIG-PIDMC-BDMM1:err}   {\ensuremath{{0.026 } } }
\vdef{bdt2011:N-EFFRHO-TRIG-PIDMC-BDMM1:tot}   {\ensuremath{{0.051 } } }
\vdef{bdt2011:N-EFFRHO-TRIG-PIDMC-BDMM1:all}   {\ensuremath{{(72.77 \pm 5.07)\times 10^{-2}} } }
\vdef{bdt2011:N-EFF-TRIG-MC-BDMM1:val}   {\ensuremath{{0.728 } } }
\vdef{bdt2011:N-EFF-TRIG-MC-BDMM1:err}   {\ensuremath{{0.017 } } }
\vdef{bdt2011:N-EFF-TRIG-MC-BDMM1:tot}   {\ensuremath{{0.047 } } }
\vdef{bdt2011:N-EFF-TRIG-MC-BDMM1:all}   {\ensuremath{{(72.77 \pm 4.70)\times 10^{-2}} } }
\vdef{bdt2011:N-EFF-CAND-BDMM1:val}   {\ensuremath{{1.000 } } }
\vdef{bdt2011:N-EFF-CAND-BDMM1:err}   {\ensuremath{{0.000 } } }
\vdef{bdt2011:N-EFF-CAND-BDMM1:tot}   {\ensuremath{{0.010 } } }
\vdef{bdt2011:N-EFF-CAND-BDMM1:all}   {\ensuremath{{(100.00 \pm 1.00)\times 10^{-2}} } }
\vdef{bdt2011:N-EFF-ANA-BDMM1:val}   {\ensuremath{{0.179 } } }
\vdef{bdt2011:N-EFF-ANA-BDMM1:err}   {\ensuremath{{0.005 } } }
\vdef{bdt2011:N-EFF-ANA-BDMM1:tot}   {\ensuremath{{0.007 } } }
\vdef{bdt2011:N-EFF-ANA-BDMM1:all}   {\ensuremath{{(17.92 \pm 0.74)\times 10^{-2}} } }
\vdef{bdt2011:N-EXP-OBS-BS1:val}   {\ensuremath{{ 2.76 } } }
\vdef{bdt2011:N-EXP-OBS-BS1:err}   {\ensuremath{{ 0.65 } } }
\vdef{bdt2011:N-EXP-OBS-BD1:val}   {\ensuremath{{ 1.35 } } }
\vdef{bdt2011:N-EXP-OBS-BD1:err}   {\ensuremath{{ 0.49 } } }
\vdef{bdt2011:N-OBS-BSMM1:val}   {\ensuremath{{4 } } }
\vdef{bdt2011:N-OBS-BDMM1:val}   {\ensuremath{{1 } } }
\vdef{bdt2011:N-OFFLO-RARE1:val}   {\ensuremath{{ 0.07 } } }
\vdef{bdt2011:N-OFFLO-RARE1:err}   {\ensuremath{{ 0.04 } } }
\vdef{bdt2011:N-OFFHI-RARE1:val}   {\ensuremath{{ 0.01 } } }
\vdef{bdt2011:N-OFFHI-RARE1:err}   {\ensuremath{{ 0.00 } } }
\vdef{bdt2011:N-PEAK-BKG-BS1:val}   {\ensuremath{{ 0.06 } } }
\vdef{bdt2011:N-PEAK-BKG-BS1:err}   {\ensuremath{{ 0.04 } } }
\vdef{bdt2011:N-PEAK-BKG-BD1:val}   {\ensuremath{{ 0.11 } } }
\vdef{bdt2011:N-PEAK-BKG-BD1:err}   {\ensuremath{{ 0.08 } } }
\vdef{bdt2011:N-TAU-BS1:val}   {\ensuremath{{ 0.18 } } }
\vdef{bdt2011:N-TAU-BS1:err}   {\ensuremath{{ 0.01 } } }
\vdef{bdt2011:N-TAU-BD1:val}   {\ensuremath{{ 0.14 } } }
\vdef{bdt2011:N-TAU-BD1:err}   {\ensuremath{{ 0.01 } } }
\vdef{bdt2011:N-OBS-OFFHI1:val}   {\ensuremath{{3 } } }
\vdef{bdt2011:N-OBS-OFFLO1:val}   {\ensuremath{{4 } } }
\vdef{bdt2011:N-EXP-SoverB1:val}   {\ensuremath{{ 0.96 } } }
\vdef{bdt2011:N-EXP-SoverSplusB1:val}   {\ensuremath{{ 0.76 } } }
\vdef{bdt2011:SgBd0:val}  {\ensuremath{{0.191 } } }
\vdef{bdt2011:SgBd0:e1}   {\ensuremath{{0.437 } } }
\vdef{bdt2011:SgBd0:e2}   {\ensuremath{{0.029 } } }
\vdef{bdt2011:SgBs0:val}  {\ensuremath{{2.772 } } }
\vdef{bdt2011:SgBs0:e1}   {\ensuremath{{1.665 } } }
\vdef{bdt2011:SgBs0:e2}   {\ensuremath{{0.416 } } }
\vdef{bdt2011:SgLo0:val}  {\ensuremath{{0.073 } } }
\vdef{bdt2011:SgLo0:e1}   {\ensuremath{{0.270 } } }
\vdef{bdt2011:SgLo0:e2}   {\ensuremath{{0.011 } } }
\vdef{bdt2011:SgHi0:val}  {\ensuremath{{0.087 } } }
\vdef{bdt2011:SgHi0:e1}   {\ensuremath{{0.295 } } }
\vdef{bdt2011:SgHi0:e2}   {\ensuremath{{0.013 } } }
\vdef{bdt2011:BdBd0:val}  {\ensuremath{{0.271 } } }
\vdef{bdt2011:BdBd0:e1}   {\ensuremath{{0.520 } } }
\vdef{bdt2011:BdBd0:e2}   {\ensuremath{{0.027 } } }
\vdef{bdt2011:BdBs0:val}  {\ensuremath{{0.109 } } }
\vdef{bdt2011:BdBs0:e1}   {\ensuremath{{0.331 } } }
\vdef{bdt2011:BdBs0:e2}   {\ensuremath{{0.011 } } }
\vdef{bdt2011:BdLo0:val}  {\ensuremath{{0.024 } } }
\vdef{bdt2011:BdLo0:e1}   {\ensuremath{{0.156 } } }
\vdef{bdt2011:BdLo0:e2}   {\ensuremath{{0.002 } } }
\vdef{bdt2011:BdHi0:val}  {\ensuremath{{0.000 } } }
\vdef{bdt2011:BdHi0:e1}   {\ensuremath{{0.000 } } }
\vdef{bdt2011:BdHi0:e2}   {\ensuremath{{0.000 } } }
\vdef{bdt2011:BgPeakLo0:val}   {\ensuremath{{0.106 } } }
\vdef{bdt2011:BgPeakLo0:e1}   {\ensuremath{{0.005 } } }
\vdef{bdt2011:BgPeakLo0:e2}   {\ensuremath{{0.075 } } }
\vdef{bdt2011:BgPeakBd0:val}   {\ensuremath{{0.369 } } }
\vdef{bdt2011:BgPeakBd0:e1}   {\ensuremath{{0.018 } } }
\vdef{bdt2011:BgPeakBd0:e2}   {\ensuremath{{0.266 } } }
\vdef{bdt2011:BgPeakBs0:val}   {\ensuremath{{0.112 } } }
\vdef{bdt2011:BgPeakBs0:e1}   {\ensuremath{{0.006 } } }
\vdef{bdt2011:BgPeakBs0:e2}   {\ensuremath{{0.080 } } }
\vdef{bdt2011:BgPeakHi0:val}   {\ensuremath{{0.011 } } }
\vdef{bdt2011:BgPeakHi0:e1}   {\ensuremath{{0.001 } } }
\vdef{bdt2011:BgPeakHi0:e2}   {\ensuremath{{0.008 } } }
\vdef{bdt2011:BgRslLo0:val}   {\ensuremath{{23.776 } } }
\vdef{bdt2011:BgRslLo0:e1}   {\ensuremath{{1.189 } } }
\vdef{bdt2011:BgRslLo0:e2}   {\ensuremath{{18.709 } } }
\vdef{bdt2011:BgRslBd0:val}   {\ensuremath{{3.704 } } }
\vdef{bdt2011:BgRslBd0:e1}   {\ensuremath{{0.185 } } }
\vdef{bdt2011:BgRslBd0:e2}   {\ensuremath{{3.888 } } }
\vdef{bdt2011:BgRslBs0:val}   {\ensuremath{{2.918 } } }
\vdef{bdt2011:BgRslBs0:e1}   {\ensuremath{{0.146 } } }
\vdef{bdt2011:BgRslBs0:e2}   {\ensuremath{{3.216 } } }
\vdef{bdt2011:BgRslHi0:val}   {\ensuremath{{0.282 } } }
\vdef{bdt2011:BgRslHi0:e1}   {\ensuremath{{0.014 } } }
\vdef{bdt2011:BgRslHi0:e2}   {\ensuremath{{0.310 } } }
\vdef{bdt2011:BgRareLo0:val}   {\ensuremath{{23.882 } } }
\vdef{bdt2011:BgRareLo0:e1}   {\ensuremath{{1.189 } } }
\vdef{bdt2011:BgRareLo0:e2}   {\ensuremath{{18.709 } } }
\vdef{bdt2011:BgRareBd0:val}   {\ensuremath{{4.073 } } }
\vdef{bdt2011:BgRareBd0:e1}   {\ensuremath{{0.186 } } }
\vdef{bdt2011:BgRareBd0:e2}   {\ensuremath{{3.897 } } }
\vdef{bdt2011:BgRareBs0:val}   {\ensuremath{{3.030 } } }
\vdef{bdt2011:BgRareBs0:e1}   {\ensuremath{{0.146 } } }
\vdef{bdt2011:BgRareBs0:e2}   {\ensuremath{{3.217 } } }
\vdef{bdt2011:BgRareHi0:val}   {\ensuremath{{0.293 } } }
\vdef{bdt2011:BgRareHi0:e1}   {\ensuremath{{0.014 } } }
\vdef{bdt2011:BgRareHi0:e2}   {\ensuremath{{0.310 } } }
\vdef{bdt2011:BgRslsLo0:val}   {\ensuremath{{3.000 } } }
\vdef{bdt2011:BgRslsLo0:e1}   {\ensuremath{{0.150 } } }
\vdef{bdt2011:BgRslsLo0:e2}   {\ensuremath{{2.361 } } }
\vdef{bdt2011:BgRslsBd0:val}   {\ensuremath{{0.467 } } }
\vdef{bdt2011:BgRslsBd0:e1}   {\ensuremath{{0.023 } } }
\vdef{bdt2011:BgRslsBd0:e2}   {\ensuremath{{0.491 } } }
\vdef{bdt2011:BgRslsBs0:val}   {\ensuremath{{0.368 } } }
\vdef{bdt2011:BgRslsBs0:e1}   {\ensuremath{{0.018 } } }
\vdef{bdt2011:BgRslsBs0:e2}   {\ensuremath{{0.406 } } }
\vdef{bdt2011:BgRslsHi0:val}   {\ensuremath{{0.036 } } }
\vdef{bdt2011:BgRslsHi0:e1}   {\ensuremath{{0.002 } } }
\vdef{bdt2011:BgRslsHi0:e2}   {\ensuremath{{0.039 } } }
\vdef{bdt2011:BgCombLo0:val}   {\ensuremath{{0.000 } } }
\vdef{bdt2011:BgCombLo0:e1}   {\ensuremath{{0.000 } } }
\vdef{bdt2011:BgCombLo0:e2}   {\ensuremath{{0.100 } } }
\vdef{bdt2011:BgCombBd0:val}   {\ensuremath{{0.000 } } }
\vdef{bdt2011:BgCombBd0:e1}   {\ensuremath{{0.000 } } }
\vdef{bdt2011:BgCombBd0:e2}   {\ensuremath{{0.100 } } }
\vdef{bdt2011:BgCombBs0:val}   {\ensuremath{{0.000 } } }
\vdef{bdt2011:BgCombBs0:e1}   {\ensuremath{{0.000 } } }
\vdef{bdt2011:BgCombBs0:e2}   {\ensuremath{{0.100 } } }
\vdef{bdt2011:BgCombHi0:val}   {\ensuremath{{0.000 } } }
\vdef{bdt2011:BgCombHi0:e1}   {\ensuremath{{0.000 } } }
\vdef{bdt2011:BgCombHi0:e2}   {\ensuremath{{0.100 } } }
\vdef{bdt2011:BgNonpLo0:val}   {\ensuremath{{3.000 } } }
\vdef{bdt2011:BgNonpLo0:e1}   {\ensuremath{{0.150 } } }
\vdef{bdt2011:BgNonpLo0:e2}   {\ensuremath{{2.363 } } }
\vdef{bdt2011:BgNonpBd0:val}   {\ensuremath{{0.467 } } }
\vdef{bdt2011:BgNonpBd0:e1}   {\ensuremath{{0.023 } } }
\vdef{bdt2011:BgNonpBd0:e2}   {\ensuremath{{0.501 } } }
\vdef{bdt2011:BgNonpBs0:val}   {\ensuremath{{0.368 } } }
\vdef{bdt2011:BgNonpBs0:e1}   {\ensuremath{{0.018 } } }
\vdef{bdt2011:BgNonpBs0:e2}   {\ensuremath{{0.418 } } }
\vdef{bdt2011:BgNonpHi0:val}   {\ensuremath{{0.036 } } }
\vdef{bdt2011:BgNonpHi0:e1}   {\ensuremath{{0.002 } } }
\vdef{bdt2011:BgNonpHi0:e2}   {\ensuremath{{0.107 } } }
\vdef{bdt2011:BgTotLo0:val}   {\ensuremath{{3.106 } } }
\vdef{bdt2011:BgTotLo0:e1}   {\ensuremath{{0.000 } } }
\vdef{bdt2011:BgTotLo0:e2}   {\ensuremath{{2.364 } } }
\vdef{bdt2011:BgTotBd0:val}   {\ensuremath{{0.836 } } }
\vdef{bdt2011:BgTotBd0:e1}   {\ensuremath{{0.000 } } }
\vdef{bdt2011:BgTotBd0:e2}   {\ensuremath{{0.567 } } }
\vdef{bdt2011:BgTotBs0:val}   {\ensuremath{{0.480 } } }
\vdef{bdt2011:BgTotBs0:e1}   {\ensuremath{{0.000 } } }
\vdef{bdt2011:BgTotBs0:e2}   {\ensuremath{{0.426 } } }
\vdef{bdt2011:BgTotHi0:val}   {\ensuremath{{0.047 } } }
\vdef{bdt2011:BgTotHi0:e1}   {\ensuremath{{0.000 } } }
\vdef{bdt2011:BgTotHi0:e2}   {\ensuremath{{0.108 } } }
\vdef{bdt2011:SgAndBgLo0:val}   {\ensuremath{{3.203 } } }
\vdef{bdt2011:SgAndBgLo0:e1}   {\ensuremath{{0.000 } } }
\vdef{bdt2011:SgAndBgLo0:e2}   {\ensuremath{{2.364 } } }
\vdef{bdt2011:SgAndBgBd0:val}   {\ensuremath{{1.298 } } }
\vdef{bdt2011:SgAndBgBd0:e1}   {\ensuremath{{0.000 } } }
\vdef{bdt2011:SgAndBgBd0:e2}   {\ensuremath{{0.770 } } }
\vdef{bdt2011:SgAndBgBs0:val}   {\ensuremath{{3.362 } } }
\vdef{bdt2011:SgAndBgBs0:e1}   {\ensuremath{{0.000 } } }
\vdef{bdt2011:SgAndBgBs0:e2}   {\ensuremath{{0.595 } } }
\vdef{bdt2011:SgAndBgHi0:val}   {\ensuremath{{0.134 } } }
\vdef{bdt2011:SgAndBgHi0:e1}   {\ensuremath{{0.000 } } }
\vdef{bdt2011:SgAndBgHi0:e2}   {\ensuremath{{0.108 } } }
\vdef{bdt2011:SgBd1:val}  {\ensuremath{{0.264 } } }
\vdef{bdt2011:SgBd1:e1}   {\ensuremath{{0.514 } } }
\vdef{bdt2011:SgBd1:e2}   {\ensuremath{{0.040 } } }
\vdef{bdt2011:SgBs1:val}  {\ensuremath{{1.194 } } }
\vdef{bdt2011:SgBs1:e1}   {\ensuremath{{1.093 } } }
\vdef{bdt2011:SgBs1:e2}   {\ensuremath{{0.179 } } }
\vdef{bdt2011:SgLo1:val}  {\ensuremath{{0.048 } } }
\vdef{bdt2011:SgLo1:e1}   {\ensuremath{{0.220 } } }
\vdef{bdt2011:SgLo1:e2}   {\ensuremath{{0.007 } } }
\vdef{bdt2011:SgHi1:val}  {\ensuremath{{0.168 } } }
\vdef{bdt2011:SgHi1:e1}   {\ensuremath{{0.410 } } }
\vdef{bdt2011:SgHi1:e2}   {\ensuremath{{0.025 } } }
\vdef{bdt2011:BdBd1:val}  {\ensuremath{{0.105 } } }
\vdef{bdt2011:BdBd1:e1}   {\ensuremath{{0.324 } } }
\vdef{bdt2011:BdBd1:e2}   {\ensuremath{{0.010 } } }
\vdef{bdt2011:BdBs1:val}  {\ensuremath{{0.062 } } }
\vdef{bdt2011:BdBs1:e1}   {\ensuremath{{0.248 } } }
\vdef{bdt2011:BdBs1:e2}   {\ensuremath{{0.006 } } }
\vdef{bdt2011:BdLo1:val}  {\ensuremath{{0.028 } } }
\vdef{bdt2011:BdLo1:e1}   {\ensuremath{{0.168 } } }
\vdef{bdt2011:BdLo1:e2}   {\ensuremath{{0.003 } } }
\vdef{bdt2011:BdHi1:val}  {\ensuremath{{0.002 } } }
\vdef{bdt2011:BdHi1:e1}   {\ensuremath{{0.050 } } }
\vdef{bdt2011:BdHi1:e2}   {\ensuremath{{0.000 } } }
\vdef{bdt2011:BgPeakLo1:val}   {\ensuremath{{0.067 } } }
\vdef{bdt2011:BgPeakLo1:e1}   {\ensuremath{{0.003 } } }
\vdef{bdt2011:BgPeakLo1:e2}   {\ensuremath{{0.048 } } }
\vdef{bdt2011:BgPeakBd1:val}   {\ensuremath{{0.106 } } }
\vdef{bdt2011:BgPeakBd1:e1}   {\ensuremath{{0.005 } } }
\vdef{bdt2011:BgPeakBd1:e2}   {\ensuremath{{0.077 } } }
\vdef{bdt2011:BgPeakBs1:val}   {\ensuremath{{0.057 } } }
\vdef{bdt2011:BgPeakBs1:e1}   {\ensuremath{{0.003 } } }
\vdef{bdt2011:BgPeakBs1:e2}   {\ensuremath{{0.041 } } }
\vdef{bdt2011:BgPeakHi1:val}   {\ensuremath{{0.008 } } }
\vdef{bdt2011:BgPeakHi1:e1}   {\ensuremath{{0.000 } } }
\vdef{bdt2011:BgPeakHi1:e2}   {\ensuremath{{0.005 } } }
\vdef{bdt2011:BgRslLo1:val}   {\ensuremath{{7.264 } } }
\vdef{bdt2011:BgRslLo1:e1}   {\ensuremath{{0.363 } } }
\vdef{bdt2011:BgRslLo1:e2}   {\ensuremath{{5.361 } } }
\vdef{bdt2011:BgRslBd1:val}   {\ensuremath{{1.402 } } }
\vdef{bdt2011:BgRslBd1:e1}   {\ensuremath{{0.070 } } }
\vdef{bdt2011:BgRslBd1:e2}   {\ensuremath{{1.437 } } }
\vdef{bdt2011:BgRslBs1:val}   {\ensuremath{{0.835 } } }
\vdef{bdt2011:BgRslBs1:e1}   {\ensuremath{{0.042 } } }
\vdef{bdt2011:BgRslBs1:e2}   {\ensuremath{{0.898 } } }
\vdef{bdt2011:BgRslHi1:val}   {\ensuremath{{0.219 } } }
\vdef{bdt2011:BgRslHi1:e1}   {\ensuremath{{0.011 } } }
\vdef{bdt2011:BgRslHi1:e2}   {\ensuremath{{0.241 } } }
\vdef{bdt2011:BgRareLo1:val}   {\ensuremath{{7.331 } } }
\vdef{bdt2011:BgRareLo1:e1}   {\ensuremath{{0.363 } } }
\vdef{bdt2011:BgRareLo1:e2}   {\ensuremath{{5.361 } } }
\vdef{bdt2011:BgRareBd1:val}   {\ensuremath{{1.508 } } }
\vdef{bdt2011:BgRareBd1:e1}   {\ensuremath{{0.070 } } }
\vdef{bdt2011:BgRareBd1:e2}   {\ensuremath{{1.439 } } }
\vdef{bdt2011:BgRareBs1:val}   {\ensuremath{{0.891 } } }
\vdef{bdt2011:BgRareBs1:e1}   {\ensuremath{{0.042 } } }
\vdef{bdt2011:BgRareBs1:e2}   {\ensuremath{{0.899 } } }
\vdef{bdt2011:BgRareHi1:val}   {\ensuremath{{0.227 } } }
\vdef{bdt2011:BgRareHi1:e1}   {\ensuremath{{0.011 } } }
\vdef{bdt2011:BgRareHi1:e2}   {\ensuremath{{0.241 } } }
\vdef{bdt2011:BgRslsLo1:val}   {\ensuremath{{2.000 } } }
\vdef{bdt2011:BgRslsLo1:e1}   {\ensuremath{{0.100 } } }
\vdef{bdt2011:BgRslsLo1:e2}   {\ensuremath{{1.476 } } }
\vdef{bdt2011:BgRslsBd1:val}   {\ensuremath{{0.312 } } }
\vdef{bdt2011:BgRslsBd1:e1}   {\ensuremath{{0.016 } } }
\vdef{bdt2011:BgRslsBd1:e2}   {\ensuremath{{0.319 } } }
\vdef{bdt2011:BgRslsBs1:val}   {\ensuremath{{0.245 } } }
\vdef{bdt2011:BgRslsBs1:e1}   {\ensuremath{{0.012 } } }
\vdef{bdt2011:BgRslsBs1:e2}   {\ensuremath{{0.264 } } }
\vdef{bdt2011:BgRslsHi1:val}   {\ensuremath{{0.024 } } }
\vdef{bdt2011:BgRslsHi1:e1}   {\ensuremath{{0.001 } } }
\vdef{bdt2011:BgRslsHi1:e2}   {\ensuremath{{0.026 } } }
\vdef{bdt2011:BgCombLo1:val}   {\ensuremath{{2.000 } } }
\vdef{bdt2011:BgCombLo1:e1}   {\ensuremath{{1.155 } } }
\vdef{bdt2011:BgCombLo1:e2}   {\ensuremath{{1.172 } } }
\vdef{bdt2011:BgCombBd1:val}   {\ensuremath{{0.667 } } }
\vdef{bdt2011:BgCombBd1:e1}   {\ensuremath{{0.385 } } }
\vdef{bdt2011:BgCombBd1:e2}   {\ensuremath{{0.391 } } }
\vdef{bdt2011:BgCombBs1:val}   {\ensuremath{{1.000 } } }
\vdef{bdt2011:BgCombBs1:e1}   {\ensuremath{{0.577 } } }
\vdef{bdt2011:BgCombBs1:e2}   {\ensuremath{{0.586 } } }
\vdef{bdt2011:BgCombHi1:val}   {\ensuremath{{3.000 } } }
\vdef{bdt2011:BgCombHi1:e1}   {\ensuremath{{1.732 } } }
\vdef{bdt2011:BgCombHi1:e2}   {\ensuremath{{1.758 } } }
\vdef{bdt2011:BgNonpLo1:val}   {\ensuremath{{4.000 } } }
\vdef{bdt2011:BgNonpLo1:e1}   {\ensuremath{{1.159 } } }
\vdef{bdt2011:BgNonpLo1:e2}   {\ensuremath{{1.885 } } }
\vdef{bdt2011:BgNonpBd1:val}   {\ensuremath{{0.978 } } }
\vdef{bdt2011:BgNonpBd1:e1}   {\ensuremath{{0.385 } } }
\vdef{bdt2011:BgNonpBd1:e2}   {\ensuremath{{0.505 } } }
\vdef{bdt2011:BgNonpBs1:val}   {\ensuremath{{1.245 } } }
\vdef{bdt2011:BgNonpBs1:e1}   {\ensuremath{{0.577 } } }
\vdef{bdt2011:BgNonpBs1:e2}   {\ensuremath{{0.472 } } }
\vdef{bdt2011:BgNonpHi1:val}   {\ensuremath{{3.024 } } }
\vdef{bdt2011:BgNonpHi1:e1}   {\ensuremath{{1.732 } } }
\vdef{bdt2011:BgNonpHi1:e2}   {\ensuremath{{1.758 } } }
\vdef{bdt2011:BgTotLo1:val}   {\ensuremath{{4.067 } } }
\vdef{bdt2011:BgTotLo1:e1}   {\ensuremath{{0.000 } } }
\vdef{bdt2011:BgTotLo1:e2}   {\ensuremath{{1.885 } } }
\vdef{bdt2011:BgTotBd1:val}   {\ensuremath{{1.084 } } }
\vdef{bdt2011:BgTotBd1:e1}   {\ensuremath{{0.000 } } }
\vdef{bdt2011:BgTotBd1:e2}   {\ensuremath{{0.510 } } }
\vdef{bdt2011:BgTotBs1:val}   {\ensuremath{{1.302 } } }
\vdef{bdt2011:BgTotBs1:e1}   {\ensuremath{{0.000 } } }
\vdef{bdt2011:BgTotBs1:e2}   {\ensuremath{{0.473 } } }
\vdef{bdt2011:BgTotHi1:val}   {\ensuremath{{3.032 } } }
\vdef{bdt2011:BgTotHi1:e1}   {\ensuremath{{0.000 } } }
\vdef{bdt2011:BgTotHi1:e2}   {\ensuremath{{1.758 } } }
\vdef{bdt2011:SgAndBgLo1:val}   {\ensuremath{{4.144 } } }
\vdef{bdt2011:SgAndBgLo1:e1}   {\ensuremath{{0.000 } } }
\vdef{bdt2011:SgAndBgLo1:e2}   {\ensuremath{{1.885 } } }
\vdef{bdt2011:SgAndBgBd1:val}   {\ensuremath{{1.454 } } }
\vdef{bdt2011:SgAndBgBd1:e1}   {\ensuremath{{0.000 } } }
\vdef{bdt2011:SgAndBgBd1:e2}   {\ensuremath{{0.606 } } }
\vdef{bdt2011:SgAndBgBs1:val}   {\ensuremath{{2.558 } } }
\vdef{bdt2011:SgAndBgBs1:e1}   {\ensuremath{{0.000 } } }
\vdef{bdt2011:SgAndBgBs1:e2}   {\ensuremath{{0.506 } } }
\vdef{bdt2011:SgAndBgHi1:val}   {\ensuremath{{3.202 } } }
\vdef{bdt2011:SgAndBgHi1:e1}   {\ensuremath{{0.000 } } }
\vdef{bdt2011:SgAndBgHi1:e2}   {\ensuremath{{1.758 } } }
\vdef{bdt2011:N-EFF-TOT-BS0:val}   {\ensuremath{{0.000608 } } }
\vdef{bdt2011:N-EFF-TOT-BS0:err}   {\ensuremath{{0.000006 } } }
\vdef{bdt2011:N-ACC-BS0:val}   {\ensuremath{{0.0084 } } }
\vdef{bdt2011:N-ACC-BS0:err}   {\ensuremath{{0.0001 } } }
\vdef{bdt2011:N-EFF-MU-PID-BS0:val}   {\ensuremath{{0.7178 } } }
\vdef{bdt2011:N-EFF-MU-PID-BS0:err}   {\ensuremath{{0.0009 } } }
\vdef{bdt2011:N-EFF-MU-PIDMC-BS0:val}   {\ensuremath{{0.6951 } } }
\vdef{bdt2011:N-EFF-MU-PIDMC-BS0:err}   {\ensuremath{{0.0011 } } }
\vdef{bdt2011:N-EFF-MU-MC-BS0:val}   {\ensuremath{{0.5040 } } }
\vdef{bdt2011:N-EFF-MU-MC-BS0:err}   {\ensuremath{{0.0028 } } }
\vdef{bdt2011:N-EFF-TRIG-PID-BS0:val}   {\ensuremath{{0.7749 } } }
\vdef{bdt2011:N-EFF-TRIG-PID-BS0:err}   {\ensuremath{{0.0012 } } }
\vdef{bdt2011:N-EFF-TRIG-PIDMC-BS0:val}   {\ensuremath{{0.7332 } } }
\vdef{bdt2011:N-EFF-TRIG-PIDMC-BS0:err}   {\ensuremath{{0.0014 } } }
\vdef{bdt2011:N-EFF-TRIG-MC-BS0:val}   {\ensuremath{{0.7491 } } }
\vdef{bdt2011:N-EFF-TRIG-MC-BS0:err}   {\ensuremath{{0.0035 } } }
\vdef{bdt2011:N-EFF-CAND-BS0:val}   {\ensuremath{{0.9964 } } }
\vdef{bdt2011:N-EFF-CAND-BS0:err}   {\ensuremath{{0.0004 } } }
\vdef{bdt2011:N-EFF-ANA-BS0:val}   {\ensuremath{{0.1933 } } }
\vdef{bdt2011:N-EFF-ANA-BS0:err}   {\ensuremath{{0.0010 } } }
\vdef{bdt2011:N-OBS-BS0:val}   {\ensuremath{{4740 } } }
\vdef{bdt2011:N-OBS-BS0:err}   {\ensuremath{{210 } } }
\vdef{bdt2011:N-EFF-TOT-BS1:val}   {\ensuremath{{0.000208 } } }
\vdef{bdt2011:N-EFF-TOT-BS1:err}   {\ensuremath{{0.000003 } } }
\vdef{bdt2011:N-ACC-BS1:val}   {\ensuremath{{0.0042 } } }
\vdef{bdt2011:N-ACC-BS1:err}   {\ensuremath{{0.0000 } } }
\vdef{bdt2011:N-EFF-MU-PID-BS1:val}   {\ensuremath{{0.6454 } } }
\vdef{bdt2011:N-EFF-MU-PID-BS1:err}   {\ensuremath{{0.0018 } } }
\vdef{bdt2011:N-EFF-MU-PIDMC-BS1:val}   {\ensuremath{{0.6773 } } }
\vdef{bdt2011:N-EFF-MU-PIDMC-BS1:err}   {\ensuremath{{0.0021 } } }
\vdef{bdt2011:N-EFF-MU-MC-BS1:val}   {\ensuremath{{0.5919 } } }
\vdef{bdt2011:N-EFF-MU-MC-BS1:err}   {\ensuremath{{0.0045 } } }
\vdef{bdt2011:N-EFF-TRIG-PID-BS1:val}   {\ensuremath{{0.6812 } } }
\vdef{bdt2011:N-EFF-TRIG-PID-BS1:err}   {\ensuremath{{0.0041 } } }
\vdef{bdt2011:N-EFF-TRIG-PIDMC-BS1:val}   {\ensuremath{{0.5371 } } }
\vdef{bdt2011:N-EFF-TRIG-PIDMC-BS1:err}   {\ensuremath{{0.0048 } } }
\vdef{bdt2011:N-EFF-TRIG-MC-BS1:val}   {\ensuremath{{0.5602 } } }
\vdef{bdt2011:N-EFF-TRIG-MC-BS1:err}   {\ensuremath{{0.0059 } } }
\vdef{bdt2011:N-EFF-CAND-BS1:val}   {\ensuremath{{0.9940 } } }
\vdef{bdt2011:N-EFF-CAND-BS1:err}   {\ensuremath{{0.0008 } } }
\vdef{bdt2011:N-EFF-ANA-BS1:val}   {\ensuremath{{0.1473 } } }
\vdef{bdt2011:N-EFF-ANA-BS1:err}   {\ensuremath{{0.0012 } } }
\vdef{bdt2011:N-OBS-BS1:val}   {\ensuremath{{1698 } } }
\vdef{bdt2011:N-OBS-BS1:err}   {\ensuremath{{48 } } }
\vdef{2011:bdt:0}     {\ensuremath{{0.290 } } }
\vdef{2011:bdtMax:0}  {\ensuremath{{10.000 } } }
\vdef{2011:mBdLo:0}   {\ensuremath{{5.200 } } }
\vdef{2011:mBdHi:0}   {\ensuremath{{5.300 } } }
\vdef{2011:mBsLo:0}   {\ensuremath{{5.300 } } }
\vdef{2011:mBsHi:0}   {\ensuremath{{5.450 } } }
\vdef{2011:etaMin:0}   {\ensuremath{{0.0 } } }
\vdef{2011:etaMax:0}   {\ensuremath{{1.4 } } }
\vdef{2011:pt:0}   {\ensuremath{{6.5 } } }
\vdef{2011:m1pt:0}   {\ensuremath{{4.5 } } }
\vdef{2011:m2pt:0}   {\ensuremath{{4.0 } } }
\vdef{2011:m1eta:0}   {\ensuremath{{1.4 } } }
\vdef{2011:m2eta:0}   {\ensuremath{{1.4 } } }
\vdef{2011:iso:0}   {\ensuremath{{0.80 } } }
\vdef{2011:chi2dof:0}   {\ensuremath{{2.2 } } }
\vdef{2011:alpha:0}   {\ensuremath{{0.050 } } }
\vdef{2011:fls3d:0}   {\ensuremath{{13.0 } } }
\vdef{2011:docatrk:0}   {\ensuremath{{0.015 } } }
\vdef{2011:closetrk:0}   {\ensuremath{{2 } } }
\vdef{2011:pvlip:0}   {\ensuremath{{100.000 } } }
\vdef{2011:pvlips:0}   {\ensuremath{{100.000 } } }
\vdef{2011:pvlip2:0}   {\ensuremath{{-100.000 } } }
\vdef{2011:pvlips2:0}   {\ensuremath{{-100.000 } } }
\vdef{2011:maxdoca:0}   {\ensuremath{{100.000 } } }
\vdef{2011:pvip:0}   {\ensuremath{{0.008 } } }
\vdef{2011:pvips:0}   {\ensuremath{{2.000 } } }
\vdef{2011:doApplyCowboyVeto:0}   {no }
\vdef{2011:fDoApplyCowboyVetoAlsoInSignal:0}   {no }
\vdef{2011:bdt:1}     {\ensuremath{{0.290 } } }
\vdef{2011:bdtMax:1}  {\ensuremath{{10.000 } } }
\vdef{2011:mBdLo:1}   {\ensuremath{{5.200 } } }
\vdef{2011:mBdHi:1}   {\ensuremath{{5.300 } } }
\vdef{2011:mBsLo:1}   {\ensuremath{{5.300 } } }
\vdef{2011:mBsHi:1}   {\ensuremath{{5.450 } } }
\vdef{2011:etaMin:1}   {\ensuremath{{1.4 } } }
\vdef{2011:etaMax:1}   {\ensuremath{{2.4 } } }
\vdef{2011:pt:1}   {\ensuremath{{8.5 } } }
\vdef{2011:m1pt:1}   {\ensuremath{{4.5 } } }
\vdef{2011:m2pt:1}   {\ensuremath{{4.2 } } }
\vdef{2011:m1eta:1}   {\ensuremath{{2.4 } } }
\vdef{2011:m2eta:1}   {\ensuremath{{2.4 } } }
\vdef{2011:iso:1}   {\ensuremath{{0.80 } } }
\vdef{2011:chi2dof:1}   {\ensuremath{{1.8 } } }
\vdef{2011:alpha:1}   {\ensuremath{{0.030 } } }
\vdef{2011:fls3d:1}   {\ensuremath{{15.0 } } }
\vdef{2011:docatrk:1}   {\ensuremath{{0.015 } } }
\vdef{2011:closetrk:1}   {\ensuremath{{2 } } }
\vdef{2011:pvlip:1}   {\ensuremath{{100.000 } } }
\vdef{2011:pvlips:1}   {\ensuremath{{100.000 } } }
\vdef{2011:pvlip2:1}   {\ensuremath{{-100.000 } } }
\vdef{2011:pvlips2:1}   {\ensuremath{{-100.000 } } }
\vdef{2011:maxdoca:1}   {\ensuremath{{100.000 } } }
\vdef{2011:pvip:1}   {\ensuremath{{0.008 } } }
\vdef{2011:pvips:1}   {\ensuremath{{2.000 } } }
\vdef{2011:doApplyCowboyVeto:1}   {no }
\vdef{2011:fDoApplyCowboyVetoAlsoInSignal:1}   {no }
\vdef{2011:bdt:0}     {\ensuremath{{0.290 } } }
\vdef{2011:bdtMax:0}  {\ensuremath{{10.000 } } }
\vdef{2011:mBdLo:0}   {\ensuremath{{5.200 } } }
\vdef{2011:mBdHi:0}   {\ensuremath{{5.300 } } }
\vdef{2011:mBsLo:0}   {\ensuremath{{5.300 } } }
\vdef{2011:mBsHi:0}   {\ensuremath{{5.450 } } }
\vdef{2011:etaMin:0}   {\ensuremath{{0.0 } } }
\vdef{2011:etaMax:0}   {\ensuremath{{1.4 } } }
\vdef{2011:pt:0}   {\ensuremath{{6.5 } } }
\vdef{2011:m1pt:0}   {\ensuremath{{4.5 } } }
\vdef{2011:m2pt:0}   {\ensuremath{{4.0 } } }
\vdef{2011:m1eta:0}   {\ensuremath{{1.4 } } }
\vdef{2011:m2eta:0}   {\ensuremath{{1.4 } } }
\vdef{2011:iso:0}   {\ensuremath{{0.80 } } }
\vdef{2011:chi2dof:0}   {\ensuremath{{2.2 } } }
\vdef{2011:alpha:0}   {\ensuremath{{0.050 } } }
\vdef{2011:fls3d:0}   {\ensuremath{{13.0 } } }
\vdef{2011:docatrk:0}   {\ensuremath{{0.015 } } }
\vdef{2011:closetrk:0}   {\ensuremath{{2 } } }
\vdef{2011:pvlip:0}   {\ensuremath{{100.000 } } }
\vdef{2011:pvlips:0}   {\ensuremath{{100.000 } } }
\vdef{2011:pvlip2:0}   {\ensuremath{{-100.000 } } }
\vdef{2011:pvlips2:0}   {\ensuremath{{-100.000 } } }
\vdef{2011:maxdoca:0}   {\ensuremath{{100.000 } } }
\vdef{2011:pvip:0}   {\ensuremath{{0.008 } } }
\vdef{2011:pvips:0}   {\ensuremath{{2.000 } } }
\vdef{2011:doApplyCowboyVeto:0}   {no }
\vdef{2011:fDoApplyCowboyVetoAlsoInSignal:0}   {no }
\vdef{2011:bdt:1}     {\ensuremath{{0.290 } } }
\vdef{2011:bdtMax:1}  {\ensuremath{{10.000 } } }
\vdef{2011:mBdLo:1}   {\ensuremath{{5.200 } } }
\vdef{2011:mBdHi:1}   {\ensuremath{{5.300 } } }
\vdef{2011:mBsLo:1}   {\ensuremath{{5.300 } } }
\vdef{2011:mBsHi:1}   {\ensuremath{{5.450 } } }
\vdef{2011:etaMin:1}   {\ensuremath{{1.4 } } }
\vdef{2011:etaMax:1}   {\ensuremath{{2.4 } } }
\vdef{2011:pt:1}   {\ensuremath{{8.5 } } }
\vdef{2011:m1pt:1}   {\ensuremath{{4.5 } } }
\vdef{2011:m2pt:1}   {\ensuremath{{4.2 } } }
\vdef{2011:m1eta:1}   {\ensuremath{{2.4 } } }
\vdef{2011:m2eta:1}   {\ensuremath{{2.4 } } }
\vdef{2011:iso:1}   {\ensuremath{{0.80 } } }
\vdef{2011:chi2dof:1}   {\ensuremath{{1.8 } } }
\vdef{2011:alpha:1}   {\ensuremath{{0.030 } } }
\vdef{2011:fls3d:1}   {\ensuremath{{15.0 } } }
\vdef{2011:docatrk:1}   {\ensuremath{{0.015 } } }
\vdef{2011:closetrk:1}   {\ensuremath{{2 } } }
\vdef{2011:pvlip:1}   {\ensuremath{{100.000 } } }
\vdef{2011:pvlips:1}   {\ensuremath{{100.000 } } }
\vdef{2011:pvlip2:1}   {\ensuremath{{-100.000 } } }
\vdef{2011:pvlips2:1}   {\ensuremath{{-100.000 } } }
\vdef{2011:maxdoca:1}   {\ensuremath{{100.000 } } }
\vdef{2011:pvip:1}   {\ensuremath{{0.008 } } }
\vdef{2011:pvips:1}   {\ensuremath{{2.000 } } }
\vdef{2011:doApplyCowboyVeto:1}   {no }
\vdef{2011:fDoApplyCowboyVetoAlsoInSignal:1}   {no }
\vdef{bdt2011:N-CSBF-TNP-BS0:val}   {\ensuremath{{0.000024 } } }
\vdef{bdt2011:N-CSBF-TNP-BS0:err}   {\ensuremath{{0.000001 } } }
\vdef{bdt2011:N-CSBF-MC-BS0:val}   {\ensuremath{{0.000025 } } }
\vdef{bdt2011:N-CSBF-MC-BS0:err}   {\ensuremath{{0.000001 } } }
\vdef{bdt2011:N-CSBF-MC-BS0:syst}   {\ensuremath{{0.000002 } } }
\vdef{bdt2011:N-CSBF-BS0:val}   {\ensuremath{{0.000025 } } }
\vdef{bdt2011:N-CSBF-BS0:err}   {\ensuremath{{0.000001 } } }
\vdef{bdt2011:N-CSBF-BS0:syst}   {\ensuremath{{0.000002 } } }
\vdef{bdt2011:bgBd2K0MuMu:loSideband0:val}   {\ensuremath{{0.000 } } }
\vdef{bdt2011:bgBd2K0MuMu:loSideband0:err}   {\ensuremath{{0.000 } } }
\vdef{bdt2011:bgBd2K0MuMu:bdRare0}   {\ensuremath{{0.000000 } } }
\vdef{bdt2011:bgBd2K0MuMu:bdRare0E}  {\ensuremath{{0.000000 } } }
\vdef{bdt2011:bgBd2K0MuMu:bsRare0}   {\ensuremath{{0.000000 } } }
\vdef{bdt2011:bgBd2K0MuMu:bsRare0E}  {\ensuremath{{0.000000 } } }
\vdef{bdt2011:bgBd2K0MuMu:hiSideband0:val}   {\ensuremath{{0.000 } } }
\vdef{bdt2011:bgBd2K0MuMu:hiSideband0:err}   {\ensuremath{{0.000 } } }
\vdef{bdt2011:bgBd2KK:loSideband0:val}   {\ensuremath{{0.002 } } }
\vdef{bdt2011:bgBd2KK:loSideband0:err}   {\ensuremath{{0.003 } } }
\vdef{bdt2011:bgBd2KK:bdRare0}   {\ensuremath{{0.001114 } } }
\vdef{bdt2011:bgBd2KK:bdRare0E}  {\ensuremath{{0.001407 } } }
\vdef{bdt2011:bgBd2KK:bsRare0}   {\ensuremath{{0.000051 } } }
\vdef{bdt2011:bgBd2KK:bsRare0E}  {\ensuremath{{0.000064 } } }
\vdef{bdt2011:bgBd2KK:hiSideband0:val}   {\ensuremath{{0.000 } } }
\vdef{bdt2011:bgBd2KK:hiSideband0:err}   {\ensuremath{{0.000 } } }
\vdef{bdt2011:bgBd2KPi:loSideband0:val}   {\ensuremath{{0.085 } } }
\vdef{bdt2011:bgBd2KPi:loSideband0:err}   {\ensuremath{{0.060 } } }
\vdef{bdt2011:bgBd2KPi:bdRare0}   {\ensuremath{{0.217428 } } }
\vdef{bdt2011:bgBd2KPi:bdRare0E}  {\ensuremath{{0.153883 } } }
\vdef{bdt2011:bgBd2KPi:bsRare0}   {\ensuremath{{0.022318 } } }
\vdef{bdt2011:bgBd2KPi:bsRare0E}  {\ensuremath{{0.015795 } } }
\vdef{bdt2011:bgBd2KPi:hiSideband0:val}   {\ensuremath{{0.001 } } }
\vdef{bdt2011:bgBd2KPi:hiSideband0:err}   {\ensuremath{{0.001 } } }
\vdef{bdt2011:bgBd2MuMuGamma:loSideband0:val}   {\ensuremath{{0.010 } } }
\vdef{bdt2011:bgBd2MuMuGamma:loSideband0:err}   {\ensuremath{{0.002 } } }
\vdef{bdt2011:bgBd2MuMuGamma:bdRare0}   {\ensuremath{{0.000683 } } }
\vdef{bdt2011:bgBd2MuMuGamma:bdRare0E}  {\ensuremath{{0.000137 } } }
\vdef{bdt2011:bgBd2MuMuGamma:bsRare0}   {\ensuremath{{0.000060 } } }
\vdef{bdt2011:bgBd2MuMuGamma:bsRare0E}  {\ensuremath{{0.000012 } } }
\vdef{bdt2011:bgBd2MuMuGamma:hiSideband0:val}   {\ensuremath{{0.000 } } }
\vdef{bdt2011:bgBd2MuMuGamma:hiSideband0:err}   {\ensuremath{{0.000 } } }
\vdef{bdt2011:bgBd2Pi0MuMu:loSideband0:val}   {\ensuremath{{0.505 } } }
\vdef{bdt2011:bgBd2Pi0MuMu:loSideband0:err}   {\ensuremath{{0.252 } } }
\vdef{bdt2011:bgBd2Pi0MuMu:bdRare0}   {\ensuremath{{0.000362 } } }
\vdef{bdt2011:bgBd2Pi0MuMu:bdRare0E}  {\ensuremath{{0.000181 } } }
\vdef{bdt2011:bgBd2Pi0MuMu:bsRare0}   {\ensuremath{{0.000000 } } }
\vdef{bdt2011:bgBd2Pi0MuMu:bsRare0E}  {\ensuremath{{0.000000 } } }
\vdef{bdt2011:bgBd2Pi0MuMu:hiSideband0:val}   {\ensuremath{{0.000 } } }
\vdef{bdt2011:bgBd2Pi0MuMu:hiSideband0:err}   {\ensuremath{{0.000 } } }
\vdef{bdt2011:bgBd2PiMuNu:loSideband0:val}   {\ensuremath{{3.584 } } }
\vdef{bdt2011:bgBd2PiMuNu:loSideband0:err}   {\ensuremath{{1.801 } } }
\vdef{bdt2011:bgBd2PiMuNu:bdRare0}   {\ensuremath{{0.099800 } } }
\vdef{bdt2011:bgBd2PiMuNu:bdRare0E}  {\ensuremath{{0.050149 } } }
\vdef{bdt2011:bgBd2PiMuNu:bsRare0}   {\ensuremath{{0.012668 } } }
\vdef{bdt2011:bgBd2PiMuNu:bsRare0E}  {\ensuremath{{0.006366 } } }
\vdef{bdt2011:bgBd2PiMuNu:hiSideband0:val}   {\ensuremath{{0.004 } } }
\vdef{bdt2011:bgBd2PiMuNu:hiSideband0:err}   {\ensuremath{{0.002 } } }
\vdef{bdt2011:bgBd2PiPi:loSideband0:val}   {\ensuremath{{0.005 } } }
\vdef{bdt2011:bgBd2PiPi:loSideband0:err}   {\ensuremath{{0.005 } } }
\vdef{bdt2011:bgBd2PiPi:bdRare0}   {\ensuremath{{0.033956 } } }
\vdef{bdt2011:bgBd2PiPi:bdRare0E}  {\ensuremath{{0.033988 } } }
\vdef{bdt2011:bgBd2PiPi:bsRare0}   {\ensuremath{{0.014331 } } }
\vdef{bdt2011:bgBd2PiPi:bsRare0E}  {\ensuremath{{0.014344 } } }
\vdef{bdt2011:bgBd2PiPi:hiSideband0:val}   {\ensuremath{{0.001 } } }
\vdef{bdt2011:bgBd2PiPi:hiSideband0:err}   {\ensuremath{{0.001 } } }
\vdef{bdt2011:bgBs2KK:loSideband0:val}   {\ensuremath{{0.012 } } }
\vdef{bdt2011:bgBs2KK:loSideband0:err}   {\ensuremath{{0.012 } } }
\vdef{bdt2011:bgBs2KK:bdRare0}   {\ensuremath{{0.104112 } } }
\vdef{bdt2011:bgBs2KK:bdRare0E}  {\ensuremath{{0.105277 } } }
\vdef{bdt2011:bgBs2KK:bsRare0}   {\ensuremath{{0.044010 } } }
\vdef{bdt2011:bgBs2KK:bsRare0E}  {\ensuremath{{0.044502 } } }
\vdef{bdt2011:bgBs2KK:hiSideband0:val}   {\ensuremath{{0.001 } } }
\vdef{bdt2011:bgBs2KK:hiSideband0:err}   {\ensuremath{{0.001 } } }
\vdef{bdt2011:bgBs2KMuNu:loSideband0:val}   {\ensuremath{{1.708 } } }
\vdef{bdt2011:bgBs2KMuNu:loSideband0:err}   {\ensuremath{{0.858 } } }
\vdef{bdt2011:bgBs2KMuNu:bdRare0}   {\ensuremath{{0.050963 } } }
\vdef{bdt2011:bgBs2KMuNu:bdRare0E}  {\ensuremath{{0.025608 } } }
\vdef{bdt2011:bgBs2KMuNu:bsRare0}   {\ensuremath{{0.007582 } } }
\vdef{bdt2011:bgBs2KMuNu:bsRare0E}  {\ensuremath{{0.003810 } } }
\vdef{bdt2011:bgBs2KMuNu:hiSideband0:val}   {\ensuremath{{0.000 } } }
\vdef{bdt2011:bgBs2KMuNu:hiSideband0:err}   {\ensuremath{{0.000 } } }
\vdef{bdt2011:bgBs2KPi:loSideband0:val}   {\ensuremath{{0.001 } } }
\vdef{bdt2011:bgBs2KPi:loSideband0:err}   {\ensuremath{{0.001 } } }
\vdef{bdt2011:bgBs2KPi:bdRare0}   {\ensuremath{{0.005890 } } }
\vdef{bdt2011:bgBs2KPi:bdRare0E}  {\ensuremath{{0.004362 } } }
\vdef{bdt2011:bgBs2KPi:bsRare0}   {\ensuremath{{0.012905 } } }
\vdef{bdt2011:bgBs2KPi:bsRare0E}  {\ensuremath{{0.009556 } } }
\vdef{bdt2011:bgBs2KPi:hiSideband0:val}   {\ensuremath{{0.000 } } }
\vdef{bdt2011:bgBs2KPi:hiSideband0:err}   {\ensuremath{{0.000 } } }
\vdef{bdt2011:bgBs2MuMuGamma:loSideband0:val}   {\ensuremath{{0.401 } } }
\vdef{bdt2011:bgBs2MuMuGamma:loSideband0:err}   {\ensuremath{{0.080 } } }
\vdef{bdt2011:bgBs2MuMuGamma:bdRare0}   {\ensuremath{{0.068853 } } }
\vdef{bdt2011:bgBs2MuMuGamma:bdRare0E}  {\ensuremath{{0.013771 } } }
\vdef{bdt2011:bgBs2MuMuGamma:bsRare0}   {\ensuremath{{0.019776 } } }
\vdef{bdt2011:bgBs2MuMuGamma:bsRare0E}  {\ensuremath{{0.003955 } } }
\vdef{bdt2011:bgBs2MuMuGamma:hiSideband0:val}   {\ensuremath{{0.000 } } }
\vdef{bdt2011:bgBs2MuMuGamma:hiSideband0:err}   {\ensuremath{{0.000 } } }
\vdef{bdt2011:bgBs2PiPi:loSideband0:val}   {\ensuremath{{0.000 } } }
\vdef{bdt2011:bgBs2PiPi:loSideband0:err}   {\ensuremath{{0.000 } } }
\vdef{bdt2011:bgBs2PiPi:bdRare0}   {\ensuremath{{0.000168 } } }
\vdef{bdt2011:bgBs2PiPi:bdRare0E}  {\ensuremath{{0.000171 } } }
\vdef{bdt2011:bgBs2PiPi:bsRare0}   {\ensuremath{{0.001565 } } }
\vdef{bdt2011:bgBs2PiPi:bsRare0E}  {\ensuremath{{0.001593 } } }
\vdef{bdt2011:bgBs2PiPi:hiSideband0:val}   {\ensuremath{{0.000 } } }
\vdef{bdt2011:bgBs2PiPi:hiSideband0:err}   {\ensuremath{{0.000 } } }
\vdef{bdt2011:bgBu2KMuMu:loSideband0:val}   {\ensuremath{{0.015 } } }
\vdef{bdt2011:bgBu2KMuMu:loSideband0:err}   {\ensuremath{{0.001 } } }
\vdef{bdt2011:bgBu2KMuMu:bdRare0}   {\ensuremath{{0.000000 } } }
\vdef{bdt2011:bgBu2KMuMu:bdRare0E}  {\ensuremath{{0.000000 } } }
\vdef{bdt2011:bgBu2KMuMu:bsRare0}   {\ensuremath{{0.000000 } } }
\vdef{bdt2011:bgBu2KMuMu:bsRare0E}  {\ensuremath{{0.000000 } } }
\vdef{bdt2011:bgBu2KMuMu:hiSideband0:val}   {\ensuremath{{0.000 } } }
\vdef{bdt2011:bgBu2KMuMu:hiSideband0:err}   {\ensuremath{{0.000 } } }
\vdef{bdt2011:bgBu2PiMuMu:loSideband0:val}   {\ensuremath{{0.828 } } }
\vdef{bdt2011:bgBu2PiMuMu:loSideband0:err}   {\ensuremath{{0.215 } } }
\vdef{bdt2011:bgBu2PiMuMu:bdRare0}   {\ensuremath{{0.001298 } } }
\vdef{bdt2011:bgBu2PiMuMu:bdRare0E}  {\ensuremath{{0.000338 } } }
\vdef{bdt2011:bgBu2PiMuMu:bsRare0}   {\ensuremath{{0.000604 } } }
\vdef{bdt2011:bgBu2PiMuMu:bsRare0E}  {\ensuremath{{0.000157 } } }
\vdef{bdt2011:bgBu2PiMuMu:hiSideband0:val}   {\ensuremath{{0.000 } } }
\vdef{bdt2011:bgBu2PiMuMu:hiSideband0:err}   {\ensuremath{{0.000 } } }
\vdef{bdt2011:bgLb2KP:loSideband0:val}   {\ensuremath{{0.001 } } }
\vdef{bdt2011:bgLb2KP:loSideband0:err}   {\ensuremath{{0.000 } } }
\vdef{bdt2011:bgLb2KP:bdRare0}   {\ensuremath{{0.001104 } } }
\vdef{bdt2011:bgLb2KP:bdRare0E}  {\ensuremath{{0.000830 } } }
\vdef{bdt2011:bgLb2KP:bsRare0}   {\ensuremath{{0.010758 } } }
\vdef{bdt2011:bgLb2KP:bsRare0E}  {\ensuremath{{0.008085 } } }
\vdef{bdt2011:bgLb2KP:hiSideband0:val}   {\ensuremath{{0.004 } } }
\vdef{bdt2011:bgLb2KP:hiSideband0:err}   {\ensuremath{{0.003 } } }
\vdef{bdt2011:bgLb2PMuNu:loSideband0:val}   {\ensuremath{{15.949 } } }
\vdef{bdt2011:bgLb2PMuNu:loSideband0:err}   {\ensuremath{{17.831 } } }
\vdef{bdt2011:bgLb2PMuNu:bdRare0}   {\ensuremath{{3.333918 } } }
\vdef{bdt2011:bgLb2PMuNu:bdRare0E}  {\ensuremath{{3.727434 } } }
\vdef{bdt2011:bgLb2PMuNu:bsRare0}   {\ensuremath{{2.757996 } } }
\vdef{bdt2011:bgLb2PMuNu:bsRare0E}  {\ensuremath{{3.083534 } } }
\vdef{bdt2011:bgLb2PMuNu:hiSideband0:val}   {\ensuremath{{0.266 } } }
\vdef{bdt2011:bgLb2PMuNu:hiSideband0:err}   {\ensuremath{{0.298 } } }
\vdef{bdt2011:bgLb2PiP:loSideband0:val}   {\ensuremath{{0.000 } } }
\vdef{bdt2011:bgLb2PiP:loSideband0:err}   {\ensuremath{{0.000 } } }
\vdef{bdt2011:bgLb2PiP:bdRare0}   {\ensuremath{{0.000332 } } }
\vdef{bdt2011:bgLb2PiP:bdRare0E}  {\ensuremath{{0.000253 } } }
\vdef{bdt2011:bgLb2PiP:bsRare0}   {\ensuremath{{0.002620 } } }
\vdef{bdt2011:bgLb2PiP:bsRare0E}  {\ensuremath{{0.001998 } } }
\vdef{bdt2011:bgLb2PiP:hiSideband0:val}   {\ensuremath{{0.004 } } }
\vdef{bdt2011:bgLb2PiP:hiSideband0:err}   {\ensuremath{{0.003 } } }
\vdef{bdt2011:bsRare0}   {\ensuremath{{0.000 } } }
\vdef{bdt2011:bsRare0E}  {\ensuremath{{0.000 } } }
\vdef{bdt2011:bdRare0}   {\ensuremath{{0.000 } } }
\vdef{bdt2011:bdRare0E}  {\ensuremath{{0.000 } } }
\vdef{bdt2011:N-CSBF-TNP-BS1:val}   {\ensuremath{{0.000030 } } }
\vdef{bdt2011:N-CSBF-TNP-BS1:err}   {\ensuremath{{0.000001 } } }
\vdef{bdt2011:N-CSBF-MC-BS1:val}   {\ensuremath{{0.000032 } } }
\vdef{bdt2011:N-CSBF-MC-BS1:err}   {\ensuremath{{0.000001 } } }
\vdef{bdt2011:N-CSBF-MC-BS1:syst}   {\ensuremath{{0.000002 } } }
\vdef{bdt2011:N-CSBF-BS1:val}   {\ensuremath{{0.000031 } } }
\vdef{bdt2011:N-CSBF-BS1:err}   {\ensuremath{{0.000001 } } }
\vdef{bdt2011:N-CSBF-BS1:syst}   {\ensuremath{{0.000002 } } }
\vdef{bdt2011:bgBd2K0MuMu:loSideband1:val}   {\ensuremath{{0.000 } } }
\vdef{bdt2011:bgBd2K0MuMu:loSideband1:err}   {\ensuremath{{0.000 } } }
\vdef{bdt2011:bgBd2K0MuMu:bdRare1}   {\ensuremath{{0.004221 } } }
\vdef{bdt2011:bgBd2K0MuMu:bdRare1E}  {\ensuremath{{0.000000 } } }
\vdef{bdt2011:bgBd2K0MuMu:bsRare1}   {\ensuremath{{0.000000 } } }
\vdef{bdt2011:bgBd2K0MuMu:bsRare1E}  {\ensuremath{{0.000000 } } }
\vdef{bdt2011:bgBd2K0MuMu:hiSideband1:val}   {\ensuremath{{0.000 } } }
\vdef{bdt2011:bgBd2K0MuMu:hiSideband1:err}   {\ensuremath{{0.000 } } }
\vdef{bdt2011:bgBd2KK:loSideband1:val}   {\ensuremath{{0.001 } } }
\vdef{bdt2011:bgBd2KK:loSideband1:err}   {\ensuremath{{0.001 } } }
\vdef{bdt2011:bgBd2KK:bdRare1}   {\ensuremath{{0.000420 } } }
\vdef{bdt2011:bgBd2KK:bdRare1E}  {\ensuremath{{0.000530 } } }
\vdef{bdt2011:bgBd2KK:bsRare1}   {\ensuremath{{0.000076 } } }
\vdef{bdt2011:bgBd2KK:bsRare1E}  {\ensuremath{{0.000095 } } }
\vdef{bdt2011:bgBd2KK:hiSideband1:val}   {\ensuremath{{0.000 } } }
\vdef{bdt2011:bgBd2KK:hiSideband1:err}   {\ensuremath{{0.000 } } }
\vdef{bdt2011:bgBd2KPi:loSideband1:val}   {\ensuremath{{0.048 } } }
\vdef{bdt2011:bgBd2KPi:loSideband1:err}   {\ensuremath{{0.034 } } }
\vdef{bdt2011:bgBd2KPi:bdRare1}   {\ensuremath{{0.058786 } } }
\vdef{bdt2011:bgBd2KPi:bdRare1E}  {\ensuremath{{0.041605 } } }
\vdef{bdt2011:bgBd2KPi:bsRare1}   {\ensuremath{{0.019412 } } }
\vdef{bdt2011:bgBd2KPi:bsRare1E}  {\ensuremath{{0.013739 } } }
\vdef{bdt2011:bgBd2KPi:hiSideband1:val}   {\ensuremath{{0.002 } } }
\vdef{bdt2011:bgBd2KPi:hiSideband1:err}   {\ensuremath{{0.001 } } }
\vdef{bdt2011:bgBd2MuMuGamma:loSideband1:val}   {\ensuremath{{0.003 } } }
\vdef{bdt2011:bgBd2MuMuGamma:loSideband1:err}   {\ensuremath{{0.001 } } }
\vdef{bdt2011:bgBd2MuMuGamma:bdRare1}   {\ensuremath{{0.000299 } } }
\vdef{bdt2011:bgBd2MuMuGamma:bdRare1E}  {\ensuremath{{0.000060 } } }
\vdef{bdt2011:bgBd2MuMuGamma:bsRare1}   {\ensuremath{{0.000045 } } }
\vdef{bdt2011:bgBd2MuMuGamma:bsRare1E}  {\ensuremath{{0.000009 } } }
\vdef{bdt2011:bgBd2MuMuGamma:hiSideband1:val}   {\ensuremath{{0.000 } } }
\vdef{bdt2011:bgBd2MuMuGamma:hiSideband1:err}   {\ensuremath{{0.000 } } }
\vdef{bdt2011:bgBd2Pi0MuMu:loSideband1:val}   {\ensuremath{{0.164 } } }
\vdef{bdt2011:bgBd2Pi0MuMu:loSideband1:err}   {\ensuremath{{0.082 } } }
\vdef{bdt2011:bgBd2Pi0MuMu:bdRare1}   {\ensuremath{{0.001651 } } }
\vdef{bdt2011:bgBd2Pi0MuMu:bdRare1E}  {\ensuremath{{0.000825 } } }
\vdef{bdt2011:bgBd2Pi0MuMu:bsRare1}   {\ensuremath{{0.000000 } } }
\vdef{bdt2011:bgBd2Pi0MuMu:bsRare1E}  {\ensuremath{{0.000000 } } }
\vdef{bdt2011:bgBd2Pi0MuMu:hiSideband1:val}   {\ensuremath{{0.000 } } }
\vdef{bdt2011:bgBd2Pi0MuMu:hiSideband1:err}   {\ensuremath{{0.000 } } }
\vdef{bdt2011:bgBd2PiMuNu:loSideband1:val}   {\ensuremath{{1.301 } } }
\vdef{bdt2011:bgBd2PiMuNu:loSideband1:err}   {\ensuremath{{0.654 } } }
\vdef{bdt2011:bgBd2PiMuNu:bdRare1}   {\ensuremath{{0.052250 } } }
\vdef{bdt2011:bgBd2PiMuNu:bdRare1E}  {\ensuremath{{0.026255 } } }
\vdef{bdt2011:bgBd2PiMuNu:bsRare1}   {\ensuremath{{0.003781 } } }
\vdef{bdt2011:bgBd2PiMuNu:bsRare1E}  {\ensuremath{{0.001900 } } }
\vdef{bdt2011:bgBd2PiMuNu:hiSideband1:val}   {\ensuremath{{0.003 } } }
\vdef{bdt2011:bgBd2PiMuNu:hiSideband1:err}   {\ensuremath{{0.001 } } }
\vdef{bdt2011:bgBd2PiPi:loSideband1:val}   {\ensuremath{{0.004 } } }
\vdef{bdt2011:bgBd2PiPi:loSideband1:err}   {\ensuremath{{0.004 } } }
\vdef{bdt2011:bgBd2PiPi:bdRare1}   {\ensuremath{{0.009850 } } }
\vdef{bdt2011:bgBd2PiPi:bdRare1E}  {\ensuremath{{0.009859 } } }
\vdef{bdt2011:bgBd2PiPi:bsRare1}   {\ensuremath{{0.006885 } } }
\vdef{bdt2011:bgBd2PiPi:bsRare1E}  {\ensuremath{{0.006891 } } }
\vdef{bdt2011:bgBd2PiPi:hiSideband1:val}   {\ensuremath{{0.001 } } }
\vdef{bdt2011:bgBd2PiPi:hiSideband1:err}   {\ensuremath{{0.001 } } }
\vdef{bdt2011:bgBs2KK:loSideband1:val}   {\ensuremath{{0.012 } } }
\vdef{bdt2011:bgBs2KK:loSideband1:err}   {\ensuremath{{0.012 } } }
\vdef{bdt2011:bgBs2KK:bdRare1}   {\ensuremath{{0.031666 } } }
\vdef{bdt2011:bgBs2KK:bdRare1E}  {\ensuremath{{0.032020 } } }
\vdef{bdt2011:bgBs2KK:bsRare1}   {\ensuremath{{0.020175 } } }
\vdef{bdt2011:bgBs2KK:bsRare1E}  {\ensuremath{{0.020401 } } }
\vdef{bdt2011:bgBs2KK:hiSideband1:val}   {\ensuremath{{0.001 } } }
\vdef{bdt2011:bgBs2KK:hiSideband1:err}   {\ensuremath{{0.001 } } }
\vdef{bdt2011:bgBs2KMuNu:loSideband1:val}   {\ensuremath{{0.567 } } }
\vdef{bdt2011:bgBs2KMuNu:loSideband1:err}   {\ensuremath{{0.285 } } }
\vdef{bdt2011:bgBs2KMuNu:bdRare1}   {\ensuremath{{0.032056 } } }
\vdef{bdt2011:bgBs2KMuNu:bdRare1E}  {\ensuremath{{0.016108 } } }
\vdef{bdt2011:bgBs2KMuNu:bsRare1}   {\ensuremath{{0.018105 } } }
\vdef{bdt2011:bgBs2KMuNu:bsRare1E}  {\ensuremath{{0.009097 } } }
\vdef{bdt2011:bgBs2KMuNu:hiSideband1:val}   {\ensuremath{{0.000 } } }
\vdef{bdt2011:bgBs2KMuNu:hiSideband1:err}   {\ensuremath{{0.000 } } }
\vdef{bdt2011:bgBs2KPi:loSideband1:val}   {\ensuremath{{0.001 } } }
\vdef{bdt2011:bgBs2KPi:loSideband1:err}   {\ensuremath{{0.001 } } }
\vdef{bdt2011:bgBs2KPi:bdRare1}   {\ensuremath{{0.002812 } } }
\vdef{bdt2011:bgBs2KPi:bdRare1E}  {\ensuremath{{0.002083 } } }
\vdef{bdt2011:bgBs2KPi:bsRare1}   {\ensuremath{{0.004407 } } }
\vdef{bdt2011:bgBs2KPi:bsRare1E}  {\ensuremath{{0.003263 } } }
\vdef{bdt2011:bgBs2KPi:hiSideband1:val}   {\ensuremath{{0.001 } } }
\vdef{bdt2011:bgBs2KPi:hiSideband1:err}   {\ensuremath{{0.000 } } }
\vdef{bdt2011:bgBs2MuMuGamma:loSideband1:val}   {\ensuremath{{0.136 } } }
\vdef{bdt2011:bgBs2MuMuGamma:loSideband1:err}   {\ensuremath{{0.027 } } }
\vdef{bdt2011:bgBs2MuMuGamma:bdRare1}   {\ensuremath{{0.022962 } } }
\vdef{bdt2011:bgBs2MuMuGamma:bdRare1E}  {\ensuremath{{0.004592 } } }
\vdef{bdt2011:bgBs2MuMuGamma:bsRare1}   {\ensuremath{{0.008396 } } }
\vdef{bdt2011:bgBs2MuMuGamma:bsRare1E}  {\ensuremath{{0.001679 } } }
\vdef{bdt2011:bgBs2MuMuGamma:hiSideband1:val}   {\ensuremath{{0.000 } } }
\vdef{bdt2011:bgBs2MuMuGamma:hiSideband1:err}   {\ensuremath{{0.000 } } }
\vdef{bdt2011:bgBs2PiPi:loSideband1:val}   {\ensuremath{{0.000 } } }
\vdef{bdt2011:bgBs2PiPi:loSideband1:err}   {\ensuremath{{0.000 } } }
\vdef{bdt2011:bgBs2PiPi:bdRare1}   {\ensuremath{{0.000147 } } }
\vdef{bdt2011:bgBs2PiPi:bdRare1E}  {\ensuremath{{0.000150 } } }
\vdef{bdt2011:bgBs2PiPi:bsRare1}   {\ensuremath{{0.000473 } } }
\vdef{bdt2011:bgBs2PiPi:bsRare1E}  {\ensuremath{{0.000481 } } }
\vdef{bdt2011:bgBs2PiPi:hiSideband1:val}   {\ensuremath{{0.000 } } }
\vdef{bdt2011:bgBs2PiPi:hiSideband1:err}   {\ensuremath{{0.000 } } }
\vdef{bdt2011:bgBu2KMuMu:loSideband1:val}   {\ensuremath{{0.016 } } }
\vdef{bdt2011:bgBu2KMuMu:loSideband1:err}   {\ensuremath{{0.001 } } }
\vdef{bdt2011:bgBu2KMuMu:bdRare1}   {\ensuremath{{0.000000 } } }
\vdef{bdt2011:bgBu2KMuMu:bdRare1E}  {\ensuremath{{0.000000 } } }
\vdef{bdt2011:bgBu2KMuMu:bsRare1}   {\ensuremath{{0.000000 } } }
\vdef{bdt2011:bgBu2KMuMu:bsRare1E}  {\ensuremath{{0.000000 } } }
\vdef{bdt2011:bgBu2KMuMu:hiSideband1:val}   {\ensuremath{{0.000 } } }
\vdef{bdt2011:bgBu2KMuMu:hiSideband1:err}   {\ensuremath{{0.000 } } }
\vdef{bdt2011:bgBu2PiMuMu:loSideband1:val}   {\ensuremath{{0.295 } } }
\vdef{bdt2011:bgBu2PiMuMu:loSideband1:err}   {\ensuremath{{0.077 } } }
\vdef{bdt2011:bgBu2PiMuMu:bdRare1}   {\ensuremath{{0.001153 } } }
\vdef{bdt2011:bgBu2PiMuMu:bdRare1E}  {\ensuremath{{0.000300 } } }
\vdef{bdt2011:bgBu2PiMuMu:bsRare1}   {\ensuremath{{0.000000 } } }
\vdef{bdt2011:bgBu2PiMuMu:bsRare1E}  {\ensuremath{{0.000000 } } }
\vdef{bdt2011:bgBu2PiMuMu:hiSideband1:val}   {\ensuremath{{0.000 } } }
\vdef{bdt2011:bgBu2PiMuMu:hiSideband1:err}   {\ensuremath{{0.000 } } }
\vdef{bdt2011:bgLb2KP:loSideband1:val}   {\ensuremath{{0.000 } } }
\vdef{bdt2011:bgLb2KP:loSideband1:err}   {\ensuremath{{0.000 } } }
\vdef{bdt2011:bgLb2KP:bdRare1}   {\ensuremath{{0.000782 } } }
\vdef{bdt2011:bgLb2KP:bdRare1E}  {\ensuremath{{0.000588 } } }
\vdef{bdt2011:bgLb2KP:bsRare1}   {\ensuremath{{0.003295 } } }
\vdef{bdt2011:bgLb2KP:bsRare1E}  {\ensuremath{{0.002476 } } }
\vdef{bdt2011:bgLb2KP:hiSideband1:val}   {\ensuremath{{0.002 } } }
\vdef{bdt2011:bgLb2KP:hiSideband1:err}   {\ensuremath{{0.001 } } }
\vdef{bdt2011:bgLb2PMuNu:loSideband1:val}   {\ensuremath{{4.555 } } }
\vdef{bdt2011:bgLb2PMuNu:loSideband1:err}   {\ensuremath{{5.093 } } }
\vdef{bdt2011:bgLb2PMuNu:bdRare1}   {\ensuremath{{1.231652 } } }
\vdef{bdt2011:bgLb2PMuNu:bdRare1E}  {\ensuremath{{1.377029 } } }
\vdef{bdt2011:bgLb2PMuNu:bsRare1}   {\ensuremath{{0.770038 } } }
\vdef{bdt2011:bgLb2PMuNu:bsRare1E}  {\ensuremath{{0.860929 } } }
\vdef{bdt2011:bgLb2PMuNu:hiSideband1:val}   {\ensuremath{{0.207 } } }
\vdef{bdt2011:bgLb2PMuNu:hiSideband1:err}   {\ensuremath{{0.231 } } }
\vdef{bdt2011:bgLb2PiP:loSideband1:val}   {\ensuremath{{0.000 } } }
\vdef{bdt2011:bgLb2PiP:loSideband1:err}   {\ensuremath{{0.000 } } }
\vdef{bdt2011:bgLb2PiP:bdRare1}   {\ensuremath{{0.000201 } } }
\vdef{bdt2011:bgLb2PiP:bdRare1E}  {\ensuremath{{0.000154 } } }
\vdef{bdt2011:bgLb2PiP:bsRare1}   {\ensuremath{{0.000910 } } }
\vdef{bdt2011:bgLb2PiP:bsRare1E}  {\ensuremath{{0.000694 } } }
\vdef{bdt2011:bgLb2PiP:hiSideband1:val}   {\ensuremath{{0.001 } } }
\vdef{bdt2011:bgLb2PiP:hiSideband1:err}   {\ensuremath{{0.001 } } }
\vdef{bdt2011:bsRare1}   {\ensuremath{{0.000 } } }
\vdef{bdt2011:bsRare1E}  {\ensuremath{{0.000 } } }
\vdef{bdt2011:bdRare1}   {\ensuremath{{0.000 } } }
\vdef{bdt2011:bdRare1E}  {\ensuremath{{0.000 } } }
\vdef{bdt2011:N-EFF-TOT-BPLUS0:val}   {\ensuremath{{0.00098 } } }
\vdef{bdt2011:N-EFF-TOT-BPLUS0:err}   {\ensuremath{{0.000003 } } }
\vdef{bdt2011:N-EFF-TOT-BPLUS0:tot}   {\ensuremath{{0.00008 } } }
\vdef{bdt2011:N-EFF-TOT-BPLUS0:all}   {\ensuremath{{(0.98 \pm 0.08)\times 10^{-3}} } }
\vdef{bdt2011:N-EFF-PRODMC-BPLUS0:val}   {\ensuremath{{0.00099 } } }
\vdef{bdt2011:N-EFF-PRODMC-BPLUS0:err}   {\ensuremath{{0.000012 } } }
\vdef{bdt2011:N-EFF-PRODMC-BPLUS0:tot}   {\ensuremath{{0.00001 } } }
\vdef{bdt2011:N-EFF-PRODMC-BPLUS0:all}   {\ensuremath{{(0.99 \pm 0.01)\times 10^{-3}} } }
\vdef{bdt2011:N-EFF-PRODTNP-BPLUS0:val}   {\ensuremath{{0.00141 } } }
\vdef{bdt2011:N-EFF-PRODTNP-BPLUS0:err}   {\ensuremath{{0.759123 } } }
\vdef{bdt2011:N-EFF-PRODTNP-BPLUS0:tot}   {\ensuremath{{0.75912 } } }
\vdef{bdt2011:N-EFF-PRODTNP-BPLUS0:all}   {\ensuremath{{(1.41 \pm 759.12)\times 10^{-3}} } }
\vdef{bdt2011:N-EFF-PRODTNPMC-BPLUS0:val}   {\ensuremath{{0.00099 } } }
\vdef{bdt2011:N-EFF-PRODTNPMC-BPLUS0:err}   {\ensuremath{{0.000012 } } }
\vdef{bdt2011:N-EFF-PRODTNPMC-BPLUS0:tot}   {\ensuremath{{0.00001 } } }
\vdef{bdt2011:N-EFF-PRODTNPMC-BPLUS0:all}   {\ensuremath{{(0.99 \pm 0.01)\times 10^{-3}} } }
\vdef{bdt2011:N-ACC-BPLUS0:val}   {\ensuremath{{0.010 } } }
\vdef{bdt2011:N-ACC-BPLUS0:err}   {\ensuremath{{0.000 } } }
\vdef{bdt2011:N-ACC-BPLUS0:tot}   {\ensuremath{{0.000 } } }
\vdef{bdt2011:N-ACC-BPLUS0:all}   {\ensuremath{{(1.03 \pm 0.04)\times 10^{-2}} } }
\vdef{bdt2011:N-EFF-MU-PID-BPLUS0:val}   {\ensuremath{{0.714 } } }
\vdef{bdt2011:N-EFF-MU-PID-BPLUS0:err}   {\ensuremath{{0.000 } } }
\vdef{bdt2011:N-EFF-MU-PID-BPLUS0:tot}   {\ensuremath{{0.029 } } }
\vdef{bdt2011:N-EFF-MU-PID-BPLUS0:all}   {\ensuremath{{(71.42 \pm 2.86)\times 10^{-2}} } }
\vdef{bdt2011:N-EFFRHO-MU-PID-BPLUS0:val}   {\ensuremath{{0.515 } } }
\vdef{bdt2011:N-EFFRHO-MU-PID-BPLUS0:err}   {\ensuremath{{0.000 } } }
\vdef{bdt2011:N-EFFRHO-MU-PID-BPLUS0:tot}   {\ensuremath{{0.000 } } }
\vdef{bdt2011:N-EFFRHO-MU-PID-BPLUS0:all}   {\ensuremath{{(51.55 \pm 0.00)\times 10^{-2}} } }
\vdef{bdt2011:N-EFF-MU-PIDMC-BPLUS0:val}   {\ensuremath{{0.690 } } }
\vdef{bdt2011:N-EFF-MU-PIDMC-BPLUS0:err}   {\ensuremath{{0.000 } } }
\vdef{bdt2011:N-EFF-MU-PIDMC-BPLUS0:tot}   {\ensuremath{{0.028 } } }
\vdef{bdt2011:N-EFF-MU-PIDMC-BPLUS0:all}   {\ensuremath{{(68.95 \pm 2.76)\times 10^{-2}} } }
\vdef{bdt2011:N-EFFRHO-MU-PIDMC-BPLUS0:val}   {\ensuremath{{0.498 } } }
\vdef{bdt2011:N-EFFRHO-MU-PIDMC-BPLUS0:err}   {\ensuremath{{0.000 } } }
\vdef{bdt2011:N-EFFRHO-MU-PIDMC-BPLUS0:tot}   {\ensuremath{{0.021 } } }
\vdef{bdt2011:N-EFFRHO-MU-PIDMC-BPLUS0:all}   {\ensuremath{{(49.77 \pm 2.06)\times 10^{-2}} } }
\vdef{bdt2011:N-EFF-MU-MC-BPLUS0:val}   {\ensuremath{{0.498 } } }
\vdef{bdt2011:N-EFF-MU-MC-BPLUS0:err}   {\ensuremath{{0.001 } } }
\vdef{bdt2011:N-EFF-MU-MC-BPLUS0:tot}   {\ensuremath{{0.020 } } }
\vdef{bdt2011:N-EFF-MU-MC-BPLUS0:all}   {\ensuremath{{(49.77 \pm 1.99)\times 10^{-2}} } }
\vdef{bdt2011:N-EFF-TRIG-PID-BPLUS0:val}   {\ensuremath{{0.771 } } }
\vdef{bdt2011:N-EFF-TRIG-PID-BPLUS0:err}   {\ensuremath{{0.000 } } }
\vdef{bdt2011:N-EFF-TRIG-PID-BPLUS0:tot}   {\ensuremath{{0.023 } } }
\vdef{bdt2011:N-EFF-TRIG-PID-BPLUS0:all}   {\ensuremath{{(77.11 \pm 2.31)\times 10^{-2}} } }
\vdef{bdt2011:N-EFFRHO-TRIG-PID-BPLUS0:val}   {\ensuremath{{0.821 } } }
\vdef{bdt2011:N-EFFRHO-TRIG-PID-BPLUS0:err}   {\ensuremath{{0.000 } } }
\vdef{bdt2011:N-EFFRHO-TRIG-PID-BPLUS0:tot}   {\ensuremath{{0.025 } } }
\vdef{bdt2011:N-EFFRHO-TRIG-PID-BPLUS0:all}   {\ensuremath{{(82.10 \pm 2.46)\times 10^{-2}} } }
\vdef{bdt2011:N-EFF-TRIG-PIDMC-BPLUS0:val}   {\ensuremath{{0.726 } } }
\vdef{bdt2011:N-EFF-TRIG-PIDMC-BPLUS0:err}   {\ensuremath{{0.001 } } }
\vdef{bdt2011:N-EFF-TRIG-PIDMC-BPLUS0:tot}   {\ensuremath{{0.022 } } }
\vdef{bdt2011:N-EFF-TRIG-PIDMC-BPLUS0:all}   {\ensuremath{{(72.61 \pm 2.18)\times 10^{-2}} } }
\vdef{bdt2011:N-EFFRHO-TRIG-PIDMC-BPLUS0:val}   {\ensuremath{{0.773 } } }
\vdef{bdt2011:N-EFFRHO-TRIG-PIDMC-BPLUS0:err}   {\ensuremath{{0.001 } } }
\vdef{bdt2011:N-EFFRHO-TRIG-PIDMC-BPLUS0:tot}   {\ensuremath{{0.023 } } }
\vdef{bdt2011:N-EFFRHO-TRIG-PIDMC-BPLUS0:all}   {\ensuremath{{(77.30 \pm 2.32)\times 10^{-2}} } }
\vdef{bdt2011:N-EFF-TRIG-MC-BPLUS0:val}   {\ensuremath{{0.773 } } }
\vdef{bdt2011:N-EFF-TRIG-MC-BPLUS0:err}   {\ensuremath{{0.001 } } }
\vdef{bdt2011:N-EFF-TRIG-MC-BPLUS0:tot}   {\ensuremath{{0.023 } } }
\vdef{bdt2011:N-EFF-TRIG-MC-BPLUS0:all}   {\ensuremath{{(77.30 \pm 2.32)\times 10^{-2}} } }
\vdef{bdt2011:N-EFF-CAND-BPLUS0:val}   {\ensuremath{{1.000 } } }
\vdef{bdt2011:N-EFF-CAND-BPLUS0:err}   {\ensuremath{{0.000 } } }
\vdef{bdt2011:N-EFF-CAND-BPLUS0:tot}   {\ensuremath{{0.010 } } }
\vdef{bdt2011:N-EFF-CAND-BPLUS0:all}   {\ensuremath{{(99.99 \pm 1.00)\times 10^{-2}} } }
\vdef{bdt2011:N-EFF-ANA-BPLUS0:val}   {\ensuremath{{0.2497 } } }
\vdef{bdt2011:N-EFF-ANA-BPLUS0:err}   {\ensuremath{{0.0005 } } }
\vdef{bdt2011:N-EFF-ANA-BPLUS0:tot}   {\ensuremath{{0.0141 } } }
\vdef{bdt2011:N-EFF-ANA-BPLUS0:all}   {\ensuremath{{(24.97 \pm 1.41)\times 10^{-2}} } }
\vdef{bdt2011:N-OBS-BPLUS0:val}   {\ensuremath{{71191 } } }
\vdef{bdt2011:N-OBS-BPLUS0:err}   {\ensuremath{{2050 } } }
\vdef{bdt2011:N-OBS-BPLUS0:tot}   {\ensuremath{{4107 } } }
\vdef{bdt2011:N-OBS-BPLUS0:all}   {\ensuremath{{71191 } } }
\vdef{bdt2011:N-OBS-CBPLUS0:val}   {\ensuremath{{72810 } } }
\vdef{bdt2011:N-OBS-CBPLUS0:err}   {\ensuremath{{277 } } }
\vdef{bdt2011:N-EFF-TOT-BPLUS1:val}   {\ensuremath{{0.00036 } } }
\vdef{bdt2011:N-EFF-TOT-BPLUS1:err}   {\ensuremath{{0.000002 } } }
\vdef{bdt2011:N-EFF-TOT-BPLUS1:tot}   {\ensuremath{{0.00004 } } }
\vdef{bdt2011:N-EFF-TOT-BPLUS1:all}   {\ensuremath{{(0.36 \pm 0.04)\times 10^{-3}} } }
\vdef{bdt2011:N-EFF-PRODMC-BPLUS1:val}   {\ensuremath{{0.00035 } } }
\vdef{bdt2011:N-EFF-PRODMC-BPLUS1:err}   {\ensuremath{{0.000006 } } }
\vdef{bdt2011:N-EFF-PRODMC-BPLUS1:tot}   {\ensuremath{{0.00001 } } }
\vdef{bdt2011:N-EFF-PRODMC-BPLUS1:all}   {\ensuremath{{(0.35 \pm 0.01)\times 10^{-3}} } }
\vdef{bdt2011:N-EFF-PRODTNP-BPLUS1:val}   {\ensuremath{{0.00044 } } }
\vdef{bdt2011:N-EFF-PRODTNP-BPLUS1:err}   {\ensuremath{{0.143284 } } }
\vdef{bdt2011:N-EFF-PRODTNP-BPLUS1:tot}   {\ensuremath{{0.14328 } } }
\vdef{bdt2011:N-EFF-PRODTNP-BPLUS1:all}   {\ensuremath{{(0.44 \pm 143.28)\times 10^{-3}} } }
\vdef{bdt2011:N-EFF-PRODTNPMC-BPLUS1:val}   {\ensuremath{{0.00035 } } }
\vdef{bdt2011:N-EFF-PRODTNPMC-BPLUS1:err}   {\ensuremath{{0.000006 } } }
\vdef{bdt2011:N-EFF-PRODTNPMC-BPLUS1:tot}   {\ensuremath{{0.00001 } } }
\vdef{bdt2011:N-EFF-PRODTNPMC-BPLUS1:all}   {\ensuremath{{(0.35 \pm 0.01)\times 10^{-3}} } }
\vdef{bdt2011:N-ACC-BPLUS1:val}   {\ensuremath{{0.005 } } }
\vdef{bdt2011:N-ACC-BPLUS1:err}   {\ensuremath{{0.000 } } }
\vdef{bdt2011:N-ACC-BPLUS1:tot}   {\ensuremath{{0.000 } } }
\vdef{bdt2011:N-ACC-BPLUS1:all}   {\ensuremath{{(0.53 \pm 0.03)\times 10^{-2}} } }
\vdef{bdt2011:N-EFF-MU-PID-BPLUS1:val}   {\ensuremath{{0.648 } } }
\vdef{bdt2011:N-EFF-MU-PID-BPLUS1:err}   {\ensuremath{{0.001 } } }
\vdef{bdt2011:N-EFF-MU-PID-BPLUS1:tot}   {\ensuremath{{0.052 } } }
\vdef{bdt2011:N-EFF-MU-PID-BPLUS1:all}   {\ensuremath{{(64.78 \pm 5.18)\times 10^{-2}} } }
\vdef{bdt2011:N-EFFRHO-MU-PID-BPLUS1:val}   {\ensuremath{{0.566 } } }
\vdef{bdt2011:N-EFFRHO-MU-PID-BPLUS1:err}   {\ensuremath{{0.001 } } }
\vdef{bdt2011:N-EFFRHO-MU-PID-BPLUS1:tot}   {\ensuremath{{0.000 } } }
\vdef{bdt2011:N-EFFRHO-MU-PID-BPLUS1:all}   {\ensuremath{{(56.65 \pm 0.00)\times 10^{-2}} } }
\vdef{bdt2011:N-EFF-MU-PIDMC-BPLUS1:val}   {\ensuremath{{0.682 } } }
\vdef{bdt2011:N-EFF-MU-PIDMC-BPLUS1:err}   {\ensuremath{{0.001 } } }
\vdef{bdt2011:N-EFF-MU-PIDMC-BPLUS1:tot}   {\ensuremath{{0.055 } } }
\vdef{bdt2011:N-EFF-MU-PIDMC-BPLUS1:all}   {\ensuremath{{(68.17 \pm 5.45)\times 10^{-2}} } }
\vdef{bdt2011:N-EFFRHO-MU-PIDMC-BPLUS1:val}   {\ensuremath{{0.596 } } }
\vdef{bdt2011:N-EFFRHO-MU-PIDMC-BPLUS1:err}   {\ensuremath{{0.001 } } }
\vdef{bdt2011:N-EFFRHO-MU-PIDMC-BPLUS1:tot}   {\ensuremath{{0.045 } } }
\vdef{bdt2011:N-EFFRHO-MU-PIDMC-BPLUS1:all}   {\ensuremath{{(59.62 \pm 4.53)\times 10^{-2}} } }
\vdef{bdt2011:N-EFF-MU-MC-BPLUS1:val}   {\ensuremath{{0.596 } } }
\vdef{bdt2011:N-EFF-MU-MC-BPLUS1:err}   {\ensuremath{{0.002 } } }
\vdef{bdt2011:N-EFF-MU-MC-BPLUS1:tot}   {\ensuremath{{0.048 } } }
\vdef{bdt2011:N-EFF-MU-MC-BPLUS1:all}   {\ensuremath{{(59.62 \pm 4.77)\times 10^{-2}} } }
\vdef{bdt2011:N-EFF-TRIG-PID-BPLUS1:val}   {\ensuremath{{0.689 } } }
\vdef{bdt2011:N-EFF-TRIG-PID-BPLUS1:err}   {\ensuremath{{0.001 } } }
\vdef{bdt2011:N-EFF-TRIG-PID-BPLUS1:tot}   {\ensuremath{{0.041 } } }
\vdef{bdt2011:N-EFF-TRIG-PID-BPLUS1:all}   {\ensuremath{{(68.87 \pm 4.13)\times 10^{-2}} } }
\vdef{bdt2011:N-EFFRHO-TRIG-PID-BPLUS1:val}   {\ensuremath{{0.764 } } }
\vdef{bdt2011:N-EFFRHO-TRIG-PID-BPLUS1:err}   {\ensuremath{{0.002 } } }
\vdef{bdt2011:N-EFFRHO-TRIG-PID-BPLUS1:tot}   {\ensuremath{{0.046 } } }
\vdef{bdt2011:N-EFFRHO-TRIG-PID-BPLUS1:all}   {\ensuremath{{(76.40 \pm 4.59)\times 10^{-2}} } }
\vdef{bdt2011:N-EFF-TRIG-PIDMC-BPLUS1:val}   {\ensuremath{{0.541 } } }
\vdef{bdt2011:N-EFF-TRIG-PIDMC-BPLUS1:err}   {\ensuremath{{0.002 } } }
\vdef{bdt2011:N-EFF-TRIG-PIDMC-BPLUS1:tot}   {\ensuremath{{0.033 } } }
\vdef{bdt2011:N-EFF-TRIG-PIDMC-BPLUS1:all}   {\ensuremath{{(54.15 \pm 3.25)\times 10^{-2}} } }
\vdef{bdt2011:N-EFFRHO-TRIG-PIDMC-BPLUS1:val}   {\ensuremath{{0.601 } } }
\vdef{bdt2011:N-EFFRHO-TRIG-PIDMC-BPLUS1:err}   {\ensuremath{{0.002 } } }
\vdef{bdt2011:N-EFFRHO-TRIG-PIDMC-BPLUS1:tot}   {\ensuremath{{0.036 } } }
\vdef{bdt2011:N-EFFRHO-TRIG-PIDMC-BPLUS1:all}   {\ensuremath{{(60.06 \pm 3.61)\times 10^{-2}} } }
\vdef{bdt2011:N-EFF-TRIG-MC-BPLUS1:val}   {\ensuremath{{0.601 } } }
\vdef{bdt2011:N-EFF-TRIG-MC-BPLUS1:err}   {\ensuremath{{0.002 } } }
\vdef{bdt2011:N-EFF-TRIG-MC-BPLUS1:tot}   {\ensuremath{{0.036 } } }
\vdef{bdt2011:N-EFF-TRIG-MC-BPLUS1:all}   {\ensuremath{{(60.06 \pm 3.61)\times 10^{-2}} } }
\vdef{bdt2011:N-EFF-CAND-BPLUS1:val}   {\ensuremath{{1.000 } } }
\vdef{bdt2011:N-EFF-CAND-BPLUS1:err}   {\ensuremath{{0.000 } } }
\vdef{bdt2011:N-EFF-CAND-BPLUS1:tot}   {\ensuremath{{0.010 } } }
\vdef{bdt2011:N-EFF-CAND-BPLUS1:all}   {\ensuremath{{(99.97 \pm 1.00)\times 10^{-2}} } }
\vdef{bdt2011:N-EFF-ANA-BPLUS1:val}   {\ensuremath{{0.1872 } } }
\vdef{bdt2011:N-EFF-ANA-BPLUS1:err}   {\ensuremath{{0.0006 } } }
\vdef{bdt2011:N-EFF-ANA-BPLUS1:tot}   {\ensuremath{{0.0106 } } }
\vdef{bdt2011:N-EFF-ANA-BPLUS1:all}   {\ensuremath{{(18.72 \pm 1.06)\times 10^{-2}} } }
\vdef{bdt2011:N-OBS-BPLUS1:val}   {\ensuremath{{21373 } } }
\vdef{bdt2011:N-OBS-BPLUS1:err}   {\ensuremath{{161 } } }
\vdef{bdt2011:N-OBS-BPLUS1:tot}   {\ensuremath{{1080 } } }
\vdef{bdt2011:N-OBS-BPLUS1:all}   {\ensuremath{{21373 } } }
\vdef{bdt2011:N-OBS-CBPLUS1:val}   {\ensuremath{{20295 } } }
\vdef{bdt2011:N-OBS-CBPLUS1:err}   {\ensuremath{{152 } } }
\vdef{bdt2011:N-EXP2-SIG-BSMM0:val}   {\ensuremath{{ 2.97 } } }
\vdef{bdt2011:N-EXP2-SIG-BSMM0:err}   {\ensuremath{{ 0.44 } } }
\vdef{bdt2011:N-EXP2-SIG-BDMM0:val}   {\ensuremath{{0.27 } } }
\vdef{bdt2011:N-EXP2-SIG-BDMM0:err}   {\ensuremath{{0.03 } } }
\vdef{bdt2011:N-OBS-BKG0:val}   {\ensuremath{{3 } } }
\vdef{bdt2011:N-EXP-BSMM0:val}   {\ensuremath{{ 0.37 } } }
\vdef{bdt2011:N-EXP-BSMM0:err}   {\ensuremath{{ 0.18 } } }
\vdef{bdt2011:N-EXP-BDMM0:val}   {\ensuremath{{ 0.46 } } }
\vdef{bdt2011:N-EXP-BDMM0:err}   {\ensuremath{{ 0.23 } } }
\vdef{bdt2011:N-LOW-BD0:val}   {\ensuremath{{5.200 } } }
\vdef{bdt2011:N-HIGH-BD0:val}   {\ensuremath{{5.300 } } }
\vdef{bdt2011:N-LOW-BS0:val}   {\ensuremath{{5.300 } } }
\vdef{bdt2011:N-HIGH-BS0:val}   {\ensuremath{{5.450 } } }
\vdef{bdt2011:N-PSS0:val}   {\ensuremath{{0.889 } } }
\vdef{bdt2011:N-PSS0:err}   {\ensuremath{{0.003 } } }
\vdef{bdt2011:N-PSS0:tot}   {\ensuremath{{0.045 } } }
\vdef{bdt2011:N-PSD0:val}   {\ensuremath{{0.273 } } }
\vdef{bdt2011:N-PSD0:err}   {\ensuremath{{0.016 } } }
\vdef{bdt2011:N-PSD0:tot}   {\ensuremath{{0.021 } } }
\vdef{bdt2011:N-PDS0:val}   {\ensuremath{{0.061 } } }
\vdef{bdt2011:N-PDS0:err}   {\ensuremath{{0.002 } } }
\vdef{bdt2011:N-PDS0:tot}   {\ensuremath{{0.004 } } }
\vdef{bdt2011:N-PDD0:val}   {\ensuremath{{0.675 } } }
\vdef{bdt2011:N-PDD0:err}   {\ensuremath{{0.017 } } }
\vdef{bdt2011:N-PDD0:tot}   {\ensuremath{{0.038 } } }
\vdef{bdt2011:N-EFF-TOT-BSMM0:val}   {\ensuremath{{0.0030 } } }
\vdef{bdt2011:N-EFF-TOT-BSMM0:err}   {\ensuremath{{0.0000 } } }
\vdef{bdt2011:N-EFF-TOT-BSMM0:tot}   {\ensuremath{{0.0003 } } }
\vdef{bdt2011:N-EFF-TOT-BSMM0:all}   {\ensuremath{{(0.30 \pm 0.04)} } }
\vdef{bdt2011:N-EFF-PRODMC-BSMM0:val}   {\ensuremath{{0.0030 } } }
\vdef{bdt2011:N-EFF-PRODMC-BSMM0:err}   {\ensuremath{{0.0000 } } }
\vdef{bdt2011:N-EFF-PRODMC-BSMM0:tot}   {\ensuremath{{0.0000 } } }
\vdef{bdt2011:N-EFF-PRODMC-BSMM0:all}   {\ensuremath{{(3.00 \pm 0.03)\times 10^{-3}} } }
\vdef{bdt2011:N-EFFRATIO-TOT-BSMM0:val}   {\ensuremath{{0.325 } } }
\vdef{bdt2011:N-EFFRATIO-TOT-BSMM0:err}   {\ensuremath{{0.003 } } }
\vdef{bdt2011:N-EFFRATIO-PRODMC-BSMM0:val}   {\ensuremath{{0.328 } } }
\vdef{bdt2011:N-EFFRATIO-PRODMC-BSMM0:err}   {\ensuremath{{0.005 } } }
\vdef{bdt2011:N-EFFRATIO-PRODTNP-BSMM0:val}   {\ensuremath{{0.324 } } }
\vdef{bdt2011:N-EFFRATIO-PRODTNP-BSMM0:err}   {\ensuremath{{185.253 } } }
\vdef{bdt2011:N-EFFRATIO-PRODTNPMC-BSMM0:val}   {\ensuremath{{0.328 } } }
\vdef{bdt2011:N-EFFRATIO-PRODTNPMC-BSMM0:err}   {\ensuremath{{0.005 } } }
\vdef{bdt2011:N-EFF-PRODTNP-BSMM0:val}   {\ensuremath{{0.0044 } } }
\vdef{bdt2011:N-EFF-PRODTNP-BSMM0:err}   {\ensuremath{{0.8467 } } }
\vdef{bdt2011:N-EFF-PRODTNP-BSMM0:tot}   {\ensuremath{{0.8467 } } }
\vdef{bdt2011:N-EFF-PRODTNP-BSMM0:all}   {\ensuremath{{(4.36 \pm 846.67)\times 10^{-3}} } }
\vdef{bdt2011:N-EFF-PRODTNPMC-BSMM0:val}   {\ensuremath{{0.0030 } } }
\vdef{bdt2011:N-EFF-PRODTNPMC-BSMM0:err}   {\ensuremath{{0.0000 } } }
\vdef{bdt2011:N-EFF-PRODTNPMC-BSMM0:tot}   {\ensuremath{{0.0000 } } }
\vdef{bdt2011:N-EFF-PRODTNPMC-BSMM0:all}   {\ensuremath{{(3.00 \pm 0.03)\times 10^{-3}} } }
\vdef{bdt2011:N-ACC-BSMM0:val}   {\ensuremath{{0.034 } } }
\vdef{bdt2011:N-ACC-BSMM0:err}   {\ensuremath{{0.000 } } }
\vdef{bdt2011:N-ACC-BSMM0:tot}   {\ensuremath{{0.001 } } }
\vdef{bdt2011:N-ACC-BSMM0:all}   {\ensuremath{{(3.36 \pm 0.12)\times 10^{-2}} } }
\vdef{bdt2011:N-EFF-MU-PID-BSMM0:val}   {\ensuremath{{0.722 } } }
\vdef{bdt2011:N-EFF-MU-PID-BSMM0:err}   {\ensuremath{{0.001 } } }
\vdef{bdt2011:N-EFF-MU-PID-BSMM0:tot}   {\ensuremath{{0.029 } } }
\vdef{bdt2011:N-EFF-MU-PID-BSMM0:all}   {\ensuremath{{(72.23 \pm 2.89)\times 10^{-2}} } }
\vdef{bdt2011:N-EFFRHO-MU-PID-BSMM0:val}   {\ensuremath{{0.486 } } }
\vdef{bdt2011:N-EFFRHO-MU-PID-BSMM0:err}   {\ensuremath{{0.001 } } }
\vdef{bdt2011:N-EFFRHO-MU-PID-BSMM0:tot}   {\ensuremath{{0.000 } } }
\vdef{bdt2011:N-EFFRHO-MU-PID-BSMM0:all}   {\ensuremath{{(48.57 \pm 0.00)\times 10^{-2}} } }
\vdef{bdt2011:N-EFF-MU-PIDMC-BSMM0:val}   {\ensuremath{{0.701 } } }
\vdef{bdt2011:N-EFF-MU-PIDMC-BSMM0:err}   {\ensuremath{{0.001 } } }
\vdef{bdt2011:N-EFF-MU-PIDMC-BSMM0:tot}   {\ensuremath{{0.028 } } }
\vdef{bdt2011:N-EFF-MU-PIDMC-BSMM0:all}   {\ensuremath{{(70.11 \pm 2.81)\times 10^{-2}} } }
\vdef{bdt2011:N-EFFRHO-MU-PIDMC-BSMM0:val}   {\ensuremath{{0.471 } } }
\vdef{bdt2011:N-EFFRHO-MU-PIDMC-BSMM0:err}   {\ensuremath{{0.001 } } }
\vdef{bdt2011:N-EFFRHO-MU-PIDMC-BSMM0:tot}   {\ensuremath{{0.019 } } }
\vdef{bdt2011:N-EFFRHO-MU-PIDMC-BSMM0:all}   {\ensuremath{{(47.14 \pm 1.94)\times 10^{-2}} } }
\vdef{bdt2011:N-EFF-MU-MC-BSMM0:val}   {\ensuremath{{0.471 } } }
\vdef{bdt2011:N-EFF-MU-MC-BSMM0:err}   {\ensuremath{{0.003 } } }
\vdef{bdt2011:N-EFF-MU-MC-BSMM0:tot}   {\ensuremath{{0.019 } } }
\vdef{bdt2011:N-EFF-MU-MC-BSMM0:all}   {\ensuremath{{(47.14 \pm 1.91)\times 10^{-2}} } }
\vdef{bdt2011:N-EFF-TRIG-PID-BSMM0:val}   {\ensuremath{{0.782 } } }
\vdef{bdt2011:N-EFF-TRIG-PID-BSMM0:err}   {\ensuremath{{0.001 } } }
\vdef{bdt2011:N-EFF-TRIG-PID-BSMM0:tot}   {\ensuremath{{0.023 } } }
\vdef{bdt2011:N-EFF-TRIG-PID-BSMM0:all}   {\ensuremath{{(78.21 \pm 2.35)\times 10^{-2}} } }
\vdef{bdt2011:N-EFFRHO-TRIG-PID-BSMM0:val}   {\ensuremath{{0.873 } } }
\vdef{bdt2011:N-EFFRHO-TRIG-PID-BSMM0:err}   {\ensuremath{{0.001 } } }
\vdef{bdt2011:N-EFFRHO-TRIG-PID-BSMM0:tot}   {\ensuremath{{0.026 } } }
\vdef{bdt2011:N-EFFRHO-TRIG-PID-BSMM0:all}   {\ensuremath{{(87.27 \pm 2.62)\times 10^{-2}} } }
\vdef{bdt2011:N-EFF-TRIG-PIDMC-BSMM0:val}   {\ensuremath{{0.740 } } }
\vdef{bdt2011:N-EFF-TRIG-PIDMC-BSMM0:err}   {\ensuremath{{0.001 } } }
\vdef{bdt2011:N-EFF-TRIG-PIDMC-BSMM0:tot}   {\ensuremath{{0.022 } } }
\vdef{bdt2011:N-EFF-TRIG-PIDMC-BSMM0:all}   {\ensuremath{{(74.05 \pm 2.23)\times 10^{-2}} } }
\vdef{bdt2011:N-EFFRHO-TRIG-PIDMC-BSMM0:val}   {\ensuremath{{0.826 } } }
\vdef{bdt2011:N-EFFRHO-TRIG-PIDMC-BSMM0:err}   {\ensuremath{{0.002 } } }
\vdef{bdt2011:N-EFFRHO-TRIG-PIDMC-BSMM0:tot}   {\ensuremath{{0.025 } } }
\vdef{bdt2011:N-EFFRHO-TRIG-PIDMC-BSMM0:all}   {\ensuremath{{(82.63 \pm 2.48)\times 10^{-2}} } }
\vdef{bdt2011:N-EFF-TRIG-MC-BSMM0:val}   {\ensuremath{{0.826 } } }
\vdef{bdt2011:N-EFF-TRIG-MC-BSMM0:err}   {\ensuremath{{0.003 } } }
\vdef{bdt2011:N-EFF-TRIG-MC-BSMM0:tot}   {\ensuremath{{0.025 } } }
\vdef{bdt2011:N-EFF-TRIG-MC-BSMM0:all}   {\ensuremath{{(82.63 \pm 2.50)\times 10^{-2}} } }
\vdef{bdt2011:N-EFF-CAND-BSMM0:val}   {\ensuremath{{1.000 } } }
\vdef{bdt2011:N-EFF-CAND-BSMM0:err}   {\ensuremath{{0.000 } } }
\vdef{bdt2011:N-EFF-CAND-BSMM0:tot}   {\ensuremath{{0.010 } } }
\vdef{bdt2011:N-EFF-CAND-BSMM0:all}   {\ensuremath{{(99.99 \pm 1.00)\times 10^{-2}} } }
\vdef{bdt2011:N-EFF-ANA-BSMM0:val}   {\ensuremath{{0.229 } } }
\vdef{bdt2011:N-EFF-ANA-BSMM0:err}   {\ensuremath{{0.001 } } }
\vdef{bdt2011:N-EFF-ANA-BSMM0:tot}   {\ensuremath{{0.007 } } }
\vdef{bdt2011:N-EFF-ANA-BSMM0:all}   {\ensuremath{{(22.92 \pm 0.70)\times 10^{-2}} } }
\vdef{bdt2011:N-EFF-TOT-BDMM0:val}   {\ensuremath{{0.0033 } } }
\vdef{bdt2011:N-EFF-TOT-BDMM0:err}   {\ensuremath{{0.0001 } } }
\vdef{bdt2011:N-EFF-TOT-BDMM0:tot}   {\ensuremath{{0.0003 } } }
\vdef{bdt2011:N-EFF-TOT-BDMM0:all}   {\ensuremath{{(0.33 \pm 0.03)} } }
\vdef{bdt2011:N-EFF-PRODMC-BDMM0:val}   {\ensuremath{{0.0033 } } }
\vdef{bdt2011:N-EFF-PRODMC-BDMM0:err}   {\ensuremath{{0.0001 } } }
\vdef{bdt2011:N-EFF-PRODMC-BDMM0:tot}   {\ensuremath{{0.0001 } } }
\vdef{bdt2011:N-EFF-PRODMC-BDMM0:all}   {\ensuremath{{(3.30 \pm 0.12)\times 10^{-3}} } }
\vdef{bdt2011:N-EFF-PRODTNP-BDMM0:val}   {\ensuremath{{0.0047 } } }
\vdef{bdt2011:N-EFF-PRODTNP-BDMM0:err}   {\ensuremath{{0.2439 } } }
\vdef{bdt2011:N-EFF-PRODTNP-BDMM0:tot}   {\ensuremath{{0.2439 } } }
\vdef{bdt2011:N-EFF-PRODTNP-BDMM0:all}   {\ensuremath{{(4.70 \pm 243.95)\times 10^{-3}} } }
\vdef{bdt2011:N-EFF-PRODTNPMC-BDMM0:val}   {\ensuremath{{0.00330 } } }
\vdef{bdt2011:N-EFF-PRODTNPMC-BDMM0:err}   {\ensuremath{{0.000110 } } }
\vdef{bdt2011:N-EFF-PRODTNPMC-BDMM0:tot}   {\ensuremath{{0.00011 } } }
\vdef{bdt2011:N-EFF-PRODTNPMC-BDMM0:all}   {\ensuremath{{(3.30 \pm 0.11)\times 10^{-3}} } }
\vdef{bdt2011:N-ACC-BDMM0:val}   {\ensuremath{{0.034 } } }
\vdef{bdt2011:N-ACC-BDMM0:err}   {\ensuremath{{0.000 } } }
\vdef{bdt2011:N-ACC-BDMM0:tot}   {\ensuremath{{0.001 } } }
\vdef{bdt2011:N-ACC-BDMM0:all}   {\ensuremath{{(3.37 \pm 0.12)\times 10^{-2}} } }
\vdef{bdt2011:N-EFF-MU-PID-BDMM0:val}   {\ensuremath{{0.718 } } }
\vdef{bdt2011:N-EFF-MU-PID-BDMM0:err}   {\ensuremath{{0.004 } } }
\vdef{bdt2011:N-EFF-MU-PID-BDMM0:tot}   {\ensuremath{{0.029 } } }
\vdef{bdt2011:N-EFF-MU-PID-BDMM0:all}   {\ensuremath{{(71.82 \pm 2.90)\times 10^{-2}} } }
\vdef{bdt2011:N-EFFRHO-MU-PID-BDMM0:val}   {\ensuremath{{0.476 } } }
\vdef{bdt2011:N-EFFRHO-MU-PID-BDMM0:err}   {\ensuremath{{0.003 } } }
\vdef{bdt2011:N-EFFRHO-MU-PID-BDMM0:tot}   {\ensuremath{{0.000 } } }
\vdef{bdt2011:N-EFFRHO-MU-PID-BDMM0:all}   {\ensuremath{{(47.57 \pm 0.00)\times 10^{-2}} } }
\vdef{bdt2011:N-EFF-MU-PIDMC-BDMM0:val}   {\ensuremath{{0.701 } } }
\vdef{bdt2011:N-EFF-MU-PIDMC-BDMM0:err}   {\ensuremath{{0.004 } } }
\vdef{bdt2011:N-EFF-MU-PIDMC-BDMM0:tot}   {\ensuremath{{0.028 } } }
\vdef{bdt2011:N-EFF-MU-PIDMC-BDMM0:all}   {\ensuremath{{(70.07 \pm 2.84)\times 10^{-2}} } }
\vdef{bdt2011:N-EFFRHO-MU-PIDMC-BDMM0:val}   {\ensuremath{{0.464 } } }
\vdef{bdt2011:N-EFFRHO-MU-PIDMC-BDMM0:err}   {\ensuremath{{0.003 } } }
\vdef{bdt2011:N-EFFRHO-MU-PIDMC-BDMM0:tot}   {\ensuremath{{0.019 } } }
\vdef{bdt2011:N-EFFRHO-MU-PIDMC-BDMM0:all}   {\ensuremath{{(46.41 \pm 1.92)\times 10^{-2}} } }
\vdef{bdt2011:N-EFF-MU-MC-BDMM0:val}   {\ensuremath{{0.464 } } }
\vdef{bdt2011:N-EFF-MU-MC-BDMM0:err}   {\ensuremath{{0.011 } } }
\vdef{bdt2011:N-EFF-MU-MC-BDMM0:tot}   {\ensuremath{{0.022 } } }
\vdef{bdt2011:N-EFF-MU-MC-BDMM0:all}   {\ensuremath{{(46.41 \pm 2.16)\times 10^{-2}} } }
\vdef{bdt2011:N-EFF-TRIG-PID-BDMM0:val}   {\ensuremath{{0.781 } } }
\vdef{bdt2011:N-EFF-TRIG-PID-BDMM0:err}   {\ensuremath{{0.004 } } }
\vdef{bdt2011:N-EFF-TRIG-PID-BDMM0:tot}   {\ensuremath{{0.024 } } }
\vdef{bdt2011:N-EFF-TRIG-PID-BDMM0:all}   {\ensuremath{{(78.12 \pm 2.38)\times 10^{-2}} } }
\vdef{bdt2011:N-EFFRHO-TRIG-PID-BDMM0:val}   {\ensuremath{{0.896 } } }
\vdef{bdt2011:N-EFFRHO-TRIG-PID-BDMM0:err}   {\ensuremath{{0.005 } } }
\vdef{bdt2011:N-EFFRHO-TRIG-PID-BDMM0:tot}   {\ensuremath{{0.027 } } }
\vdef{bdt2011:N-EFFRHO-TRIG-PID-BDMM0:all}   {\ensuremath{{(89.59 \pm 2.74)\times 10^{-2}} } }
\vdef{bdt2011:N-EFF-TRIG-PIDMC-BDMM0:val}   {\ensuremath{{0.739 } } }
\vdef{bdt2011:N-EFF-TRIG-PIDMC-BDMM0:err}   {\ensuremath{{0.005 } } }
\vdef{bdt2011:N-EFF-TRIG-PIDMC-BDMM0:tot}   {\ensuremath{{0.023 } } }
\vdef{bdt2011:N-EFF-TRIG-PIDMC-BDMM0:all}   {\ensuremath{{(73.90 \pm 2.28)\times 10^{-2}} } }
\vdef{bdt2011:N-EFFRHO-TRIG-PIDMC-BDMM0:val}   {\ensuremath{{0.848 } } }
\vdef{bdt2011:N-EFFRHO-TRIG-PIDMC-BDMM0:err}   {\ensuremath{{0.006 } } }
\vdef{bdt2011:N-EFFRHO-TRIG-PIDMC-BDMM0:tot}   {\ensuremath{{0.026 } } }
\vdef{bdt2011:N-EFFRHO-TRIG-PIDMC-BDMM0:all}   {\ensuremath{{(84.75 \pm 2.61)\times 10^{-2}} } }
\vdef{bdt2011:N-EFF-TRIG-MC-BDMM0:val}   {\ensuremath{{0.848 } } }
\vdef{bdt2011:N-EFF-TRIG-MC-BDMM0:err}   {\ensuremath{{0.012 } } }
\vdef{bdt2011:N-EFF-TRIG-MC-BDMM0:tot}   {\ensuremath{{0.028 } } }
\vdef{bdt2011:N-EFF-TRIG-MC-BDMM0:all}   {\ensuremath{{(84.75 \pm 2.80)\times 10^{-2}} } }
\vdef{bdt2011:N-EFF-CAND-BDMM0:val}   {\ensuremath{{1.000 } } }
\vdef{bdt2011:N-EFF-CAND-BDMM0:err}   {\ensuremath{{0.000 } } }
\vdef{bdt2011:N-EFF-CAND-BDMM0:tot}   {\ensuremath{{0.010 } } }
\vdef{bdt2011:N-EFF-CAND-BDMM0:all}   {\ensuremath{{(100.00 \pm 1.00)\times 10^{-2}} } }
\vdef{bdt2011:N-EFF-ANA-BDMM0:val}   {\ensuremath{{0.249 } } }
\vdef{bdt2011:N-EFF-ANA-BDMM0:err}   {\ensuremath{{0.005 } } }
\vdef{bdt2011:N-EFF-ANA-BDMM0:tot}   {\ensuremath{{0.009 } } }
\vdef{bdt2011:N-EFF-ANA-BDMM0:all}   {\ensuremath{{(24.87 \pm 0.89)\times 10^{-2}} } }
\vdef{bdt2011:N-EXP-OBS-BS0:val}   {\ensuremath{{ 3.64 } } }
\vdef{bdt2011:N-EXP-OBS-BS0:err}   {\ensuremath{{ 0.48 } } }
\vdef{bdt2011:N-EXP-OBS-BD0:val}   {\ensuremath{{ 1.3 } } }
\vdef{bdt2011:N-EXP-OBS-BD0:err}   {\ensuremath{{ 0.2 } } }
\vdef{bdt2011:N-OBS-BSMM0:val}   {\ensuremath{{4 } } }
\vdef{bdt2011:N-OBS-BDMM0:val}   {\ensuremath{{3 } } }
\vdef{bdt2011:N-OFFLO-RARE0:val}   {\ensuremath{{ 0.11 } } }
\vdef{bdt2011:N-OFFLO-RARE0:err}   {\ensuremath{{ 0.06 } } }
\vdef{bdt2011:N-OFFHI-RARE0:val}   {\ensuremath{{ 0.01 } } }
\vdef{bdt2011:N-OFFHI-RARE0:err}   {\ensuremath{{ 0.00 } } }
\vdef{bdt2011:N-PEAK-BKG-BS0:val}   {\ensuremath{{ 0.11 } } }
\vdef{bdt2011:N-PEAK-BKG-BS0:err}   {\ensuremath{{ 0.08 } } }
\vdef{bdt2011:N-PEAK-BKG-BD0:val}   {\ensuremath{{ 0.36 } } }
\vdef{bdt2011:N-PEAK-BKG-BD0:err}   {\ensuremath{{ 0.26 } } }
\vdef{bdt2011:N-TAU-BS0:val}   {\ensuremath{{ 0.12 } } }
\vdef{bdt2011:N-TAU-BS0:err}   {\ensuremath{{ 0.00 } } }
\vdef{bdt2011:N-TAU-BD0:val}   {\ensuremath{{ 0.15 } } }
\vdef{bdt2011:N-TAU-BD0:err}   {\ensuremath{{ 0.01 } } }
\vdef{bdt2011:N-OBS-OFFHI0:val}   {\ensuremath{{0 } } }
\vdef{bdt2011:N-OBS-OFFLO0:val}   {\ensuremath{{3 } } }
\vdef{bdt2011:N-EXP-SoverB0:val}   {\ensuremath{{ 8.12 } } }
\vdef{bdt2011:N-EXP-SoverSplusB0:val}   {\ensuremath{{ 1.62 } } }
\vdef{bdt2011:N-EXP2-SIG-BSMM1:val}   {\ensuremath{{ 1.28 } } }
\vdef{bdt2011:N-EXP2-SIG-BSMM1:err}   {\ensuremath{{ 0.19 } } }
\vdef{bdt2011:N-EXP2-SIG-BDMM1:val}   {\ensuremath{{0.11 } } }
\vdef{bdt2011:N-EXP2-SIG-BDMM1:err}   {\ensuremath{{0.01 } } }
\vdef{bdt2011:N-OBS-BKG1:val}   {\ensuremath{{7 } } }
\vdef{bdt2011:N-EXP-BSMM1:val}   {\ensuremath{{ 1.24 } } }
\vdef{bdt2011:N-EXP-BSMM1:err}   {\ensuremath{{ 0.62 } } }
\vdef{bdt2011:N-EXP-BDMM1:val}   {\ensuremath{{ 0.98 } } }
\vdef{bdt2011:N-EXP-BDMM1:err}   {\ensuremath{{ 0.49 } } }
\vdef{bdt2011:N-LOW-BD1:val}   {\ensuremath{{5.200 } } }
\vdef{bdt2011:N-HIGH-BD1:val}   {\ensuremath{{5.300 } } }
\vdef{bdt2011:N-LOW-BS1:val}   {\ensuremath{{5.300 } } }
\vdef{bdt2011:N-HIGH-BS1:val}   {\ensuremath{{5.450 } } }
\vdef{bdt2011:N-PSS1:val}   {\ensuremath{{0.716 } } }
\vdef{bdt2011:N-PSS1:err}   {\ensuremath{{0.005 } } }
\vdef{bdt2011:N-PSS1:tot}   {\ensuremath{{0.036 } } }
\vdef{bdt2011:N-PSD1:val}   {\ensuremath{{0.315 } } }
\vdef{bdt2011:N-PSD1:err}   {\ensuremath{{0.021 } } }
\vdef{bdt2011:N-PSD1:tot}   {\ensuremath{{0.026 } } }
\vdef{bdt2011:N-PDS1:val}   {\ensuremath{{0.158 } } }
\vdef{bdt2011:N-PDS1:err}   {\ensuremath{{0.004 } } }
\vdef{bdt2011:N-PDS1:tot}   {\ensuremath{{0.009 } } }
\vdef{bdt2011:N-PDD1:val}   {\ensuremath{{0.537 } } }
\vdef{bdt2011:N-PDD1:err}   {\ensuremath{{0.023 } } }
\vdef{bdt2011:N-PDD1:tot}   {\ensuremath{{0.035 } } }
\vdef{bdt2011:N-EFF-TOT-BSMM1:val}   {\ensuremath{{0.0020 } } }
\vdef{bdt2011:N-EFF-TOT-BSMM1:err}   {\ensuremath{{0.0000 } } }
\vdef{bdt2011:N-EFF-TOT-BSMM1:tot}   {\ensuremath{{0.0002 } } }
\vdef{bdt2011:N-EFF-TOT-BSMM1:all}   {\ensuremath{{(0.20 \pm 0.02)} } }
\vdef{bdt2011:N-EFF-PRODMC-BSMM1:val}   {\ensuremath{{0.0020 } } }
\vdef{bdt2011:N-EFF-PRODMC-BSMM1:err}   {\ensuremath{{0.0000 } } }
\vdef{bdt2011:N-EFF-PRODMC-BSMM1:tot}   {\ensuremath{{0.0000 } } }
\vdef{bdt2011:N-EFF-PRODMC-BSMM1:all}   {\ensuremath{{(1.96 \pm 0.02)\times 10^{-3}} } }
\vdef{bdt2011:N-EFFRATIO-TOT-BSMM1:val}   {\ensuremath{{0.183 } } }
\vdef{bdt2011:N-EFFRATIO-TOT-BSMM1:err}   {\ensuremath{{0.002 } } }
\vdef{bdt2011:N-EFFRATIO-PRODMC-BSMM1:val}   {\ensuremath{{0.181 } } }
\vdef{bdt2011:N-EFFRATIO-PRODMC-BSMM1:err}   {\ensuremath{{0.004 } } }
\vdef{bdt2011:N-EFFRATIO-PRODTNP-BSMM1:val}   {\ensuremath{{0.339 } } }
\vdef{bdt2011:N-EFFRATIO-PRODTNP-BSMM1:err}   {\ensuremath{{119.311 } } }
\vdef{bdt2011:N-EFFRATIO-PRODTNPMC-BSMM1:val}   {\ensuremath{{0.181 } } }
\vdef{bdt2011:N-EFFRATIO-PRODTNPMC-BSMM1:err}   {\ensuremath{{0.004 } } }
\vdef{bdt2011:N-EFF-PRODTNP-BSMM1:val}   {\ensuremath{{0.0013 } } }
\vdef{bdt2011:N-EFF-PRODTNP-BSMM1:err}   {\ensuremath{{0.1770 } } }
\vdef{bdt2011:N-EFF-PRODTNP-BSMM1:tot}   {\ensuremath{{0.1770 } } }
\vdef{bdt2011:N-EFF-PRODTNP-BSMM1:all}   {\ensuremath{{(1.30 \pm 176.96)\times 10^{-3}} } }
\vdef{bdt2011:N-EFF-PRODTNPMC-BSMM1:val}   {\ensuremath{{0.0020 } } }
\vdef{bdt2011:N-EFF-PRODTNPMC-BSMM1:err}   {\ensuremath{{0.0000 } } }
\vdef{bdt2011:N-EFF-PRODTNPMC-BSMM1:tot}   {\ensuremath{{0.0000 } } }
\vdef{bdt2011:N-EFF-PRODTNPMC-BSMM1:all}   {\ensuremath{{(1.96 \pm 0.03)\times 10^{-3}} } }
\vdef{bdt2011:N-ACC-BSMM1:val}   {\ensuremath{{0.024 } } }
\vdef{bdt2011:N-ACC-BSMM1:err}   {\ensuremath{{0.000 } } }
\vdef{bdt2011:N-ACC-BSMM1:tot}   {\ensuremath{{0.001 } } }
\vdef{bdt2011:N-ACC-BSMM1:all}   {\ensuremath{{(2.39 \pm 0.12)\times 10^{-2}} } }
\vdef{bdt2011:N-EFF-MU-PID-BSMM1:val}   {\ensuremath{{0.626 } } }
\vdef{bdt2011:N-EFF-MU-PID-BSMM1:err}   {\ensuremath{{0.002 } } }
\vdef{bdt2011:N-EFF-MU-PID-BSMM1:tot}   {\ensuremath{{0.050 } } }
\vdef{bdt2011:N-EFF-MU-PID-BSMM1:all}   {\ensuremath{{(62.63 \pm 5.01)\times 10^{-2}} } }
\vdef{bdt2011:N-EFFRHO-MU-PID-BSMM1:val}   {\ensuremath{{0.631 } } }
\vdef{bdt2011:N-EFFRHO-MU-PID-BSMM1:err}   {\ensuremath{{0.002 } } }
\vdef{bdt2011:N-EFFRHO-MU-PID-BSMM1:tot}   {\ensuremath{{0.000 } } }
\vdef{bdt2011:N-EFFRHO-MU-PID-BSMM1:all}   {\ensuremath{{(63.11 \pm 0.00)\times 10^{-2}} } }
\vdef{bdt2011:N-EFF-MU-PIDMC-BSMM1:val}   {\ensuremath{{0.658 } } }
\vdef{bdt2011:N-EFF-MU-PIDMC-BSMM1:err}   {\ensuremath{{0.002 } } }
\vdef{bdt2011:N-EFF-MU-PIDMC-BSMM1:tot}   {\ensuremath{{0.053 } } }
\vdef{bdt2011:N-EFF-MU-PIDMC-BSMM1:all}   {\ensuremath{{(65.77 \pm 5.27)\times 10^{-2}} } }
\vdef{bdt2011:N-EFFRHO-MU-PIDMC-BSMM1:val}   {\ensuremath{{0.663 } } }
\vdef{bdt2011:N-EFFRHO-MU-PIDMC-BSMM1:err}   {\ensuremath{{0.002 } } }
\vdef{bdt2011:N-EFFRHO-MU-PIDMC-BSMM1:tot}   {\ensuremath{{0.051 } } }
\vdef{bdt2011:N-EFFRHO-MU-PIDMC-BSMM1:all}   {\ensuremath{{(66.27 \pm 5.05)\times 10^{-2}} } }
\vdef{bdt2011:N-EFF-MU-MC-BSMM1:val}   {\ensuremath{{0.663 } } }
\vdef{bdt2011:N-EFF-MU-MC-BSMM1:err}   {\ensuremath{{0.004 } } }
\vdef{bdt2011:N-EFF-MU-MC-BSMM1:tot}   {\ensuremath{{0.053 } } }
\vdef{bdt2011:N-EFF-MU-MC-BSMM1:all}   {\ensuremath{{(66.27 \pm 5.32)\times 10^{-2}} } }
\vdef{bdt2011:N-EFF-TRIG-PID-BSMM1:val}   {\ensuremath{{0.512 } } }
\vdef{bdt2011:N-EFF-TRIG-PID-BSMM1:err}   {\ensuremath{{0.006 } } }
\vdef{bdt2011:N-EFF-TRIG-PID-BSMM1:tot}   {\ensuremath{{0.031 } } }
\vdef{bdt2011:N-EFF-TRIG-PID-BSMM1:all}   {\ensuremath{{(51.18 \pm 3.13)\times 10^{-2}} } }
\vdef{bdt2011:N-EFFRHO-TRIG-PID-BSMM1:val}   {\ensuremath{{0.901 } } }
\vdef{bdt2011:N-EFFRHO-TRIG-PID-BSMM1:err}   {\ensuremath{{0.010 } } }
\vdef{bdt2011:N-EFFRHO-TRIG-PID-BSMM1:tot}   {\ensuremath{{0.055 } } }
\vdef{bdt2011:N-EFFRHO-TRIG-PID-BSMM1:all}   {\ensuremath{{(90.10 \pm 5.50)\times 10^{-2}} } }
\vdef{bdt2011:N-EFF-TRIG-PIDMC-BSMM1:val}   {\ensuremath{{0.413 } } }
\vdef{bdt2011:N-EFF-TRIG-PIDMC-BSMM1:err}   {\ensuremath{{0.004 } } }
\vdef{bdt2011:N-EFF-TRIG-PIDMC-BSMM1:tot}   {\ensuremath{{0.025 } } }
\vdef{bdt2011:N-EFF-TRIG-PIDMC-BSMM1:all}   {\ensuremath{{(41.33 \pm 2.52)\times 10^{-2}} } }
\vdef{bdt2011:N-EFFRHO-TRIG-PIDMC-BSMM1:val}   {\ensuremath{{0.727 } } }
\vdef{bdt2011:N-EFFRHO-TRIG-PIDMC-BSMM1:err}   {\ensuremath{{0.008 } } }
\vdef{bdt2011:N-EFFRHO-TRIG-PIDMC-BSMM1:tot}   {\ensuremath{{0.044 } } }
\vdef{bdt2011:N-EFFRHO-TRIG-PIDMC-BSMM1:all}   {\ensuremath{{(72.74 \pm 4.43)\times 10^{-2}} } }
\vdef{bdt2011:N-EFF-TRIG-MC-BSMM1:val}   {\ensuremath{{0.727 } } }
\vdef{bdt2011:N-EFF-TRIG-MC-BSMM1:err}   {\ensuremath{{0.004 } } }
\vdef{bdt2011:N-EFF-TRIG-MC-BSMM1:tot}   {\ensuremath{{0.044 } } }
\vdef{bdt2011:N-EFF-TRIG-MC-BSMM1:all}   {\ensuremath{{(72.74 \pm 4.39)\times 10^{-2}} } }
\vdef{bdt2011:N-EFF-CAND-BSMM1:val}   {\ensuremath{{1.000 } } }
\vdef{bdt2011:N-EFF-CAND-BSMM1:err}   {\ensuremath{{0.000 } } }
\vdef{bdt2011:N-EFF-CAND-BSMM1:tot}   {\ensuremath{{0.010 } } }
\vdef{bdt2011:N-EFF-CAND-BSMM1:all}   {\ensuremath{{(99.99 \pm 1.00)\times 10^{-2}} } }
\vdef{bdt2011:N-EFF-ANA-BSMM1:val}   {\ensuremath{{0.170 } } }
\vdef{bdt2011:N-EFF-ANA-BSMM1:err}   {\ensuremath{{0.001 } } }
\vdef{bdt2011:N-EFF-ANA-BSMM1:tot}   {\ensuremath{{0.005 } } }
\vdef{bdt2011:N-EFF-ANA-BSMM1:all}   {\ensuremath{{(16.99 \pm 0.52)\times 10^{-2}} } }
\vdef{bdt2011:N-EFF-TOT-BDMM1:val}   {\ensuremath{{0.0020 } } }
\vdef{bdt2011:N-EFF-TOT-BDMM1:err}   {\ensuremath{{0.0001 } } }
\vdef{bdt2011:N-EFF-TOT-BDMM1:tot}   {\ensuremath{{0.0002 } } }
\vdef{bdt2011:N-EFF-TOT-BDMM1:all}   {\ensuremath{{(0.20 \pm 0.02)} } }
\vdef{bdt2011:N-EFF-PRODMC-BDMM1:val}   {\ensuremath{{0.0020 } } }
\vdef{bdt2011:N-EFF-PRODMC-BDMM1:err}   {\ensuremath{{0.0001 } } }
\vdef{bdt2011:N-EFF-PRODMC-BDMM1:tot}   {\ensuremath{{0.0001 } } }
\vdef{bdt2011:N-EFF-PRODMC-BDMM1:all}   {\ensuremath{{(1.96 \pm 0.09)\times 10^{-3}} } }
\vdef{bdt2011:N-EFF-PRODTNP-BDMM1:val}   {\ensuremath{{0.0013 } } }
\vdef{bdt2011:N-EFF-PRODTNP-BDMM1:err}   {\ensuremath{{0.0450 } } }
\vdef{bdt2011:N-EFF-PRODTNP-BDMM1:tot}   {\ensuremath{{0.0450 } } }
\vdef{bdt2011:N-EFF-PRODTNP-BDMM1:all}   {\ensuremath{{(1.29 \pm 44.95)\times 10^{-3}} } }
\vdef{bdt2011:N-EFF-PRODTNPMC-BDMM1:val}   {\ensuremath{{0.00196 } } }
\vdef{bdt2011:N-EFF-PRODTNPMC-BDMM1:err}   {\ensuremath{{0.000103 } } }
\vdef{bdt2011:N-EFF-PRODTNPMC-BDMM1:tot}   {\ensuremath{{0.00010 } } }
\vdef{bdt2011:N-EFF-PRODTNPMC-BDMM1:all}   {\ensuremath{{(1.96 \pm 0.10)\times 10^{-3}} } }
\vdef{bdt2011:N-ACC-BDMM1:val}   {\ensuremath{{0.023 } } }
\vdef{bdt2011:N-ACC-BDMM1:err}   {\ensuremath{{0.000 } } }
\vdef{bdt2011:N-ACC-BDMM1:tot}   {\ensuremath{{0.001 } } }
\vdef{bdt2011:N-ACC-BDMM1:all}   {\ensuremath{{(2.30 \pm 0.12)\times 10^{-2}} } }
\vdef{bdt2011:N-EFF-MU-PID-BDMM1:val}   {\ensuremath{{0.627 } } }
\vdef{bdt2011:N-EFF-MU-PID-BDMM1:err}   {\ensuremath{{0.006 } } }
\vdef{bdt2011:N-EFF-MU-PID-BDMM1:tot}   {\ensuremath{{0.051 } } }
\vdef{bdt2011:N-EFF-MU-PID-BDMM1:all}   {\ensuremath{{(62.75 \pm 5.06)\times 10^{-2}} } }
\vdef{bdt2011:N-EFFRHO-MU-PID-BDMM1:val}   {\ensuremath{{0.613 } } }
\vdef{bdt2011:N-EFFRHO-MU-PID-BDMM1:err}   {\ensuremath{{0.006 } } }
\vdef{bdt2011:N-EFFRHO-MU-PID-BDMM1:tot}   {\ensuremath{{0.000 } } }
\vdef{bdt2011:N-EFFRHO-MU-PID-BDMM1:all}   {\ensuremath{{(61.35 \pm 0.00)\times 10^{-2}} } }
\vdef{bdt2011:N-EFF-MU-PIDMC-BDMM1:val}   {\ensuremath{{0.670 } } }
\vdef{bdt2011:N-EFF-MU-PIDMC-BDMM1:err}   {\ensuremath{{0.007 } } }
\vdef{bdt2011:N-EFF-MU-PIDMC-BDMM1:tot}   {\ensuremath{{0.054 } } }
\vdef{bdt2011:N-EFF-MU-PIDMC-BDMM1:all}   {\ensuremath{{(66.95 \pm 5.40)\times 10^{-2}} } }
\vdef{bdt2011:N-EFFRHO-MU-PIDMC-BDMM1:val}   {\ensuremath{{0.655 } } }
\vdef{bdt2011:N-EFFRHO-MU-PIDMC-BDMM1:err}   {\ensuremath{{0.007 } } }
\vdef{bdt2011:N-EFFRHO-MU-PIDMC-BDMM1:tot}   {\ensuremath{{0.049 } } }
\vdef{bdt2011:N-EFFRHO-MU-PIDMC-BDMM1:all}   {\ensuremath{{(65.46 \pm 4.95)\times 10^{-2}} } }
\vdef{bdt2011:N-EFF-MU-MC-BDMM1:val}   {\ensuremath{{0.655 } } }
\vdef{bdt2011:N-EFF-MU-MC-BDMM1:err}   {\ensuremath{{0.015 } } }
\vdef{bdt2011:N-EFF-MU-MC-BDMM1:tot}   {\ensuremath{{0.054 } } }
\vdef{bdt2011:N-EFF-MU-MC-BDMM1:all}   {\ensuremath{{(65.46 \pm 5.45)\times 10^{-2}} } }
\vdef{bdt2011:N-EFF-TRIG-PID-BDMM1:val}   {\ensuremath{{0.501 } } }
\vdef{bdt2011:N-EFF-TRIG-PID-BDMM1:err}   {\ensuremath{{0.024 } } }
\vdef{bdt2011:N-EFF-TRIG-PID-BDMM1:tot}   {\ensuremath{{0.038 } } }
\vdef{bdt2011:N-EFF-TRIG-PID-BDMM1:all}   {\ensuremath{{(50.07 \pm 3.82)\times 10^{-2}} } }
\vdef{bdt2011:N-EFFRHO-TRIG-PID-BDMM1:val}   {\ensuremath{{0.838 } } }
\vdef{bdt2011:N-EFFRHO-TRIG-PID-BDMM1:err}   {\ensuremath{{0.040 } } }
\vdef{bdt2011:N-EFFRHO-TRIG-PID-BDMM1:tot}   {\ensuremath{{0.064 } } }
\vdef{bdt2011:N-EFFRHO-TRIG-PID-BDMM1:all}   {\ensuremath{{(83.84 \pm 6.40)\times 10^{-2}} } }
\vdef{bdt2011:N-EFF-TRIG-PIDMC-BDMM1:val}   {\ensuremath{{0.435 } } }
\vdef{bdt2011:N-EFF-TRIG-PIDMC-BDMM1:err}   {\ensuremath{{0.015 } } }
\vdef{bdt2011:N-EFF-TRIG-PIDMC-BDMM1:tot}   {\ensuremath{{0.030 } } }
\vdef{bdt2011:N-EFF-TRIG-PIDMC-BDMM1:all}   {\ensuremath{{(43.46 \pm 3.03)\times 10^{-2}} } }
\vdef{bdt2011:N-EFFRHO-TRIG-PIDMC-BDMM1:val}   {\ensuremath{{0.728 } } }
\vdef{bdt2011:N-EFFRHO-TRIG-PIDMC-BDMM1:err}   {\ensuremath{{0.026 } } }
\vdef{bdt2011:N-EFFRHO-TRIG-PIDMC-BDMM1:tot}   {\ensuremath{{0.051 } } }
\vdef{bdt2011:N-EFFRHO-TRIG-PIDMC-BDMM1:all}   {\ensuremath{{(72.77 \pm 5.07)\times 10^{-2}} } }
\vdef{bdt2011:N-EFF-TRIG-MC-BDMM1:val}   {\ensuremath{{0.728 } } }
\vdef{bdt2011:N-EFF-TRIG-MC-BDMM1:err}   {\ensuremath{{0.017 } } }
\vdef{bdt2011:N-EFF-TRIG-MC-BDMM1:tot}   {\ensuremath{{0.047 } } }
\vdef{bdt2011:N-EFF-TRIG-MC-BDMM1:all}   {\ensuremath{{(72.77 \pm 4.70)\times 10^{-2}} } }
\vdef{bdt2011:N-EFF-CAND-BDMM1:val}   {\ensuremath{{1.000 } } }
\vdef{bdt2011:N-EFF-CAND-BDMM1:err}   {\ensuremath{{0.000 } } }
\vdef{bdt2011:N-EFF-CAND-BDMM1:tot}   {\ensuremath{{0.010 } } }
\vdef{bdt2011:N-EFF-CAND-BDMM1:all}   {\ensuremath{{(100.00 \pm 1.00)\times 10^{-2}} } }
\vdef{bdt2011:N-EFF-ANA-BDMM1:val}   {\ensuremath{{0.179 } } }
\vdef{bdt2011:N-EFF-ANA-BDMM1:err}   {\ensuremath{{0.005 } } }
\vdef{bdt2011:N-EFF-ANA-BDMM1:tot}   {\ensuremath{{0.007 } } }
\vdef{bdt2011:N-EFF-ANA-BDMM1:all}   {\ensuremath{{(17.92 \pm 0.74)\times 10^{-2}} } }
\vdef{bdt2011:N-EXP-OBS-BS1:val}   {\ensuremath{{ 2.86 } } }
\vdef{bdt2011:N-EXP-OBS-BS1:err}   {\ensuremath{{ 0.65 } } }
\vdef{bdt2011:N-EXP-OBS-BD1:val}   {\ensuremath{{ 1.36 } } }
\vdef{bdt2011:N-EXP-OBS-BD1:err}   {\ensuremath{{ 0.49 } } }
\vdef{bdt2011:N-OBS-BSMM1:val}   {\ensuremath{{4 } } }
\vdef{bdt2011:N-OBS-BDMM1:val}   {\ensuremath{{1 } } }
\vdef{bdt2011:N-OFFLO-RARE1:val}   {\ensuremath{{ 0.07 } } }
\vdef{bdt2011:N-OFFLO-RARE1:err}   {\ensuremath{{ 0.04 } } }
\vdef{bdt2011:N-OFFHI-RARE1:val}   {\ensuremath{{ 0.01 } } }
\vdef{bdt2011:N-OFFHI-RARE1:err}   {\ensuremath{{ 0.00 } } }
\vdef{bdt2011:N-PEAK-BKG-BS1:val}   {\ensuremath{{ 0.06 } } }
\vdef{bdt2011:N-PEAK-BKG-BS1:err}   {\ensuremath{{ 0.04 } } }
\vdef{bdt2011:N-PEAK-BKG-BD1:val}   {\ensuremath{{ 0.10 } } }
\vdef{bdt2011:N-PEAK-BKG-BD1:err}   {\ensuremath{{ 0.08 } } }
\vdef{bdt2011:N-TAU-BS1:val}   {\ensuremath{{ 0.18 } } }
\vdef{bdt2011:N-TAU-BS1:err}   {\ensuremath{{ 0.01 } } }
\vdef{bdt2011:N-TAU-BD1:val}   {\ensuremath{{ 0.14 } } }
\vdef{bdt2011:N-TAU-BD1:err}   {\ensuremath{{ 0.01 } } }
\vdef{bdt2011:N-OBS-OFFHI1:val}   {\ensuremath{{3 } } }
\vdef{bdt2011:N-OBS-OFFLO1:val}   {\ensuremath{{4 } } }
\vdef{bdt2011:N-EXP-SoverB1:val}   {\ensuremath{{ 1.03 } } }
\vdef{bdt2011:N-EXP-SoverSplusB1:val}   {\ensuremath{{ 0.80 } } }
\vdef{bdt2011:SgBd0:val}  {\ensuremath{{0.204 } } }
\vdef{bdt2011:SgBd0:e1}   {\ensuremath{{0.452 } } }
\vdef{bdt2011:SgBd0:e2}   {\ensuremath{{0.031 } } }
\vdef{bdt2011:SgBs0:val}  {\ensuremath{{2.965 } } }
\vdef{bdt2011:SgBs0:e1}   {\ensuremath{{1.722 } } }
\vdef{bdt2011:SgBs0:e2}   {\ensuremath{{0.445 } } }
\vdef{bdt2011:SgLo0:val}  {\ensuremath{{0.078 } } }
\vdef{bdt2011:SgLo0:e1}   {\ensuremath{{0.279 } } }
\vdef{bdt2011:SgLo0:e2}   {\ensuremath{{0.012 } } }
\vdef{bdt2011:SgHi0:val}  {\ensuremath{{0.093 } } }
\vdef{bdt2011:SgHi0:e1}   {\ensuremath{{0.305 } } }
\vdef{bdt2011:SgHi0:e2}   {\ensuremath{{0.014 } } }
\vdef{bdt2011:BdBd0:val}  {\ensuremath{{0.271 } } }
\vdef{bdt2011:BdBd0:e1}   {\ensuremath{{0.520 } } }
\vdef{bdt2011:BdBd0:e2}   {\ensuremath{{0.027 } } }
\vdef{bdt2011:BdBs0:val}  {\ensuremath{{0.109 } } }
\vdef{bdt2011:BdBs0:e1}   {\ensuremath{{0.331 } } }
\vdef{bdt2011:BdBs0:e2}   {\ensuremath{{0.011 } } }
\vdef{bdt2011:BdLo0:val}  {\ensuremath{{0.024 } } }
\vdef{bdt2011:BdLo0:e1}   {\ensuremath{{0.156 } } }
\vdef{bdt2011:BdLo0:e2}   {\ensuremath{{0.002 } } }
\vdef{bdt2011:BdHi0:val}  {\ensuremath{{0.000 } } }
\vdef{bdt2011:BdHi0:e1}   {\ensuremath{{0.000 } } }
\vdef{bdt2011:BdHi0:e2}   {\ensuremath{{0.000 } } }
\vdef{bdt2011:BgPeakLo0:val}   {\ensuremath{{0.105 } } }
\vdef{bdt2011:BgPeakLo0:e1}   {\ensuremath{{0.005 } } }
\vdef{bdt2011:BgPeakLo0:e2}   {\ensuremath{{0.075 } } }
\vdef{bdt2011:BgPeakBd0:val}   {\ensuremath{{0.364 } } }
\vdef{bdt2011:BgPeakBd0:e1}   {\ensuremath{{0.018 } } }
\vdef{bdt2011:BgPeakBd0:e2}   {\ensuremath{{0.262 } } }
\vdef{bdt2011:BgPeakBs0:val}   {\ensuremath{{0.109 } } }
\vdef{bdt2011:BgPeakBs0:e1}   {\ensuremath{{0.005 } } }
\vdef{bdt2011:BgPeakBs0:e2}   {\ensuremath{{0.078 } } }
\vdef{bdt2011:BgPeakHi0:val}   {\ensuremath{{0.011 } } }
\vdef{bdt2011:BgPeakHi0:e1}   {\ensuremath{{0.001 } } }
\vdef{bdt2011:BgPeakHi0:e2}   {\ensuremath{{0.008 } } }
\vdef{bdt2011:BgRslLo0:val}   {\ensuremath{{23.000 } } }
\vdef{bdt2011:BgRslLo0:e1}   {\ensuremath{{1.150 } } }
\vdef{bdt2011:BgRslLo0:e2}   {\ensuremath{{17.946 } } }
\vdef{bdt2011:BgRslBd0:val}   {\ensuremath{{3.556 } } }
\vdef{bdt2011:BgRslBd0:e1}   {\ensuremath{{0.178 } } }
\vdef{bdt2011:BgRslBd0:e2}   {\ensuremath{{3.728 } } }
\vdef{bdt2011:BgRslBs0:val}   {\ensuremath{{2.799 } } }
\vdef{bdt2011:BgRslBs0:e1}   {\ensuremath{{0.140 } } }
\vdef{bdt2011:BgRslBs0:e2}   {\ensuremath{{3.084 } } }
\vdef{bdt2011:BgRslHi0:val}   {\ensuremath{{0.271 } } }
\vdef{bdt2011:BgRslHi0:e1}   {\ensuremath{{0.014 } } }
\vdef{bdt2011:BgRslHi0:e2}   {\ensuremath{{0.298 } } }
\vdef{bdt2011:BgRareLo0:val}   {\ensuremath{{23.105 } } }
\vdef{bdt2011:BgRareLo0:e1}   {\ensuremath{{1.150 } } }
\vdef{bdt2011:BgRareLo0:e2}   {\ensuremath{{17.946 } } }
\vdef{bdt2011:BgRareBd0:val}   {\ensuremath{{3.920 } } }
\vdef{bdt2011:BgRareBd0:e1}   {\ensuremath{{0.179 } } }
\vdef{bdt2011:BgRareBd0:e2}   {\ensuremath{{3.737 } } }
\vdef{bdt2011:BgRareBs0:val}   {\ensuremath{{2.907 } } }
\vdef{bdt2011:BgRareBs0:e1}   {\ensuremath{{0.140 } } }
\vdef{bdt2011:BgRareBs0:e2}   {\ensuremath{{3.085 } } }
\vdef{bdt2011:BgRareHi0:val}   {\ensuremath{{0.282 } } }
\vdef{bdt2011:BgRareHi0:e1}   {\ensuremath{{0.014 } } }
\vdef{bdt2011:BgRareHi0:e2}   {\ensuremath{{0.298 } } }
\vdef{bdt2011:BgRslsLo0:val}   {\ensuremath{{3.000 } } }
\vdef{bdt2011:BgRslsLo0:e1}   {\ensuremath{{0.150 } } }
\vdef{bdt2011:BgRslsLo0:e2}   {\ensuremath{{2.341 } } }
\vdef{bdt2011:BgRslsBd0:val}   {\ensuremath{{0.464 } } }
\vdef{bdt2011:BgRslsBd0:e1}   {\ensuremath{{0.023 } } }
\vdef{bdt2011:BgRslsBd0:e2}   {\ensuremath{{0.486 } } }
\vdef{bdt2011:BgRslsBs0:val}   {\ensuremath{{0.365 } } }
\vdef{bdt2011:BgRslsBs0:e1}   {\ensuremath{{0.018 } } }
\vdef{bdt2011:BgRslsBs0:e2}   {\ensuremath{{0.402 } } }
\vdef{bdt2011:BgRslsHi0:val}   {\ensuremath{{0.035 } } }
\vdef{bdt2011:BgRslsHi0:e1}   {\ensuremath{{0.002 } } }
\vdef{bdt2011:BgRslsHi0:e2}   {\ensuremath{{0.039 } } }
\vdef{bdt2011:BgCombLo0:val}   {\ensuremath{{0.000 } } }
\vdef{bdt2011:BgCombLo0:e1}   {\ensuremath{{0.000 } } }
\vdef{bdt2011:BgCombLo0:e2}   {\ensuremath{{0.100 } } }
\vdef{bdt2011:BgCombBd0:val}   {\ensuremath{{0.000 } } }
\vdef{bdt2011:BgCombBd0:e1}   {\ensuremath{{0.000 } } }
\vdef{bdt2011:BgCombBd0:e2}   {\ensuremath{{0.100 } } }
\vdef{bdt2011:BgCombBs0:val}   {\ensuremath{{0.000 } } }
\vdef{bdt2011:BgCombBs0:e1}   {\ensuremath{{0.000 } } }
\vdef{bdt2011:BgCombBs0:e2}   {\ensuremath{{0.100 } } }
\vdef{bdt2011:BgCombHi0:val}   {\ensuremath{{0.000 } } }
\vdef{bdt2011:BgCombHi0:e1}   {\ensuremath{{0.000 } } }
\vdef{bdt2011:BgCombHi0:e2}   {\ensuremath{{0.100 } } }
\vdef{bdt2011:BgNonpLo0:val}   {\ensuremath{{3.000 } } }
\vdef{bdt2011:BgNonpLo0:e1}   {\ensuremath{{0.150 } } }
\vdef{bdt2011:BgNonpLo0:e2}   {\ensuremath{{2.343 } } }
\vdef{bdt2011:BgNonpBd0:val}   {\ensuremath{{0.464 } } }
\vdef{bdt2011:BgNonpBd0:e1}   {\ensuremath{{0.023 } } }
\vdef{bdt2011:BgNonpBd0:e2}   {\ensuremath{{0.496 } } }
\vdef{bdt2011:BgNonpBs0:val}   {\ensuremath{{0.365 } } }
\vdef{bdt2011:BgNonpBs0:e1}   {\ensuremath{{0.018 } } }
\vdef{bdt2011:BgNonpBs0:e2}   {\ensuremath{{0.414 } } }
\vdef{bdt2011:BgNonpHi0:val}   {\ensuremath{{0.035 } } }
\vdef{bdt2011:BgNonpHi0:e1}   {\ensuremath{{0.002 } } }
\vdef{bdt2011:BgNonpHi0:e2}   {\ensuremath{{0.107 } } }
\vdef{bdt2011:BgTotLo0:val}   {\ensuremath{{3.105 } } }
\vdef{bdt2011:BgTotLo0:e1}   {\ensuremath{{0.000 } } }
\vdef{bdt2011:BgTotLo0:e2}   {\ensuremath{{2.344 } } }
\vdef{bdt2011:BgTotBd0:val}   {\ensuremath{{0.828 } } }
\vdef{bdt2011:BgTotBd0:e1}   {\ensuremath{{0.000 } } }
\vdef{bdt2011:BgTotBd0:e2}   {\ensuremath{{0.561 } } }
\vdef{bdt2011:BgTotBs0:val}   {\ensuremath{{0.474 } } }
\vdef{bdt2011:BgTotBs0:e1}   {\ensuremath{{0.000 } } }
\vdef{bdt2011:BgTotBs0:e2}   {\ensuremath{{0.422 } } }
\vdef{bdt2011:BgTotHi0:val}   {\ensuremath{{0.046 } } }
\vdef{bdt2011:BgTotHi0:e1}   {\ensuremath{{0.000 } } }
\vdef{bdt2011:BgTotHi0:e2}   {\ensuremath{{0.108 } } }
\vdef{bdt2011:SgAndBgLo0:val}   {\ensuremath{{3.208 } } }
\vdef{bdt2011:SgAndBgLo0:e1}   {\ensuremath{{0.000 } } }
\vdef{bdt2011:SgAndBgLo0:e2}   {\ensuremath{{2.344 } } }
\vdef{bdt2011:SgAndBgBd0:val}   {\ensuremath{{1.3 } } }
\vdef{bdt2011:SgAndBgBd0:e1}   {\ensuremath{{0.000 } } }
\vdef{bdt2011:SgAndBgBd0:e2}   {\ensuremath{{0.8 } } }
\vdef{bdt2011:SgAndBgBs0:val}   {\ensuremath{{3.6 } } }
\vdef{bdt2011:SgAndBgBs0:e1}   {\ensuremath{{0.000 } } }
\vdef{bdt2011:SgAndBgBs0:e2}   {\ensuremath{{0.6 } } }
\vdef{bdt2011:SgAndBgHi0:val}   {\ensuremath{{0.140 } } }
\vdef{bdt2011:SgAndBgHi0:e1}   {\ensuremath{{0.000 } } }
\vdef{bdt2011:SgAndBgHi0:e2}   {\ensuremath{{0.108 } } }
\vdef{bdt2011:SgBd1:val}  {\ensuremath{{0.283 } } }
\vdef{bdt2011:SgBd1:e1}   {\ensuremath{{0.532 } } }
\vdef{bdt2011:SgBd1:e2}   {\ensuremath{{0.042 } } }
\vdef{bdt2011:SgBs1:val}  {\ensuremath{{1.277 } } }
\vdef{bdt2011:SgBs1:e1}   {\ensuremath{{1.130 } } }
\vdef{bdt2011:SgBs1:e2}   {\ensuremath{{0.192 } } }
\vdef{bdt2011:SgLo1:val}  {\ensuremath{{0.052 } } }
\vdef{bdt2011:SgLo1:e1}   {\ensuremath{{0.228 } } }
\vdef{bdt2011:SgLo1:e2}   {\ensuremath{{0.008 } } }
\vdef{bdt2011:SgHi1:val}  {\ensuremath{{0.180 } } }
\vdef{bdt2011:SgHi1:e1}   {\ensuremath{{0.424 } } }
\vdef{bdt2011:SgHi1:e2}   {\ensuremath{{0.027 } } }
\vdef{bdt2011:BdBd1:val}  {\ensuremath{{0.105 } } }
\vdef{bdt2011:BdBd1:e1}   {\ensuremath{{0.324 } } }
\vdef{bdt2011:BdBd1:e2}   {\ensuremath{{0.010 } } }
\vdef{bdt2011:BdBs1:val}  {\ensuremath{{0.062 } } }
\vdef{bdt2011:BdBs1:e1}   {\ensuremath{{0.248 } } }
\vdef{bdt2011:BdBs1:e2}   {\ensuremath{{0.006 } } }
\vdef{bdt2011:BdLo1:val}  {\ensuremath{{0.028 } } }
\vdef{bdt2011:BdLo1:e1}   {\ensuremath{{0.168 } } }
\vdef{bdt2011:BdLo1:e2}   {\ensuremath{{0.003 } } }
\vdef{bdt2011:BdHi1:val}  {\ensuremath{{0.002 } } }
\vdef{bdt2011:BdHi1:e1}   {\ensuremath{{0.050 } } }
\vdef{bdt2011:BdHi1:e2}   {\ensuremath{{0.000 } } }
\vdef{bdt2011:BgPeakLo1:val}   {\ensuremath{{0.067 } } }
\vdef{bdt2011:BgPeakLo1:e1}   {\ensuremath{{0.003 } } }
\vdef{bdt2011:BgPeakLo1:e2}   {\ensuremath{{0.048 } } }
\vdef{bdt2011:BgPeakBd1:val}   {\ensuremath{{0.105 } } }
\vdef{bdt2011:BgPeakBd1:e1}   {\ensuremath{{0.005 } } }
\vdef{bdt2011:BgPeakBd1:e2}   {\ensuremath{{0.076 } } }
\vdef{bdt2011:BgPeakBs1:val}   {\ensuremath{{0.056 } } }
\vdef{bdt2011:BgPeakBs1:e1}   {\ensuremath{{0.003 } } }
\vdef{bdt2011:BgPeakBs1:e2}   {\ensuremath{{0.040 } } }
\vdef{bdt2011:BgPeakHi1:val}   {\ensuremath{{0.008 } } }
\vdef{bdt2011:BgPeakHi1:e1}   {\ensuremath{{0.000 } } }
\vdef{bdt2011:BgPeakHi1:e2}   {\ensuremath{{0.005 } } }
\vdef{bdt2011:BgRslLo1:val}   {\ensuremath{{7.038 } } }
\vdef{bdt2011:BgRslLo1:e1}   {\ensuremath{{0.352 } } }
\vdef{bdt2011:BgRslLo1:e2}   {\ensuremath{{5.144 } } }
\vdef{bdt2011:BgRslBd1:val}   {\ensuremath{{1.346 } } }
\vdef{bdt2011:BgRslBd1:e1}   {\ensuremath{{0.067 } } }
\vdef{bdt2011:BgRslBd1:e2}   {\ensuremath{{1.377 } } }
\vdef{bdt2011:BgRslBs1:val}   {\ensuremath{{0.800 } } }
\vdef{bdt2011:BgRslBs1:e1}   {\ensuremath{{0.040 } } }
\vdef{bdt2011:BgRslBs1:e2}   {\ensuremath{{0.861 } } }
\vdef{bdt2011:BgRslHi1:val}   {\ensuremath{{0.210 } } }
\vdef{bdt2011:BgRslHi1:e1}   {\ensuremath{{0.010 } } }
\vdef{bdt2011:BgRslHi1:e2}   {\ensuremath{{0.231 } } }
\vdef{bdt2011:BgRareLo1:val}   {\ensuremath{{7.104 } } }
\vdef{bdt2011:BgRareLo1:e1}   {\ensuremath{{0.352 } } }
\vdef{bdt2011:BgRareLo1:e2}   {\ensuremath{{5.144 } } }
\vdef{bdt2011:BgRareBd1:val}   {\ensuremath{{1.451 } } }
\vdef{bdt2011:BgRareBd1:e1}   {\ensuremath{{0.068 } } }
\vdef{bdt2011:BgRareBd1:e2}   {\ensuremath{{1.379 } } }
\vdef{bdt2011:BgRareBs1:val}   {\ensuremath{{0.856 } } }
\vdef{bdt2011:BgRareBs1:e1}   {\ensuremath{{0.040 } } }
\vdef{bdt2011:BgRareBs1:e2}   {\ensuremath{{0.862 } } }
\vdef{bdt2011:BgRareHi1:val}   {\ensuremath{{0.218 } } }
\vdef{bdt2011:BgRareHi1:e1}   {\ensuremath{{0.010 } } }
\vdef{bdt2011:BgRareHi1:e2}   {\ensuremath{{0.231 } } }
\vdef{bdt2011:BgRslsLo1:val}   {\ensuremath{{2.000 } } }
\vdef{bdt2011:BgRslsLo1:e1}   {\ensuremath{{0.100 } } }
\vdef{bdt2011:BgRslsLo1:e2}   {\ensuremath{{1.462 } } }
\vdef{bdt2011:BgRslsBd1:val}   {\ensuremath{{0.309 } } }
\vdef{bdt2011:BgRslsBd1:e1}   {\ensuremath{{0.015 } } }
\vdef{bdt2011:BgRslsBd1:e2}   {\ensuremath{{0.316 } } }
\vdef{bdt2011:BgRslsBs1:val}   {\ensuremath{{0.243 } } }
\vdef{bdt2011:BgRslsBs1:e1}   {\ensuremath{{0.012 } } }
\vdef{bdt2011:BgRslsBs1:e2}   {\ensuremath{{0.262 } } }
\vdef{bdt2011:BgRslsHi1:val}   {\ensuremath{{0.024 } } }
\vdef{bdt2011:BgRslsHi1:e1}   {\ensuremath{{0.001 } } }
\vdef{bdt2011:BgRslsHi1:e2}   {\ensuremath{{0.026 } } }
\vdef{bdt2011:BgCombLo1:val}   {\ensuremath{{2.000 } } }
\vdef{bdt2011:BgCombLo1:e1}   {\ensuremath{{1.155 } } }
\vdef{bdt2011:BgCombLo1:e2}   {\ensuremath{{1.172 } } }
\vdef{bdt2011:BgCombBd1:val}   {\ensuremath{{0.667 } } }
\vdef{bdt2011:BgCombBd1:e1}   {\ensuremath{{0.385 } } }
\vdef{bdt2011:BgCombBd1:e2}   {\ensuremath{{0.391 } } }
\vdef{bdt2011:BgCombBs1:val}   {\ensuremath{{1.000 } } }
\vdef{bdt2011:BgCombBs1:e1}   {\ensuremath{{0.577 } } }
\vdef{bdt2011:BgCombBs1:e2}   {\ensuremath{{0.586 } } }
\vdef{bdt2011:BgCombHi1:val}   {\ensuremath{{3.000 } } }
\vdef{bdt2011:BgCombHi1:e1}   {\ensuremath{{1.732 } } }
\vdef{bdt2011:BgCombHi1:e2}   {\ensuremath{{1.758 } } }
\vdef{bdt2011:BgNonpLo1:val}   {\ensuremath{{4.000 } } }
\vdef{bdt2011:BgNonpLo1:e1}   {\ensuremath{{1.159 } } }
\vdef{bdt2011:BgNonpLo1:e2}   {\ensuremath{{1.874 } } }
\vdef{bdt2011:BgNonpBd1:val}   {\ensuremath{{0.976 } } }
\vdef{bdt2011:BgNonpBd1:e1}   {\ensuremath{{0.385 } } }
\vdef{bdt2011:BgNonpBd1:e2}   {\ensuremath{{0.503 } } }
\vdef{bdt2011:BgNonpBs1:val}   {\ensuremath{{1.243 } } }
\vdef{bdt2011:BgNonpBs1:e1}   {\ensuremath{{0.577 } } }
\vdef{bdt2011:BgNonpBs1:e2}   {\ensuremath{{0.470 } } }
\vdef{bdt2011:BgNonpHi1:val}   {\ensuremath{{3.023 } } }
\vdef{bdt2011:BgNonpHi1:e1}   {\ensuremath{{1.732 } } }
\vdef{bdt2011:BgNonpHi1:e2}   {\ensuremath{{1.758 } } }
\vdef{bdt2011:BgTotLo1:val}   {\ensuremath{{4.067 } } }
\vdef{bdt2011:BgTotLo1:e1}   {\ensuremath{{0.000 } } }
\vdef{bdt2011:BgTotLo1:e2}   {\ensuremath{{1.874 } } }
\vdef{bdt2011:BgTotBd1:val}   {\ensuremath{{1.081 } } }
\vdef{bdt2011:BgTotBd1:e1}   {\ensuremath{{0.000 } } }
\vdef{bdt2011:BgTotBd1:e2}   {\ensuremath{{0.508 } } }
\vdef{bdt2011:BgTotBs1:val}   {\ensuremath{{1.299 } } }
\vdef{bdt2011:BgTotBs1:e1}   {\ensuremath{{0.000 } } }
\vdef{bdt2011:BgTotBs1:e2}   {\ensuremath{{0.472 } } }
\vdef{bdt2011:BgTotHi1:val}   {\ensuremath{{3.031 } } }
\vdef{bdt2011:BgTotHi1:e1}   {\ensuremath{{0.000 } } }
\vdef{bdt2011:BgTotHi1:e2}   {\ensuremath{{1.758 } } }
\vdef{bdt2011:SgAndBgLo1:val}   {\ensuremath{{4.147 } } }
\vdef{bdt2011:SgAndBgLo1:e1}   {\ensuremath{{0.000 } } }
\vdef{bdt2011:SgAndBgLo1:e2}   {\ensuremath{{1.874 } } }
\vdef{bdt2011:SgAndBgBd1:val}   {\ensuremath{{1.5 } } }
\vdef{bdt2011:SgAndBgBd1:e1}   {\ensuremath{{0.000 } } }
\vdef{bdt2011:SgAndBgBd1:e2}   {\ensuremath{{0.6 } } }
\vdef{bdt2011:SgAndBgBs1:val}   {\ensuremath{{2.6 } } }
\vdef{bdt2011:SgAndBgBs1:e1}   {\ensuremath{{0.000 } } }
\vdef{bdt2011:SgAndBgBs1:e2}   {\ensuremath{{0.5 } } }
\vdef{bdt2011:SgAndBgHi1:val}   {\ensuremath{{3.214 } } }
\vdef{bdt2011:SgAndBgHi1:e1}   {\ensuremath{{0.000 } } }
\vdef{bdt2011:SgAndBgHi1:e2}   {\ensuremath{{1.758 } } }
\vdef{bdt2011:N-EFF-TOT-BS0:val}   {\ensuremath{{0.000608 } } }
\vdef{bdt2011:N-EFF-TOT-BS0:err}   {\ensuremath{{0.000006 } } }
\vdef{bdt2011:N-ACC-BS0:val}   {\ensuremath{{0.0084 } } }
\vdef{bdt2011:N-ACC-BS0:err}   {\ensuremath{{0.0001 } } }
\vdef{bdt2011:N-EFF-MU-PID-BS0:val}   {\ensuremath{{0.7178 } } }
\vdef{bdt2011:N-EFF-MU-PID-BS0:err}   {\ensuremath{{0.0009 } } }
\vdef{bdt2011:N-EFF-MU-PIDMC-BS0:val}   {\ensuremath{{0.6951 } } }
\vdef{bdt2011:N-EFF-MU-PIDMC-BS0:err}   {\ensuremath{{0.0011 } } }
\vdef{bdt2011:N-EFF-MU-MC-BS0:val}   {\ensuremath{{0.5040 } } }
\vdef{bdt2011:N-EFF-MU-MC-BS0:err}   {\ensuremath{{0.0028 } } }
\vdef{bdt2011:N-EFF-TRIG-PID-BS0:val}   {\ensuremath{{0.7749 } } }
\vdef{bdt2011:N-EFF-TRIG-PID-BS0:err}   {\ensuremath{{0.0012 } } }
\vdef{bdt2011:N-EFF-TRIG-PIDMC-BS0:val}   {\ensuremath{{0.7332 } } }
\vdef{bdt2011:N-EFF-TRIG-PIDMC-BS0:err}   {\ensuremath{{0.0014 } } }
\vdef{bdt2011:N-EFF-TRIG-MC-BS0:val}   {\ensuremath{{0.7491 } } }
\vdef{bdt2011:N-EFF-TRIG-MC-BS0:err}   {\ensuremath{{0.0035 } } }
\vdef{bdt2011:N-EFF-CAND-BS0:val}   {\ensuremath{{0.9964 } } }
\vdef{bdt2011:N-EFF-CAND-BS0:err}   {\ensuremath{{0.0004 } } }
\vdef{bdt2011:N-EFF-ANA-BS0:val}   {\ensuremath{{0.1933 } } }
\vdef{bdt2011:N-EFF-ANA-BS0:err}   {\ensuremath{{0.0010 } } }
\vdef{bdt2011:N-OBS-BS0:val}   {\ensuremath{{4740 } } }
\vdef{bdt2011:N-OBS-BS0:err}   {\ensuremath{{210 } } }
\vdef{bdt2011:N-EFF-TOT-BS1:val}   {\ensuremath{{0.000208 } } }
\vdef{bdt2011:N-EFF-TOT-BS1:err}   {\ensuremath{{0.000003 } } }
\vdef{bdt2011:N-ACC-BS1:val}   {\ensuremath{{0.0042 } } }
\vdef{bdt2011:N-ACC-BS1:err}   {\ensuremath{{0.0000 } } }
\vdef{bdt2011:N-EFF-MU-PID-BS1:val}   {\ensuremath{{0.6454 } } }
\vdef{bdt2011:N-EFF-MU-PID-BS1:err}   {\ensuremath{{0.0018 } } }
\vdef{bdt2011:N-EFF-MU-PIDMC-BS1:val}   {\ensuremath{{0.6773 } } }
\vdef{bdt2011:N-EFF-MU-PIDMC-BS1:err}   {\ensuremath{{0.0021 } } }
\vdef{bdt2011:N-EFF-MU-MC-BS1:val}   {\ensuremath{{0.5919 } } }
\vdef{bdt2011:N-EFF-MU-MC-BS1:err}   {\ensuremath{{0.0045 } } }
\vdef{bdt2011:N-EFF-TRIG-PID-BS1:val}   {\ensuremath{{0.6812 } } }
\vdef{bdt2011:N-EFF-TRIG-PID-BS1:err}   {\ensuremath{{0.0041 } } }
\vdef{bdt2011:N-EFF-TRIG-PIDMC-BS1:val}   {\ensuremath{{0.5371 } } }
\vdef{bdt2011:N-EFF-TRIG-PIDMC-BS1:err}   {\ensuremath{{0.0048 } } }
\vdef{bdt2011:N-EFF-TRIG-MC-BS1:val}   {\ensuremath{{0.5602 } } }
\vdef{bdt2011:N-EFF-TRIG-MC-BS1:err}   {\ensuremath{{0.0059 } } }
\vdef{bdt2011:N-EFF-CAND-BS1:val}   {\ensuremath{{0.9940 } } }
\vdef{bdt2011:N-EFF-CAND-BS1:err}   {\ensuremath{{0.0008 } } }
\vdef{bdt2011:N-EFF-ANA-BS1:val}   {\ensuremath{{0.1473 } } }
\vdef{bdt2011:N-EFF-ANA-BS1:err}   {\ensuremath{{0.0012 } } }
\vdef{bdt2011:N-OBS-BS1:val}   {\ensuremath{{1698 } } }
\vdef{bdt2011:N-OBS-BS1:err}   {\ensuremath{{48 } } }
\vdef{2011:ulObs:val:mu_s}	{\ensuremath{{6.1}}}
\vdef{2011:ulObs:exponent:mu_s}	{\ensuremath{{-9}}}
\vdef{2011:ulSM:val:mu_s}	{\ensuremath{{5.9}}}
\vdef{2011:ulSM:errHi:mu_s}	{\ensuremath{{1.7}}}
\vdef{2011:ulSM:errLo:mu_s}	{\ensuremath{{1.5}}}
\vdef{2011:ulSM:exponent:mu_s}	{\ensuremath{{-9}}}
\vdef{2011:ulBkg:val:mu_s}	{\ensuremath{{2.2}}}
\vdef{2011:ulBkg:errHi:mu_s}	{\ensuremath{{0.8}}}
\vdef{2011:ulBkg:errLo:mu_s}	{\ensuremath{{0.7}}}
\vdef{2011:ulBkg:exponent:mu_s}	{\ensuremath{{-9}}}
\vdef{2011:ulObs:val:mu_d}	{\ensuremath{{8.9}}}
\vdef{2011:ulObs:exponent:mu_d}	{\ensuremath{{-10}}}
\vdef{2011:ulSM:val:mu_d}	{\ensuremath{{9.0}}}
\vdef{2011:ulSM:errHi:mu_d}	{\ensuremath{{3.6}}}
\vdef{2011:ulSM:errLo:mu_d}	{\ensuremath{{4.0}}}
\vdef{2011:ulSM:exponent:mu_d}	{\ensuremath{{-10}}}
\vdef{2011:ulBkg:val:mu_d}	{\ensuremath{{8.3}}}
\vdef{2011:ulBkg:errHi:mu_d}	{\ensuremath{{2.6}}}
\vdef{2011:ulBkg:errLo:mu_d}	{\ensuremath{{3.4}}}
\vdef{2011:ulBkg:exponent:mu_d}	{\ensuremath{{-10}}}
\vdef{20112012:ulObs:val:mu_s}	{\ensuremath{{3.6}}}
\vdef{20112012:ulObs:exponent:mu_s}	{\ensuremath{{-9}}}
\vdef{20112012:ulSM:val:mu_s}	{\ensuremath{{6.0}}}
\vdef{20112012:ulSM:errHi:mu_s}	{\ensuremath{{1.8}}}
\vdef{20112012:ulSM:errLo:mu_s}	{\ensuremath{{1.7}}}
\vdef{20112012:ulSM:exponent:mu_s}	{\ensuremath{{-9}}}
\vdef{20112012:ulBkg:val:mu_s}	{\ensuremath{{2.4}}}
\vdef{20112012:ulBkg:errHi:mu_s}	{\ensuremath{{1.1}}}
\vdef{20112012:ulBkg:errLo:mu_s}	{\ensuremath{{0.8}}}
\vdef{20112012:ulBkg:exponent:mu_s}	{\ensuremath{{-9}}}
\vdef{20112012:ulObs:val:mu_d}	{\ensuremath{{1.1}}}
\vdef{20112012:ulObs:exponent:mu_d}	{\ensuremath{{-9}}}
\vdef{20112012:ulSM:val:mu_d}	{\ensuremath{{7.6}}}
\vdef{20112012:ulSM:errHi:mu_d}	{\ensuremath{{3.4}}}
\vdef{20112012:ulSM:errLo:mu_d}	{\ensuremath{{2.3}}}
\vdef{20112012:ulSM:exponent:mu_d}	{\ensuremath{{-10}}}
\vdef{20112012:ulBkg:val:mu_d}	{\ensuremath{{6.4}}}
\vdef{20112012:ulBkg:errHi:mu_d}	{\ensuremath{{2.8}}}
\vdef{20112012:ulBkg:errLo:mu_d}	{\ensuremath{{2.0}}}
\vdef{20112012:ulBkg:exponent:mu_d}	{\ensuremath{{-10}}}
\vdef{20112012:fitComb:val:bdmm:barrel}	{\ensuremath{{9.669}}}
\vdef{20112012:fitPeak:val:bdmm:barrel}	{\ensuremath{{0.535}}}
\vdef{20112012:fitSig:val:bdmm:barrel}	{\ensuremath{{8.458}}}
\vdef{20112012:fitComb:val:bsmm:barrel}	{\ensuremath{{10.516}}}
\vdef{20112012:fitPeak:val:bsmm:barrel}	{\ensuremath{{0.143}}}
\vdef{20112012:fitSig:val:bsmm:barrel}	{\ensuremath{{5.050}}}
\vdef{20112012:fitComb:val:bdmm:endcap}	{\ensuremath{{5.392}}}
\vdef{20112012:fitPeak:val:bdmm:endcap}	{\ensuremath{{0.104}}}
\vdef{20112012:fitSig:val:bdmm:endcap}	{\ensuremath{{2.435}}}
\vdef{20112012:fitComb:val:bsmm:endcap}	{\ensuremath{{6.495}}}
\vdef{20112012:fitPeak:val:bsmm:endcap}	{\ensuremath{{0.052}}}
\vdef{20112012:fitSig:val:bsmm:endcap}	{\ensuremath{{1.698}}}
\vdef{20112012:intObs:val:mu_s}	{\ensuremath{{1.2}}}
\vdef{20112012:intObs:errHi:mu_s}	{\ensuremath{{1.3}}}
\vdef{20112012:intObs:errLo:mu_s}	{\ensuremath{{0.9}}}
\vdef{20112012:intObs:exponent:mu_s}	{\ensuremath{{-9}}}
\vdef{20112012:intSM:val:mu_s}	{\ensuremath{{3.2}}}
\vdef{20112012:intSM:errHi:mu_s}	{\ensuremath{{1.1}}}
\vdef{20112012:intSM:errLo:mu_s}	{\ensuremath{{1.3}}}
\vdef{20112012:intSM:exponent:mu_s}	{\ensuremath{{-9}}}
\vdef{20112012:intObs:val:mu_d}	{\ensuremath{{6.9}}}
\vdef{20112012:intObs:errHi:mu_d}	{\ensuremath{{4.2}}}
\vdef{20112012:intObs:errLo:mu_d}	{\ensuremath{{3.7}}}
\vdef{20112012:intObs:exponent:mu_d}	{\ensuremath{{-10}}}
\vdef{20112012:intSM:val:mu_d}	{\ensuremath{{1.0}}}
\vdef{20112012:intSM:errHi:mu_d}	{\ensuremath{{2.8}}}
\vdef{20112012:intSM:errLo:mu_d}	{\ensuremath{{1.0}}}
\vdef{20112012:intSM:exponent:mu_d}	{\ensuremath{{-10}}}

\cmsNoteHeader{BPH-13-004} 
\title{\texorpdfstring{Measurement of the $\PBzs \rightarrow \mu^+ \mu^-$ branching fraction and search for $\PBz \rightarrow \mu^+ \mu^-$ with the CMS experiment}{Measurement of the Bs to mu mu branching fraction and search for B0 to mu mu with the CMS Experiment}}

\date{\today}

\abstract{Results are presented from a search for the rare
decays $\PBzs \rightarrow \mu^+ \mu^-$ and $\PBz \rightarrow \mu^+
\mu^-$ in $\Pp\Pp$ collisions at $\sqrt{s}=7$ and $8\TeV$, with data
samples corresponding to integrated luminosities of \lumiseven\ and
\lumieight\fbinv, respectively, collected by the CMS experiment at
the LHC. An unbinned maximum-likelihood fit to the dimuon invariant
mass distribution gives a branching fraction $\cbf(\bsmm) =
\bsResultBF$, where the uncertainty includes both statistical and
systematic contributions. An excess of \bsmm\ events  with
respect to background is observed with a significance of \bssignif\ standard deviations. For the decay \bdmm\
an upper limit of $\cbf(\bdmm) < \vuse{20112012:ulObs:val:mu_d} \times
10^{\vuse{20112012:ulObs:exponent:mu_d} }$ at the 95\% confidence
level is determined. Both results are in agreement with the expectations from the
standard model.}

\hypersetup{%
pdfauthor={CMS Collaboration},%
pdftitle={Measurement of Bs to mu mu branching fraction and search for B0 to mu mu with the CMS Experiment},%
pdfsubject={CMS},%
pdfkeywords={CMS, physics}}

\maketitle 

In the standard model (SM) of particle physics, tree-level diagrams do
not
contribute to flavor-changing neutral-current (FCNC) decays.  However,
FCNC decays may proceed through higher-order loop diagrams, and this
opens up the possibility for contributions from non-SM particles. In
the SM, the rare FCNC decays $\PBzs$ $(\PBz) \to \Pgmp\Pgmm$ have
small branching fractions of $\cbf(\bsmm)=(3.57\pm
0.30)\times10^{-9}$, corresponding to the decay-time integrated
branching fraction, and
$\cbf(\bdmm)=(1.07\pm0.10)\times10^{-10}$~\cite{Buras:2012ru,DeBruyn:2012wk}.
Charge conjugation is implied throughout this Letter.  Several
extensions of the SM, such as supersymmetric models with non-universal
Higgs boson masses~\cite{Ellis:2006jy}, specific models containing
leptoquarks~\cite{Davidson:2010uu}, and the minimal supersymmetric
standard model with large
$\tan\beta$~\cite{Choudhury:2005rz,Parry:2005fp}, predict enhancements
to the branching fractions for these rare decays.  The decay rates can
also be suppressed for specific choices of model
parameters~\cite{Ellis:2007kb}.  Over the past 30 years, significant
progress in sensitivity has been made, with exclusion limits on the
branching fractions improving by five orders of magnitude.  The ARGUS~\cite{Albrecht:1987rj}, UA1~\cite{Albajar:1991ct}, %
CLEO~\cite{Bergfeld:2000ui}, Belle~\cite{Chang:2003yy},
BaBar~\cite{Aubert:2007hb}, CDF~\cite{Aaltonen:2013as},
D0~\cite{Abazov:2013wjb}, ATLAS~\cite{Aad:2012pn},
CMS~\cite{Chatrchyan:2012rg}, and LHCb~\cite{Aaij:2011rj} experiments
have all published limits on these decays.  The LHCb experiment has
subsequently shown evidence, with 3.5 standard deviation significance,
for the decay \bsmm with $\cbf(\bsmm) = (3.2^{+1.5}_{-1.2}) \times
10^{-9}$ \cite{aaij:2012nna}.

This Letter reports a measurement of $\cbf(\PBzs \rightarrow \mu^+ \mu^-)$
based on a simultaneous search for \bsmm\ and \bdmm\ decays using a
data sample of $\Pp\Pp$ collisions corresponding to integrated
luminosities of \lumiseven\ $\fbinv$ at $\sqrt{s} =7\TeV$ and
\lumieight $\fbinv$ at $8\TeV$ collected by the Compact Muon Solenoid
(CMS) experiment at the Large Hadron Collider (LHC).  For these data,
the peak luminosity varied from $3.5\times10^{30}$ to
$7.7\times10^{33} \textrm{cm}^{-2}\textrm{s}^{-1}$.  The average
number of interactions per bunch crossing (pileup) was 9 (21) at
$\sqrt{s} = 7(8)\TeV$.

The search for the $\PB\to\mu^+\mu^-$ signal, where \PB\ denotes $\PBzs$
or $\PBz$, is performed in the dimuon invariant mass regions around
the \PBzs\ and \Bz\ masses.  To avoid possible biases, the signal
region $5.20 < m_{\mu\mu} < 5.45\gev$ was kept blind until all
selection criteria were established. For the $7\TeV$ data, this Letter
reports a re-analysis of the data used in the previous
result~\cite{Chatrchyan:2012rg}, where the data were re-blinded.  The
combinatorial dimuon background, mainly from semileptonic decays of separate
\PB\ mesons, is evaluated by extrapolating the data in
nearby mass sidebands into the signal region. Monte Carlo (MC)
simulations are used to account for backgrounds from \PB\ and
$\Lambda_{\cPqb}$ decays. These background samples consist of
$\PB\to \cPhx \mu\nu$, $\PB\to \cPhx \mu\mu$, and $\Lambda_\cPqb\to
\Pp\mu\nu$ decays, as well as ``peaking'' decays of the type $\PB\to \cPhx \cPhx'$,
where \cPhx, $\cPhx'$ are charged hadrons misidentified as
muons, which give a dimuon invariant mass distribution that peaks in
the signal region. The MC simulation event samples are generated using {\PYTHIA} (version
{6.424} for 7\TeV, version {6.426} for
$8\TeV$)~\cite{Sjostrand:2006za}, with the underlying event simulated
with the Z2 tune~\cite{Z2}, unstable particles decayed via \textsc{
EvtGen}~\cite{Lange:2001uf}, and the detector response simulated
with {\GEANTfour}~\cite{Ivanchenko:2003xp}.  A normalization sample of
$\bupsik\to \Pgmp\Pgmm\PKp$ decays is used to minimize uncertainties
related to the $\bbbar$ production cross section and the integrated
luminosity.  A control sample of $\bspsiphi\to\Pgmp\Pgmm\PKp\PKm$
decays is used to validate the MC simulation and to evaluate potential
effects from differences in fragmentation between \PBp\ and \PBzs.
The efficiencies of all samples, including detector acceptances,
are determined with MC simulation studies.

A detailed description of the CMS apparatus can be found in
Ref.~\cite{:2008zzk}. The CMS experiment uses a right-handed
coordinate system, with the origin at the nominal interaction point,
the $x$ axis pointing to the center of the LHC ring, the $y$ axis
pointing up, and the $z$ axis along the counterclockwise-beam
direction. The polar angle $\theta$ is measured from the positive $z$
axis and the azimuthal angle $\phi$ is measured in the $x$-$y$ plane.
The main subdetectors used in this analysis are the silicon tracker
and the muon detectors.  Muons are tracked within the pseudorapidity
region $\abs{\eta} < 2.4$, where $\eta = -\ln[\tan(\theta/2)]$.  A
transverse momentum (\pt)\ resolution of about 1.5\% is obtained for
muons in this analysis~\cite{Khachatryan:2010pw}.

The events are selected with a two-level trigger system.  The first
level only requires two muon candidates in the muon detectors.
The high-level trigger (HLT) uses additional information from the
silicon tracker to provide essentially a full event reconstruction.
The dimuon invariant mass was
required to satisfy $4.8 < m_{\mu\mu} < 6.0\gev$.  For the $7\TeV$
data set, the HLT selection required two muons, each
with $\pt>4.0\gev$, and a dimuon $\pt^{\mu\mu}>3.9\gev$. For events
having at least one muon with $\abs{\eta}>1.5$, $\pt^{\mu\mu}>5.9\gev$ was
required.  For the $8\TeV$ data set, the \pt\ criterion on the muon
with lower \pt\ was loosened to $\pt>3.0\gev$, with
$\pt^{\mu\mu}>4.9\gev$. For events containing at least one muon with
$\abs{\eta}>1.8$, the muons were each required to have $\pt>4.0\gev$,
$\pt^{\mu\mu}>7.0\gev$, and the dimuon vertex fit $p$-value ${>}0.5\%$.

For the normalization and control samples the HLT selection required
the following: two muons, each with $\pt > 4\gev$ and $\abs{\eta}<2.2$; $\pt^{\mu\mu} >
6.9\gev$; $2.9 < m_{\mu\mu} <3.3\gev$; and the dimuon
vertex fit $p$-value $> 15\%$.  Two additional requirements were
imposed in the transverse plane: (i) the pointing angle $\alpha_{xy}$
between the dimuon momentum and the vector from the average
interaction point to the dimuon vertex had to fulfill $\cos\alpha_{xy}
> 0.9$, and (ii) the flight length significance
$\ell_{xy}/\sigma(\ell_{xy})$ must be greater than 3, where
$\ell_{xy}$ is the two-dimensional distance between the average
interaction point and the dimuon vertex, and $\sigma(\ell_{xy})$ is
its uncertainty.  The signal, normalization, and control triggers required
the three-dimensional (3D) distance of closest approach (\dca) between
the two muons to satisfy $\dca < 0.5\cm$.  The average trigger
efficiency for events in the signal and normalization samples, as
determined from MC simulation and calculated after all other selection
criteria are applied, is in the range 39--85\%, depending on the
running period and detector region.  The uncertainty in the ratio of
trigger efficiencies (muon identification efficiencies) for the signal
and normalization samples is estimated to be 3--6\% (1--4\%) by
comparing simulation and data.

The $\PB \to \Pgmp\Pgmm$ candidates are constructed from two
oppositely charged ``tight'' muons as described in
Ref.~\cite{misid:2010}.  Both muons must have $\pt>4\gev$ and be consistent
in direction and \pt\ with
the muons that triggered the event. A boosted decision
tree (BDT) constructed within the {\sc TMVA}
framework~\cite{Hocker:2007ht} is trained to further separate genuine
muons from those arising from misidentified charged hadrons.  The
variables used in the BDT can be divided into four classes: basic
kinematic quantities, silicon-tracker fit information, combined
silicon and muon track fit information, and muon detector information.
The BDT is trained on MC simulation samples of \PB-meson decays to
kaons and muons. Compared to the ``tight'' muons, the BDT working
point used to select muons for this analysis reduces the
hadron-to-muon misidentification probability by 50\% while retaining
90\% of true muons. The probability to misidentify a charged hadron as
a muon because of decay in flight or detector punch-through is
measured in data from samples of well-identified pions, kaons, and
protons. This probability ranges from
(0.5--1.3$)\times 10^{-3}$, (0.8--2.2)$\times 10^{-3}$, and
(0.4--1.5$)\times 10^{-3}$, for pions, kaons, and protons, respectively, depending on whether the
particle is in the barrel or endcap, the running period, and the
momentum. Each of these probabilities is ascribed an uncertainty of
50\%, based on differences between data and MC simulation.

Candidates are kept for further analysis if they have $4.9 <
m_{\mu\mu} < 5.9\gev$, after constraining the tracks to a common
vertex.  The $\PB$-candidate momentum and vertex position are used to
choose a primary vertex based on the distance of closest approach along
the beamline.  Since the background level and mass resolution depend
significantly on $\eta_{\mu\mu}$, where $\eta_{\mu\mu}$ is the
pseudorapidity of the \PB-meson candidate, the events are separated into
two categories: the ``barrel channel'' with candidates where both
muons have $\abs{\eta} < 1.4$, and the ``endcap channel'' containing those
where at least one muon has $\abs{\eta} > 1.4$.  The $m_{\mu\mu}$
resolution, as determined from simulated signal events, ranges from
$32\MeV$ for $\eta_{\mu\mu}\approx0$ to $75\MeV$ for $\abs{\eta_{\mu\mu}}
> 1.8$.

Four isolation variables are defined. (1)~$I =
\pt^{\mu\mu}/(\pt^{\mu\mu} + \sum_{\mathrm{trk}}\pt)$, where
$\sum_{\text{trk}}\pt$ is the sum of \pt\ of all tracks, other than
muon candidates, satisfying $\Delta R = \sqrt{(\Delta\eta)^2 +
(\Delta\phi)^2} < 0.7$, with $\Delta\eta$ and $\Delta\phi$ as the
differences in $\eta$ and $\phi$ between a charged track and the
direction of the \PB\ candidate.  The sum includes all tracks with
$\pt>0.9\GeV$ that are (i)~consistent with originating from the same
primary vertex as the $\PB$ candidate or (ii)~have a \dca\ with
respect to the $\PB$ vertex ${<}0.05\cm$ and are not associated with
any other primary vertex. (2)~$I_\mu$ is the isolation variable of each muon,
calculated as for the \PB\ candidate, but with respect to the muon track.
A cone size of $\Delta R = 0.5$ around the muon and tracks with
$\pt > 0.5\GeV$ and $\dca < 0.1\cm$ from the muon are used.
(3)~\closetrk\ is defined as
the number of tracks with $\pt>0.5\GeV$ and \dca\ with respect to the
$\PB$ vertex less than $<0.03\cm$. (4)~\docatrk\ is defined as the
smallest \dca\ to the $\PB$ vertex, considering all tracks in the
event that are either associated with the same primary vertex as the
$\PB$ candidate or not associated with any primary vertex.

The final selection is performed with BDTs trained to distinguish
between signal and background event candidates. For the training, \bsmm\ MC
simulation
samples are used for the signal, and candidates from the data dimuon
mass sidebands after a loose preselection
for the background. The preselection retains at least 10,000 events
dominated by combinatorial background for each BDT.
To avoid any selection bias, the
data background events are randomly split into three sets, such that
the training and testing of the BDT is performed on sets independent
of its application. Studies with sideband events and signal MC
simulation samples
with shifted \PB\ mass show that the BDT response is independent of
mass.  Separate BDTs are trained for each of the four combinations of
7 and 8\TeV data and the barrel and endcap regions of the
detector. For each BDT, a number of variables is considered and only
those found to be effective are included. Each of the following twelve
variables, shown to be independent of pileup, are used in at least one
of the BDTs: $I$; $I_\mu$; \closetrk; \docatrk; $\pt^{\mu\mu}$;
$\eta_{\mu\mu}$; the $\PB$-vertex fit $\chi^2$ per degree of freedom
(dof); the \dca\ between the two muon tracks; the 3D pointing angle
$\alpha_{3\mathrm{D}}$; the 3D flight length significance
$\ell_{3\mathrm{D}}/\sigma(\ell_{3\mathrm{D}})$; the 3D impact
parameter $\ip$ of the $\PB$ candidate; and its significance $\ips$,
where $\sigma(\delta_\mathrm{3D})$ is the uncertainty on $\ip$.  The last
four variables are computed with respect to the primary vertex.  Good
agreement between data and MC simulation is observed for these variables. In
total, including the division into three sets, 12 BDTs are trained.

The output discriminant $b$ of the BDT is used in two ways for further
analysis. (1)~In the 1D-BDT method, a minimum requirement on $b$ per
channel is used to define the final selection. The requirement on $b$
is optimized for best $S/\sqrt{S+B}$ (where $S$ is the expected signal
and $B$ the background) on statistically independent data control samples.
The optimization gives $b > 0.29$ for both barrel and endcap in the
$\sqrt{s} = 7\TeV$ data, and $b>0.36$ (0.38) in the barrel (endcap)
for the $\sqrt{s} = 8\TeV$ sample. The 1D-BDT method is used for the
determination of the upper limit on $\cbf(\PBz \to \Pgmp\Pgmm)$.
The signal efficiencies
$\varepsilon_{\text{tot}}$ for method (1) are provided in
Table~\ref{t:allTheNumbers}, together with the expected number of
events (signal and signal plus background) for the \PBz\ signal region
$5.20 < m < 5.30\GeV$ and the \PBzs\ signal region $5.30 < m <
5.45\GeV$. (2)~In the categorized-BDT method, the discriminant $b$ is
used to define twelve
event categories with different signal-to-background ratios. For the
$\sqrt{s} = 7\TeV$ data in the barrel (endcap) channel, the
two categories have boundaries of 0.10, 0.31, 1.00 (0.10, 0.26, 1.00). For
the $\sqrt{s} = 8\TeV$ sample in the barrel (endcap) channel, the
corresponding boundaries for the four categories
are 0.10, 0.23, 0.33, 0.44, 1.00 (0.10, 0.22,
0.33, 0.45, 1.00). This binning is chosen to give the same expected
signal yield in each bin. The dimuon invariant mass distributions for the
twelve categories of events are fitted simultaneously to obtain the final
results. Method (2) has higher expected sensitivity
and thus provides the main methodology for the extraction of
$\cbf(\PBzs \to \Pgmp\Pgmm)$.

The $\bupsik\to\Pgmp\Pgmm \PKp$ ($\bspsiphi\to\Pgmp\Pgmm\PKp\PKm$)
selection requires two oppositely charged muons with $3.0 < m_{\mu\mu}
< 3.2\gev$ and $\pt^{\mu\mu} > 7\gev$, combined with one or two tracks,
assumed to be kaons, fulfilling $\pt > 0.5 \gev$ and $\abs{\eta} < 2.4 $
($\abs{\eta} < 2.1$ in the $8\TeV$ data).  The distance of closest
approach between all pairs among the three (four) tracks is required
to be less than $0.1\cm$.  For \bspsiphi\ candidates the two assumed
kaon tracks must have invariant mass $0.995 < m_{\PK\PK} < 1.045\gev$
and $\Delta R < 0.25$.  The $\PB$ vertex is fitted from the three
(four) tracks; a candidate is accepted if the resulting invariant mass
is in the range 4.8--6.0\GeV. The final selection is
achieved using the same BDT as for the signal, with the following
modifications: the \PB-vertex $\chi^2/\mathrm{dof}$ is determined
from the dimuon vertex fit, and for the calculation of the isolation
variables all \PB-candidate decay tracks are neglected.

\renewcommand{\base}{bdt2012}
\newcommand{\StrutSmall}{\rule[-1.0mm]{0pt}{4.7mm}}
\newcommand{\StrutLarge}{\rule[-2.3mm]{0pt}{6.3mm}}

\renewcommand{\sample}{SgData-SgMc}
\renewcommand{\channel}{A}

\begin{table*}[!tb]
  \begin{center}
    \topcaption{The signal selection efficiencies
      $\varepsilon_{\text{tot}}$, the predicted number of SM signal
      events $N_{\text{signal}}^{\text{exp}}$,
      the expected number of signal and background events
      $N_{\text{total}}^{\text{exp}}$, and the number of observed
      events $N_{\text{obs}}$ in the barrel and endcap channels for the 7 and $8\TeV$ data
      using the 1D-BDT method. The event numbers refer to the \PBz\
      and \PBzs\ signal regions, respectively.}
    \label{t:allTheNumbers}

    \begin{scotch}{clcccc}
        & & $\varepsilon_{\text{tot}} [10^{-2}]$ & $N_{\text{signal}}^{\text{exp}}$ & $N_{\mathrm{total}}^{\text{exp}}$ &$N_{\text{obs}}$\StrutSmall \\
      \hline
      \multirow{5}{*}{$7\TeV$ }
      & \PBz\ Barrel
      &  $\vuse{bdt2011:N-EFF-TOT-BDMM0:all} $
      & $\vuse{bdt2011:N-EXP2-SIG-BDMM0:val} \pm \vuse{bdt2011:N-EXP2-SIG-BDMM0:err}  $
      &$\vuse{bdt2011:SgAndBgBd0:val} \pm  \vuse{bdt2011:SgAndBgBd0:e2} $
      &$\vuse{bdt2011:N-OBS-BDMM0:val} $\StrutSmall
      \\
      & \PBzs\ Barrel
      & $\vuse{bdt2011:N-EFF-TOT-BSMM0:all} $
      & $\vuse{bdt2011:N-EXP2-SIG-BSMM0:val} \pm \vuse{bdt2011:N-EXP2-SIG-BSMM0:err}  $
      &$\vuse{bdt2011:SgAndBgBs0:val} \pm  \vuse{bdt2011:SgAndBgBs0:e2} $
      &$\vuse{bdt2011:N-OBS-BSMM0:val} $\StrutSmall \\
      & \PBz\ Endcap
      &  $\vuse{bdt2011:N-EFF-TOT-BDMM1:all} $
      & $\vuse{bdt2011:N-EXP2-SIG-BDMM1:val} \pm  \vuse{bdt2011:N-EXP2-SIG-BDMM1:err}  $
      &$\vuse{bdt2011:SgAndBgBd1:val} \pm  \vuse{bdt2011:SgAndBgBd1:e2} $
      &$\vuse{bdt2011:N-OBS-BDMM1:val} $\StrutSmall \\
      &\PBzs\ Endcap
      & $\vuse{bdt2011:N-EFF-TOT-BSMM1:all} $
      & $\vuse{bdt2011:N-EXP2-SIG-BSMM1:val} \pm  \vuse{bdt2011:N-EXP2-SIG-BSMM1:err}  $
      &$\vuse{bdt2011:SgAndBgBs1:val} \pm  \vuse{bdt2011:SgAndBgBs1:e2} $
      &$\vuse{bdt2011:N-OBS-BSMM1:val} $\StrutSmall \\
      \hline
      \multirow{5}{*}{$8\TeV$ }
      & \PBz\ Barrel
      & $\vuse{bdt2012:N-EFF-TOT-BDMM0:all} $
      & $\vuse{bdt2012:N-EXP2-SIG-BDMM0:val} \pm      \vuse{bdt2012:N-EXP2-SIG-BDMM0:err}  $
      &$\vuse{bdt2012:SgAndBgBd0:val} \pm  \vuse{bdt2012:SgAndBgBd0:e2} $
      &$\vuse{bdt2012:N-OBS-BDMM0:val} $\StrutSmall \\
      &\PBzs\ Barrel
      & $\vuse{bdt2012:N-EFF-TOT-BSMM0:all} $
      & $\vuse{bdt2012:N-EXP2-SIG-BSMM0:val} \pm \vuse{bdt2012:N-EXP2-SIG-BSMM0:err}  $
      &$\vuse{bdt2012:SgAndBgBs0:val} \pm  \vuse{bdt2012:SgAndBgBs0:e2} $
      &$\vuse{bdt2012:N-OBS-BSMM0:val} $\StrutSmall \\
      & \PBz\ Endcap
      & $\vuse{bdt2012:N-EFF-TOT-BDMM1:all} $
      & $\vuse{bdt2012:N-EXP2-SIG-BDMM1:val} \pm \vuse{bdt2012:N-EXP2-SIG-BDMM1:err}  $
      &$\vuse{bdt2012:SgAndBgBd1:val} \pm  \vuse{bdt2012:SgAndBgBd1:e2} $
      &$\vuse{bdt2012:N-OBS-BDMM1:val} $\StrutSmall \\
      & \PBzs\ Endcap
      & $\vuse{bdt2012:N-EFF-TOT-BSMM1:all} $
      & $\vuse{bdt2012:N-EXP2-SIG-BSMM1:val} \pm \vuse{bdt2012:N-EXP2-SIG-BSMM1:err}  $
      &$\vuse{bdt2012:SgAndBgBs1:val} \pm  \vuse{bdt2012:SgAndBgBs1:e2} $
      &$\vuse{bdt2012:N-OBS-BSMM1:val} $\StrutSmall \\
    \end{scotch}
  \end{center}
\end{table*}

The total efficiency to reconstruct with the 1D-BDT method a
$\bupsik\to\Pgmp\Pgmm\PKp$ decay, including the detector acceptance,
is $\varepsilon_{\text{tot}}^{\PBp} =
\vuse{bdt2011:N-EFF-TOT-BPLUS0:all} $ and
$\vuse{bdt2011:N-EFF-TOT-BPLUS1:all} $, respectively, for the barrel
and endcap channels in the $7\TeV$ analysis, and
$\vuse{bdt2012:N-EFF-TOT-BPLUS0:all} $ and
$\vuse{bdt2012:N-EFF-TOT-BPLUS1:all} $ for the $8\TeV$ analysis, where
statistical and systematic uncertainties are combined in quadrature.
The distributions of $b$ for the normalization and control samples are
found to agree well between data and MC simulation, with residual
differences used to estimate systematic uncertainties.  No dependence
of the selection efficiency on pileup is observed.  The systematic
uncertainty in the acceptance is estimated by comparing the values
obtained with different \bbbar\ production mechanisms (gluon
splitting, flavor excitation, and flavor creation).  The uncertainty
in the event selection efficiency for the \bupsik\ normalization
sample is evaluated from differences between measured and simulated
\bupsik\ events.  The uncertainty in the \bsmm\ and \bdmm\ signal
efficiencies (3--10\%, depending on the channel and $\sqrt{s}$) is
evaluated using the \bspsiphi\ control sample.

The yields for the normalization (control) sample in each category are
fitted with a double (single) Gaussian function.  The backgrounds
under the normalization and control sample peaks are described with an
exponential (plus an error function for the normalization
sample). Additional functions are included, with shape templates fixed
from simulation, to account for backgrounds from \bupsipi\ (Gaussian
function) for the normalization sample, and $\PB^0\rightarrow\cPJgy
\PK^{*0}$ (Landau function) for the control sample.  In the $7\TeV$ data,
the observed number of \bupsik\ candidates in the barrel
is $(71.2\pm4.1)\times10^3$ and $(21.4\pm1.1)\times10^3$ in the
endcap channel.
For the $8\TeV$ sample the corresponding yields are
$(309\pm16)\times10^3$ (barrel) and
$(69.3\pm3.5)\times10^3$ (endcap).
The uncertainties include a systematic component estimated from simulated events by considering
alternative fitting functions.

The \bsmm\ branching fraction is measured using
\begin{equation}
  \cbf(\bsmm)
  =  \frac{N_\mathrm{S}}{N_{\text{obs}}^{\PBp}} \,
  \frac{f_\cPqu}{f_\cPqs} \,
  \frac{\varepsilon_{\text{tot}}^{\PBp}}{\varepsilon_{\text{tot}}} \,
  \cbf(\PBp)\label{eq:schema},
\end{equation}
and analogously for the \bdmm\ case, where $N_\mathrm{S}$
($N_{\text{obs}}^{\PBp}$) is the number of reconstructed \bsmm\
(\bupsik) decays, $\varepsilon_\text{tot}$
($\varepsilon_{\text{tot}}^{\PBp}$) is the total signal (\PBp)
efficiency, $\cbf(\PBp) = (6.0\pm0.2)\times10^{-5}$~\cite{Nakamura:2010zzi} is the branching fraction for
$\bupsik\to\Pgmp\Pgmm \PKp$, and $\fufs$ is the ratio of the \PBp\
and \PBzs\ fragmentation fractions.  The value $\fsfu=0.256\pm0.020$,
as measured by LHCb~\cite{Aaij:2013jhep}, is used and  an additional systematic uncertainty of
5\% is assigned to account for possible pseudorapidity and $\pt^{\mu\mu}$ dependence of
this ratio. Studies based on the $\PBp \to
\JPsi\PK$ and $\PBzs \to \JPsi \phi$ control samples reveal no
discernible pseudorapidity or $\pt^{\mu\mu}$ dependence of this ratio in the
kinematic region used in the analysis.

An unbinned maximum-likelihood fit to the $m_{\mu\mu}$ distribution is
used to extract the signal and background yields.  Events in the
signal window can result from genuine signal, combinatorial
background, background from semileptonic \cPqb-hadron decays, and the
peaking background.  The probability density functions (PDFs) for the
signal, semileptonic, and peaking backgrounds are obtained from fits
to MC simulation.  The $\PBzs$ and $\PBz$ signal shapes are modeled by
Crystal Ball functions~\cite{bib-crystalball}.  The peaking background
is modeled with the sum of Gaussian and Crystal Ball functions (with a
common mean).  The semileptonic background is modeled with a Gaussian
kernels method~\cite{kernel1,kernel2}.  The PDF for the combinatorial
background is modeled with a first-degree polynomial. Since the dimuon
mass resolution $\sigma$, determined on an event-by-event basis from
the dimuon mass fit, varies significantly, the PDFs described above
are combined as a conditional product with the PDF for the per-event
mass resolution, such that the Crystal Ball function width correctly reflects
the resolution on a per-event basis.  To avoid any effect of the
correlation between $\sigma$ and the candidate mass, we divide the
invariant mass uncertainty by the mass to obtain a ``reduced'' mass
uncertainty, $\sigma_r = \sigma/m_{\mu\mu}$, which is used in the fit.

\begin{figure}[htbp]
  \begin{centering}
    \includegraphics[width=\figwid]{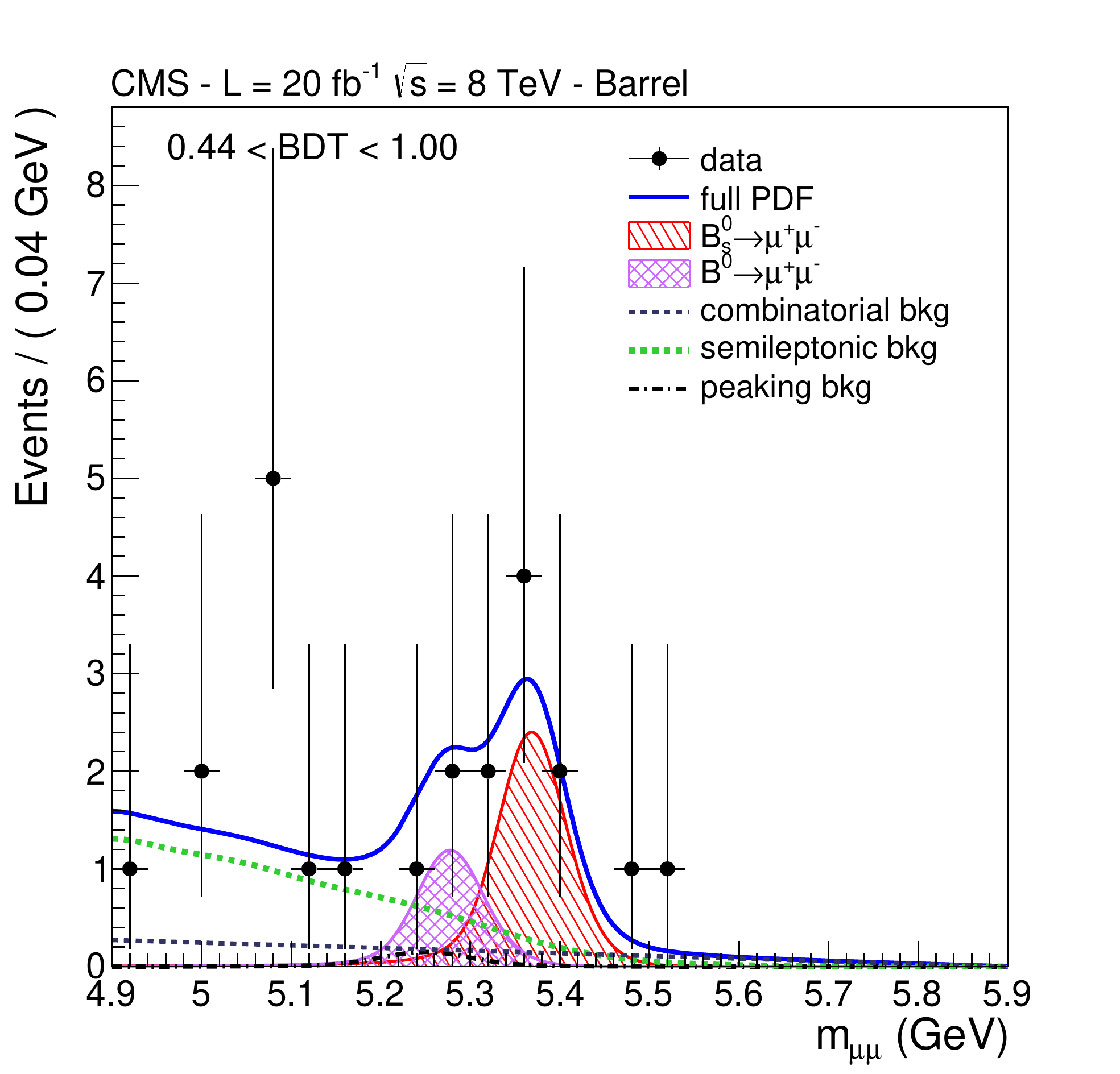}
    \includegraphics[width=\figwid]{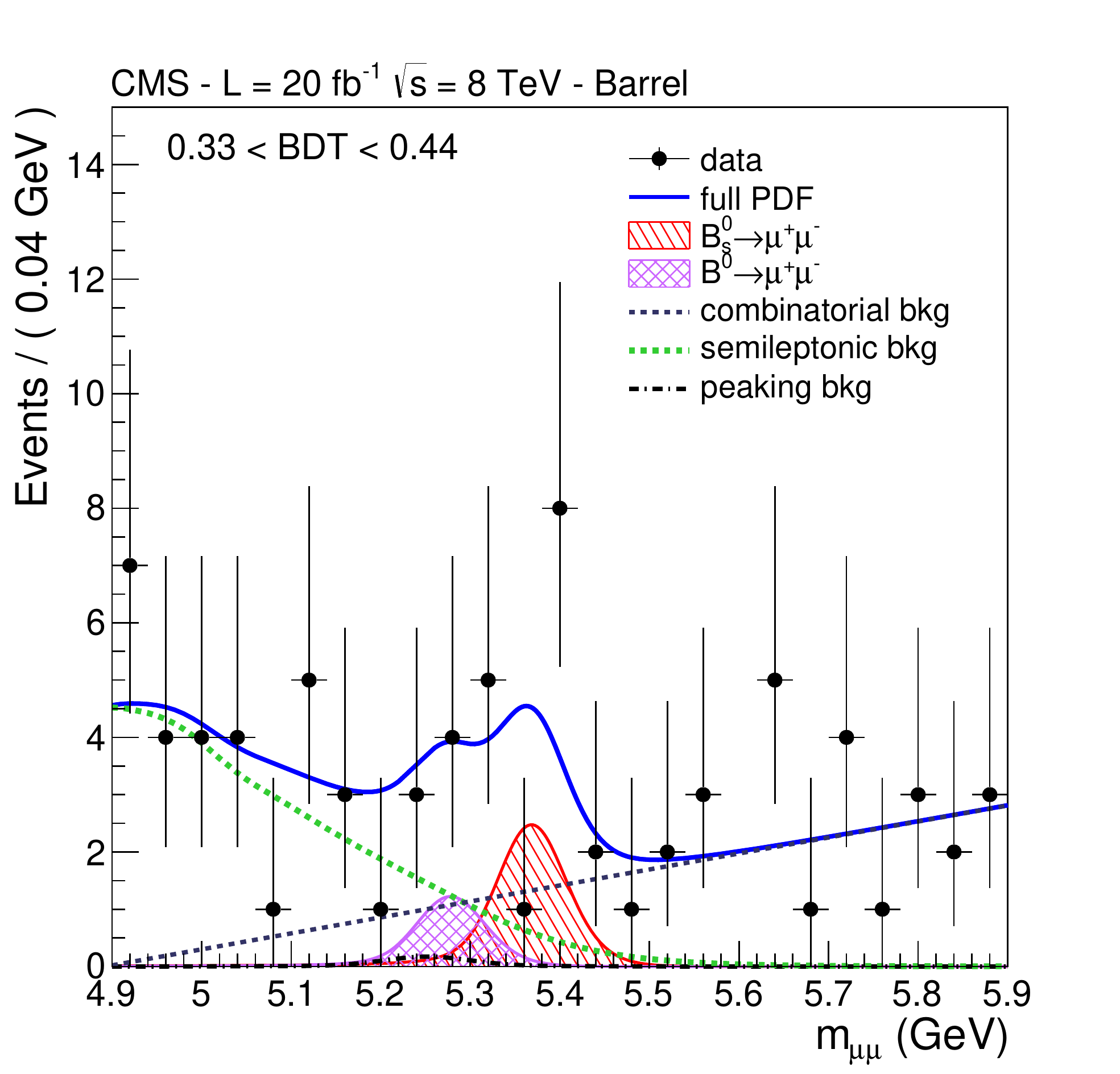}
    \includegraphics[width=\figwid]{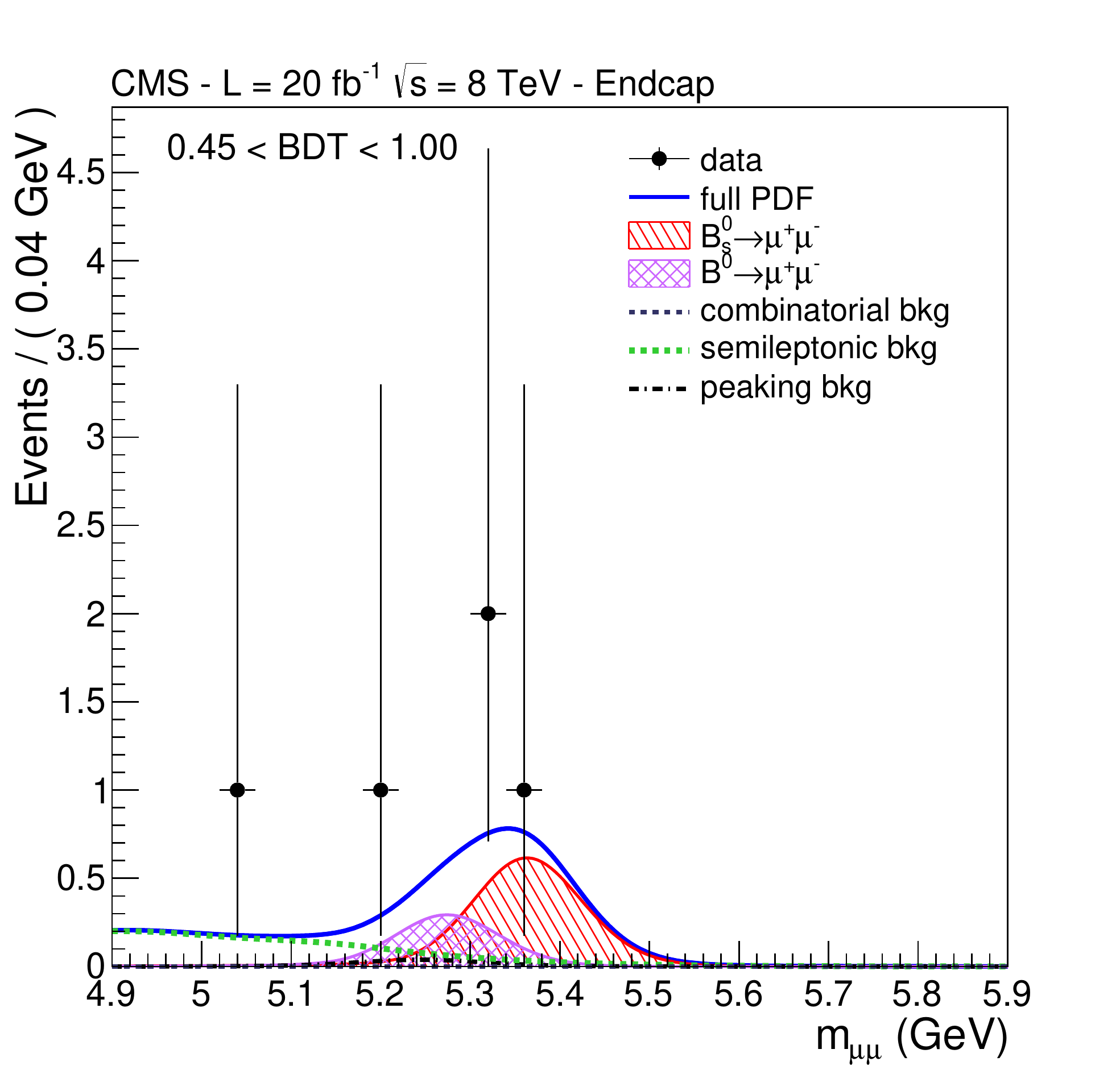}
    \includegraphics[width=\figwid]{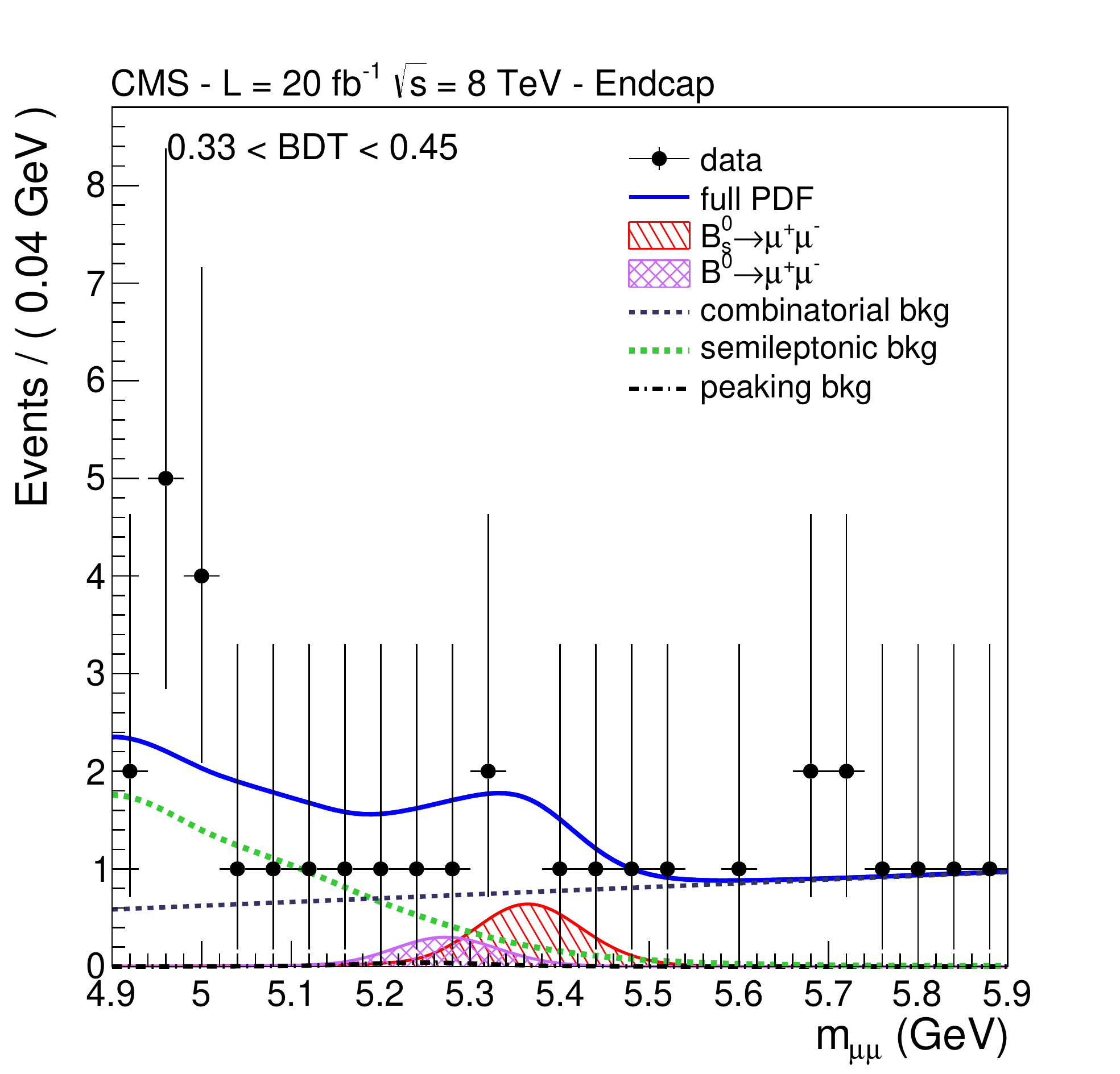}
   \caption{Results from the categorized-BDT method of the fit to the dimuon invariant mass
   distributions for the $\sqrt{s}=8\TeV$ data in the barrel (top) and endcap (bottom)
   for the BDT bins with the highest (left) and second-highest (right) signal-to-background ratio.}
   \label{fig:result}
  \end{centering}
\end{figure}

The dimuon mass distributions for the four channels (barrel and endcap
in 7 and $8\TeV$ data), further divided into categories corresponding
to different bins in the BDT parameter $b$, are fitted simultaneously.
The results are illustrated for the most sensitive categories in
Fig.~\ref{fig:result}. The fits for all twelve categories are shown
in \suppMaterial\ifthenelse{\boolean{cms@external}}{ showing additional plots of the mass fits}{}.
Pseudo-experiments, done with MC simulated
events, confirm the robustness and accuracy of the fitting procedure.

Systematic uncertainties are constrained with Gaussian PDFs with the
standard deviations of the constraints set equal to the uncertainties.
Sources of systematic uncertainty arise from the hadron-to-muon
misidentification probability, the branching fraction uncertainties
(dominated by 100\% for $\Lambda_\cPqb\to \Pp\mu\nu$), and the
normalization of the peaking background.  The $\PB\to \cPhx \cPhx'$
and semileptonic backgrounds are estimated by normalizing to the
observed $\PBp\to\cPJgy\PKp$ yield.  The peaking background yield is
constrained in the fit with log-normal PDFs with r.m.s.~parameters set
to the mean 1-standard-deviation uncertainties.  The absolute level of
peaking background has been studied on an independent data sample,
obtained with single-muon triggers, and is found to agree with the
expectation described above. The shape parameters for the
peaking and the semileptonic backgrounds and for the signals are fixed
to the expectation.  The mass scale uncertainty at the \PB-meson mass
is 6\MeV (7\MeV) for the barrel (endcap) channels, as determined with
charmonium and bottomium decays to dimuon final states.

\begin{figure}[htbp]
  \begin{centering}
    \includegraphics[width=\cmsFigWidth]{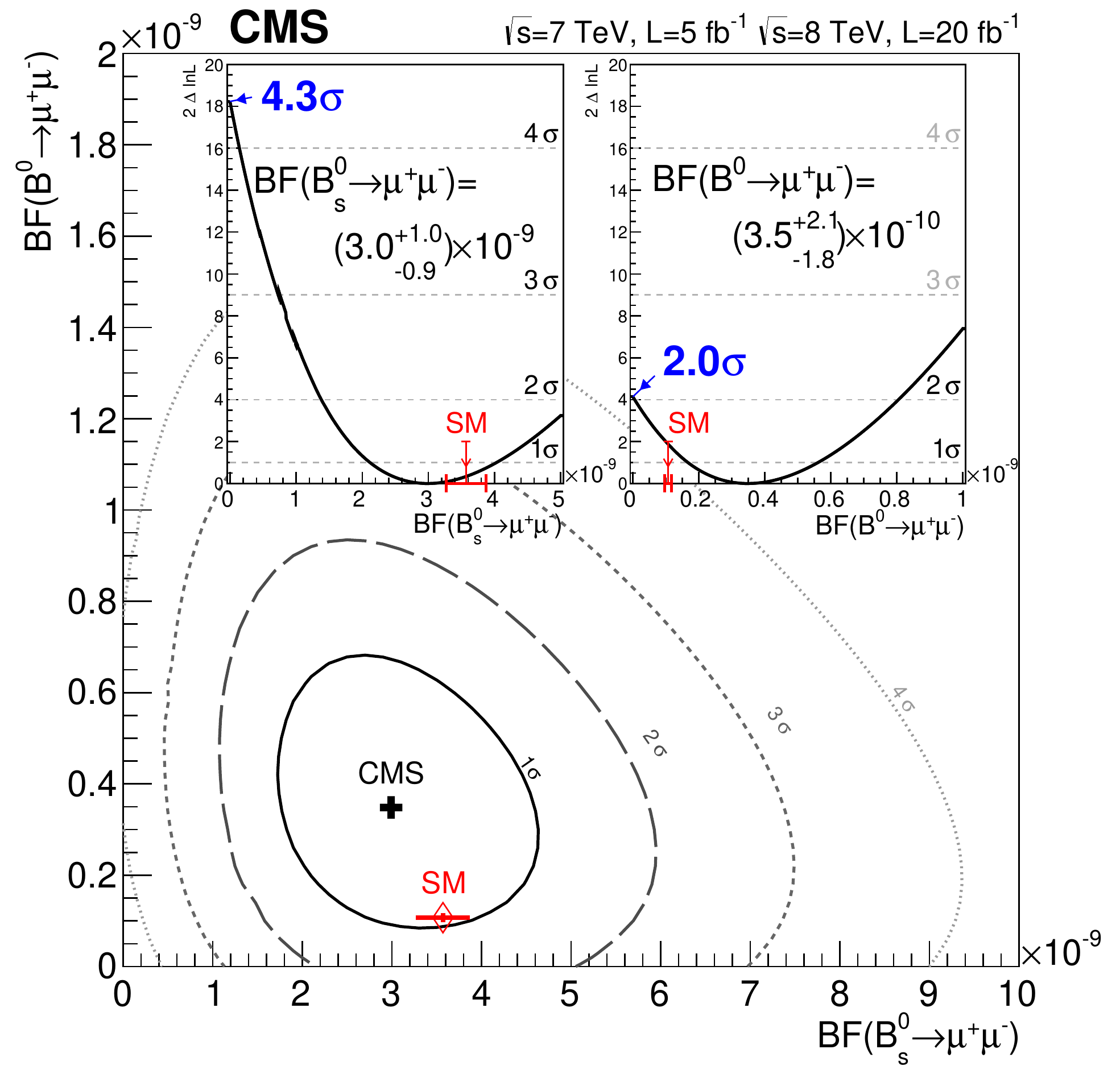}
    \includegraphics[width=\cmsFigWidth]{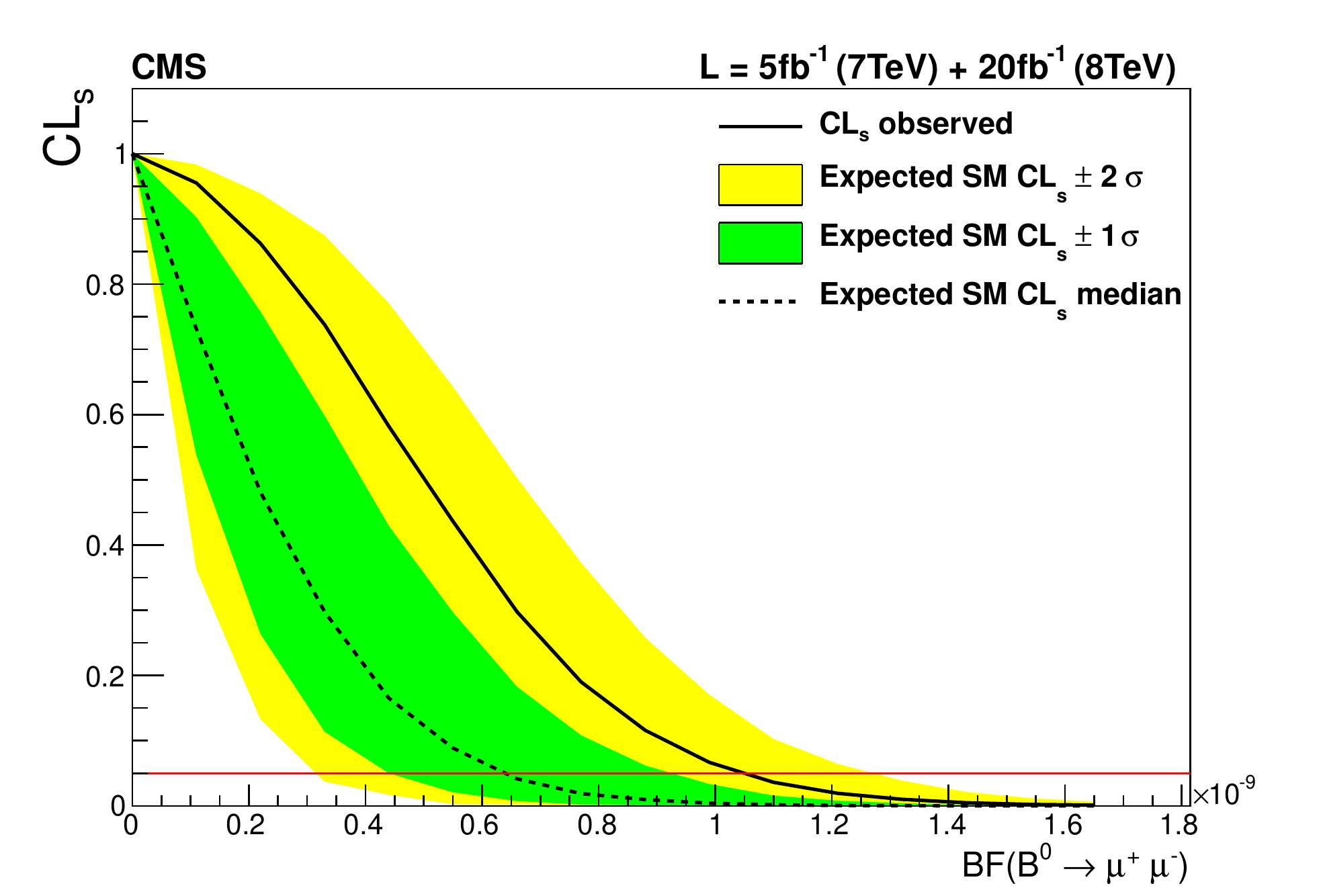}
   \caption{\cmsLeft, scan of the ratio of the joint likelihood for $\cbf(\bsmm)$ and
  $\cbf(\bdmm)$. As insets, the likelihood ratio scan for each of the
  branching fractions when the other is profiled together with other
  nuisance parameters; the significance at which the background-only
  hypothesis is rejected is also shown. \cmsRight, observed and expected  CL$\mathrm{_S}$ for $\PBz \to \mu^+ \mu^-$ as a function of the   assumed branching fraction.  }
   \label{fig:l2d}
  \end{centering}
\end{figure}

\begin{figure}[htbp]
  \begin{centering}
    \includegraphics[width=\figwid]{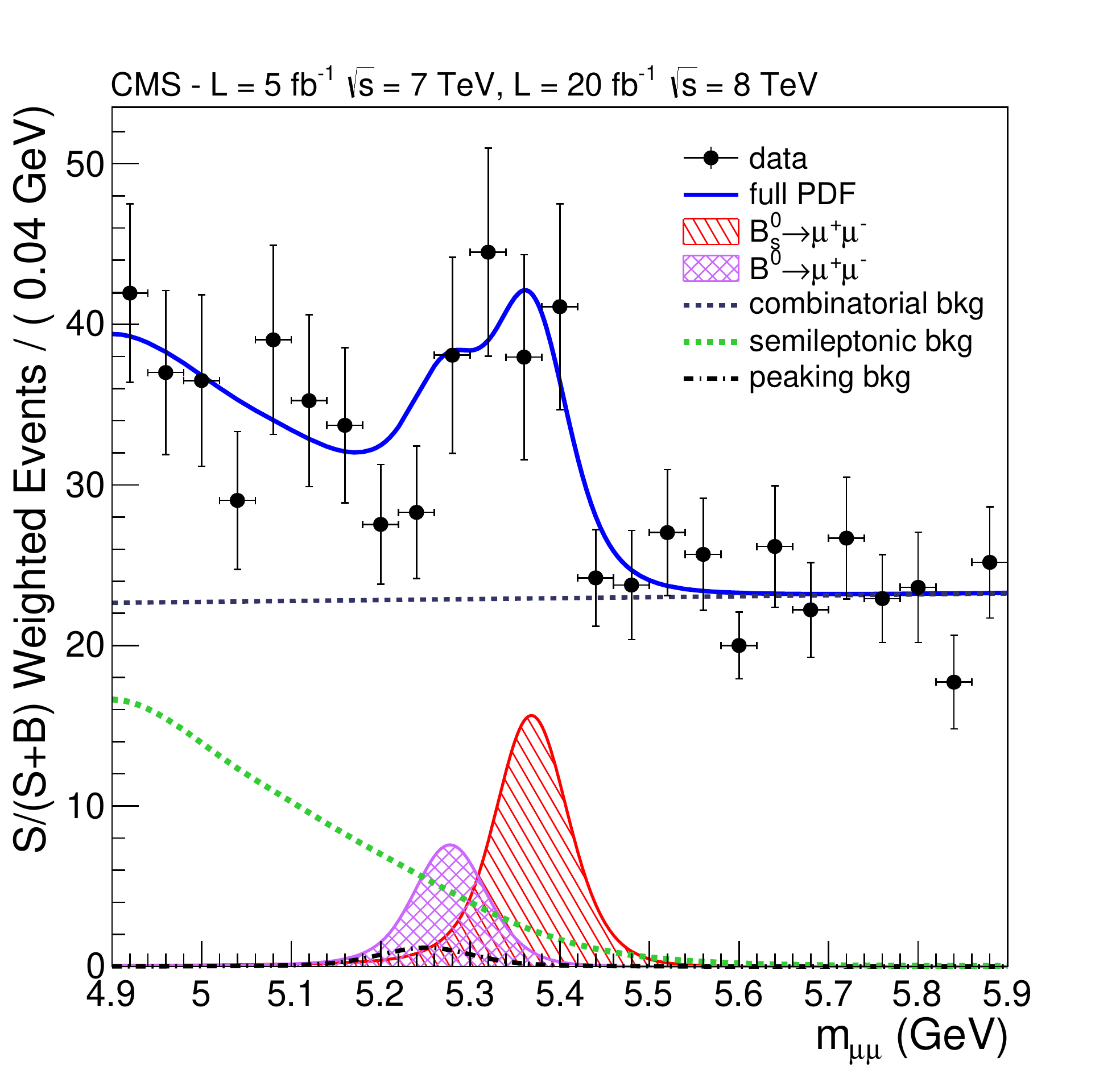}
    \includegraphics[width=\figwid]{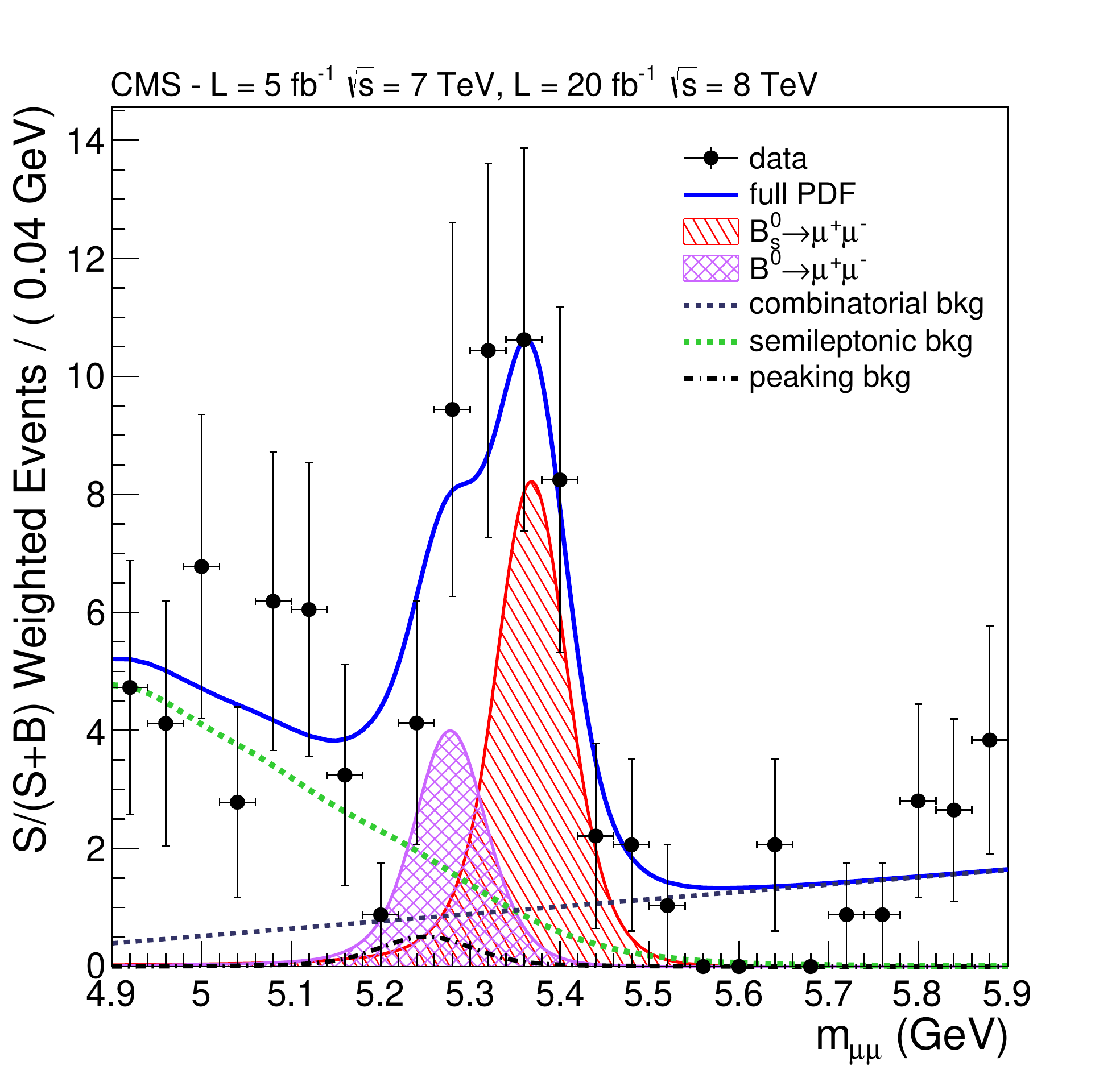}
   \caption{Plots illustrating  the combination of all
   categories used in the categorized-BDT method (left) and the 1D-BDT
   method (right). For these plots, the individual categories are
   weighted with $S/(S+B)$, where $S$ ($B$) is the signal (background)
   determined at the \PBzs\ peak position. The overall normalization is
   set such that the fitted \PBzs\ signal corresponds to the total yield of the
   individual contributions. These distributions are
   for illustrative purposes only and were not used in obtaining the
   final results. }
   \label{fig:cookedResult}
  \end{centering}
\end{figure}

An excess of $\PBzs \rightarrow \mu^+ \mu^-$ decays is observed above
the background predictions. The measured decay-time integrated
branching fraction from the fit is $\cbf(\bsmm) = \bsResultBF$, where
the uncertainty includes both the statistical and systematic
components, but is dominated by the statistical uncertainties.
The observed (expected median) significance of the excess
is \bssignif\ (4.8) standard deviations and is determined by
evaluating the ratio of the likelihood value for the hypothesis with
no signal, divided by the likelihood with $\mathcal{B}(\PBzs \rightarrow
\mu^+ \mu^-)$ floating.  For this determination, $\mathcal{B}(\PBz
\rightarrow \mu^+ \mu^-)$ is allowed to float and is treated as a
nuisance parameter in the fit (see the left plot in Fig.~\ref{fig:l2d}). The measured
branching fraction is consistent with the expectation from the
SM. With the 1D-BDT method, the observed (expected median)
significance is 4.8 (4.7) standard
deviations. Figure~\ref{fig:cookedResult} shows the combined
mass distributions weighted by $S/(S+B)$ for the categorized-BDT (left) and the 1D-BDT (right)
methods. However, these distributions are illustrative only and were not used to obtain the
final results.

No significant excess is observed for $\PBz \rightarrow \mu^+ \mu^-$,
and the upper limit $\mathcal{B}(\PBz \rightarrow \mu^+\mu^-) <
\vuse{20112012:ulObs:val:mu_d} \times
10^{\vuse{20112012:ulObs:exponent:mu_d} } $ ($9.2\times10^{-10}$) at
95\% (90\%) confidence level (CL) is determined with the
CL$\mathrm{_S}$ approach~\cite{Read2002,Junk1999}, based on the
observed numbers of events in the signal and sideband regions with the 1D-BDT method
as summarized in Table~\ref{t:allTheNumbers}.  The expected
95\% CL upper limit for $\mathcal{B}(\PBz \rightarrow \mu^+\mu^-)$ in the
presence of SM signal plus background (background only) is
$6.3\times10^{-10}$ ($5.4\times10^{-10}$), where the statistical and systematic
uncertainties are considered.  The right plot in Fig.~\ref{fig:l2d}
shows the observed and expected CL$\mathrm{_S}$ curves versus the assumed
$\cbf(\bdmm)$. From the fit, the branching fraction for this decay is
determined to be $\cbf(\bdmm) = \bdResultBF$. The significance of this
measurement is \bdsignif\ standard deviations. The dimuon invariant mass
distributions with the 1D-BDT method for the four channels are shown in
\ifthenelse{\boolean{cms@external}}{the supplemental material.}{Fig.~\ref{fig:bdt2} in the Appendix.}

In summary, a search for the rare decays $\PBzs \to \mu^+ \mu^-$ and
$\PBz \to \mu^+ \mu^-$ has been performed on a data sample of $\Pp\Pp$
collisions at $\sqrt{s}=7$ and $8\TeV$ corresponding to integrated
luminosities of $\lumiseven$ and $\lumieight\fbinv$, respectively.  No
significant evidence is observed for $\PBz \to \mu^+ \mu^-$ and an upper limit of
$\mathcal{B}(\PBz \rightarrow \mu^+\mu^-) < \bdUL$ is established at 95\% CL. For
$\PBzs \to \mu^+ \mu^-$, an excess of events with a
significance of \bssignif\ standard deviations is observed, and
a branching fraction of $\cbf(\bsmm) = \bsResultBF$ is determined, in agreement with
the standard model expectations.

We congratulate our colleagues in the CERN accelerator departments for
the excellent performance of the LHC and thank the technical and
administrative staffs at CERN and at other CMS institutes for their
contributions to the success of the CMS effort. In addition, we
gratefully acknowledge the computing centers and personnel of the
Worldwide LHC Computing Grid for delivering so effectively the
computing infrastructure essential to our analyses. Finally, we
acknowledge the enduring support for the construction and operation of
the LHC and the CMS detector provided by the following funding
agencies: BMWF and FWF (Austria); FNRS and FWO (Belgium); CNPq, CAPES,
FAPERJ, and FAPESP (Brazil); MES (Bulgaria); CERN; CAS, MoST, and NSFC
(China); COLCIENCIAS (Colombia); MSES (Croatia); RPF (Cyprus); MoER,
SF0690030s09 and ERDF (Estonia); Academy of Finland, MEC, and HIP
(Finland); CEA and CNRS/IN2P3 (France); BMBF, DFG, and HGF (Germany);
GSRT (Greece); OTKA and NKTH (Hungary); DAE and DST (India); IPM
(Iran); SFI (Ireland); INFN (Italy); NRF and WCU (Republic of Korea);
LAS (Lithuania); CINVESTAV, CONACYT, SEP, and UASLP-FAI (Mexico); MBIE
(New Zealand); PAEC (Pakistan); MSHE and NSC (Poland); FCT (Portugal);
JINR (Dubna); MON, RosAtom, RAS and RFBR (Russia); MESTD (Serbia);
SEIDI and CPAN (Spain); Swiss Funding Agencies (Switzerland); NSC
(Taipei); ThEPCenter, IPST, STAR and NSTDA (Thailand); TUBITAK and
TAEK (Turkey); NASU (Ukraine); STFC (United Kingdom); DOE and NSF
(USA).

\bibliography{auto_generated}   
\ifthenelse{\boolean{cms@external}}{}{
\clearpage
\appendix
\newlength\figwidX
\ifthenelse{\boolean{cms@external}}{\setlength\figwidX{0.23\textwidth}}{\setlength\figwidX{0.49\textwidth}}
\section{Additional plots of mass fits\label{app:suppMat}}
\begin{sidewaysfigure}[htbp]
  \begin{centering}
    \includegraphics[width=0.95\linewidth]{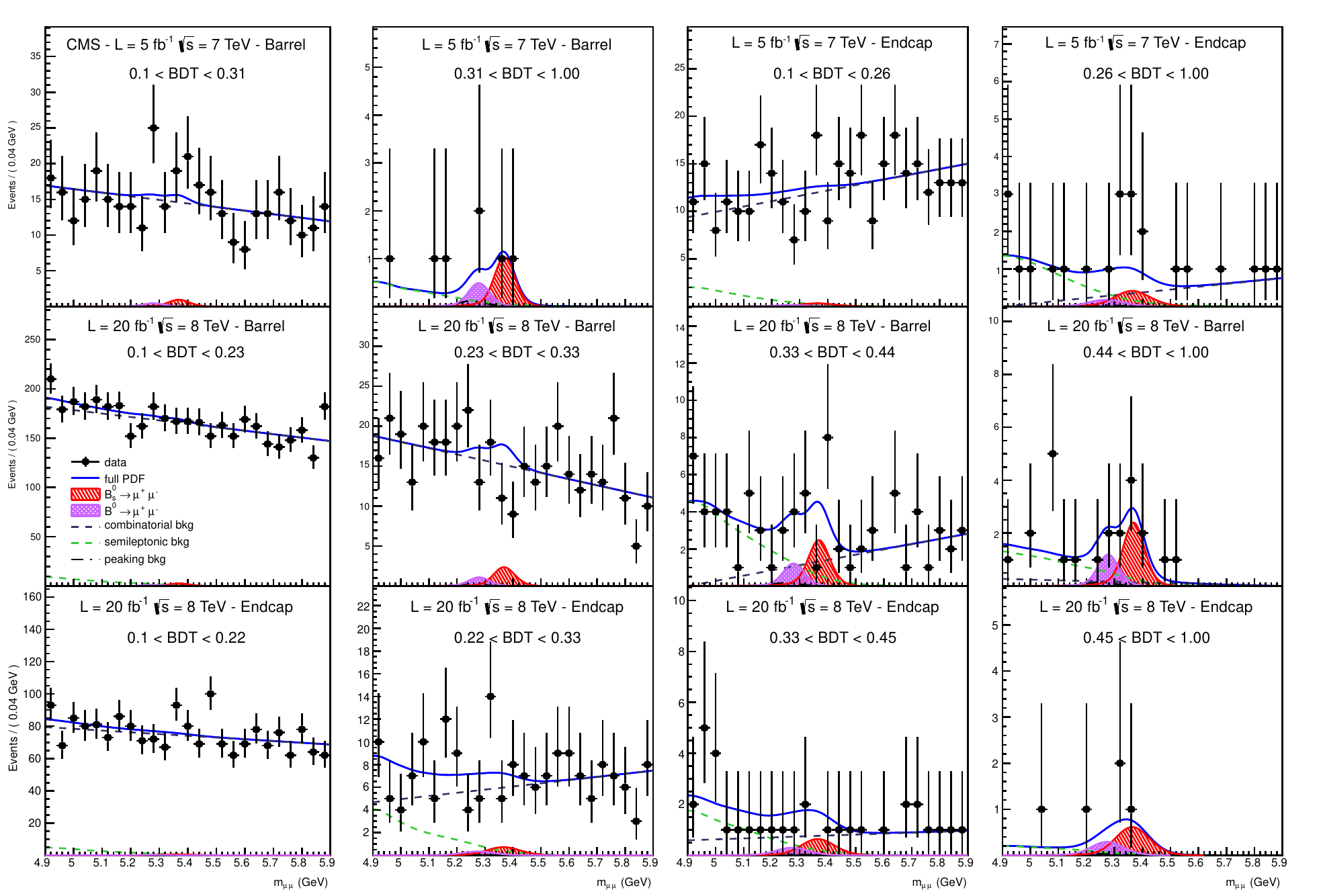}
   \caption{Results of the fit to the dimuon invariant mass
   distributions for all BDT bins in the data with the categorized-BDT
   method. The points are the data, the solid line is the result of the
   fit, the shaded areas are the two \PB\ signals, and the different dotted
   lines are the backgrounds.}
   \label{fig:twelveplots}
  \end{centering}
\end{sidewaysfigure}

\begin{figure}[htbp]
  \begin{centering}
    \includegraphics[width=\figwidX]{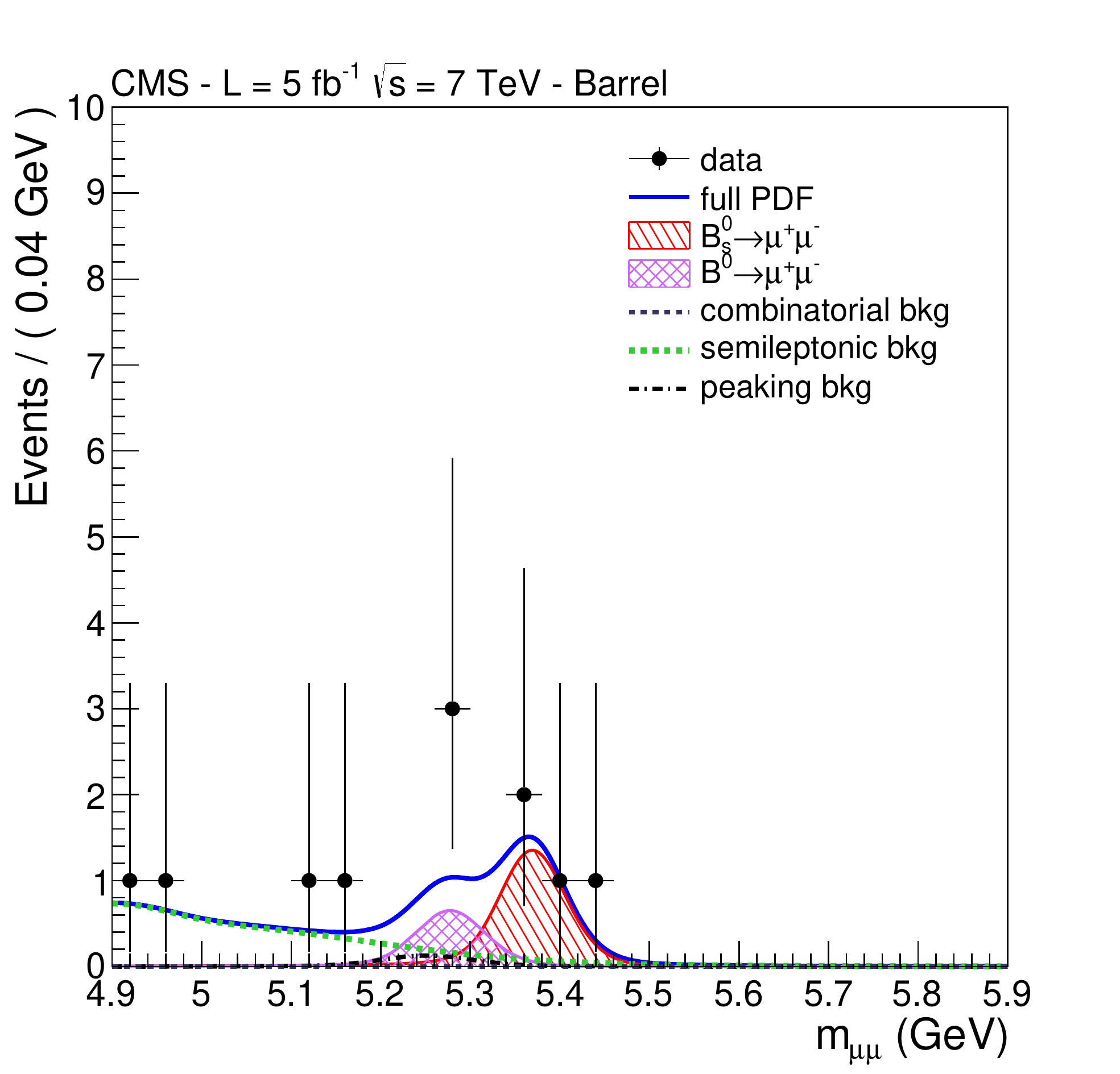}
    \includegraphics[width=\figwidX]{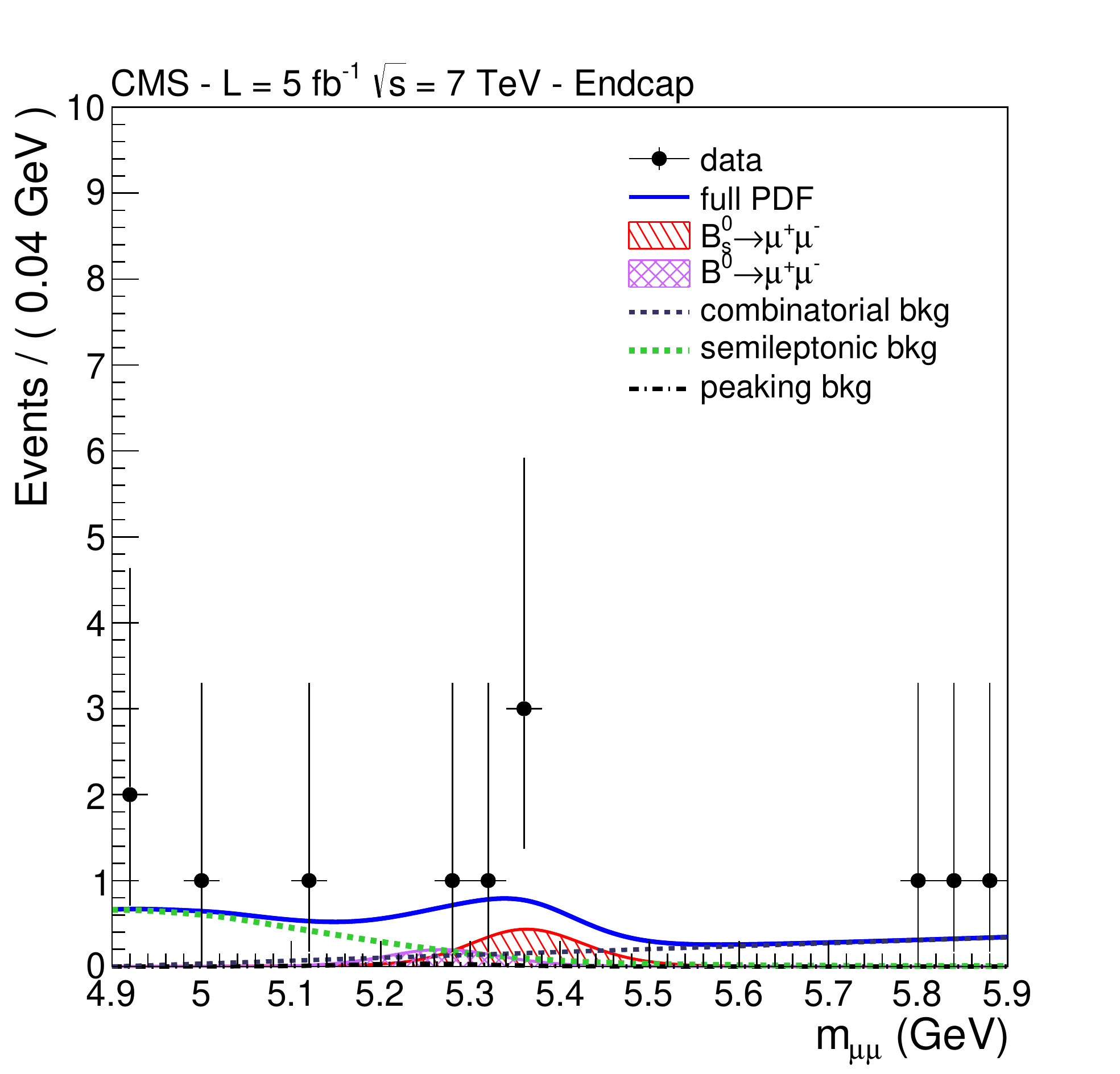}
    \includegraphics[width=\figwidX]{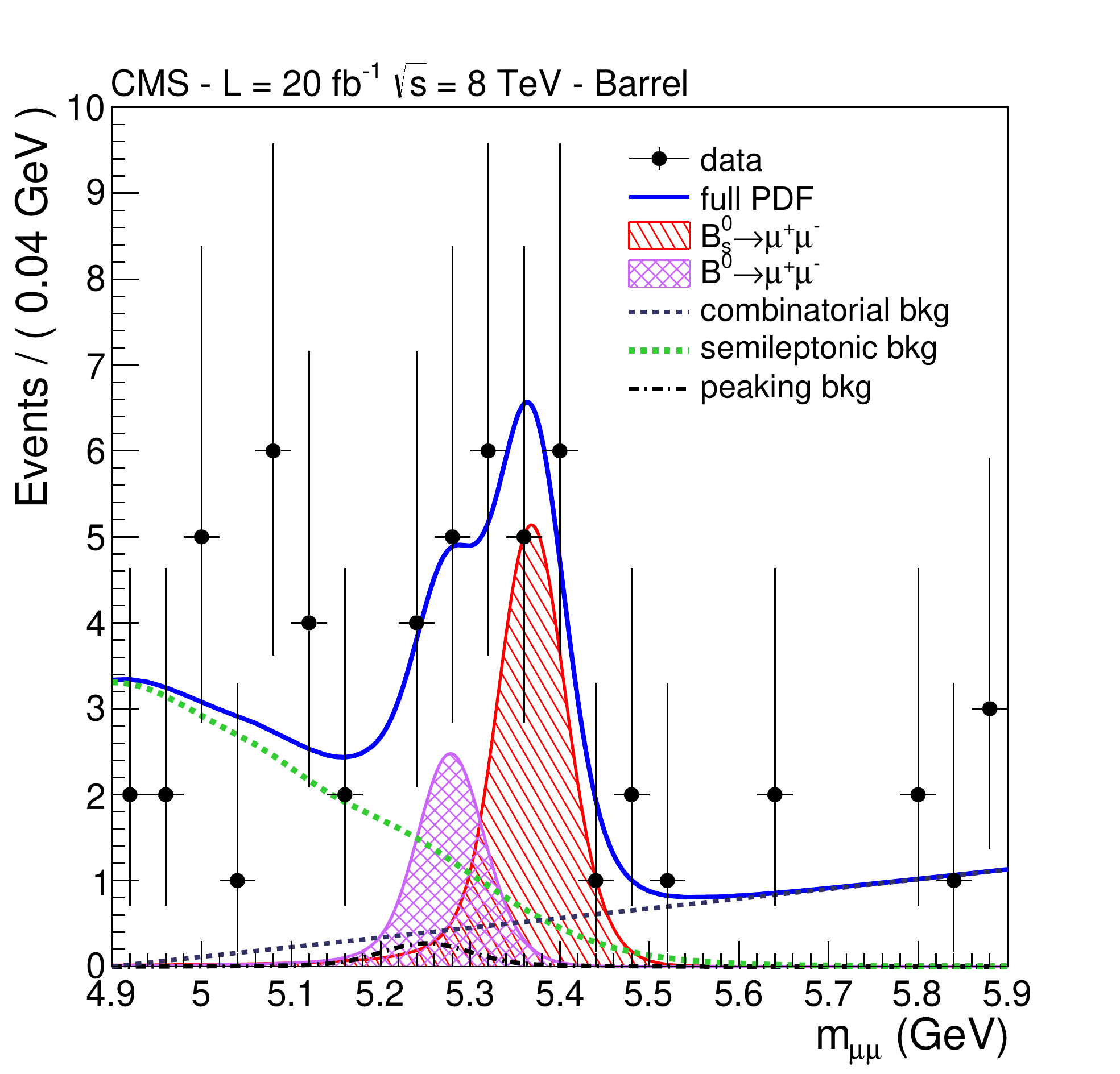}
    \includegraphics[width=\figwidX]{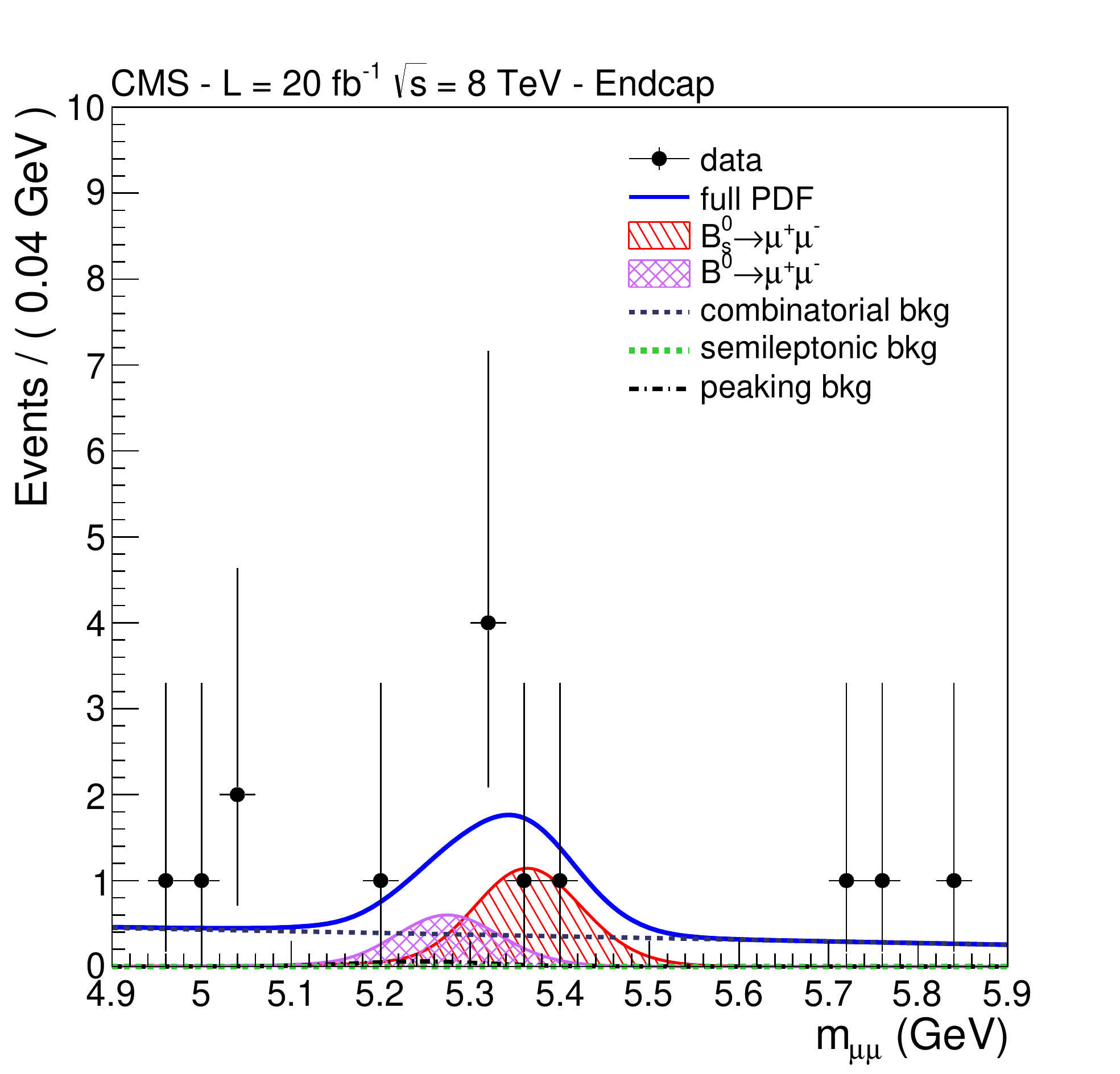}
   \caption{Results of the fit to the dimuon invariant mass
   distributions with the 1D-BDT method for the barrel (left) and endcap (right)
   from the 7\TeV (top) and 8\TeV (bottom) data samples. The points are
   the data, the solid line is the result of the fit, the shaded areas are the
   two \PB\ signals, and the different dotted lines are the backgrounds.}
   \label{fig:bdt2}
  \end{centering}
\end{figure}

}
\cleardoublepage \section{The CMS Collaboration \label{app:collab}}\begin{sloppypar}\hyphenpenalty=5000\widowpenalty=500\clubpenalty=5000\textbf{Yerevan Physics Institute,  Yerevan,  Armenia}\\*[0pt]
S.~Chatrchyan, V.~Khachatryan, A.M.~Sirunyan, A.~Tumasyan
\vskip\cmsinstskip
\textbf{Institut f\"{u}r Hochenergiephysik der OeAW,  Wien,  Austria}\\*[0pt]
W.~Adam, T.~Bergauer, M.~Dragicevic, J.~Er\"{o}, C.~Fabjan\cmsAuthorMark{1}, M.~Friedl, R.~Fr\"{u}hwirth\cmsAuthorMark{1}, V.M.~Ghete, N.~H\"{o}rmann, J.~Hrubec, M.~Jeitler\cmsAuthorMark{1}, W.~Kiesenhofer, V.~Kn\"{u}nz, M.~Krammer\cmsAuthorMark{1}, I.~Kr\"{a}tschmer, D.~Liko, I.~Mikulec, D.~Rabady\cmsAuthorMark{2}, B.~Rahbaran, C.~Rohringer, H.~Rohringer, R.~Sch\"{o}fbeck, J.~Strauss, A.~Taurok, W.~Treberer-Treberspurg, W.~Waltenberger, C.-E.~Wulz\cmsAuthorMark{1}
% ...
\vskip\cmsinstskip
% ...
1:~~Also at Vienna University of Technology, Vienna, Austria\\
2:~~Also at CERN, European Organization for Nuclear Research, Geneva, Switzerland\\
% ...

\end{sloppypar}
\end{document}